\documentclass[twoside,12pt]{article}
\usepackage{epsfig}
\usepackage{amsmath}
\usepackage{amssymb}
\def\Journal#1#2#3#4{{#1} {#2} (#4) #3}

\def\EPJC{{\em Eur.~Phys.~J.} C}

\def\IJMPA{{\em Int. J. Mod. Phys.} A}

\def\JHEP{{\em J.~High.~E.~Phys.}}
\def\MPLA{{\em Mod.~Phys.~Lett.} A}

\def\NPB{{\em Nucl.~Phys.} B}
\def\NJP{{\em New~J.~Phys.}}

\def\PLB{{\em Phys. Lett.} B}

\def\PRL{\em Phys. Rev. Lett.}

\def\PREP{\em Phys. Rep.}

\def\PRD{{\em Phys. Rev.} D}

\def\PRO{{\em Prog. Theor. Phys.}}

\def\ZPC{{\em Z. Phys.} C}

\def\etal{{\em et al.}}

\newcommand{\be}{\begin{equation}}
\newcommand{\ee}{\end{equation}}
\newcommand{\bea}{\begin{eqnarray}}
\newcommand{\eea}{\end{eqnarray}}

\newcommand{\ba}{\begin{array}}
\newcommand{\ea}{\end{array}} 

\newcommand{\lsim}{\raisebox{-1.5mm}{$\:\stackrel{\textstyle{<}}{\textstyle{\sim}}\:$}}
\newcommand{\gsim}{\raisebox{-0.5mm}{$\stackrel{>}{\scriptstyle{\sim}}$}}
\def\eV{\hbox{$\;\hbox{\rm eV}$}}
\def\GeV{\hbox{$\;\hbox{\rm GeV}$}}

\def\TeV{\hbox{$\;\hbox{\rm TeV}$}}

\newcommand{\pb}{\mbox{{\rm ~pb}~}}
\newcommand{\fb}{\mbox{{\rm ~fb}~}}
\newcommand{\LV}{\mbox{$\not \hspace{-0.13cm} L$}}
\newcommand{\BV}{\mbox{$\not \hspace{-0.13cm} B$}}
\newcommand{\Rp}{\mbox{$\not \hspace{-0.15cm} R_p$}}

\newcommand{\LLE}      {$ {\rm {LL \bar E}}$}
\newcommand{\LQD}      {$ {\rm {LQ \bar D}}$}
\newcommand{\UDD}      {$ {\rm {\bar U \bar D \bar D}}$}

\newcommand{\Xo}[1]    {\mbox {$\tilde{\chi}_{#1}^0                  $}}
\newcommand{\Xpm}[1]   {\mbox {$\tilde{\chi}_{#1}^{\pm}              $}}

\newcommand{\sel}      {\mbox{$ \tilde{e}                            $}}
\newcommand{\selL}     {\mbox{$ \tilde{e_L}                          $}}
\newcommand{\selR}     {\mbox{$ \tilde{e_R}                          $}}
\newcommand{\smu}      {\mbox{$ \tilde{\mu}                          $}}
\newcommand{\smuL}     {\mbox{$ \tilde{\mu}_L                        $}}
\newcommand{\smuR}     {\mbox{$ \tilde{\mu}_R                        $}}
\newcommand{\stau}     {\mbox{$ \tilde{\tau}                         $}}
\newcommand{\stauL}    {\mbox{$ \tilde{\tau}_L                       $}}
\newcommand{\stauR}    {\mbox{$ \tilde{\tau}_R                       $}}
\newcommand{\snu}      {\mbox{$ \tilde{\nu}                          $}}
\newcommand{\snue}     {\mbox{$ \tilde{\nu}_{{\rm e}L}               $}}
\newcommand{\snumu}    {\mbox{$ \tilde{\nu}_{{\mu}L}                 $}}
\newcommand{\snutau}   {\mbox{$ \tilde{\nu}_{{\tau}L}                $}}
\newcommand{\sqrk}     {\mbox{$\tilde{q}                             $}}

\newcommand{\sqrkuL}   {\mbox{$\tilde{u}_L                           $}}
\newcommand{\sqrkuR}   {\mbox{$\tilde{u}_R                           $}}

\newcommand{\sqrkdL}   {\mbox{$\tilde{d}_L                           $}}
\newcommand{\sqrkdR}   {\mbox{$\tilde{d}_R                           $}}

\newcommand{\sqrkcL}   {\mbox{$\tilde{c}_L                           $}}
\newcommand{\sqrkcR}   {\mbox{$\tilde{c}_R                           $}}

\newcommand{\sqrksL}   {\mbox{$\tilde{s}_L                           $}}
\newcommand{\sqrksR}   {\mbox{$\tilde{s}_R                           $}}

\newcommand{\sqrkt}[1] {\mbox{$\tilde{t}_{#1}                        $}}

\newcommand{\sqrkb}[1] {\mbox{$\tilde{b}_{#1}                        $}}

\def \Eslash {E \kern-.5em\slash }
\def \pslash {p \kern-.5em\slash }
\def \kslash {k \kern-.5em\slash }
\newcommand{\Emiss}{\( \not \! {E} \) }

\newcommand{\master}{.}
\begin{document}
%\special{papersize=21.0cm,29.7cm}

%\begin{flushleft}
%{\bf  Draft 0.00 \hspace{5mm} \today}
%\end{flushleft}
%\vspace*{1.5cm}

%
%\title{
\title{\begin{flushright}{\normalsize DESY--02--165}\\
\vspace*{-0.3cm}
{\normalsize November 2002}  \end{flushright}
\vspace*{1cm}
Search for Particles and Forces Beyond the Standard Model 
       at HERA $ep$ and Tevatron $p\bar{p}$ Colliders\footnote{
submitted for publication in ``Progress in Particle and Nuclear Physics''
under the title: ``Search for Particles and Forces Beyond the Standard Model
in High Energy Lepton-Hadron and Hadron-Hadron Collisions''}}
%       in High Energy Lepton-Hadron and Hadron-Hadron Collisions}
\author{M.\ Kuze$^{1}$ and Y.\ Sirois$^{2}$\\
\\
$^1$KEK, Institute of Particle and Nuclear Studies, Japan \\
$^2$LLR, Ecole Polytechnique and IN2P3-CNRS, France}

\date{}
\maketitle\thispagestyle{empty}
\begin{abstract}
A review of searches for physics beyond the Standard Model carried out at 
high energy lepton-hadron and hadron-hadron facilities is presented, 
with emphasis on topics of interest for future data taking at the 
upgraded Tevatron $p{\bar p}$ and HERA $ep$ colliders.
The status and discovery prospects are discussed for leptoquarks, 
Technicolour and supersymmetry, forbidden lepton and quark flavour-changing 
processes, extra gauge bosons, excited states of composite fermions, 
generic contact interactions and extra compactified dimensions.
\end{abstract}

\newpage
\pagestyle{empty}
\tableofcontents
\newpage
\pagestyle{plain}
\setcounter{page}{1}

%=======================Include TeX of the Sections ===========================

%%%%%%%%%%%%%%%%%%%%%%%%%%%%%%%%%%%%%%%%%%%%%%%%%%%%%%%%%%%%%%%%%%%%%%%%%%%%%%
\section{Introduction}
\label{sec:intro}
%%%%%%%%%%%%%%%%%%%%%%%%%%%%%%%%%%%%%%%%%%%%%%%%%%%%%%%%%%%%%%%%%%%%%%%%%%%%%%

Although remarkably confirmed at the phenomenological level over the past
quarter of century, in particular most recently at high-energy colliders, 
the Standard $SU(3) \times SU(2) \times U(1)$ Model of strong, 
electromagnetic and weak forces remains incomplete and unsatisfactory. 
There are reasons to believe that the search for new physics could be 
fruitful at existing colliders in the years to come, hopefully providing 
a deeper understanding of elementary forces in Nature. 

The Standard Model is incomplete and unsatisfactory because, first of
all, it only offers a partial ``unification'' of the electroweak and 
strong forces. Quarks are assumed to carry flavour degrees of freedom and
colour quantum numbers somehow independently . The electroweak interactions
couple only to flavours and are indifferent to colours. The strong 
interaction of coloured quarks and gluons described by the $SU(3)$ 
quantum chromodynamics (QCD) gauge field theory remains separate. 
A ``Grand Unification'' of carriers of fundamental interactions in, for
instance, a larger local gauge theory is simply postponed.
Needless to say, no connection is made at an even grander level with an 
eventual quantum theory of gravity.

The Standard Model's predictive power moreover suffers from a large 
number of arbitrary parameters. For instance, the particle masses 
are not predicted and must be measured
experimentally. These masses are assumed to originate solely from the 
electroweak sector. A fundamental scalar field, the Higgs-boson field,
is assumed to pervade the universe and to possess, through 
self-interaction, a non-zero field strength of 
$v = (\sqrt{2} G_F)^{-1/2} \simeq 246 \GeV$ of the ground state.
This non-zero vacuum expectation value induces a breaking of the
electroweak $SU(2)_L \times U(1)_Y$ symmetry down to the electromagnetic
$U(1)_{EM}$ symmetry. This ``Higgs mechanism'', which gives masses to the 
$W^{\pm}$ and $Z$ bosons and leaves the photon massless, remains unproven.
The mass of the Higgs boson itself is not predicted by the Standard Model
but an upper bound must nevertheless be imposed to preserve the internal 
consistency of the model. 
It is the Yukawa interactions, of arbitrary strengths, of fermions with the 
Higgs field that are assumed to be responsible for the fermion masses 
after electroweak symmetry breaking.
In contrast, the local gauge symmetries of the strong interactions remain
unbroken at all levels of the theory.
The masses of protons and neutrons, which themselves contribute to more 
than 99\% of the mass of ordinary cold matter, are understood to 
originate from the dynamics of colour confinement in QCD.

In the Standard Model there are no direct couplings between quark
and lepton families and the theory is consistent with a separate and
exact conservation of lepton and baryon numbers in all processes.
The viability of the Standard Model rests on a somewhat empirical 
similarity between lepton and quark sectors. 
Disastrous anomalies that would prevent renormalizability of the theory 
are avoided by an exact cancellation between contributions of lepton and 
quark fields. No deeper understanding is provided for this exact 
cancellation, which happens thanks to the special arrangement of fermion 
multiplets in the model and the fact that quarks have the additional 
colour degree of freedom. At a more fundamental level, the structure
of the leptonic and quarkonic sectors could, for instance, imply the 
existence of new bosonic carriers of lepton and baryon numbers.

Finally, the Standard Model incorporates an apparent threefold ``replica''
of fermion generations which remain unexplained.
The electroweak interaction Lagrangian is simply constructed separately
for each of the lepton and quark generations, with anomalies cancelled
within each generation. There are no direct couplings between different 
lepton families while, intriguingly, three quark families (at least)
are needed if quark mixing is to be the cause of all observed
electroweak CP violation.
The existence of the fermion generations could be hinting that more 
elementary constituents exist which form the known quarks and leptons. 
\begin{table}
   \caption{Main contemporary collider facilities. These are listed 
            together with their beam particles and the 
	    available centre-of-mass energies. Also given are the 
	    integrated luminosities accumulated (or expected) per
	    experiment. The multiplicative factor after the 
	    $\otimes$ sign denotes the number of multi-purpose 
	    collider experiments operating simultaneously at each 
	    of the facilities.}
   \label{tab:colliders}
   \begin{center}
   \begin{tabular}{c|c|r|c|c}
   \hline
     Collider & Beams  & $\sqrt{s}$ \,\,\, & $\int {\cal{L}} dt$ & Years \\
   \hline
    LEP$_I$   & $e^+ e^- $ & $M_Z$ 
                              & $\sim 160 \pb^{-1} \otimes 4 $ & 1989-95 \\ 
    LEP$_{II}$ & $e^+ e^- $  & $> 2 \times M_W $ 
                             & $\sim 620 \pb^{-1} \otimes 4 $ & 1996-00 \\ 
   \hline
    HERA$_{Ia}$ & $e^{-} p$ & $300 \GeV$ 
                               & ${\cal{O}} (1 \pb^{-1}) \otimes 2$ & 1992-93  \\ 
    HERA$_{Ib}$ & $e^{\pm} p$ & $\lsim 320 \GeV$ 
                     & ${\cal{O}} (100 \pb^{-1}) \otimes 2$ & 1994-00 \\ 
   \hline
    Tevatron$_{Ia}$ & $p \bar{p}$ & $1.8 \TeV$  
                             & ${\cal{O}} (10 \pb^{-1})$   & 1987-89  \\
    Tevatron$_{Ib}$ & $p \bar{p}$ & $1.8 \TeV$  
                    & ${\cal{O}} (100 \pb^{-1}) \otimes 2$ & 1992-96  \\ 
   \hline \hline 
    HERA$_{II}$ & $e^{\pm}_{L,R} p$ & $\sim 320 \GeV$ 
                            & $\sim 1 \fb^{-1} \otimes 2$  & 2002-06 \\
   \hline
    Tevatron$_{IIa}$ & $p \bar{p}$ & $\sim 2.0 \TeV$  & $\sim 2 \fb^{-1} \otimes 2$
                                                           & 2002-05 \\
   \hline
    Tevatron$_{IIb}$ & $p \bar{p}$ & $\sim 2.0 \TeV$  & 
                       ${\cal{O}} (10 \fb^{-1}) \otimes 2$ & 2005-08 \\
   \hline
   \end{tabular}
   \end{center}
\end{table}

The dissatisfaction with the Standard Model makes the search for new 
physics a central duty in experiments at colliders. Existing or planned 
lepton-hadron and hadronic colliders could provide the required discovery 
reach. 
This is motivated on the theoretical side, where there exists in various
more-or-less predictive models a strong prejudice for new physics 
``close to'' electroweak unification scale.
For example, supersymmetric models like the Minimal Supersymmetric 
Standard Model or minimal Supergravity would most naturally yield
a rich phenomenology at the ${\cal{O}}(\TeV)$ scale.
This is because the mass difference between ordinary particles and their 
supersymmetric partners must not be too large if such models are to be 
useful to avoid excessive ``fine tuning'' while preserving 
the masses in the Higgs scalar sector from quadratically divergent 
renormalization corrections. 
Another example is provided by theories which attempt to unify all known 
interactions including gravity.
It has been realized recently that a relevant scale for quantum-gravity 
models with ``large'' compactified extra dimensions could be as low 
as ${\cal{O}}(1$ to $10 \TeV)$ with possibly observable effects at 
colliders from the propagation of fields in the extra dimensions.
Technicolour models in which new composite scalar fields are 
responsible for the electroweak symmetry breaking possibly also 
yield a rich spectrum of new composite states accessible at 
colliders.
On the experimental side, the recent observation of neutrino 
oscillations~\cite{Fukuda98} could be a guiding sign towards physics
beyond the Standard Model. 
In the forthcoming years, scales from $1$ to $10 \TeV$ will be best probed 
in complementary facilities such as the HERA $ep$ and the Tevatron 
$p\bar{p}$ upgraded colliders (Table~\ref{tab:colliders}).

These $ep$ and $p\bar{p}$ colliders are well suited to search for
new physics affecting lepton-quark couplings. For $ep$ colliders this is
obvious given the quantum numbers available in the initial state, which
allow for contributions to the process $e q \rightarrow e q$ via $s$-channel 
resonant production or $u$-channel virtual exchange of new bosons 
coupling to lepton-quark pairs. In $p\bar{p}$ collisions, such new bosons, 
if pair produced, could be easily recognized via their decay, possibly leading 
to final states involving lepton pairs. Furthermore, the $t$-channel exchange
of such a boson could contribute to the Drell-Yan-like process
$q \bar{q} \rightarrow l \bar{l}$.
The status of the searches for leptoquark production will be 
discussed in section~\ref{sec:leptoq}.
Searches motivated by theories possessing new composite or elementary 
scalar fields are then discussed in section~\ref{sec:ewbreaking}.
We review collider constraints 
on Technicolour models, which are models designed 
to provide an alternative to the Higgs mechanism in the Standard Model 
and which contain leptoquark-like particles.

The phenomenology and searches for supersymmetric (SUSY) particles are 
discussed in section~\ref{sec:SUSY}, with some emphasis on $R$-parity-violating
theories. New Yukawa couplings to lepton-quark pairs appear 
in such theories where they connect 
the SUSY scalar partners of known fermions to ordinary matter
via lepton-number violating interactions. 
In view of existing indirect constraints, the collider facilities appear 
particularly competitive for couplings involving heavy quark flavours.
Particular attention is given to searches and constraints on stop squarks.
New bosons couplings to lepton-quark pairs are one of the various possible 
contributions beyond the Standard Model to flavour-changing neutral 
currents. The sensitivity of collider experiments to such currents
is compared in the top sector using an effective Lagrangian 
approach in section~\ref{sec:FLAVOURS}. 
The search for lepton-flavour-violation processes in an effective and 
generic approach is also discussed in this section~\ref{sec:FLAVOURS}.

Searches for new vector gauge bosons or new scalar Higgs bosons predicted
by theories incorporating an extension of the electroweak gauge symmetries
are discussed in section~\ref{sec:extendedSM}.

Direct searches for excited fermions are discussed in section~\ref{sec:FSTAR}.
A comparison of the sensitivity of existing colliders to contact interactions 
is presented in section~\ref{sec:contact} in the context of compositeness 
and leptoquark models. The possible effects on inclusive measurements of
the exchange of gravitons which are allowed 
to propagate in the extra compactified dimensions
in ($4+n$)-dimensional string theory
are discussed in section~\ref{sec:xtra}.

A summary and conclusions on future discovery prospects are presented
in section~\ref{sec:conclusion}.

  % Introduction

% Symmetry between quarks and leptons, Grand Unification
\clearpage
%%%%%%%%%%%%%%%%%%%%%%%%%%%%%%%%%%%%%%%%%%%%%%%%%%%%%%%%%%%%%%%%%%%%%%%%%%%%%%
\section{Leptoquarks}
\label{sec:leptoq}
%%%%%%%%%%%%%%%%%%%%%%%%%%%%%%%%%%%%%%%%%%%%%%%%%%%%%%%%%%%%%%%%%%%%%%%%%%%%%%

%
\subsection{Introduction}

An intriguing property of the Standard Model is the apparent symmetry
between the lepton and quark sectors.
This symmetry is manifest in their assignment to singlets and doublets
of the weak interaction, with their ``replica'' over three fermion generations.
This symmetry is furthermore essential in achieving an exact cancellation 
of chiral (triangular) anomalies. The cancellation demands that the sum 
of the electric charges is exactly neutralized in each generation, which 
incidentally requires three quark colours.
This could possibly be an indication that leptons and quarks are fundamentally 
connected through a new ``lepto-quark'' interaction.

Leptoquarks (LQs) are colour-triplet scalar (S) or vector (V) bosons carrying
lepton and baryon numbers, and a fractional electromagnetic charge,
Q$_{em}$. 
They appear naturally in Grand Unified Theories (GUT) for electroweak
and strong interactions of both the ``Georgi-Glashow type''~\cite{GUT-GG} 
(based on simple gauge groups with a superheavy unifying mass scale) or 
of the ``Pati-Salam  type''~\cite{GUT-PS} (with flavour-colour and left-right 
symmetric semi-simple gauge groups with intermediate or low unifying mass 
scale), as well as in superstring-``inspired'' $E_6$ models~\cite{Hewett89}.
They also appear as mediators between quark and lepton doublets in 
horizontal-symmetry schemes~\cite{Pakvasa87},
in Technicolour theories addressing the issue of electroweak symmetry 
breaking (see section~\ref{sec:technic}), in strongly coupled
weak-interaction models attempting to reconcile the conceptual differences 
between the weak and strong sectors~\cite{Abbott81}, 
and in some matter-compositeness theories~\cite{Schremp85}
attempting to provide an explanation for the three generations of fermions.
Actual searches at colliders have been mostly carried out in the context of 
effective models.

\subsection{Effective Interactions, Models and Nomenclature}
\label{sec:lqmodels}

A most general effective Lagrangian for leptoquark interactions
with SM fermion pairs was proposed by Buchm\"uller, R\"uckl and 
Wyler~\cite{Buchmuller87} under the assumptions that leptoquarks:
\, {\bf i)} have renormalizable interactions;
\, {\bf ii)} have interactions invariant under Standard Model
          $SU(3) \times SU(2) \times U(1)$ gauge groups;
\, {\bf iii)} couple only to Standard Model fermions and gauge bosons.
Furthermore, unacceptable instability of the proton is avoided by 
imposing that leptoquarks:
\, {\bf iv)} conserve leptonic number $L_l$ and baryonic number 
             $B_q$ separately.
Such leptoquarks carry a fermionic number $F = 3B_q + L_l$ of either
$|F|=0$ or $2$ and have interactions with lepton-quark pairs
described by~\cite{Buchmuller87}
$$
 {\cal L} = {\cal L}_{|F| =2}+{\cal L}_{F=0} 
$$
with
\begin{eqnarray*}
{\cal L}_{|F| =2} & = &
  (g_{1L}\bar{q}^c_L i \tau_2l_L + g_{1R}\bar{u}^c_L e^-_R)S_0
  + \tilde{g}_{1R}\bar{d}^c_Re^-_R\tilde{S_0}
  + g_{3L}\bar{q}^c_L i \tau_2 \tau l_L S_1 \\
  & + & (g_{2L}\bar{d}^c_R\gamma^\mu l_L + g_{2R}\bar{q}^c_L\gamma^\mu
  e^-_R)V_{1/2\mu} +
  \tilde{g}_{2L}\bar{u}^c_R\gamma^\mu l_L\tilde{V}_{1/2\mu} + h.c. \\
{\cal L}_{F=0} & = & 
  (h_{1L}\bar{q}_L\gamma^\mu l_L +
  h_{1R}\bar{d}_R\gamma^\mu e^-_R)V_{0\mu}
  + \tilde{h}_{1R}\bar{u}_R\gamma^\mu e^-_R \tilde{V}_{o\mu} + h_{3L}\bar{q}_L
  \tau \gamma^\mu l_L V_{1\mu} \\
  & + &(h_{2L}\bar{u}_Rl_L + h_{2R}\bar{q}_L i \tau_2 e^-_R )S_{1/2} 
  + \tilde{h}_{2L}\bar{d}_Rl_L \tilde{S}_{1/2} + h.c.,
\end{eqnarray*}
where $q_L$ and $l_L$ are the $SU(2)_L$ left-handed quark and lepton 
doublets and $e_R, d_R$ and $u_R$ denote the corresponding 
right-handed singlets for leptons, $d$-type and $u$-type quarks.
The $\psi^c$ are the charge conjugate of the fermion fields with the
convention $\psi^c \equiv C\bar{\psi}^T$. The indices $L$ and $R$ appended
to the coupling constants correspond to the chirality of the lepton
involved. For simplicity, the colour and generational indices have been
suppressed. 

Having chosen for the leptoquark interactions with lepton-quark pairs the
above effective Lagrangian which preserves the symmetries of the Standard Model,
the possible representations of the leptoquarks with respect to the gauge groups
and the couplings to the gauge bosons are in principle completely determined.
This is strictly true for scalars. However, for vector leptoquarks interacting
with gauge bosons ($g$), the cross-section that depends on trilinear $gVV$ and 
quartic $ggVV$ couplings might require damping by the introduction 
of anomalous couplings.
These will be necessary for instance if the vector leptoquarks
are composite low-energy manifestations of a more fundamental theory
at higher energy scales. Four independent anomalous couplings 
$\kappa_{\gamma}$, $\kappa_{Z}$, $\lambda_{\gamma}$ and $\lambda_{Z}$ are 
introduced for the electroweak sector. A theory with pure Yang-Mills 
couplings is recovered by setting $\kappa_{\gamma,Z} = \lambda_{\gamma,Z} = 0$.
Models with ``minimal vector couplings'' are obtained by setting
$\kappa_{\gamma,Z} = 1$ and $\lambda_{\gamma,Z} = 0$.
A discussion of the leptoquark interactions with $\gamma$ and $Z$ bosons in 
a model-independent effective Lagrangian approach can be found in 
Refs.~\cite{Blumlein93,Blumlein97}. 
Two anomalous couplings $\kappa_{g}$ and $\lambda_{g}$ are 
introduced for the strong sector. A general effective Lagrangian for
the interactions with gluons can be found in~Ref.~\cite{Blumlein97b}.

Two further restrictions can be imposed to cope with the existing 
low-energy constraints~\cite{Davidson94,Hewett97b} in what will be 
henceforward called the ``minimal Buchm\"uller-R\"uckl-Wyler 
effective model'' (mBRW).
In the mBRW model, leptoquarks:
\, {\bf v)} each couple to a single lepton-quark generation $i$ 
            with $i=1,2$ or $3$;
\, {\bf vi)} each has pure chiral couplings to SM fermions.
With the restrictions imposed to the mBRW model, it will be sufficient to
use the generic symbol $\lambda$ for the different Yukawa couplings
$g, \tilde{g}, h$ and $\tilde{h}$.
The restriction $\lambda^i \times \lambda^j \simeq 0$ ($i \neq j$)
on inter-generational connections avoids possibly large tree-level 
flavour-changing neutral currents and flavour-universality violations. 
The last restriction $\lambda_L \times \lambda_R \simeq 0$ avoids direct 
contributions to chirally suppressed meson decays such as the process 
$\pi \rightarrow e \nu$ as well as for instance virtual-loop contributions 
to the $g-2$ of the muon.

For each fermion generation $i$, the mBRW model allows for the existence 
of five different weak-isospin families (iso-singlets, iso-doublets
and iso-triplets) for both scalar and vector leptoquarks. 
These are listed in Table~\ref{tab:brwlqs}.
% ------------------ TABLE : Scalar Leptoquarks  -------------------------
\begin{table*}[htb]
  \renewcommand{\doublerulesep}{0.4pt}
  \renewcommand{\arraystretch}{1.2}
 \vspace{-0.1cm}

\begin{center}
    \begin{tabular}{||c|c|r|c|c||c|c|r|c|c||}
      \hline \hline
%
% -> Scalar LQ :
     \multicolumn{5}{||c||}{$|\rm F|$=2 Leptoquarks} & \multicolumn{5}{c||}{F=0 Leptoquarks} \\ \hline     
     LQ    & $Q_{em}$ & $T_3$ & LQ   & $\beta$ & LQ   &  $Q_{em}$ & $T_3$ & LQ    & $\beta$ \\
     Type  &          &       &Decay &         & Type &           &       & Decay &         \\ \hline

  $S_{0,L}$  &$-1/3$&$0$& $l^-_L u_L$    & $1/2$ & $V_{0,L}$  &$-2/3$&$0$ & $l^-_L \bar{d}_R$  & $1/2$ \\
             &      &   & $\nu_L d_L$    & $1/2$ &            &      &    & $\nu_L \bar{u}_R$  & $1/2$ \\
  $S_{0,R}$  &      &   & $l^-_R u_R$    & $1$   & $V_{0,R}$  &      &    & $l^-_R \bar{d}_L$  & $1$   \\ \hline
 $\tilde{S}_{0,R}$ &$-4/3$&$0$& $l^-_R d_R$ & $1$ & $\tilde{V}_{0,R}$ &$-5/3$&$0$& $l^-_R \bar{u}_L$ &  $1$  \\ \hline  
  $S_{1,L}$ &$-4/3$&$-1$& $l^-_L d_L$  & $1$   & $V_{1,L}$    &$-5/3$&$-1$& $l^-_L \bar{u}_R$ &  $1$  \\
            &$-1/3$&$0$ & $l^-_L u_L$  & $1/2$ &              &$-2/3$&$0$ & $l^-_L \bar{d}_R$ & $1/2$ \\
            &      &    & $\nu_L d_L$  & $1/2$ &              &       &   & $\nu_L \bar{u}_R$ & $1/2$ \\ 
            &$+2/3$&$+1$& $\nu_L u_L$  & $1$ &              &$+1/3$&$+1$& $\nu_L \bar{d}_R$ & $1$ \\ \hline
  $V_{1/2,L}$ &$-4/3$&$-1/2$& $l^-_L d_R$  & $1$ & $S_{1/2,L}$  &$-5/3$&$-1/2$& $l^-_L \bar{u}_L$ & $1$ \\
  $V_{1/2,R}$ &$-4/3$&      & $l^-_R d_L$  & $1$ & $S_{1/2,R}$  &$-5/3$&      & $l^-_R \bar{u}_R$ & $1$ \\
              &$-1/3$&$+1/2$& $l^-_R u_L$  & $1$ &            &$-2/3$&$+1/2$& $l^-_R \bar{d}_R$ & $1$ \\ \hline
 $\tilde{V}_{1/2,L}$ &$-1/3$&$-1/2$& $l^-_L u_R$ & $1$
                                                 &$\tilde{S}_{1/2,L}$&$-2/3$&$-1/2$& $l^-_L \bar{d}_L$ & $1$ \\
                    &$+2/3$&$+1/2$& $\nu_L u_R$ & $1$ 
		                               &              &$+1/3$&$+1/2$& $\nu_L \bar{d}_L$ & $1$ \\ \hline
      \hline
    \end{tabular}
    \caption {\small \label{tab:brwlqs}
               Leptoquarks with fermionic number $|F|=2$ ({\it left column}) 
	       and $F=0$ ({\it right column}) in the Buchm\"uller-R\"uckl-Wyler
	       (BRW) effective model~\cite{Buchmuller87}. 
	       The scalar (S) and vector (V) leptoquarks (LQ) are grouped into 
	       weak-isospin families (subscript index). In the minimal BRW model, 
	       leptoquarks coupling to fermions (i.e. in lepton ($l$)-quark ($q$) 
	       pairs) of different chiralities are assumed independent. Here, by 
	       convention~\cite{Kohler89}, the leptoquark types are
	       distinguished by the chirality ($L,R$ index) of the 
	       coupled lepton. Also given for each leptoquark is the electric 
	       charge $Q_{em}$, the third component $T_3$ of the weak isospin, their 
	       allowed decay modes and the corresponding branching fractions $\beta$. 
	       For simplicity, the same symbols are often used to designate 
	       both leptoquarks and anti-leptoquarks. Thus, for
	       example, the $S_{1,L}$ is used implicitely in the text for the
	       leptoquark $Q_{em} = -4/3$ ($Q_{em} = -1/3$) involved in the 
	       production process $e^-_L d_L \rightarrow S_{1,L}$ 
	       ($e^-_L u_L \rightarrow S_{1,L}$) or in the conjugate processes
	       $e^+_R \bar{d}_R \rightarrow S_{1,L}$ 
	       ($e^+_R \bar{u}_R \rightarrow S_{1,L}$). Note for instance that 
	       this $S_{1,L}$ cannot be produced in the $Q_{em} = +2/3$ state 
	       in a $e q$ collision. Compared to the original BRW 
	       nomenclature~\cite{Buchmuller87}, the 
	       ``Aachen notations''~\cite{Kohler89} adopted here have the 
	       following correspondence: 
	       $S_{0}             \leftrightarrow S_{1}^{BRW}$;
	       $\tilde{S}_{0}     \leftrightarrow \tilde{S}_{1}^{BRW}$;
	       $S_{1}           \leftrightarrow S_{3}^{BRW}$;
	       $V_{1/2}         \leftrightarrow V_{2}^{BRW}$;
               $\tilde{V}_{1/2} \leftrightarrow \tilde{V}_{2}^{BRW}$;
	       $V_{0}             \leftrightarrow U_{1}^{BRW}$;
	       $\tilde{V}_{0}     \leftrightarrow \tilde{U}_{1}^{BRW}$;
	       $V_{1}             \leftrightarrow U_{3}^{BRW}$;
	       $S_{1/2}           \leftrightarrow R_2^{BRW}$;
	       $\tilde{S}_{1/2}   \leftrightarrow \tilde{R}_2^{BRW}$.}

\end{center}
\end{table*}
% ------------------------------------------------------------------------

For experimental searches, mass degeneracy is generally assumed within each 
isospin family.
This is motivated theoretically since one would expect all leptoquarks 
within a given $SU(2)_L$ representation to be degenerate apart 
from loop corrections. 
Hence, for simplicity, the same symbol represents any of the 
various states of different electric charges within a family.
For instance, the $S_{1/2,L}$ designates both the scalar leptoquark $S_{1/2}$ 
states of electric charge $-5/3$ and $-2/3$ coupling to a left-handed lepton.
In total, one distinguishes fourteen types of leptoquarks; 
seven scalars with either $|F|=2$ 
($S_{0,L}$, $S_{0,R}$, $\tilde{S}_{0,R}$, $S_{1,L}$) 
or $F=0$ 
($S_{1/2,L}$, $S_{1/2,R}$, $\tilde{S}_{1/2,L}$),
and seven vectors with either $|F|=2$ 
($V_{1/2,L}$, $V_{1/2,R}$, $\tilde{V}_{1/2,L}$)
or $F=0$ 
($V_{0,L}$, $V_{0,R}$, $\tilde{V}_{0,R}$, $V_{1,L}$).
By construction, the decay branching ratios $\beta(LQ \rightarrow l q)$ 
of each of these leptoquarks into a final state with a charged lepton 
$l$ are fixed by the model to $0$, $1/2$ or $1$.

Generally, only a subset of the allowed BRW-leptoquark states are 
predicted by a specific fundamental model.
For instance, 
the scalar leptoquark corresponding to the $S_{0,L}$ of 
Table~\ref{tab:brwlqs} is the one present in superstring-inspired $E_6$ 
theories~\cite{Hewett89}. 
A light scalar iso-doublet of leptoquarks corresponding to the
anti-$\tilde{S}_{1/2}$ of Table~\ref{tab:brwlqs} has been 
proposed~\cite{Murayama92} in a model that attempts to reconcile 
SU(5) GUT theories with the existing constraint on the proton lifetime 
and the observed $\sin^2 \theta_W$. Light colour-exotic scalars appear
to be a generic feature in such models~\cite{Rizzo92}.
In contrast, a weak-isospin singlet vector leptoquark of hypercharge 2/3 
and corresponding to the $V_{0}$ appears in the Pati-Salam GUT 
model~\cite{GUT-PS}.
Interestingly, all possible fourteen states appear in a GUT theory based
on the SU(15) gauge group~\cite{Frampton90a, Frampton90b}.

Enriched phenomenology appears in leptoquark models that depart from the 
assumptions of the BRW model~\cite{Hewett97}. 
Instead of relying on a specific model, searches at colliders can be performed 
in what will be henceforward called  ``generic models''; models in which 
$\beta(LQ \rightarrow l q)$ is simply left as a free parameter. 
This is assumed to be made possible by (e.g.) dropping some of the 
above constraints. 
It might be for instance reasonable to assume, contrary to assumption (iii), 
that leptoquarks also couple to other (unspecified) new fields. 
Alternatively, relaxing the assumptions (iv) or (v) in the lepton sector 
could open new lepton-flavour violating (LFV) decays.
The low-energy constraints and the discovery reach at colliders in this 
particular case will be discussed in section~\ref{sec:flavours}.
Squarks in $R$-parity violating supersymmetry (see section~\ref{sec:SUSY})
can fall, from a phenomenological point of view, into the ``free $\beta$'' 
category of the ``LQ'' phenomenology. 
This is because they might possess leptoquark-like decay modes through 
Yukawa couplings in addition to their normal decay modes through gauge 
couplings. The $\tilde{u}$-like and $\tilde{d}$-like squarks  
can have lepton-quark couplings resembling those of the $\tilde{S}_{1/2}$ 
and $S_{0}$ leptoquarks, respectively.
For instance, the $\tilde{u}_L$ (the superpartner of the left-handed $u$ 
quark) may couple to an $e^++d$ pair via a Yukawa coupling $\lambda'_{111}$ 
in a way similar to the coupling of the first generation $\tilde{S}_{1/2,L}$ 
leptoquark of charge $|{\rm Q}_{em}| = 2/3$.
Via the same coupling, the $\tilde{d}_R$ (the superpartner of the 
right-handed $d$ quark) couples to $e^-+u$ or $\nu_e+d$ pairs like the 
first-generation $S_{0,L}$ of charge $|{\rm Q}_{em}| = 1/3$.
As a general consequence, it will be possible to translate constraints
on the $\lambda$ couplings of leptoquarks into constraints on the 
$\lambda'_{ijk}$ couplings of squarks in $R$-parity violating 
supersymmetry. However, as will be discussed in section~\ref{sec:SUSY}, 
additional constraints will affect the $\lambda'_{ijk}$ couplings 
since they also induce decays of other supersymmetric particles and,
in contrast to LQ couplings, enter into explicit lepton-number 
violating procesess. 

\subsection{Phenomenology at Colliders}

\begin{figure}[htb]
  \begin{center}                                                                
  \begin{tabular}{ccc}
  \vspace*{-0.2cm}
    
  \hspace*{-1.0cm} \epsfig{file=\master/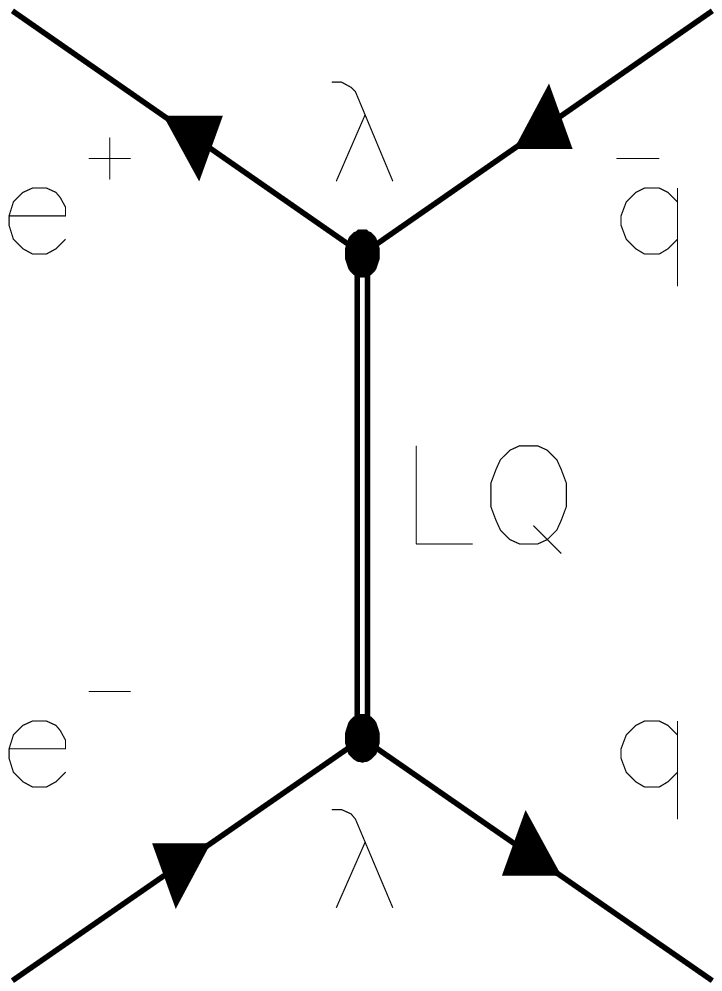,width=0.42\textwidth}
&
  \hspace*{-2.2cm} \epsfig{file=\master/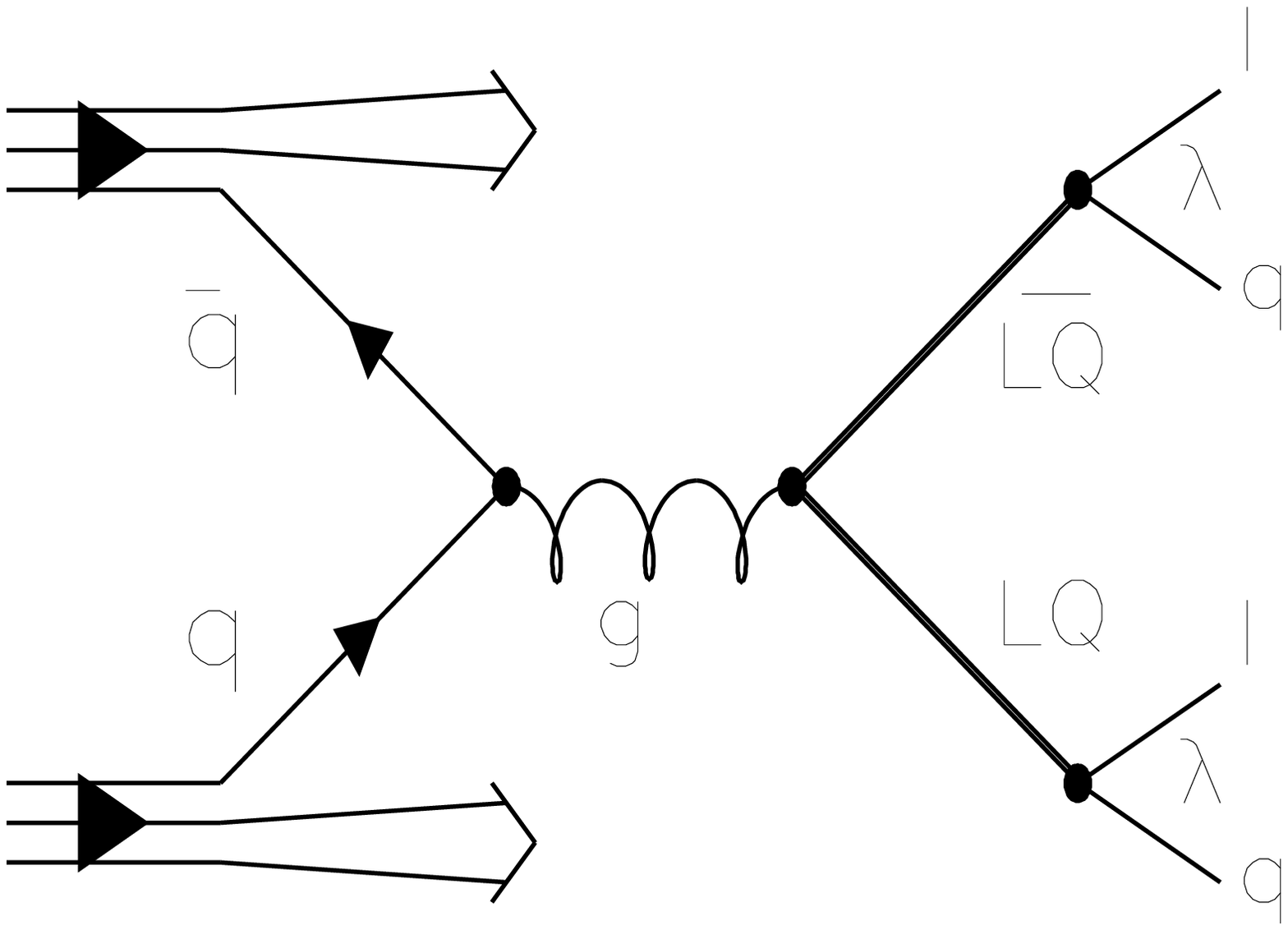,width=0.40\textwidth}
&
  \hspace*{-1.2cm} \epsfig{file=\master/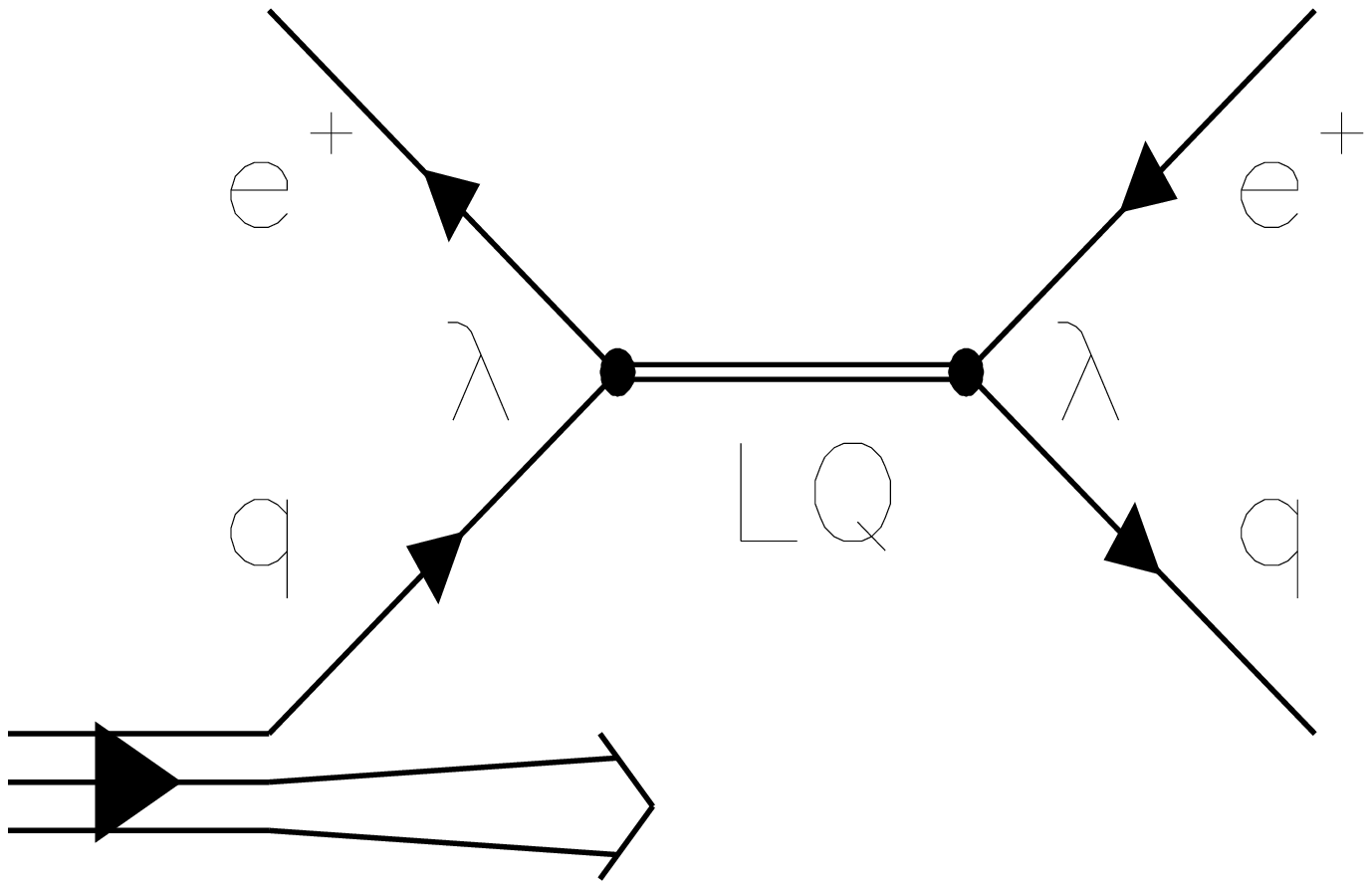,width=0.40\textwidth}

  \end{tabular}
  \end{center}
  \vspace*{-2.5cm}
 
  \hspace*{3.0cm} (a) \hspace*{5.0cm} (b) \hspace*{6.0cm} (c) \\ 

  \vspace*{-0.5cm}   
  
 \caption[]{ \label{fig:diaglq}
            Typical diagrams for leptoquark production at colliders;
            a) $t$-channel exchange in 
	                 $e^+ e^- \rightarrow q \bar{q}$; 
            b) pair-production in 
	                   $p \bar{p} \rightarrow LQ \bar{LQ} + X
			   \rightarrow l \bar{l} q \bar{q} + X$;    
	    c) $s$-channel resonance in 
	                   $e p \rightarrow LQ + X \rightarrow e  q + X$.} 
\end{figure}

Diagrams for the production or exchange of leptoquarks at $e^+ e^-$,
$p \bar{p}$ and $e p$ colliders are shown in Fig.~\ref{fig:diaglq}.
%----------------------------------------------------------------------------
%-----------------------------------------------------------------------------
% Leptoquarks at e+e- colliders: 
 
\subsubsection{Leptoquarks at {\boldmath $e^+e^-$} Colliders}

% Production

Leptoquarks of all three generations can be pair-produced at an 
$e^+e^-$ collider through $s$-channel $\gamma$ and $Z^0$ exchange,
and in addition, leptoquarks of the first generation can be pair-produced 
through $t$-channel quark exchange.
Furthermore, they can be exchanged virtually in 
$t$-channel to contribute to the process $e^+e^- \rightarrow q\bar q$,
as shown in Fig.~\ref{fig:diaglq}(a).
Single real production of leptoquarks is only possible via higher order 
processes, with a main contribution coming from the fusion of an incoming 
$e$ beam particle with a $q$ from the resolved component of a (quasi-real) 
$\gamma^*$ radiated off the other $e$ beam~\cite{OPALlq}. 
However, the sensitivity of LEP$_{II}$ for leptoquarks
in this production mode is smaller than that of HERA.

% Cross-sections

Detailed expressions of the total and differential cross-sections for
leptoquark production at $e^+e^-$ colliders can be found 
in~\cite{Blumlein93}.
The total pair-production cross-sections for the various leptoquark species 
strongly depend on their specific $SU(2)_L \times U(1)_Y$ quantum numbers. 
For collider centre-of-mass energies $\sqrt{s_{ee}} \gg M_Z$ and a leptoquark 
mass $M_{LQ} \lsim 1/2 \times \sqrt{s_{ee}}$, they can vary by an order of 
magnitude among scalar or vector species and are systematically larger for 
vectors. 
The largest cross-sections for vector leptoquarks are expected for 
a ``Yang-Mills'' model (i.e. $\kappa_{\gamma,Z} = \lambda_{\gamma,Z} = 0$,
see section~\ref{sec:lqmodels}). 
In contrast, the set of anomalous-coupling values that minimizes the 
total cross-section (``Minimal $\sigma_{LQ}$ scenario''), thus leading to
most conservative constraints, depends~\cite{Blumlein97} on the kinematic 
factor $\beta = \sqrt{ 1 - 4 M^2_{LQ} / s_{ee}}$.
It does not in general coincide with the ``Minimal vector couplings''
scenario (i.e. $\kappa_{\gamma,Z} = 1$ and $\lambda_{\gamma,Z} = 0$).

The $t$-channel quark exchange contributes significantly to leptoquark
pair production only if the Yukawa interaction is of electromagnetic 
strength (i.e. if $\lambda$ approaches $\sqrt{4 \pi \alpha}$). 
It interferes with the $s$-channel pair production.

% Signals

Pair-produced scalar and vector leptoquarks can be distinguished by their 
angular distributions $(1/\sigma) d \sigma / d \cos \theta$, where $\theta$ 
is the polar angle of the leptoquark relative to the incident electron. 
In the $s$-channel, scalar leptoquarks are produced with an approximate 
$\sin^2 \theta$ distribution while vector leptoquarks are produced 
approximately flat in $\cos \theta$.

The virtual $t$-channel exchange of leptoquarks can be detected by a 
$q\bar{q}$ production cross-section departing from Standard Model expectation
and in jet-charge asymmetry measurements.
Since it is a virtual exchange, the analysis in this channel is sensitive
to leptoquark masses much higher than $\sqrt{s_{ee}}$.
More detailed discussions in this ``contact interaction''-type analysis
appear in section~\ref{sec:contact}.

\subsubsection{Leptoquarks at {\boldmath $p \bar{p}$ and $p p$} Colliders}

% Production

The dominant production processes for leptoquarks at hadronic machines
such as the Tevatron $p \bar{p}$ collider are pair-production via 
gauge couplings in $q \bar{q}$ annihilation and $g g$ fusion.
Leptoquarks of all three generations can be thus produced.
An example diagram is shown in Fig.~\ref{fig:diaglq}(b).
In addition, leptoquarks of the first generation can be exchanged singly 
in $t$-channel virtual processes.

% Cross-sections

For scalar leptoquarks, the total pair-production cross-section
is essentially parameter free. 
For vector leptoquarks, additional anomalous-coupling parameters $\kappa_{g}$ 
and $\lambda_{g}$ are introduced (see section~\ref{sec:lqmodels}) and treated as 
independent~\footnote{As discussed in~\cite{Blumlein97b}, the coupling parameters 
                      $\kappa_{g}$ and $\lambda_{g}$ can be related through the 
		      anomalous `magnetic' moment and `electric' quadrupole 
		      moment of the vector leptoquark in the colour field.}. 
The production cross-section is generally larger for vector leptoquarks but can
vary by one or two orders of magnitude depending on the specific choices of 
anomalous-coupling values~\cite{Blumlein97b,Hewett97b}. 
The relative contributions of the $q \bar{q}$ and $g g$ partonic processes
depends on the fraction $\xi$ of the $p p$ or $p \bar{p}$ centre-of-mass energy
($\sqrt{s_{pp}}$) required in the partonic subprocess, with $g g$ always 
dominating at small $\xi$ values and $q \bar{q}$ dominating at 
$p \bar{p}$ colliders for large enough $\xi$ values (e.g. above 
$\xi \sim 10^{-2}$).

% Signals

Depending on whether each of the leptoquarks decays to a charged lepton
or a neutrino, the final state either consists of a lepton pair and two jets
($lljj$), one lepton, missing momentum and two jets ($l\nu jj$) or
missing momentum and two jets ($\nu\nu jj$), each of which requires a 
different background-reduction strategy.  
Also, specific analysis strategies are taken depending on the generation 
of the leptoquarks.

In contrast to the case at $e^+ e^-$ colliders, pair-produced scalar and
vector leptoquarks at hadronic colliders cannot be distinguished by their 
angular distributions, given only the very slight spin-related differences 
expected~\cite{Blumlein97b,Hewett97b}.

The virtual $t$-channel exchange of leptoquarks is investigated by 
searching for deviations from Standard Model expectations for 
Drell-Yan $e^+ e^-$ production and is sensitive to leptoquark masses 
well above $\sqrt{s_{pp}}$.
Results from this type of analysis will be discussed in the context of
``contact interactions'' in section~\ref{sec:contact}.
		              
\subsubsection{Leptoquarks at {\boldmath $ep$} Colliders}

First-generation leptoquarks can be resonantly produced at the HERA $ep$ 
collider by the fusion of an $e$ beam particle with a $q$ from the 
proton, or exchanged in the $u$-channel. 
An example diagram is shown in Fig.~\ref{fig:diaglq}(c).
Since valence quarks dominate the parton distribution function (PDF)
at the large Bjorken-$x$ values needed to produce high-mass leptoquarks, $e^+p$
collision is most sensitive to $F=0$ leptoquarks and $e^-p$ for
$|F|=2$ leptoquarks.

The leptoquark processes interfere with $t$-channel electroweak-boson 
exchange. Thus, LQ searches at HERA 
involve the analysis of event signatures indistinguishable from Standard
Model deep inelastic scattering (DIS) at high squared momentum transfer, $Q^2$.
However, different angular distributions (or $y$ distributions,
$y=Q^2/xs_{ep}$) can be used to separate the signal from
background.  The $y$ variable (inelasticity) is related to the decay
angle $\theta^*$ in the leptoquark rest frame by $\cos\theta^* = 1-2y$
in the quark-parton model.  While the neutral current DIS
shows a $1/y^2$ fall-off at fixed $x$, scalar leptoquarks
show flat $y$ distributions and vector leptoquarks have a $(1-y)^2$
dependence, which is more enhanced than the SM background at large $y$.

For small enough Yukawa couplings ($\lambda \ll 1$) and leptoquark masses
not too close to the kinematical limit, the narrow-width approximation
for the dominant $s$-channel resonance
gives a good description of the production cross-section:
\begin{equation*}
\sigma_{LQ} = \frac{\pi}{4s_{ep}}\lambda^2 \cdot q(x=\frac{M^2_{LQ}}{s_{ep}}, Q^2=M^2_{LQ}),
\end{equation*}
where $q(x,Q^2)$ is the PDF evaluated at the resonance pole for the quark
flavour corresponding to the $SU(2)$ multiplet member in Table~\ref{tab:brwlqs}.
When the leptoquark mass approaches the kinematical limit and $\lambda$
becomes large, the effect of interference with the SM diagram (photon and $Z$
exchange) and the $u$-channel diagram becomes non-negligible and this
simple $\lambda^2$ dependence of the cross-section no longer holds.

The experimental search is made by looking for a mass resonance in the electron-jet
final state at large $y$.  Also a resonance search in the neutrino-jet
system is possible, with the assumption that only one neutrino escapes
detection and accounts for the missing momentum.  In this case, the
dominant SM background is charged current DIS.
HERA experiments are also able to detect such leptoquark signals with
high efficiency and small background, in contrast to the $\nu \nu j j $
analyses at Tevatron whose sensitivity becomes degraded compared to
the $eejj$ channel because of the harsh QCD background.

\subsection{Search Results and Prospects}
%

% Collider results (BRW model)

Early searches in ALEPH, DELPHI, L3 and OPAL experiments at LEP$_{I}$ 
concentrated on pair production in $Z^0$ decays~\cite{ADLO9193lq}. 
Leptoquarks of all types and of each generation were considered.
Direct searches for singly and pair-produced LQ as well as indirect 
searches from virtual exchange have been recently performed~\cite{ADLO9899lq} 
at LEP$_{II}$.
Early searches for pair production of scalar and vector leptoquarks
of all three generations have been carried out by the CDF and D$\emptyset$
experiments~\cite{CDFD09397lq} and recently 
updated~\cite{CDFLQa,D0LQa,CDFLQb,D0LQb,CDFLQbb,CDFLQc,D0LQc,TeVLQa} 
to consider all available data from Tevatron$_{I}$.
Searches by the H1 and ZEUS experiments using early $e^-p$ data
from 1993-94 were discussed in Refs.~\cite{HERA9394lq}.
Results based on $e^+p$ data up to 1997 were discussed
in Refs.~\cite{H19497lq,ZEUS9497lq,ZEUSLQa}.

\begin{figure}[htb]
  \begin{center}
  \begin{tabular}{cc}
  \hspace*{-0.5cm} \epsfig{file=\master/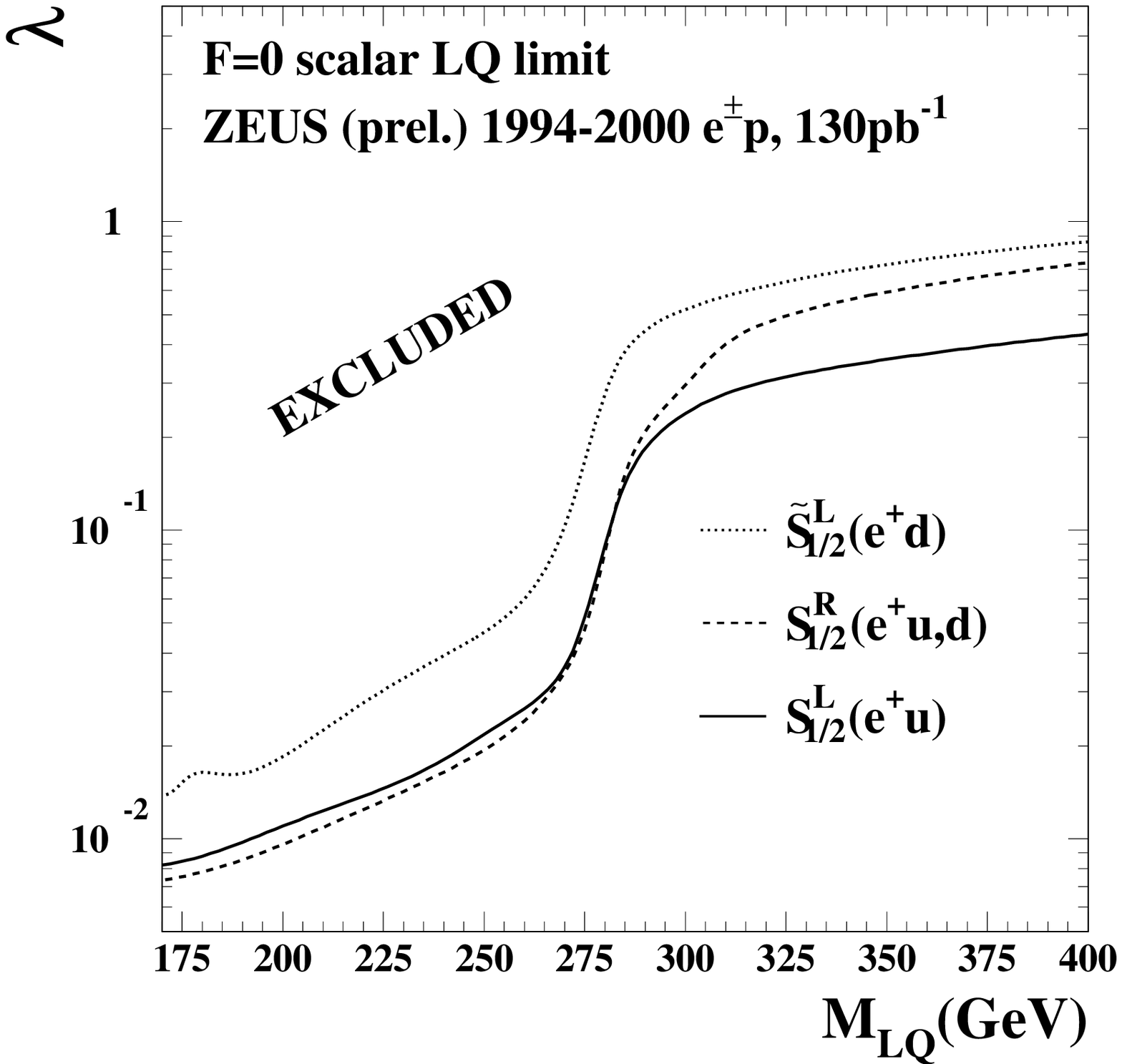,width=0.50\textwidth}
&
  \hspace*{-0.5cm} \epsfig{file=\master/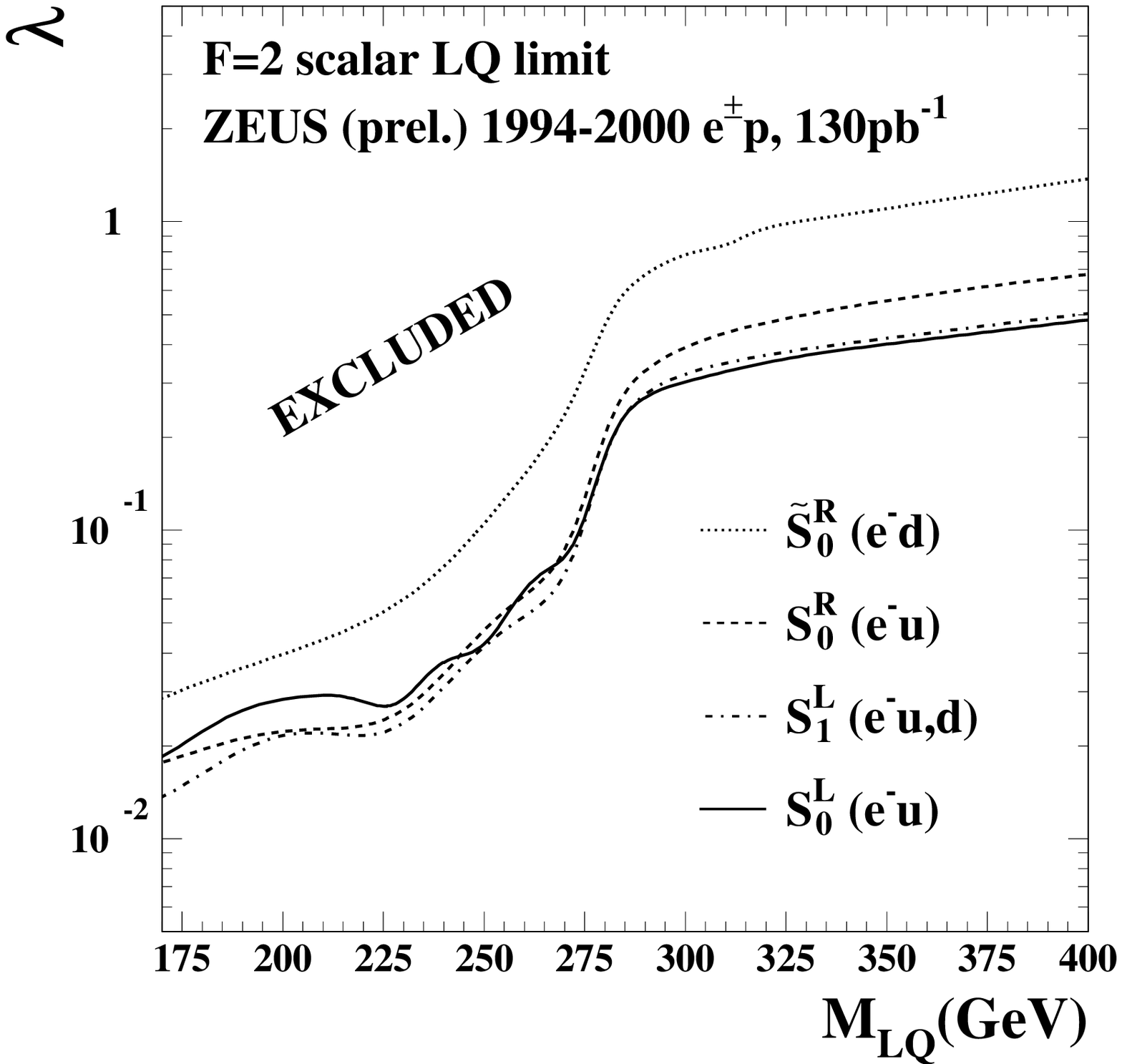,width=0.50\textwidth}
\\
  \hspace*{-0.5cm} \epsfig{file=\master/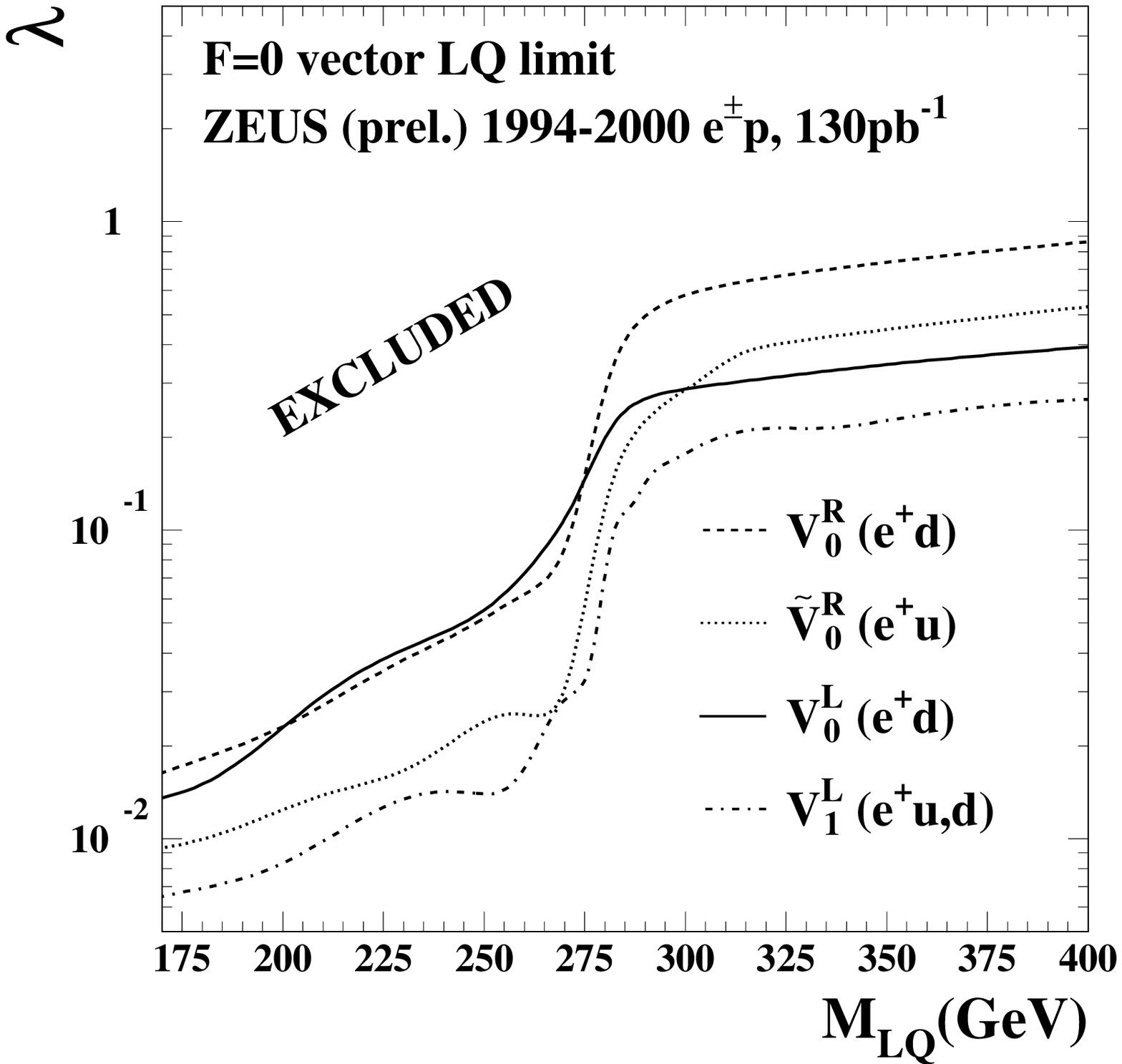,width=0.50\textwidth}
&
  \hspace*{-0.5cm} \epsfig{file=\master/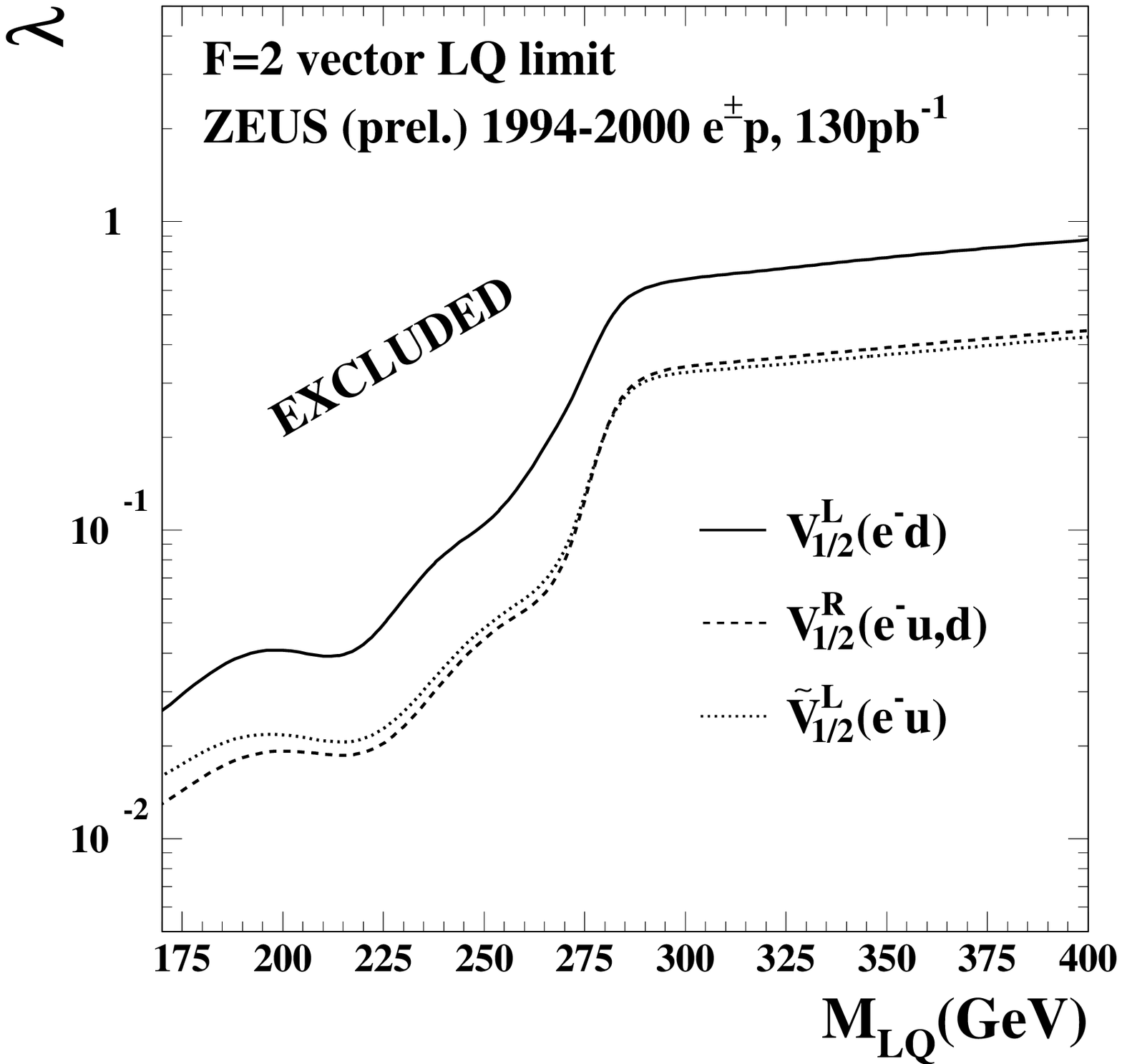,width=0.50\textwidth}
  \end{tabular}
  \end{center}
  
 \caption[]{\label{fig:lqbrwlim}
            Exclusion limits obtained~\cite{ZEUSLQc} at the HERA$_I$ 
	    collider in the $\lambda$ {\it vs.} $M_{LQ}$ plane
	    for leptoquarks of the BRW model.
	    Other recent leptoquark results from the H1 and ZEUS
	    experiments using all available $e^{\pm}p$ data from
	    HERA$_I$ can be found in 
	    Refs.~\cite{H1LQa,H1LQb,ZEUSLQb,ZEUSLQc}.} 
\end{figure}

Recent H1 and ZEUS results combining most or all available 
$e^{\pm}p$ HERA$_I$ data taken from 1994 to 2000 are discussed in 
Refs.~\cite{H1LQa,H1LQb,ZEUSLQb,ZEUSLQc}.
The exclusion limits thus obtained by ZEUS for first-generation leptoquarks
in the framework of the BRW model are shown for all leptoquark types 
in Fig.~\ref{fig:lqbrwlim}.
%---------------------------------------------------------------------------
%---------------------------------------------------------------------------- 

\begin{figure}[htb]
  \begin{center}                                                                

  \epsfig{file=\master/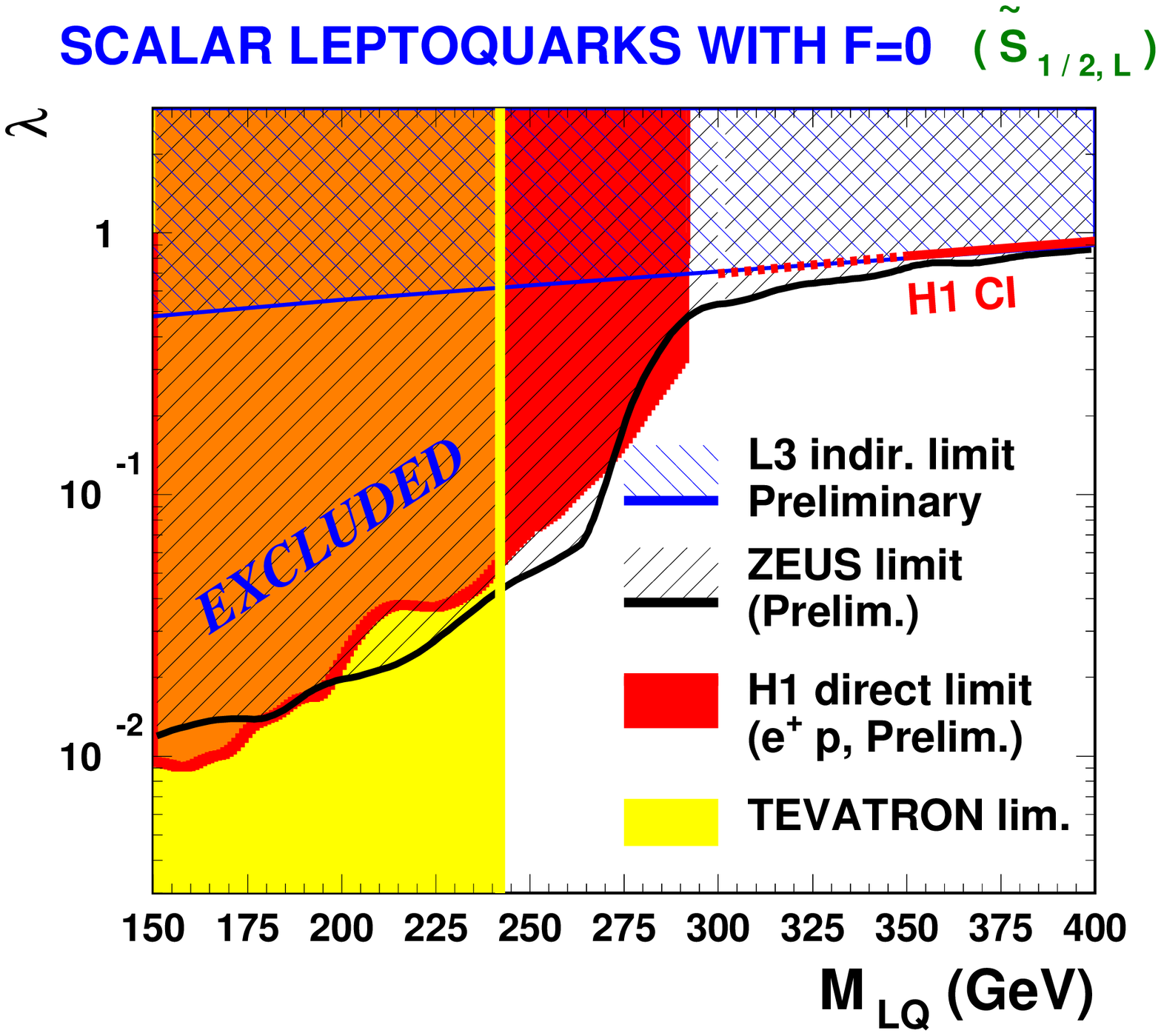,width=0.60\textwidth}

  \end{center}

 \vspace*{-3.5cm}
  
 \caption[]{ \label{fig:lqslimits}
             Existing collider constraints on a typical scalar leptoquark 
	     obtained at HERA, LEP (from L3~\cite{l3ams}) and Tevatron colliders in the Yukawa 
	     coupling {\it vs.} mass plane.} 
\end{figure}

The sensitivities of the collider searches for first-generation leptoquarks
of the BRW model are compared in Fig.~\ref{fig:lqslimits} for a typical
scalar with $F=0$,
namely the $\tilde{S}_{1/2,L}$ for which
$\beta_{eq} \equiv \beta (LQ \rightarrow e^+ q) = 1.0$.
%----------------------------------------------------------------------------
The Tevatron$_{I}$ experiments exclude leptoquark masses up to $242 \GeV$
independently of $\lambda$ for a scalar carrying the quantum numbers of 
this $\tilde{S}_{1/2,L}$.
For a $S_{0,L}$ ($\beta = 0.5$), the exclusion limit decreases to $204 \GeV$.
For an interaction stronger than the electromagnetic interaction
(i.e. $\lambda^2 / 4 \pi \alpha > 1$), virtual LQ exchange at HERA$_{I}$
and LEP$_{II}$ provide comparable exclusion limits.
For smaller values of $\lambda$, in the mass range beyond the reach of 
Tevatron$_{I}$ and below $\sim 300 \GeV$, a discovery domain 
remains open for HERA$_{II}$. This domain will be ultimately covered
independently of $\lambda$ at Tevatron$_{II}$.  

% Indirect Constraints

The allowed domain for a possible discovery of leptoquarks at colliders 
is furthermore restricted by severe and utterly unavoidable constraints 
from low-energy experiments. These indirect constraints for leptoquarks
of the mBRW model have been studied in detail in 
Refs~\cite{Davidson94,Hewett97,Leurer94}.
The most stringent bounds originate from measurements of Atomic Parity 
Violation and from the universality in leptonic $\pi$ decays. 
Lower limits in the TeV range on the ratio $M / \lambda$ are found 
for all leptoquark types of the first generation 
(see section~\ref{sec:contact}). 
Thus, leptoquarks allowed in the $200$ to $300 \GeV$ range must have interactions 
with lepton-quark pairs much weaker than the electromagnetic 
interaction (i.e. $\lambda \ll 0.3$).

\begin{figure}[htb]
  \begin{center}                                                                

  \epsfig{file=\master/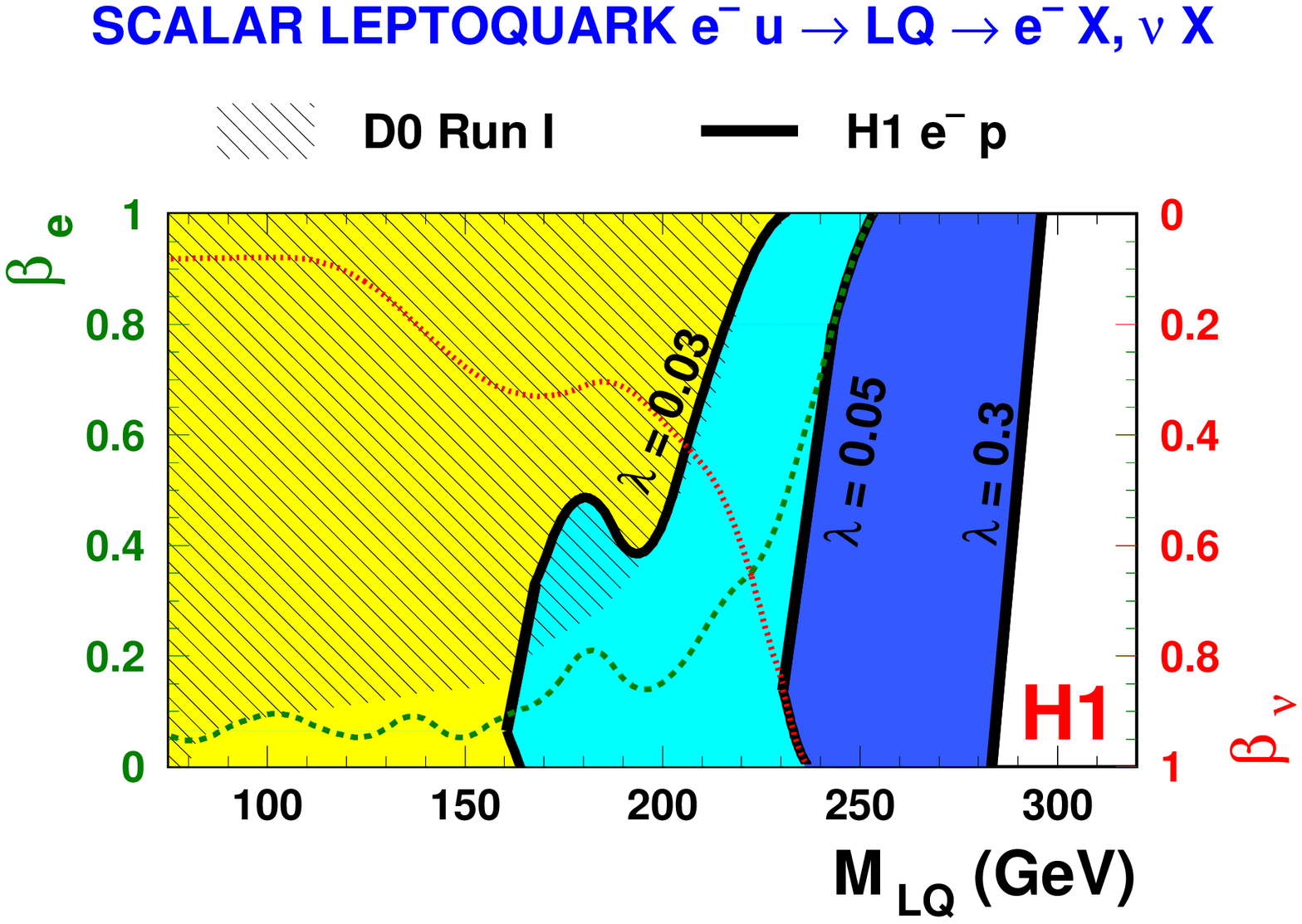,width=0.60\textwidth}

  \end{center}

 \vspace*{-0.5cm}
  
 \caption[]{ \label{fig:lqbtavsm}
             Comparison of the HERA and Tevatron bounds for generic scalar
	     leptoquarks in the branching ratio {\it vs.} mass plane.
             The HERA bounds (from H1) are shown for three assumptions on
             Yukawa coupling $\lambda$.  For $\lambda=0.05$, limits using
             only $eq$ final state (dashed line) and only $\nu q$
             final state (dotted line) are also shown.} 
\end{figure}

In generic models with an arbitrarily small branching ratio 
$\beta (LQ \rightarrow e q)$, the chances of a discovery at HERA 
increase as $\lambda$ grows,
as can be inferred from the actual HERA$_{I}$
and Tevatron$_{I}$ constraints shown in 
Fig.~\ref{fig:lqbtavsm}~\cite{H1LQb}.
%----------------------------------------------------------------------------
%-----------------------------------------------------------------------------

%
% ----------  TABLE : Scalar and Vector Leptoquarks   1rst Generation --------------
\begin{table*}[htb]
  \renewcommand{\doublerulesep}{0.4pt}
  \renewcommand{\arraystretch}{1.2}
 \vspace{-0.1cm}
\begin{center}
  \begin{tabular}{||c||c|c|c||c|c||}
  \hline
   \multicolumn{6}{||c||}{COLLIDER CONSTRAINTS on {\boldmath $1^{st}$} GENERATION LEPTOQUARKS} \\
  \hline \hline
   \multicolumn{6}{||l||}{SCALARS} \\
  \hline
  $\beta_e$& \multicolumn{3}{||c||}{Lower Mass Limits (in $\GeV$) at 95\%CL for} 
                                                                   & Assumptions & Experiment \\
           & {\;\;\;\;\;any $\lambda_{lq}$ values\;\;\;\;\;\;} 
	         & $\lambda_{lq} \ge 0.1$ 
	               & $\lambda_{lq} \ge 0.3$  &             &            \\
 \hline 
      1    & {242} 
                 &  -  &  -  & $p \bar{p} \rightarrow eeqq + X$ 
	                                       & CDF$\oplus$D$\emptyset$~\cite{TeVLQa} \\
           & { 213 } &  -  &  -  &         & CDF~\cite{CDFLQa}           \\
           & { 225 } &  -  &  -  &         & D$\emptyset$~\cite{D0LQa}   \\
           & { - } 
	         & 282 & 298 & $e^+ u \rightarrow LQ^{F=0} \rightarrow eq$ & H1~\cite{H1LQa}  \\
           & { - } 
	         & 268 & 282 & $e^+ u \rightarrow LQ^{F=0} \rightarrow eq$ & ZEUS~\cite{ZEUSLQa} \\
           & { - } 
	         & 246 & 270 & $e^+ d \rightarrow LQ^{F=0} \rightarrow eq$ & ZEUS~\cite{ZEUSLQa} \\
           & { - } 
	         & 273 & 296 & $e^- u \rightarrow LQ^{F=2} \rightarrow eq$ & H1~\cite{H1LQb}  \\
           & { - } 
	         & 276 & 295 & $e^- u \rightarrow LQ^{F=2} \rightarrow eq$ & ZEUS~\cite{ZEUSLQb} \\
           & { - } 
	         & 243 & 276 & $e^- d \rightarrow LQ^{F=2} \rightarrow eq$ & H1~\cite{H1LQb}  \\
           & { - } 
                 & 249 & 278 & $e^- d \rightarrow LQ^{F=2} \rightarrow eq$ & ZEUS~\cite{ZEUSLQb} \\

 \hline 
    1/2    & {204} 
                 &  -  &   - & $p \bar{p} \rightarrow e{\nu}qq (eeqq; {\nu}{\nu}qq) + X$ 
		                                                  & D$\emptyset$~\cite{D0LQa} \\
           & { - } 
	         & 275 & 292 & $e^+ u \rightarrow LQ^{F=0} \rightarrow eq$ & H1~\cite{H1LQa}     \\
           & { - } 
	         & 261 & 278 & $e^+ u \rightarrow LQ^{F=0} \rightarrow eq$ & ZEUS~\cite{ZEUSLQa} \\
           & { - } 
	         & 235 & 265 & $e^+ d \rightarrow LQ^{F=0} \rightarrow eq, \nu q$ 
		                                                           & ZEUS~\cite{ZEUSLQa} \\
           & { - } 
	         & 262 & 289 & $e^- u \rightarrow LQ^{F=2} \rightarrow eq, \nu q$ 
		                                                           & H1~\cite{H1LQb}     \\
           & { - } 
	         & 271 & 294 & $e^- u \rightarrow LQ^{F=2} \rightarrow eq, \nu q$
                                                                           & ZEUS~\cite{ZEUSLQb} \\
           & { - } 
	         & 230 & 270 & $e^- d \rightarrow LQ^{F=2} \rightarrow eq$ & H1~\cite{H1LQb}     \\
           & { - } 
	         & 231 & 271 & $e^- d \rightarrow LQ^{F=2} \rightarrow eq$ & ZEUS~\cite{ZEUSLQb} \\
 \hline 
     0     & { 98} 
                 &  -  &  -  & $p \bar{p} \rightarrow {\nu}{\nu}qq + X$ 
	                                             & D$\emptyset$~\cite{D0LQa}  \\
           & { - } 
	         & 237 & 262 & $e^+ d \rightarrow LQ^{F=0} \rightarrow \nu q$
		                                                   & ZEUS~\cite{ZEUSLQa} \\
           & { - } 
	         & 262 & 282 & $e^- u \rightarrow LQ^{F=2} \rightarrow \nu q$
		                                                   & H1~\cite{H1LQb} \\
           & { - } 
	         & 268 & 293 & $e^- u \rightarrow LQ^{F=2} \rightarrow \nu q$
		                                                   & ZEUS~\cite{ZEUSLQb} \\
 \hline
 \hline
  \end{tabular}
  \caption {\small \label{tab:lqlimit1s}
             Lower mass limits (95\%CL) on first-generation scalar leptoquarks from direct 
	     searches at colliders for different decay branching fraction $\beta_e$.
             For $\beta_e = 1/2$ limits, when both $eq$ and $\nu q$ decays are used, 
             $\beta_e + \beta_\nu = 1$ is assumed.
	     The results from H1 and ZEUS experiments given here were derived in the 
	     context of generic models (with arbitrary $\beta_e$) and depend on the Yukawa 
	     coupling $\lambda_{lq}$ to lepton-quark pairs. Other results obtained in the
	     strict context of the minimal BRW model are available from HERA (see text).}              
\end{center}
\end{table*}
%
% VECTOR TABLE
%
\begin{table*}[htb]
  \renewcommand{\doublerulesep}{0.4pt}
  \renewcommand{\arraystretch}{1.2}
 \vspace{-0.1cm}
\begin{center}
  \begin{tabular}{||c||c|c|c|c||c|c||}
  \hline
   \multicolumn{7}{||c||}{COLLIDER CONSTRAINTS on {\boldmath $1^{st}$} GENERATION LEPTOQUARKS} \\
  \hline \hline
  \multicolumn{7}{||l||}{VECTORS} \\
 \hline
  $\beta_e$ & \multicolumn{4}{||c||}{Lower Mass Limits (in $\GeV$) at 95\%CL for} 
                                                                    & Assumptions & Experiment \\
            & \multicolumn{2}{||c|}{LQ $\leftrightarrow$ boson couplings:} 
	          & $\lambda_{lq} \ge 0.1$ 
	                & $\lambda_{lq} \ge 0.3$  &            &            \\
            & Min. Vec. 
	           & Yang-Mills       &      &    &           &            \\
 \hline 
      1     & 292  & 345 &  -  &  -   & $p \bar{p} \rightarrow eeqq + X$ & D$\emptyset$~\cite{D0LQa} \\
            &  -   &  -  & 272 & 283  & $e^+ u \rightarrow LQ^{F=0} \rightarrow eq$ 	                                                               & ZEUS~\cite{ZEUSLQa} \\
            &  -   &  -  & 264 & 292  & $e^+ d \rightarrow LQ^{F=0} \rightarrow eq$   & H1~\cite{H1LQa} \\
            &  -   &  -  & 241 & 271  & $e^+ d \rightarrow LQ^{F=0} \rightarrow eq$ 	                                                               & ZEUS~\cite{ZEUSLQa} \\
            &  -   &  -  & 275 & 295  & $e^- u \rightarrow LQ^{F=2} \rightarrow eq$	                                                               & ZEUS~\cite{ZEUSLQb} \\
            &  -   &  -  & 246 & 277  & $e^- d \rightarrow LQ^{F=2} \rightarrow eq$	                                                               & ZEUS~\cite{ZEUSLQb} \\
 \hline 
    1/2     & 282  & 337 &  -  &  -   & $p \bar{p} \rightarrow e{\nu}qq 
                                        (eeqq; {\nu}{\nu}qq) + X$      & D$\emptyset$~\cite{D0LQa} \\
            &  -   &  -  & 266 & 281  & $e^+ u \rightarrow LQ^{F=0} \rightarrow eq$ 	                                                               & ZEUS~\cite{ZEUSLQa} \\
            &  -   &  -  & 260 & 290  & $e^+ d \rightarrow LQ^{F=0} \rightarrow eq, \nu q$	                                                               & H1~\cite{H1LQa} \\
            &  -   &  -  & 239 & 267  & $e^+ d \rightarrow LQ^{F=0} \rightarrow eq, \nu q$ 	                                                               & ZEUS~\cite{ZEUSLQa} \\
            &  -   &  -  & 276 & 295  & $e^- u \rightarrow LQ^{F=2} \rightarrow eq, \nu q$ 	                                                               & ZEUS~\cite{ZEUSLQb} \\
            &  -   &  -  & 230 & 271  & $e^- d \rightarrow LQ^{F=2} \rightarrow eq$ 	                                                               & ZEUS~\cite{ZEUSLQb} \\
  \hline 
     0      & 238  & 298 &  -  & -    & $p \bar{p} \rightarrow {\nu}{\nu}qq + X$ 
                                                                       & D$\emptyset$~\cite{D0LQa} \\
            &  -   &  -  & 268 & 300  & $e^+ d \rightarrow LQ^{F=0} \rightarrow \nu q$	                                                               & H1~\cite{H1LQa} \\
            &  -   &  -  & 243 & 267  & $e^+ d \rightarrow LQ^{F=0} \rightarrow \nu q$	                                                               & ZEUS~\cite{ZEUSLQa} \\
            &  -   &  -  & 280 & 295  & $e^- u \rightarrow LQ^{F=2} \rightarrow \nu q$                                                               & ZEUS~\cite{ZEUSLQb} \\
  \hline
  \hline
  \end{tabular}
  \caption {\small \label{tab:lqlimit1v}
             Lower mass limits (95\%CL) on first-generation vector leptoquarks from direct searches 
	     at colliders for different decay branching fraction $\beta_e$.
	     The results from CDF and D$\emptyset$ experiments
             depend on anomalous couplings to gauge bosons 
	     (see text) and are given here for ``Yang-Mills'' or ``Minimal Vector'' models.
             For $\beta_e = 1/2$ limits, when both $eq$ and $\nu q$
decays are used, $\beta_e + \beta_\nu = 1$ is assumed.
	     The results from H1 and ZEUS experiments given here were derived in the 
	     context of generic models (with arbitrary $\beta_e$) and depend on the Yukawa 
	     coupling $\lambda_{lq}$ to lepton-quark pairs. Other results obtained in the
	     strict context of the minimal BRW model are available from HERA (see text).}              
\end{center}
\end{table*}
%
%
% ----------  TABLE : Scalar and Vector Leptoquarks   2nd and 3rd Generation --------------
\begin{table*}[htb]
  \renewcommand{\doublerulesep}{0.4pt}
  \renewcommand{\arraystretch}{1.2}
 \vspace{-0.1cm}
\begin{center}
  \begin{tabular}{||c||c|c|c||c|c||}
  \hline
  \multicolumn{6}{||c||}{COLLIDER CONSTRAINTS on LEPTOQUARKS of HIGHER GENERATIONS} \\
  \hline  
  \multicolumn{6}{||c||}{ {\boldmath $2^{nd}$} {\bf Generation}} \\
  \hline 
  $\beta_{\mu}$ & \multicolumn{3}{||c||}{Lower Limits on LQ Mass}  & Assumptions & Experiment \\
            & \multicolumn{3}{||c||}{(95\%CL; in $\GeV$) for}  &  &  \\
            & SCALARS   & \multicolumn{2}{||c||}{VECTORS}      &  &  \\ 
            &           & \multicolumn{2}{||c|}{LQ $\leftrightarrow$ boson couplings:} 
	                                                       &  &  \\
            &  & Min. Vec. & Yang-Mills              &  &  \\	    
  \hline 
      1     & 202 &  -  &  -  & $p \bar{p} \rightarrow {\mu}{\mu}qq + X$ & CDF~\cite{CDFLQb} \\
            & 200 & 275 & 325 & $p \bar{p} \rightarrow {\mu}{\mu}qq + X$ & D$\emptyset$~\cite{D0LQb} \\
     1/2    & 160 &  -  &  -  & $p \bar{p} \rightarrow {\mu}{\mu}qq + X$ & CDF~\cite{CDFLQb} \\
            & 180 & 260 & 310 
	                      & $p \bar{p} \rightarrow {\mu}{\nu}qq ({\mu}{\mu}qq; {\nu}{\nu}qq)  + X$
						                     & D$\emptyset$~\cite{D0LQb} \\
      0     & 123 & 171 & 222 & $p \bar{p} \rightarrow {\nu}{\nu}cc + X$  & CDF~\cite{CDFLQbb} \\
            &  98 & 238 & 298 & $p \bar{p} \rightarrow {\nu}{\nu}qq + X$ & D$\emptyset$~\cite{D0LQa} \\
  \hline \hline
  \multicolumn{6}{||c||}{ {\boldmath  $3^{rd}$} {\bf Generation}} \\
  \hline 
  $\beta_{\tau}$ & \multicolumn{3}{||c||}{Lower Limits on LQ Mass}  & Assumptions & Experiment \\
            & \multicolumn{3}{||c||}{(95\%CL; in $\GeV$) for}  &  &  \\
            & SCALARS   & \multicolumn{2}{||c||}{VECTORS}      &  &  \\ 
            &           & \multicolumn{2}{||c|}{LQ $\leftrightarrow$ boson couplings:} 
	                                                       &  &  \\
            &           & Min. Vec.  & Yang-Mills              &  &  \\	    
  \hline
      1     &  99 & 170 & 225 & $p \bar{p} \rightarrow {\tau}{\tau}qq + X$ & CDF~\cite{CDFLQc} \\
      0     & 148 & 199 & 250 & $p \bar{p} \rightarrow {\nu}{\nu}bb + X$& CDF~\cite{CDFLQbb} \\
            &  94 & 148$^*$ & 216 & $p \bar{p} \rightarrow {\nu}{\nu}bb +
X, b\rightarrow \mu+X'$& D$\emptyset$~\cite{D0LQc} \\

  \hline
  \hline
  \end{tabular}
  \caption {\small \label{tab:lqlimit23}
             Lower mass limits (95\%CL) on second- and third-generation leptoquarks from 
	     direct searches at colliders for different decay branching fraction $\beta_{\mu}$
	     and $\beta_{\tau}$.
	     The results in the case of vector leptoquarks possibly depend on anomalous 
	     couplings to gauge bosons (see text) and are given here for 
	     ``Yang-Mills'' or ``Minimal Vector'' models
	     (expect for the result marked with a $^*$ which was obtained for anomalous 
	     couplings leading to a minimal cross-section). 
             For $\beta_\mu = 1/2$ limits, when both $\mu q$ and $\nu q$ decays are used, 
             $\beta_\mu + \beta_\nu = 1$ is assumed.
             The limits for the third-generation quoted here assume no decays to top.}
\end{center}
\end{table*}
% ----------------------------------------------------------------------
%     
The constraints on first-, second- and third-generation leptoquarks
obtained from Tevatron and HERA experiments are summarized in
Tables~\ref{tab:lqlimit1s}, \ref{tab:lqlimit1v} and~\ref{tab:lqlimit23}.
Tevatron experiments offer the best opportunity to search for second- and 
third-generation leptoquarks.
At Tevatron$_{I}$, masses below $202 \GeV$ ($99 \GeV$) are excluded for
second(third)-generation scalar leptoquarks with 
$\beta (LQ \rightarrow \mu q) = 1.0$ ($\beta (LQ \rightarrow \tau q) = 1.0$).
Above these excluded domains, HERA has access to leptoquarks of higher
generations only in cases where lepton-flavour violating processes are 
allowed. These are discussed in detail in section~\ref{sec:flavours}.
Striking event topologies could result from $s$- or 
$u$-channel exchange of leptoquarks if 
$\lambda_{e q} \times \lambda_{\mu q} \neq 0$   
or $\lambda_{e q} \times \lambda_{\tau q} \neq 0$.

\begin{figure}[htb]
  \begin{center}                                                                
 
 \epsfig{file=\master/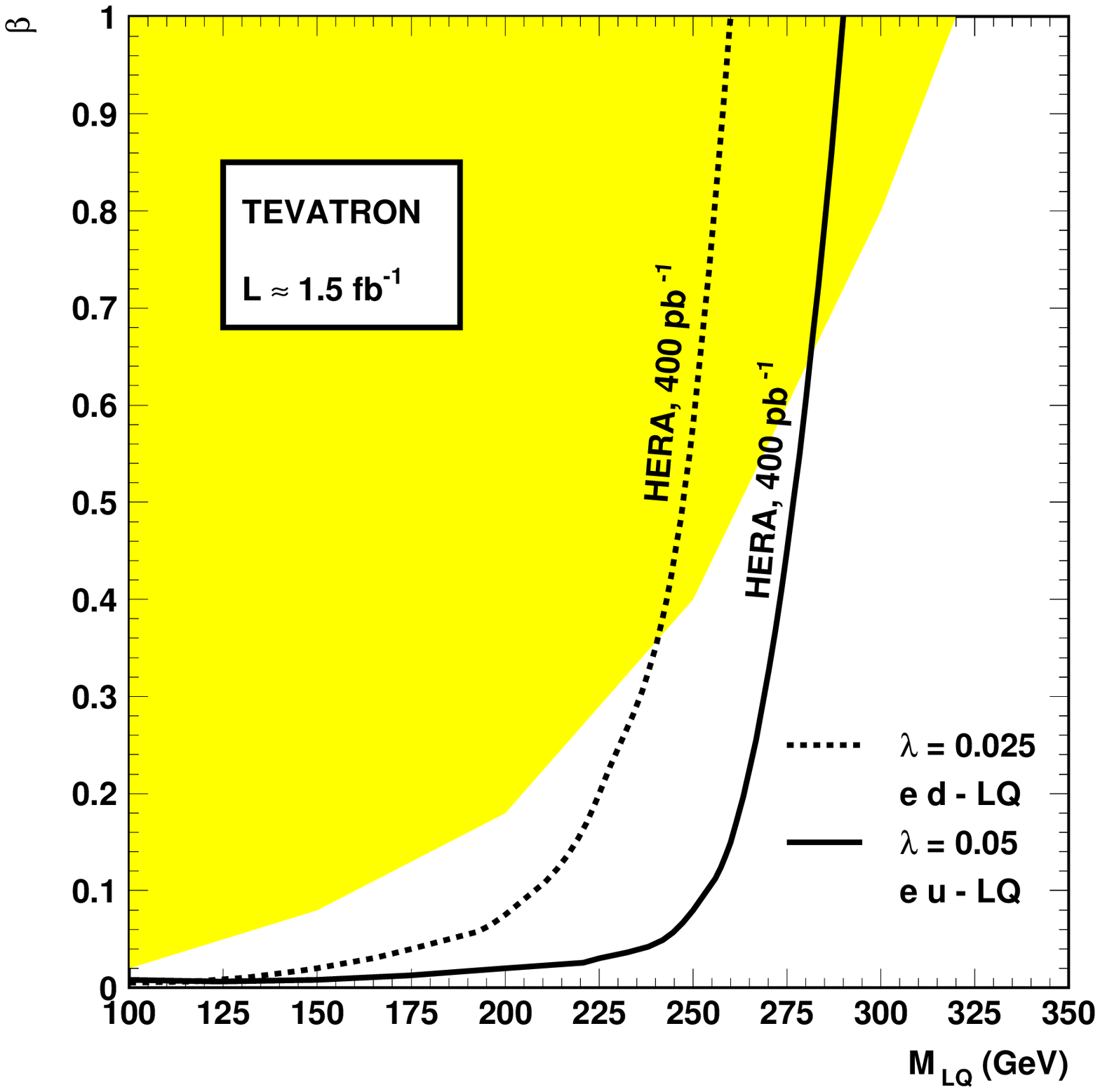,width=0.60\textwidth}

  \end{center}

 \vspace*{-0.5cm}
  
 \caption[]{ \label{fig:lqprospect}
             Prospect of mass-dependent sensitivities on the branching
             ratio $\beta$ of a leptoquark decaying into $eq$, for
             1.5~fb$^{-1}$ of Tevatron$_{II}$ data and 400~pb$^{-1}$
             of HERA$_{II}$ data.}
\end{figure}

Future prospects for HERA$_{II}$ and Tevatron$_{II}$ are illustated in
Fig.~\ref{fig:lqprospect}~\cite{Perez00} in the case of a first-generation scalar 
leptoquark decaying into $eq$. Tevatron$_{II}$ will offer a better mass
reach for $\beta (LQ \rightarrow e q) \simeq 1$ while the sensitivity will
be best at HERA$_{II}$ for $\beta (LQ \rightarrow e q) \lsim 0.5$ even for
interaction strengths two orders of magnitude weaker than the electromagnetic
interaction strength. 
%----------------------------------------------------------------------------
%-----------------------------------------------------------------------------

\hfill \\
\noindent
{\bf Other lepton-parton exoticas:} \\
For completeness, it should be mentioned that other exotic lepton-parton
resonances have been discussed~\cite{buchmuller87b} in the context of 
$ep$ colliders. Most prominent among these are leptogluons, which appear
as colour-octet partners of the known (colour-singlet) leptons in composite
models~\cite{Fritzsch81} in which the leptons are bound states of some
coloured constituents. The leptoquark resonance-search results have
been re-interpreted to establish constraints on leptogluon masses depending
on a composite scale $\Lambda$ in early search papers at 
HERA~\cite{HERA9394lq} but the subjet has not been revisited recently.
   % Search for Leptoquarks

% Alternative Theories for Electroweak Symmetry Breaking:
\clearpage
%%%%%%%%%%%%%%%%%%%%%%%%%%%%%%%%%%%%%%%%%%%%%%%%%%%%%%%%%%%%%%%%%%%%%%%%%%%%%%
\section{Alternative Theories for Electroweak Symmetry Breaking}
\label{sec:ewbreaking}
%%%%%%%%%%%%%%%%%%%%%%%%%%%%%%%%%%%%%%%%%%%%%%%%%%%%%%%%%%%%%%%%%%%%%%%%%%%%%%

The origin of ordinary particle masses remains a mystery. 
It is nevertheless a common belief that an
electroweak symmetry breaking mechanism characterized by one 
(or more) scalar particles is responsible. 
Such particles could be elementary as in the Standard Model 
or in supersymmetric theories like Supergravity.
Alternatively, our parametrization in terms of scalar couplings may in
fact represent a low-energy manifestation of more fundamental dynamics,
with additional particles and interactions. This is the underlying 
assumption of Technicolour or compositeness theories. 
Searches carried out in the framework of Technicolour theories, where
specific dynamical assumptions are made, are discussed in 
subsection~\ref{sec:technic}. The motivations for prospective studies
carried out in the framework of the BESS (Breaking Electroweak Symmetry 
Strongly) model, which possesses new composite bosonic states, are 
discussed in subsection~\ref{sec:bess}.

\subsection{Technicolour}
\label{sec:technic}
%%%%%%%%%%%%%%%%%%%

The Technicolour theory was originally motivated by the premise that
any fundamental energy scale, such as the scale of electroweak 
symmetry breaking, should have a dynamical origin.
Thus, a dynamical electroweak symmetry breaking mechanism is implemented,
in which a r\^ole similar to that of the Higgs boson in the Standard Model 
is now played by multiplets of technihadrons composed of fundamental 
techniquarks bound by a new Technicolour force.

The simplest Technicolour theories~\cite{Weinberg79,Susskind79} did
not address the flavour problem and failed to explain lepton and quark 
masses. Moreover, such theories have now been excluded in particular by 
LEP$_{I}$ constraints~\cite{Erler99} on contributions to vacuum-polarization 
amplitudes~\cite{Peskin92}.
In the Extended Technicolour (ETC) model~\cite{Eichten80}, a new
gauge interaction is introduced to couple ordinary quarks and leptons
to technifermions. Thus, quarks and leptons acquire masses
$ m_{q,l} \simeq \Lambda^3_{TC} / M^2_{ETC} $, with  
$\Lambda_{TC}$ of ${\cal{O}} (10^{2-3}) \GeV$,
the characteristic scale of the new strong gauge interaction, and 
$M_{ETC}$ of ${\cal{O}} (10^{5}) \GeV$, the scale at which the ETC
gauge group breaks down to flavour, colour and technicolour.
But the ETC model in turn has severe problems with unwanted
flavour-changing neutral current (FCNC) interactions.
%---------------------------------------------------------------------------
\begin{figure}[htb]
  \begin{center}                                                                
 \vspace*{-1.5cm}
  \begin{tabular}{cc}
  \epsfig{file=\master/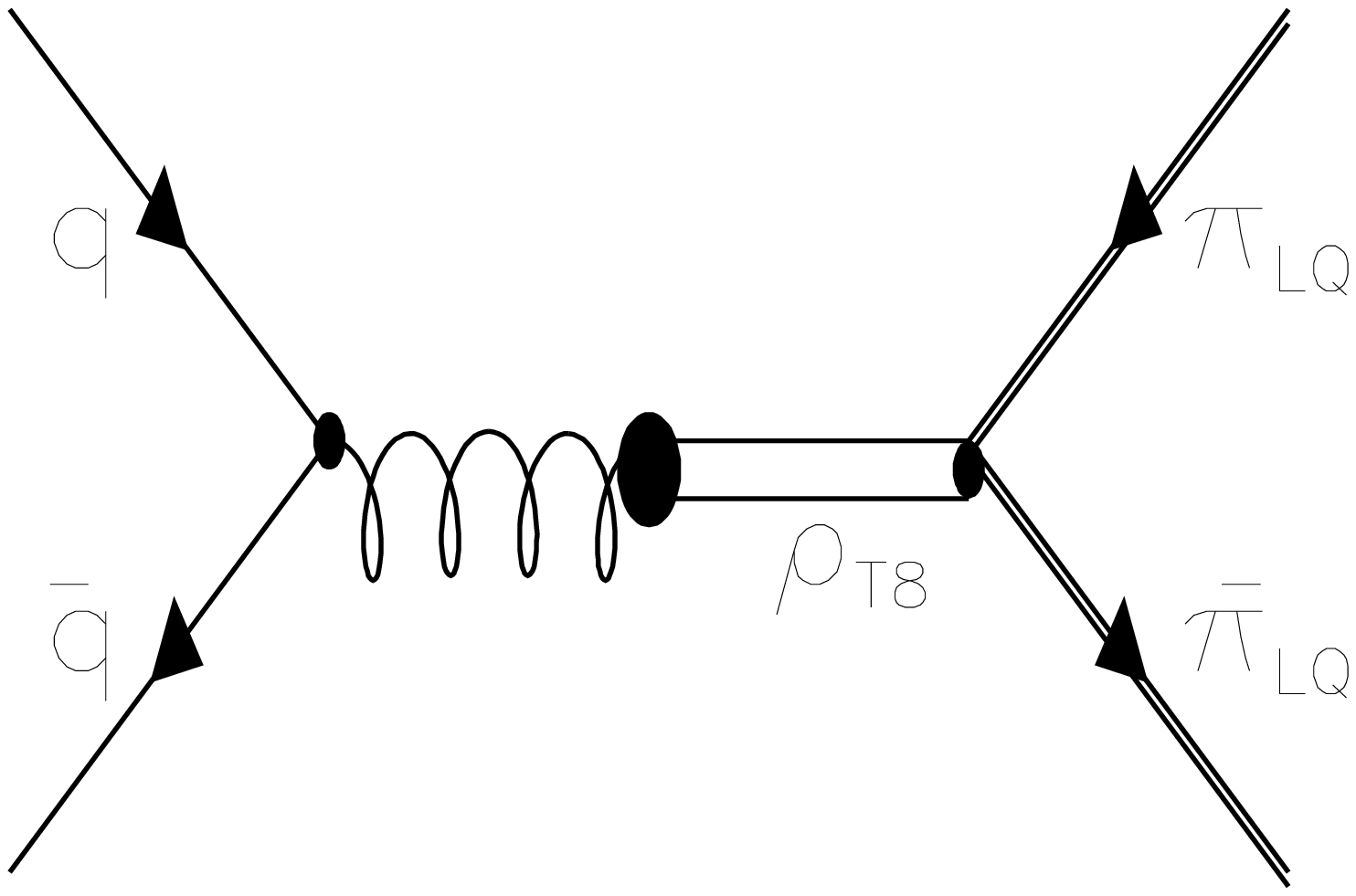,width=0.45\textwidth}
&
  \epsfig{file=\master/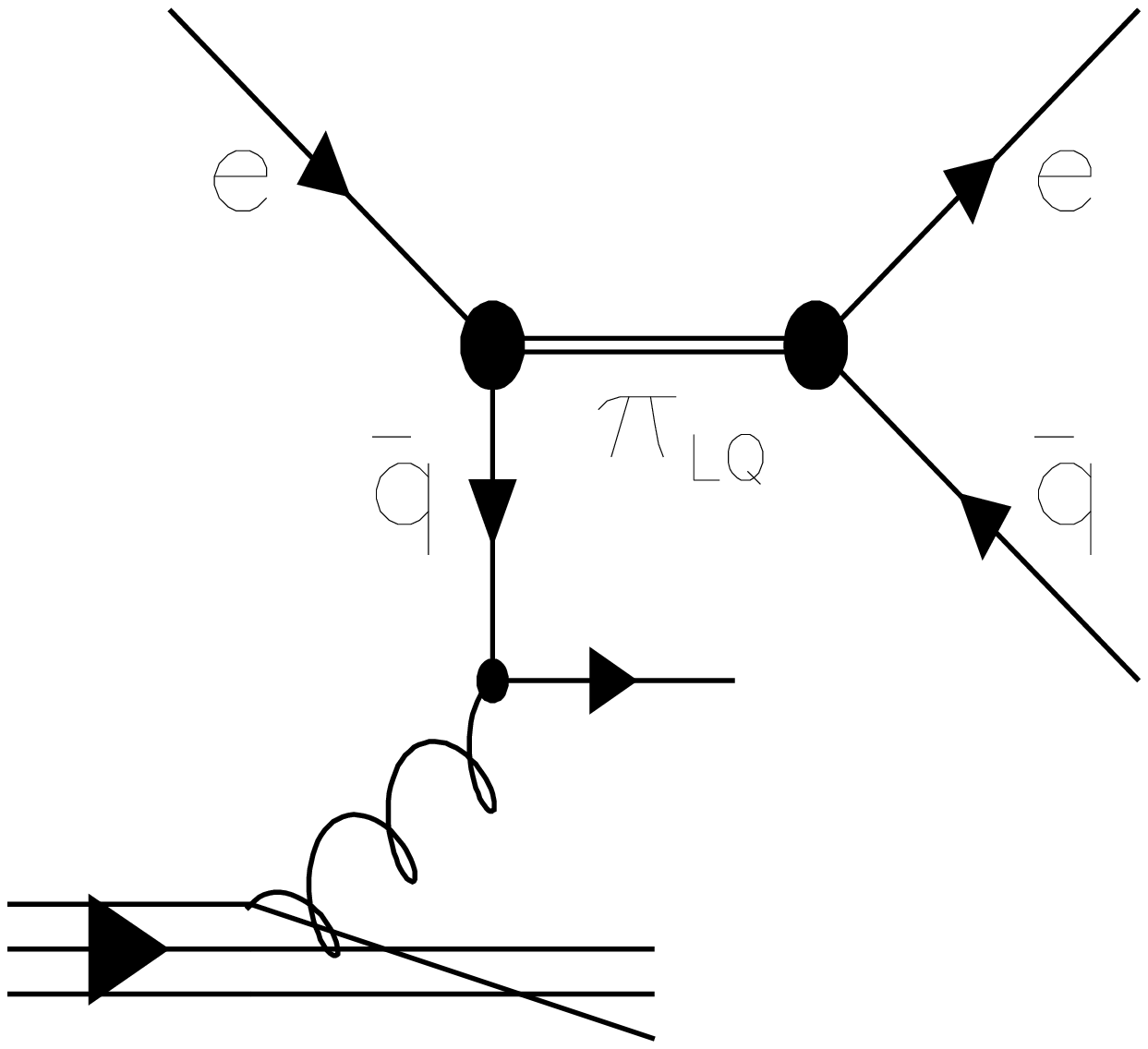,width=0.45\textwidth}
  \end{tabular}
  \end{center}
 \vspace*{-2.5cm}
 
 \hspace*{5.0cm} (a) \hspace*{9.0cm} (b) \\
 
 \vspace*{-0.5cm}
  
 \caption[]{\label{fig:tcdiag}
            a) Diagram for production of technipion (``leptoquark'')
                         pairs in hadronic collisions via $s$-channel  
			 production of a technirho.
	    b) Diagram for single production of leptoquarks
	                 involving heavy quarks in lepton-hadron collisions.} 
\end{figure}
  
%----------------------------------------------------------------------------
The FCNC problems of the ETC model are avoided in recent and more involved 
''Walking Technicolour'' models~\cite{Eichten96,Eichten97,Lane99}. 
This is achieved, at the expense of a loss in predictive power, 
by departing from the original QCD analogy and imposing that, in the presence 
of a large number of technifermions, the Technicolour gauge coupling runs much 
more slowly.
The slow running of the coupling permits ordinary quark and lepton
masses below ${\cal{O}} (1) \GeV$ to be generated from ETC interactions at
${\cal{O}}(10^5) \GeV$.
Walking Technicolour cannot be fully tested by precision experiments  
but it implies the existence of numerous Goldstone-boson bound states of
the technifermions which should appear at masses of ${\cal{O}} (10^{2-3}) \GeV$ 
and can be searched for at colliders.
These include colour-singlet mesons (e.g. scalar technipions 
$\pi_{T}^{\pm,0}$), colour-triplets (e.g. $\pi_{LQ}$ leptoquarks) and 
colour-octets (e.g. $\pi_{T8}^{\pm,0}$ technipions or $\rho_{T8}$ 
technirhos).
In contrast to the leptoquarks of the BRW model discussed in 
section~\ref{sec:leptoq}, the $\pi_{LQ}$'s have Higgs-like couplings 
to ordinary fermions. 

Searches for colour-non-singlet technimesons in the context of Walking
Technicolour have been performed at the 
Tevatron$_{I}$~\cite{CDFtechni99, CDFLQbb} collider based on the
model assumptions of Lane and Ramana~\cite{Lane91}.
The constraints thus established appear particularly relevant in view
of future data taking at Tevatron$_{II}$ and HERA$_{II}$. These are
reviewed in the following.
Otherwise, general reviews of existing bounds on technihadrons 
at colliders can be found in literature~\cite{Lane00,Womersley97}. 

Resonant production of colour-octet technirhos can proceed in $p \bar{p}$
collisions through 
$q \bar{q} \, , \, gg \rightarrow  ( g \leftrightarrow \rho_{T8} ) $
followed by the decay of the $\rho_{T8}$ via 
$\rho_{T8} \rightarrow \pi_{T8} \bar{\pi}_{T8}, \pi_{LQ} \bar{\pi}_{LQ}$
or via e.g. $\rho_{T8} \rightarrow q \bar{q}, g g'$.
An example diagram is shown in Fig.~\ref{fig:tcdiag}(a).
%---------------------------------------------------------------------------
\begin{figure}[htb]
  \begin{center}
  \begin{tabular}{cc}
  \hspace*{-0.5cm} \epsfig{file=\master/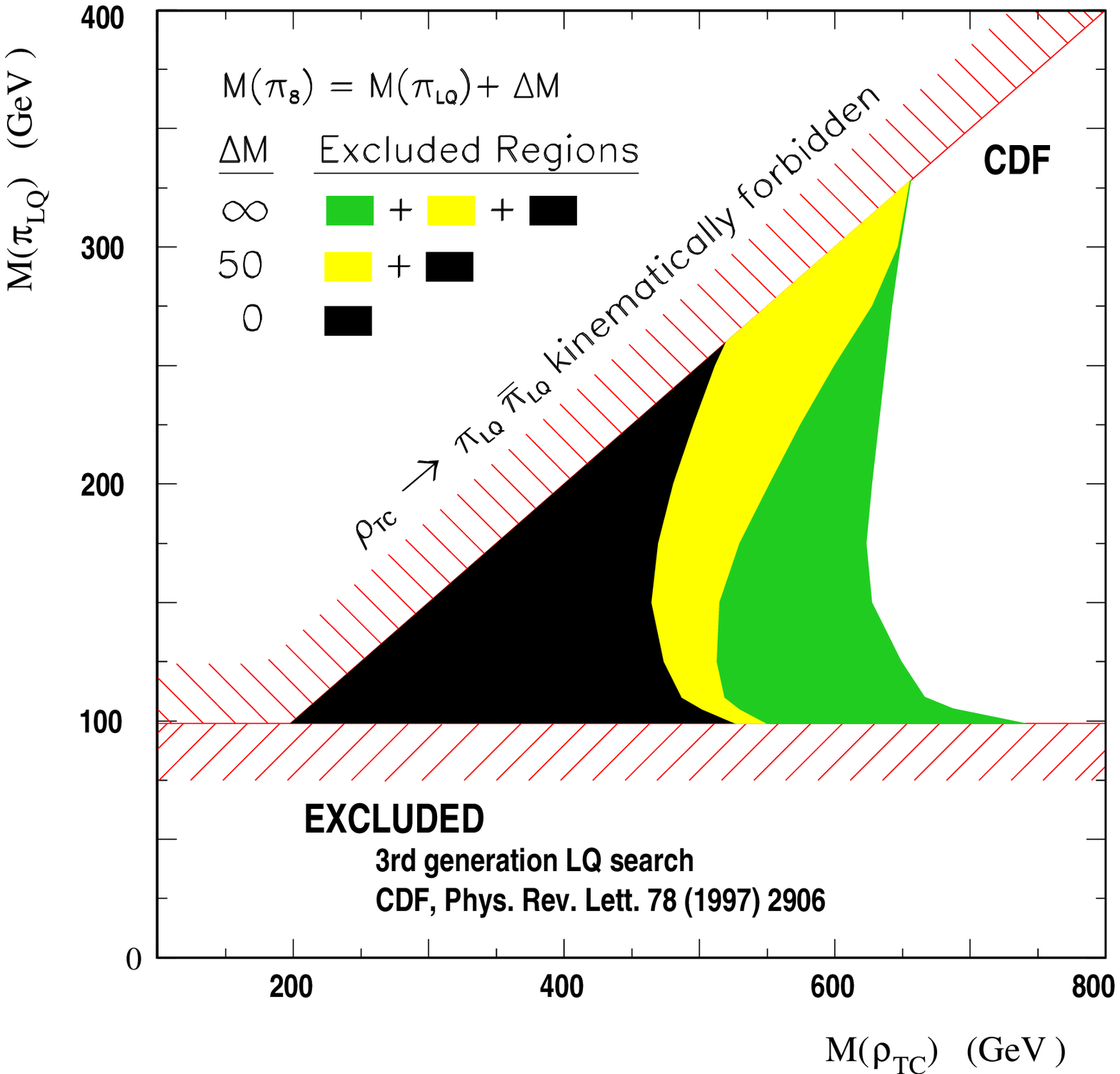,width=0.50\textwidth}
&
  \hspace*{-0.5cm} \epsfig{file=\master/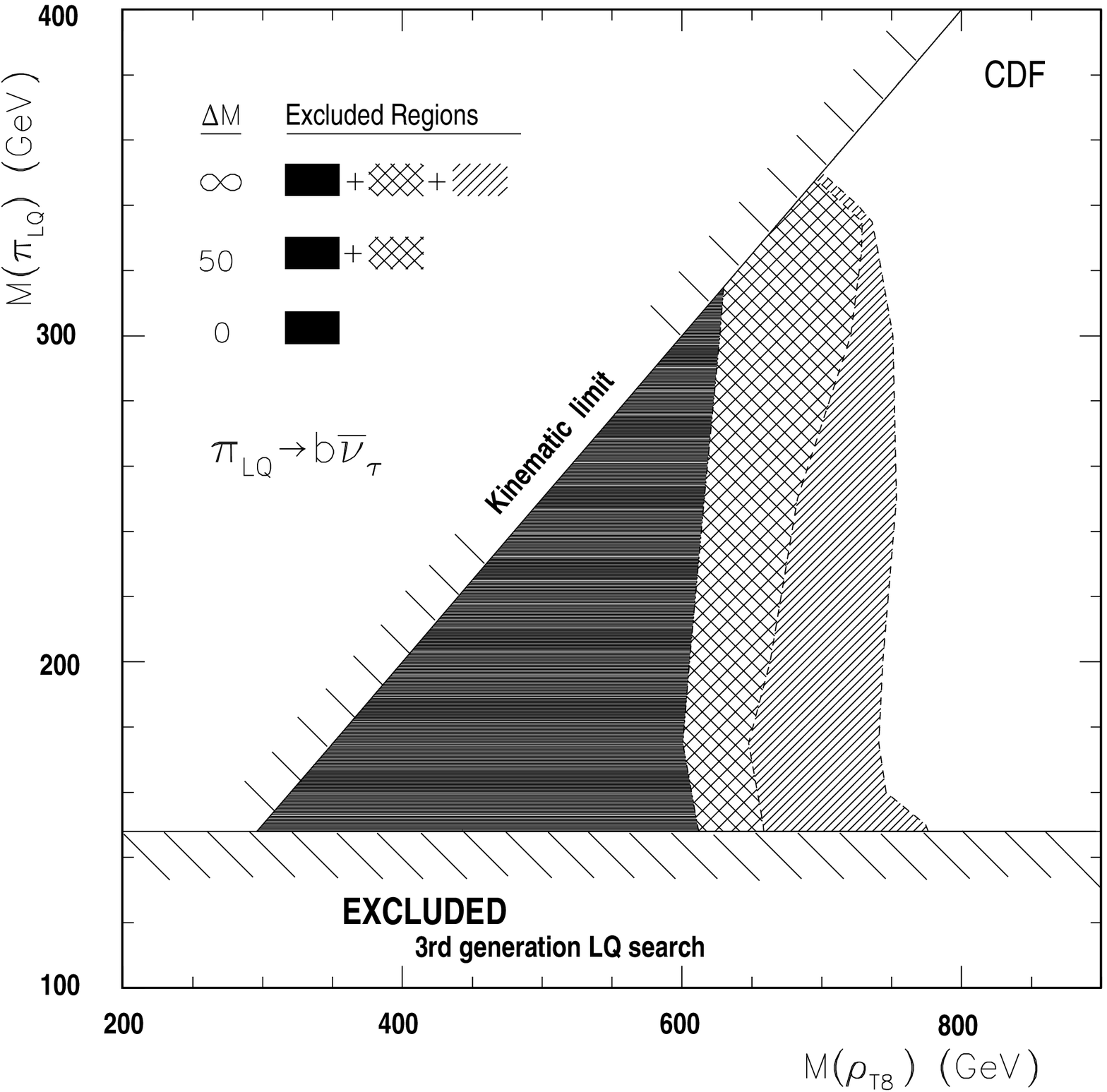,width=0.50\textwidth}
  \end{tabular}
  \end{center}
  
 \vspace*{-0.5cm}
 
 \hspace*{5.0cm} (a) \hspace*{7.5cm} (b) \\
 
 \vspace*{-0.5cm}
  
 \caption[]{\label{fig:tclim}
            Technicolour constraints (95\% CL) from the CDF experiment at the
            Tevatron. 
	    a) Excluded regions for various 
	    $M(\pi_{T8}) - M(\pi_{LQ})$ assumptions from a search for
	    $\rho_{T8} \rightarrow \pi_{LQ} \bar{\pi_{LQ}}
	               \rightarrow \tau^+ \tau^- j j$;
		       (from~\cite{CDFtechni99}).
	    b) Excluded regions for various 
	    $M(\pi_{T8}) - M(\pi_{LQ})$ assumptions from a search for
	    $\rho_{T8} \rightarrow \pi_{LQ} \bar{\pi_{LQ}}
	               \rightarrow b \bar{b} \nu \bar{\nu} $;
		       (from~\cite{CDFLQbb}).} 
\end{figure}
  
%----------------------------------------------------------------------------
The CDF experiment has searched for $\rho_{T8}$ production followed by the
decay $\rho_{T8} \rightarrow \pi_{LQ} \bar{\pi}_{LQ}$, taking into account
the (unobserved) branching fraction of 
$\rho_{T8} \rightarrow \pi_{T8} \bar{\pi}_{T8}$. 
The leptoquark technipions were assumed to decay either via
$\pi_{LQ} \rightarrow \tau^+ b$, in which case the analysis has 
imposed no $b$ tagging (Fig.~\ref{fig:tclim}a) 
or via $\pi_{LQ}  \rightarrow \nu b$ in which case $b$ 
tagging is imposed (Fig.~\ref{fig:tclim}b).
Thus, the analysis is common to that of searches of generic third-generation
leptoquarks produced in pairs (section~\ref{sec:leptoq})
which sets a lower bound on the mass $M (\pi_{LQ})$ independent of
$M (\rho_{T8})$. The constraints from $\rho_{T8}$ production which 
extends beyond this lower bound are shown in Fig.~\ref{fig:tclim}
in the $M (\pi_{LQ})$ {\it vs.} $M (\rho_{T8})$ plane for various
$\Delta M = M(\pi_{T8}) - M(\pi_{LQ})$ assumptions.
Provided that $M(\pi_{T8}) > 2 \times M (\pi_{LQ})$, colour-octet
technirhos are excluded at 95\% CL for masses up to 
$M (\rho_{T8}) < 600 {\rm GeV}$ independently of $\Delta M$.

Leptoquark technipions could be singly produced in $ep$ collisions via a
$t$-channel diagram as shown in Fig.~\ref{fig:tcdiag}(b), preferably
involving a heavy quark $b$ generated by $g \rightarrow b \bar{b}$
splitting. It should be noted that such a process, which requires at 
production an $(eb)$-type of lepton-quark coupling~\cite{Djouadi90},
could turn out to be strongly suppressed in Technicolour theories 
by interfamily mixing parameters. Possible production and decay modes
(including lepton-flavour violating processes such as
    $e + p \rightarrow e + \pi_{LQ} + \bar{b} + X;  
  \pi_{LQ} \rightarrow \tau + b$)
were discussed for $ep$ colliders in the context of early Technicolour 
theories some 20 years ago in~\cite{Cashmore85}. Unfortunately the
topic has not been revisited. Lepton-flavour conserving $(eb)$-type
of leptoquarks~\cite{Djouadi90} would be confronted to the stringent 
Tevatron constraints obtained for first-generation leptoquarks unless
their dominant decay were to be into $\nu_e t$. 

In any case, the Technicolour constraints of Fig.~\ref{fig:tclim} from 
the Tevatron incidentally push leptoquark technipions beyond the reach 
of HERA, unless, as is likely in Walking Technicolour, the 
$\rho_{T8} \rightarrow \pi_{LQ} \bar{\pi}_{LQ}$ decay is kinematically
not allowed. Even then, for most models, very stringent bounds can be 
deduced from Tevatron data; for instance from the absence of dijet resonances
if the $\rho_{T8}$ decays dominantly into $q \bar{q}$ or $g g'$ pairs.
It is nevertheless possible that the dijet rate itself could be 
depleted~\cite{Lane00} if, for instance, the $\rho_{T8}$ decays dominantly 
through $\rho_{T8} \rightarrow g \pi_{T8} ; \pi_{T8} \rightarrow q \bar{q}$.
In such a case, the relevant bound from CDF is the third-generation
leptoquark bound at $99 \GeV$ (95\% CL), beyond which HERA$_{II}$
could have a sensitivity for $t$-channel $\pi_{LQ}$ 
production~\cite{Djouadi90,Cashmore85}. 
A richer set of other possible signals are being explored for 
Tevatron$_{II}$.

\subsection{The BESS Model}
\label{sec:bess}
%%%%%%%%%%%%%%%%

On the basis of unitarity arguments, it is widely believed that, in
the absence of a light Higgs boson or other low-lying scalar resonances,
the interaction among electroweak gauge bosons must become strong
at high energies. In other words, in the absence of an elementary Higgs boson,
new physics should in any case become manifest in the gauge-boson
sector at the electroweak symmetry breaking characteristic scale
of typically $\Lambda_{EWSB} = 4 \pi v \simeq 3 \TeV$.  

Avoiding the difficult task of constructing a viable dynamical scheme, 
the idea of a strongly interacting sector as responsible for electroweak 
symmetry breaking can be tested through an effective-Lagrangian approach. 
This is the motivation for the searches carried out in the framework of the
so-called BESS model.

In its minimal version~\cite{Casalb87}, the BESS model contains a
triplet of new vector resonances $V^{\pm,0}$ similar to the $\rho$
or techni-$\rho$ of Technicolour models. These new vector bosons mix
with the electroweak gauge bosons. 
The mixing depends on the ratio $g/g''$ where $g$ is the $SU(2)_L$ 
Standard Model coupling and $g''$ is the new gauge coupling entering
the self-coupling of the $V^{\pm,0}$.
The coupling of the $V^{\pm,0}$ to ordinary fermions is fixed
by introducing a parameter $b$.
Besides $g''$ and $b$, the model also requires a characteristic mass scale
$M$ which might be taken as the mass $M_V$ of the new strongly interacting
bosons. 
The Standard Model is recovered in the limit $g'' \rightarrow 0$ and
$b \rightarrow 0$. Specific versions of the BESS model can be made to
mimic Technicolour models.

Constraints on the parameters $g/g''$ and $b$ of the BESS model have been 
established by combining precision electroweak data from the LEP and SLC colliders
with $M_{top}$ and $M_W$ measurements from Tevatron 
experiments~\cite{Anichini95}. 
The prospects for direct $V$ resonant production via $l^+l^-$ or 
$q \bar{q}$ annihilation (through $b$ or due to mixing) and via
$WW$ fusion have been studied for future multi-TeV colliders
in the case of a minimal BESS model in Ref.~\cite{Deandrea94}.

A particular BESS model that has received attention for collider
physics is the so-called degenerate-BESS model~\cite{Casalb87b,Casalb95} 
(d-BESS) which requires two new triplets of gauge bosons $L^{\pm,0}$ 
and $R^{\pm,0}$ quasi-degenerate in mass. 
A main property of the d-BESS model is that all deviations at low
energy [i.e. ${\cal{O}}(M_Z)$] from Standard Model expectations 
are completely suppressed. 
Thus, a sensitivity to new strongly interacting bosons could even be
possible at existing colliders despite the constraints of precision 
electroweak tests~\cite{Casalb87b,Casalb95}.

A full description of the effective Lagrangian for the d-BESS model
relevant at colliders can be found in Ref~\cite{Casalb95}. 
A gold-plated signature of the model at hadronic colliders would be a 
pair of leptons originating from the decay of new bosons resonantly 
produced in $q \bar{q}'$ annihilation processes. 
In the charged channel for instance, the process
$p \bar{p} \rightarrow L^{\pm} + X \rightarrow e {\nu}_e (\mu {\nu}_{\mu}) + X$
would lead to a Jacobian peak in the transverse-mass distribution of
final state leptons lying on a background continuum from standard 
Drell-Yan processes with $W$ propagator.
In the neutral channel, the cleanest signal would be provided by
the process $p \bar{p} \rightarrow L^{0} (R^{0}) + X \rightarrow e^+ e^- + X$.
The signal would appear as a narrow resonance in the invariant-mass
distribution of final state electrons lying on a background 
continuum from $\gamma^*$ and $Z^*$ production. 
It has been shown~\cite{Casalb00} that experiments at Tevatron$_{II}$ will be 
sensitive to an unexplored range of $g/g''$ values for characteristic mass 
scales $M$ in the range $200 \GeV < M < 1 \TeV$ via $L^{\pm}$ searches.
In contrast, the HERA experiments cannot improve on existing constraints,
as was examined in Ref.~\cite{Casalb91}.
  % Search for Technicolour

% Supersymmetry
\clearpage
%%%%%%%%%%%%%%%%%%%%%%%%%%%%%%%%%%%%%%%%%%%%%%%%%%%%%%%%%%%%%%%%%%%%%%%%%%%%%%
\section{Supersymmetry}
\label{sec:SUSY}
%%%%%%%%%%%%%%%%%%%%%%%%%%%%%%%%%%%%%%%%%%%%%%%%%%%%%%%%%%%%%%%%%%%%%%%%%%%%%%

\subsection{Introduction: Supersymmetric Matter and Model Parameters}

The search for supersymmetry (SUSY) has constituted one of the central
themes in theoretical and experimental high energy physics over the past 
decades. 
Supersymmetry was originally introduced~\cite{WessBagger} in the framework 
of relativistic field theories as the only possible remaining non-trivial 
extension of the Poincar\'e group (which contains space-time translations 
and Lorentz transformations), relating fermionic with bosonic fields
through its algebraic structure.
Supersymmetric models provide a consistent framework for the unification of 
gauge interactions at some Grand Unification (GUT) scale while resolving the
``hierarchy problem'', i.e. explaining the stability of the electroweak energy 
scale, ${\cal{O}}(10^{2}) \GeV$, relative to the GUT scale 
($M_{GUT} \simeq 10^{16} \GeV$) in the presence of quantum corrections.
It furthermore stabilizes a low Higgs mass against radiative corrections,
provided that the characteristic mass scale of SUSY is below 
${\cal{O}}(1) \TeV$. 

Yet, the existence of any particular realisation of SUSY in Nature remains 
to be proven (or falsified~!). 
Excellent and comprehensive review articles have been written on 
this topic recently, exploring in particular the aftermath 
of LEP precision data~\cite{AfterLEPa,AfterLEPb,AfterLEPc}. 
Our goal here is more modest as we shall review those aspects most 
relevant for experimental searches at the HERA$_{II}$ and Tevatron$_{II}$ 
colliders, following a brief introduction on some essential aspects of 
SUSY models.

\hfill \\
\noindent
{\bf Minimal Sparticle Spectrum:} \\

An immediate consequence of the algebraic structure of supersymmetry is
that particles belonging to the same supermultiplet must have the same
mass. However, the experimental constraints clearly forbid the existence of,
say, a scalar with a mass equal to that of the electron. Known bosons
cannot be made to be superpartners of known fermions. They do not appear
to have very much in common, with different gauge-symmetry properties 
and a number of known degrees of freedom significantly larger for fermions. 
Hence supersymmetry, if it exists in the physical world, must be broken. 
It requires the introduction of new heavy bosonic (fermionic) partners that ought 
to be associated with each of the ordinary quarks and leptons (gauge bosons
and the Higgs bosons). 
A supersymmetric extension of the Standard Model with minimal new 
particle content requires in addition the introduction of two Higgs-field 
doublets, $H_1$ and~$H_2$.

Ultimately, the number of free parameters of the new theory and the details 
of the observable phenomenology will depend on the supersymmetry-breaking 
mechanism chosen by Nature (see below) and on the absence or 
presence of $R$-parity-violating interactions (see below).
However, the minimal particle content of supersymmetric extensions of the Standard Model 
is essentially common to all models and summarised in Table~\ref{tab:susymat}.
% ............ TABLE  x: Supersymmetric matter ....................................
%
\begin{table*}[htb]
   \renewcommand{\doublerulesep}{0.4pt}
 \begin{center}
   \begin{tabular}{||c|c|c||c|c|c||}
   \hline \hline
     \multicolumn{6}{||c||}{{\it SUSY Physical States After Electroweak Symmetry Breaking}} \\
     \multicolumn{3}{||c||}{ $R_P = +1$ }                  & \multicolumn{3}{c||}{ $R_P = -1$ }                \\
     {\bf Name}    &  {\bf  Symbol}     & {\bf  Spin} & {\bf Name} &  {\bf  Symbol}             & {\bf Spin} \\
   \hline
     \multicolumn{3}{||c||}{{\it Ordinary Standard Model Matter}}
                                                           & \multicolumn{3}{c||}{{\it Supersymmetric Matter}} \\
     leptons       & $e, \nu_e$         &    1/2      & sleptons   & \selL, \selR, \snue        &  0    \\
                   & $\mu, \nu_{\mu}$   &    1/2      &            & \smuL, \smuR, \snumu       &  0    \\
                   & $\tau, \nu_{\tau}$ &    1/2      &            & \stauL, \stauR, \snutau    &  0    \\
     quarks        & $u, d$             &    1/2      & squarks    & \sqrkuL, \sqrkuR, 
                                                                                 \sqrkdL, \sqrkdR &  0    \\
                   & $c, s$             &    1/2      &            & \sqrkcL, \sqrkcR, 
		                                                                 \sqrksL, \sqrksR &  0    \\
                   & $t, b$             &    1/2      &            & \sqrkt{1}, \sqrkt{2}, 
		                                                             \sqrkb{1}, \sqrkb{2} &  0    \\  
   \hline
     gluons            &   $g$        &   1           & gluino      &   $\tilde{g}$                  &   1/2 \\
     photon            & $\gamma$     &   1           &             &                                &       \\
     electroweak bosons         & $Z^{0}$    &   1           & neutralinos & \Xo{1}, \Xo{2}, \Xo{3}, \Xo{4} &   1/2 \\
                       & $W^{\pm}$      &   1           & charginos   &  \Xpm{1}, \Xpm{2}              &   1/2 \\
     \multicolumn{3}{||c||}{{\it Higgs Sector}}       &             &                                &       \\ 
     $CP$-even scalars  &  $h^0$,$H^0$ &   0          &             &                                &       \\ 
     $CP$-odd pseudoscalars &  $A^0$   &   0          &             &                                &       \\ 
     charged scalars    &  $H^{\pm}$   &   0          &             &                                &       \\ 
   \hline \hline
  \end{tabular}
  \caption {{\it Minimal particle content of Supersymmetric Standard Models.}} 
  \label{tab:susymat}
\end{center}
\end{table*}
%
%...................................................................................

To a given fermion, $f$, corresponds a superpartner for each of its chirality states.
These $\tilde{f}_L$ and $\tilde{f}_R$ are fundamentally independent fields. 
They are expected to mix significantly only in the third generation, leading to the
mass eigenstates \sqrkt{1}, \sqrkt{2}, \sqrkb{1} and \sqrkb{2}.

The ``neutralinos'' \Xo{1,2,3,4} (``charginos'' \Xpm{1,2}) are the mass eigenstates 
resulting from the mixing of the non-strongly interacting gauginos, $\tilde{\gamma}$
and $\tilde{Z}$ ($\tilde{W}^{\pm}$), with the higgsinos,
$\tilde{H}^0_1$ and $\tilde{H}^0_2$ ($\tilde{H}^{\pm}$).

The two Higgs doublets lead to five physical Higgs bosons, two CP-even neutral 
scalars ($h^0$,$H^0$), a CP-odd neutral pseudo-scalar ($A^0$) and a pair of
charged scalars ($H^{\pm}$).

Needless to say, searches for the vast number of new ``sparticles'' and for the
other indirect effects predicted by supersymmetric models have constituted 
a major analysis activity at high-energy colliders over the past decade. 
This is likely to remain so at the HERA$_{II}$ and Tevatron$_{II}$ colliders. 
If realized at low energies, and not yet found by then, it is widely believed 
that Nature cannot hide SUSY beyond the expected experimental sensitivity 
of the future LHC collider.

\hfill \\
\noindent
{\bf Supersymmetry Breaking Mechanisms and Model Parameters:} \\

How supersymmetry is broken and in which way this breaking is communicated
to the particles remains an open question on the theoretical side.
The phenomenology at colliders will depend strongly on the chosen answer.

In the so-called Minimal Supersymmetric Standard Model (MSSM), $R$-parity is 
assumed to be conserved and the SUSY breaking is simply parametrized by 
introducing explicitly ``soft'' terms in the effective Lagrangian~\cite{Haber85}. 
For phenomenological studies, the sfermion masses are generally treated as free 
parameters while the masses of the neutralinos and charginos,
as well as the gauge couplings between any two sparticles and a standard fermion
or boson, are determined by a set of five parameters: the three soft-breaking
parameters $M_1$, $M_2$ and $M_3$ for the $U(1)$, $SU(2)$ and $SU(3)$ gauginos,
the ratio $\tan \beta$ of the vacuum expectation values of the two neutral
Higgs bosons,
and the ``mass'' term $\mu$ which mixes the Higgs superfields.
The SUSY-breaking soft terms also contain in general bilinear ($B$) and 
trilinear ($A_{ijk}$) couplings. 

However, one could expect that, if SUSY is a fundamental symmetry of Nature, then it
should preferably be an exact symmetry which is spontaneously broken. In other words,
the ultimate theory should have a Lagrangian density that is invariant under SUSY
but a vacuum that isn't. The symmetry would be hidden at low (collider) energies
in a way analogous to the fate of electroweak symmetry in the ordinary Standard
Model. Three prominent schemes have been extensively considered: Minimal 
Supergravity (mSUGRA), Gauge Mediated (GMSB) and Anomaly Mediated (AMSB) SUSY
Breaking Models.

In Supergravity theories~\cite{Nilles84}, the supersymmetry is broken in a 
``hidden'' sector and the breaking is transmitted to the ``visible'' sector by 
gravitational interactions.
In mSUGRA, masses and couplings must obey unification conditions at the GUT scale,
$M_{GUT} \simeq 10^{16} \GeV$. The gaugino masses $M_1$, $M_2$ and $M_3$ unify 
to a common mass $m_{1/2}$; scalar particles (sfermions and Higgs bosons) have
a common mass $m_0$ and all trilinear coupling parameters $A_{ijk}$ have a common
value $A_0$. Furthermore, the mass mixing parameter $\mu$ can be expressed as a
function of the other parameters via equations corresponding to the minimization
of the Higgs potential when invoking radiative electroweak symmetry breaking, so
that only the sign of $\mu$ remains free. Thus, in mSUGRA, one is left with five
unknowns (four free parameters and a sign), namely $m_{1/2}$, $m_0$, $A_0$,
$\tan \beta$ and the sign of $\mu$. The measurements of the sparticle spectrum 
and SUSY phenomenology at collider energies could provide an overconstrained
determination of these parameters at the GUT scale through renormalization
group equations. In large domains of the parameter space, the lightest neutralino
acts as the LSP (lightest supersymmetric particle)
appearing at the end of the gauge-decay chains of other SUSY 
particles.

In GMSB models~\cite{Giudice99b}, the SUSY breaking is transmitted through gauge
interactions via some messenger states of mass $M \ll M_{Planck}$. These models 
depend on the free parameters $M_{mess}$, $N_{mess}$, $\Lambda$ and the sign 
of $\mu$ where $M_{mess}$ is the messenger scale, $N_{mess}$
is an index depending on the chosen structure of the messenger sector, and 
$\Lambda$ is a universal SUSY-breaking scale. If the SUSY breaking occurs at
relatively low energy scales, e.g. $\Lambda \sim {\cal{O}}(10^{1-2}) \TeV$,
then a very light gravitino ($\tilde{G}$) will act as the LSP appearing
at the end of the gauge-decay chains of other SUSY particles. 

In AMSB models~\cite{AMSBtheory}, the SUSY breaking is not transmitted directly 
from the ``hidden'' to the ``visible'' sector. The gaugino masses are rather 
generated at one loop as a consequence of the ``super-Weyl'' anomaly and depend 
on the mass parameter $m_{3/2}$. In minimal models, a universal scalar mass $m_0$ 
is introduced at the GUT scale, leaving four unknowns (three free parameters 
and a sign), $m_0$, $m_{3/2}$, $\tan \beta$ and the sign of $\mu$.
The wino acts as the LSP and the lightest chargino and neutralino are nearly
degenerate in mass.

\hfill \\
\noindent
{\bf $R$-parity and the Phenomenology:} \\

In the Standard Model, the conservation of the baryon and lepton number is 
an automatic consequence of the gauge invariance and renormalizability.
In contrast, baryon and lepton number are no longer protected in 
supersymmetric extensions of the Standard Model. The introduction of a
scalar (fermionic) partner for each ordinary fermion (boson) allows in 
general for new interactions that do not preserve baryon or lepton number. 
Such interactions can be avoided in an {\it ad hoc} manner by the introduction of 
a discrete symmetry implying the conservation of the quantum number 
$R_p$ ($R$-parity) which distinguishes ordinary particles ($R_p = +1$) 
from supersymmetric particles ($R_p = -1$), and is defined as
$R_p \equiv (-1)^{3B+L+2S}$ with $S$ denoting the particle spin, 
$B$ the baryon number and $L$ the lepton number.

Whether or not $R_p$ is conserved in supersymmetric models has dramatic 
observable consequences. If $R_p$ is exactly conserved then sparticles 
can only be produced in pairs and the
LSP is absolutely stable. The LSP is then a natural 
candidate for Cold Dark Matter in cosmology. At collider experiments
the (cascade) decays of pair-produced heavy sparticles would always leave 
a pair of LSPs escaping detection, thus leading to a characteristic
``missing energy-momenta'' signal.

Reviews of the phenomenology relevant for collider physics from 
$R_p$-conserving supersymmetry can be found in the case of the MSSM/mSUGRA
models in Refs.~\cite{AfterLEPa, AfterLEPb, AfterLEPc}, for GMSB models in Refs.~\cite{GMSBpheno}
and for AMSB models in Refs.~\cite{AMSBpheno}. 

If $R_p$-violating ($\Rp$) interactions are allowed, then all supersymmetric 
matter becomes intrinsically unstable, sparticles can be singly produced and 
spectacular processes with lepton- or baryon-number violation could be
observed at colliders.

\begin{figure}[htb]
  \begin{center}                                                                
  \begin{tabular}{cccc}
  \vspace*{-0.5cm}
    
  \hspace*{-1.2cm} \epsfig{file=\master/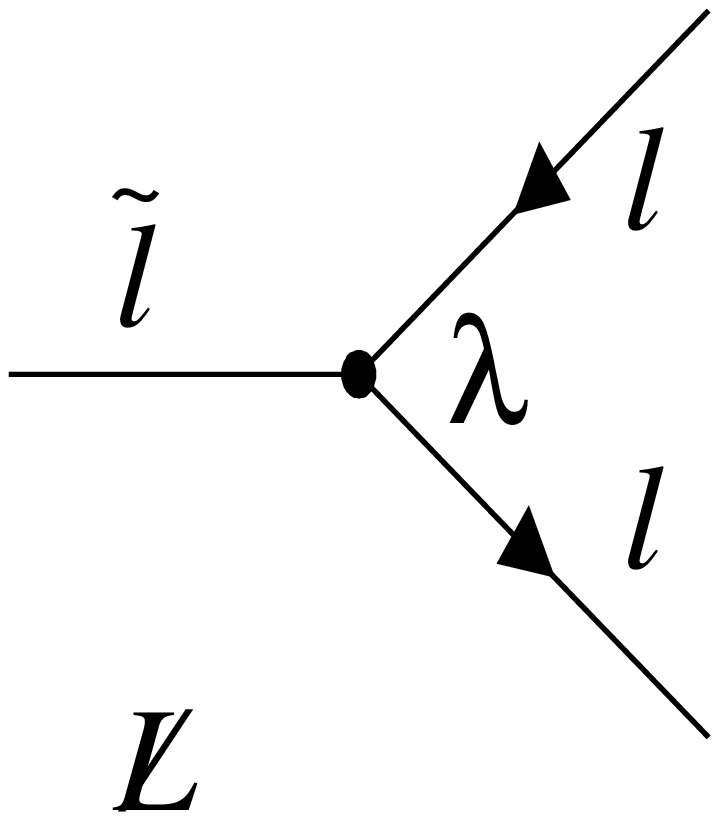,width=0.40\textwidth}
&
  \hspace*{-3.0cm} \epsfig{file=\master/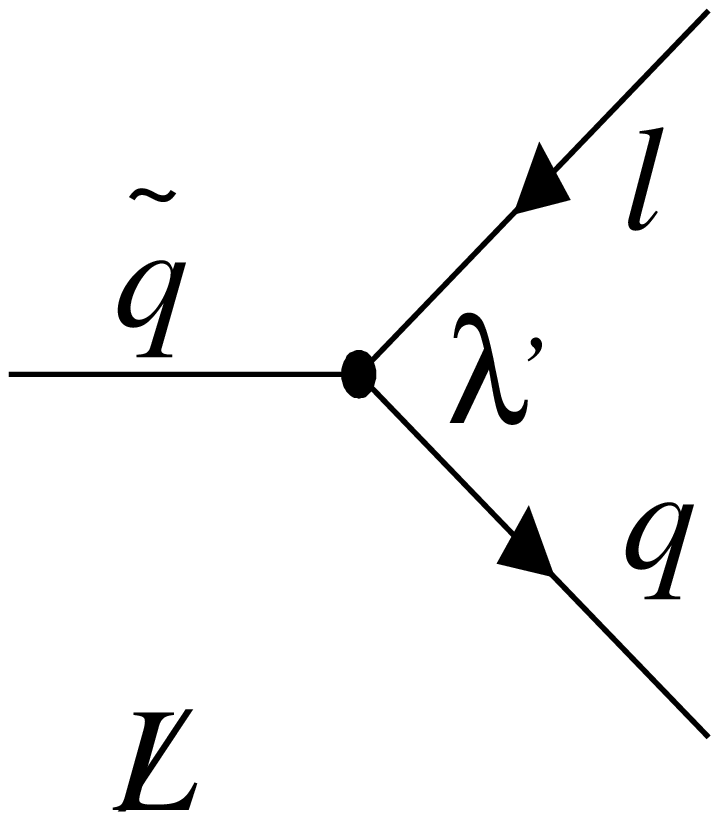,width=0.40\textwidth}
&
  \hspace*{-4.2cm} \epsfig{file=\master/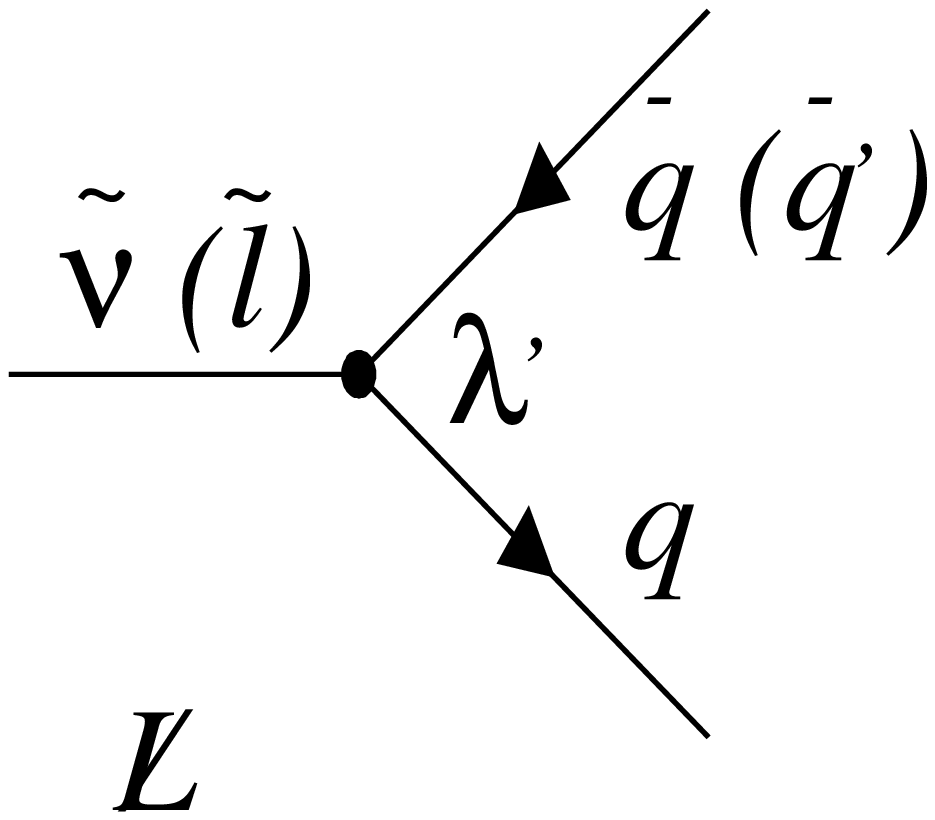,width=0.40\textwidth}
&
  \hspace*{-2.7cm} \epsfig{file=\master/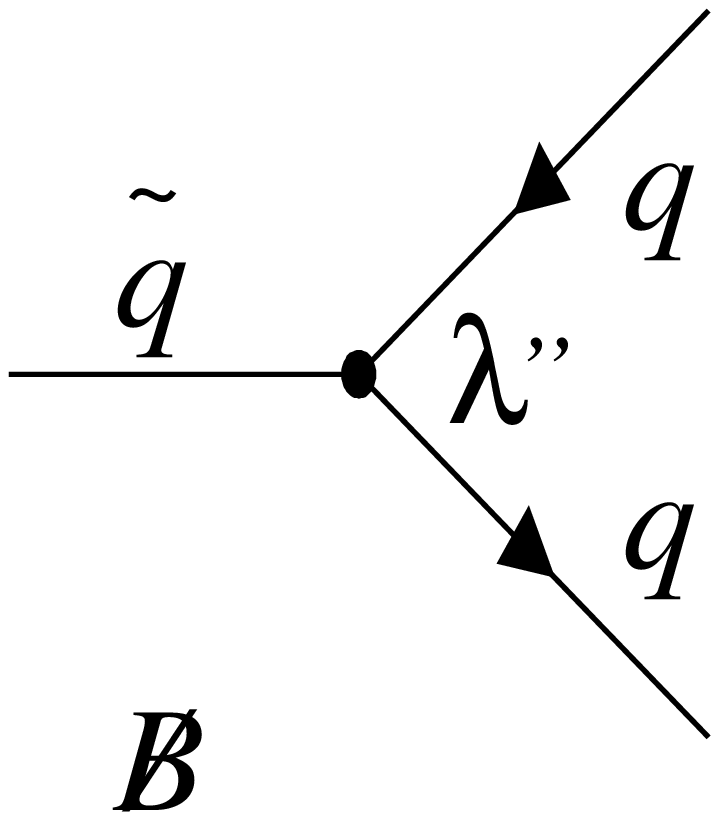,width=0.40\textwidth}

  \end{tabular}
  \end{center}
 \vspace*{-1.8cm}
  
 \caption[]{ \label{fig:diagrpv}
            Basic tree diagrams for sfermion-fermion-fermion $\Rp$ Yukawa
	    interactions via $\lambda, \lambda'$ and $\lambda''$. } 
\end{figure}

The most general renormalizable $R_p$-violating superpotential 
consistent with the gauge symmetry and field content of the MSSM 
contains bilinear and trilinear terms: 
\begin{equation*}
  W_{\Rp}\ =\ \mu_i\, H_u L_i\
  +\ \frac{1}{2}\, \lambda_{ijk}\, L_i L_j E^c_k\
  +\ \lambda'_{ijk}\, L_i Q_j D^c_k\
  +\ \frac{1}{2}\, \lambda''_{ijk}\, U^c_i D^c_j D^c_k \ ,
\label{eq:W_Rp_odd_app}  
\end{equation*}
where an implicit summation over the generation indices,
$i,j,k = 1,2,3$, and over gauge indices is understood. 
The $\mu_i$ associated to fermion bilinears are dimensionful 
mixing parameters and the $\lambda, \lambda'$, and $\lambda''$ 
are Yukawa-like couplings which are trilinear in the fields.
The corresponding Lagrangian expanded in terms of four-component Dirac 
spinors is written as
%------------
\begin{eqnarray*}
{\cal L}_{L_{i}L_{j}E^c_{k}}= - \frac{1}{2} \lambda_{ijk}
\bigg ( \tilde  \nu_{iL}\bar l_{kR}l_{jL} +
\tilde l_{jL}\bar l_{kR}\nu_{iL} + \tilde l^\star _{kR}\bar \nu^c_{iR}
l_{jL}-(i \leftrightarrow j) \bigg ) + \ \mbox{h.c.},
\label{eq:laglambda_app}
\end{eqnarray*} 
where for instance $\bar \nu^c_{iR}=\overline{(\nu^c_i)_R}$.
Similarly,
%------------
\begin{eqnarray*}
{\cal L}_{L_i Q_j D^c_k}= 
- \lambda '_{ijk} && \bigg ( \tilde  \nu_{iL}\bar d_{kR}d_{jL} +
\tilde d_{jL}\bar d_{kR}\nu_{iL} + \tilde d^\star _{kR}\bar \nu^c_{iR}
d_{jL} \cr && -\tilde  l_{iL}\bar d_{kR}u_{jL} -
\tilde u_{jL}\bar d_{kR}l_{iL} - \tilde d^\star _{kR}\bar l^c_{iR} u_{jL}
\bigg ) + \ \mbox{h.c.},
\label{eq:laglambdap_app}
\end{eqnarray*} 
%------------
and
%------------ 
\begin{eqnarray*}
{\cal L}_{U_i^c D_j^c D_k^c}= - \frac{1}{2} {\lambda ''}_{ijk} 
\bigg (\tilde  u^\star _{i R}\bar d_{j R}d^c_{k L} +
\tilde  d^\star _{j R}\bar u_{i R}d^c_{k L} +
\tilde  d^\star _{k R}\bar u_{i R}d^c_{j L} \bigg ) 
+ \ \mbox{h.c.} \ .
\label{eq:laglambdapp_app}
\end{eqnarray*} 
%------------
Now gauge invariance enforces antisymmetry of the $\lambda_{ijk}$ 
couplings in their first two indices 
(i.e. $\lambda_{ijk} = - \lambda_{jik}$), 
and antisymmetry of the $\lambda''_{ijk}$ couplings in their last two indices 
(i.e. $\lambda''_{ijk} = - \lambda''_{ikj}$).
Hence, altogether there exist 45 dimensionless Yukawa couplings, 
9 $\lambda_{ijk}$ plus 27 $\lambda'_{ijk}$ which break the conservation
of lepton number and 9 $\lambda''_{ijk}$ which break baryon-number 
conservation.
Basic tree diagrams illustrating the three types of Yukawa couplings
are shown in Fig.~\ref{fig:diagrpv}.
%----------------------------------------------------------------------------
%-----------------------------------------------------------------------------

The lepton-number violating (\LV) term \LLE\ couples sleptons 
and leptons through $\lambda$.
The \LV\ term \LQD\ couples squarks to lepton-quark pairs 
and sleptons to quark pairs through $\lambda'$.
The baryon-number violating (\BV) term \UDD\ couples squarks to 
quark pairs through $\lambda''$. 
Testing $2^{45}-1$ possible combinations of $\lambda, \lambda'$ and $\lambda''$ 
couplings of comparable size would obviously be an insurmountable task.
The problem is partially reduced when taking into account constraints 
already established. The preservation of proton stability imposes for instance
very stringent constraints on the co-existence of \LV\ and \BV\ couplings.
The coupling products $\lambda \times \lambda'$ and $\lambda' \times \lambda''$ 
must essentially vanish. 
It is moreover reasonable to assume that there could be a strong hierarchy among
the couplings, not unlike that observed in the Yukawa sector of the Standard Model.
Thus, in actual searches it is generally assumed (conservatively !) that a single 
$\lambda, \lambda'$ or $\lambda''$ coupling dominates.

The discovery mass reach for sfermions at colliders can be considerably enlarged
by single sparticle real production or virtual exchange involving $\Rp$ Yukawa couplings.
For sfermions masses below the available centre-of-mass energies
($\sqrt{s}$) at a given collider, 
$s$-channel resonant processes are allowed with a production rate scaling with the 
Yukawa coupling squared.
The list of the $s$-channel processes allowed at lowest order in 
$e^+e^-$, $ep$ and $p\bar{p}$ collisions is given in Table~\ref{tab:rpvres}.
% ... TABLE  x: Rp-violating production processes .......................
%
\begin{table*}[htb]
  \renewcommand{\doublerulesep}{0.4pt}
  \renewcommand{\arraystretch}{1.2}
  \begin{center}
  \begin{tabular}{||c||c|c|lr||}
  \hline \hline
  \multicolumn{5}{||c||}{Resonant Production of Sfermions at Colliders} \\
  \multicolumn{5}{||c||}{(lowest-order processes)} \\
   \hline
   Collider  &  Coupling     &  Sfermion           & \multicolumn{2}{|c||}{Elementary Process} \\
             &               &  Type               & \multicolumn{2}{|c||}{} \\
   \hline
    $e^+e^-$ &$\lambda_{1j1}$
                             & $\tilde{\nu}_{\mu}, \tilde{\nu}_{\tau}$ & 
               $l^+_i l^-_k \rightarrow \tilde{\nu}_j$ & $i=k=1 \, , \, j=2,3$ \\
   \hline
  $p \bar{p}$ &$\lambda'_{ijk}$
                    & $\tilde{\nu}_{e}, \tilde{\nu}_{\mu}, \tilde{\nu}_{\tau}$
		    & $d_k \bar{d}_j \rightarrow \ \tilde{\nu}_i$ & $i,j,k = 1,\ldots,3$ \\
              &     & $\tilde{e}, \tilde{\mu}, \tilde{\tau}$
	 & $u_j \bar{d}_k \rightarrow \ \tilde{l}_{iL}$ & $i,k = 1,\ldots,3 \, , \, j = 1,\ldots,2$ \\
              & $\lambda''_{ijk}$
                    & $\tilde{d}, \tilde{s}, \tilde{b}$
	 & $\bar{u}_i \bar{d}_j \rightarrow \ \tilde{d}_{k}$ & $i,j,k = 1,\ldots,3 \, , \, j \neq k$\\
              &     & $\tilde{u}, \tilde{c}, \tilde{t}$
	 & $\bar{d}_j \bar{d}_k \rightarrow \ \tilde{u}_{i}$ & $i,j,k = 1,\ldots,3 \, , \, j \neq k$\\
   \hline
  $e p$ & $\lambda'_{1jk}$
                    & $\tilde{d}_R, \tilde{s}_R, \tilde{b}_R$
	     & $l^-_1 u_j \rightarrow \tilde{d}_{kR}$ & $j=1,2$ \\
        & $\lambda'_{1jk}$
                    & $\tilde{u}_L, \tilde{c}_L, \tilde{t}_L$
		    & $l^+_1 d_k \rightarrow \tilde{u}_{jL}$ & $i,j,k = 1,\ldots,3$\\
   \hline \hline
  \end{tabular}
  \caption {{\it
         Sfermions $s$-channel resonant production at colliders. Charge conjugate 
	 processes (not listed here) are also possible. Real $\tilde{\nu}$ production 
	 at an $e^+e^-$ collider can only proceed via $\lambda_{121}$ or $\lambda_{131}$
	 while virtual $\tilde{\nu}$ exchange in the $t$-channel is possible for any
	 $\lambda_{ijk}$ provided either $i, j$ or $k=1$. Real $\tilde{q}$ production
	 at an $ep$ collider is possible via any of the nine $\lambda'_{1jk}$ couplings.}}
  \label{tab:rpvres}	 
\end{center}
\end{table*}
%
%......................................................................
The \LV\ coupling $\lambda$ allows for $s$-channel resonant production 
of $\tilde{\nu}$ at $l^+l^-$ colliders.
The \LV\ coupling $\lambda'$ allows for $s$-channel resonant production of 
$\tilde{q}$ at $ep$ colliders and of $\tilde{\nu}$ or $\tilde{l}^{\pm}$ 
at $p\bar{p}$ colliders.
The \BV\ coupling $\lambda''$ allows for $s$-channel resonant production of 
$\tilde{q}$ at $pp$ colliders.

Virtual exchange involving $\Rp$ couplings provides a sensitivity to sfermions 
with masses above the available $\sqrt{s}$ at colliders.
In the simplest process, the virtual exchange will provide a contribution
to fermion pair production with a cross-section depending on the square of the
Yukawa coupling squared. For $M_{\tilde{f}} \gg \sqrt{s}$, this will effectively
contract to a four-fermion contact interaction (see section~\ref{sec:contact}).
At an $e^+e^-$ collider, lepton-pair production can receive a contribution
involving $\lambda$ from $s$-channel exchange of $\tilde{\nu}_{\mu}$ and 
$\tilde{\nu}_{\tau}$ and from $t$-channel exchange of $\tilde{\nu}_{e}$, 
$\tilde{\nu}_{\mu}$ and $\tilde{\nu}_{\tau}$.
Quark-pair production can receive a contribution involving $\lambda'$ from
$t$-channel exchange of $\tilde{u}_L$-like or $\tilde{d}_R$-like squarks.
At an $ep$ collider, $s$- and $t$-channel exchange of $\tilde{u}_L$-like or 
$\tilde{d}_R$-like squarks involving $\lambda'$ can contribute to lepton-quark
pair production.
At a $p\bar{p}$ collider, quark pair production can receive a contribution
involving $\lambda'$ from $\tilde{\nu}$ or $\tilde{l}$ exchange in the 
$s$-channel. It can also receive a contribution involving $\lambda''$
from $\tilde{q}$ exchange in the $s$- or $t$-channel.

The presence of an $\Rp$ coupling will open new decay modes for sparticles.
``Direct'' $\Rp$ decays will compete with ``indirect'' decays initiated by
gauge couplings and in which the $\Rp$ couplings enter at a later stage in
the decay chain.
The coupling $\lambda$ allows for direct decays of sleptons into lepton
pairs, and of gaugino-higgsinos into three leptons.
The coupling $\lambda'$ allows for direct decays of sleptons into quark
pairs, of squarks into lepton-quark pairs, and of gaugino-higgsinos into
a lepton plus a quark pair.
The coupling $\lambda''$ allows for direct decays of squarks into quark
pairs, and of gaugino-higgsinos into three quarks.

\subsection{Status of Supersymmetry Searches at Colliders}
\label{sec:statsusy}

% Precision measurements, extended Higgs sector:

The precision measurements at LEP have left, as a legacy, at least indirect 
hints suggesting that supersymmetry could hide just above existing direct 
constraints. 
One such hint comes from the precisely measured gauge couplings, which are 
found to be consistent with a supersymmetric GUT provided that the 
sparticle masses are less than ${\cal{O}}(1) \TeV$.
Another hint comes from the precision electroweak data which suggest
the existence of a relatively light neutral Higgs boson.
The existence of a light neutral Higgs boson is a strong requirement 
in all supersymmetric models.
A Standard Model fit~\cite{LEPWG2001} to these electroweak precision 
data provides
an upper indirect limit of $M_H \le 193 \GeV$ (95\% CL). 
The direct search at LEP$_{II}$ gives a lower limit of 
$M_H \ge 114.4 \GeV$ (95\% CL)~\cite{LEPSMH}.
The limits also apply at small $\tan \beta$, in the framework 
of the MSSM, to the light CP-even neutral Higgs boson $h^0$.
Scanning over the parameter space of the MSSM gives~\cite{LEPSUSYH} a 
conservative lower limit from LEP$_{II}$ of $M_{h^0} > 91.0$~GeV.

% Sparticle searches:

No direct evidence for the existence of supersymmetric matter has yet been 
found at colliders and considerable efforts went into the derivation of
constraints on supersymmetric models. Yet, trying to establish universal
bounds is a formidable task given the flexibility of general formulations
of supersymmetry. Hence two main avenues have been followed in a 
complementary manner by the experiments. 
% ............ TABLE  x: Constraints on Sparticle Masses .....................
%
\begin{table*}[htb]
   \renewcommand{\doublerulesep}{0.4pt}
 \begin{center}
 \begin{tabular}{||c|c|c|c|c||}
 \hline \hline
   \multicolumn{5}{||c||}{UTTERLY UNAVOIDABLE CONSTRAINTS ON SPARTICLE MASSES} \\

   {\bf Sparticle}& {\bf  Model} & {\bf 95\% CL}  & {\bf Applicability} &{\bf Experimental}\\
   {\bf Type}     &              & {\bf $M_{low}$ Limits}&                  &{\bf Ressources} \\
                  &              & (in $\GeV$)&                     &                 \\
   \hline
   \multicolumn{5}{||l||}{{\it Gauginos-Higgsinos (EW sector)}}  \\
%                              ------------------------------
   \hline
% neutralinos:
  \Xo{i}  &  MSSM      &  37     & $i=1$; $\forall (\tan \beta , m_0)$;
                                        LSP $\equiv \tilde{\chi}_1^0$;
                                        GUT rel.    
		                      & LEP$_{II}$~\cite{LEPMSSM} \\
                   & $\Rp$-SUSY & 35-40 & \LLE, \UDD; $\forall$ MSSM    
		                      & LEP$_{II}$~\cite{LEPRPV02} \\
                   &            &  30 & \LQD; $\forall$ MSSM   
		                      & LEP$_{II}$~\cite{LEPRPV02} \\
                   &  GMSB      &  77 & LSP/NLSP $\equiv \tilde{G} / \tilde{\chi}_1^0$;
                                $\beta(\tilde{\chi}_1^0 \rightarrow \gamma \tilde{G}) = 100 \%$
		                      & Tevatron$_{I}$~\cite{D0gx98} \\
                   &            &  91 & $\tilde{G}$ is LSP, \stau$_1$ is NLSP;
		                        $M_{\tilde{G}} < 1 \eV$           
		                      & LEP$_{II}$~\cite{DELPHIgs} \\
% charginos:
  \Xpm{i} &  MSSM  &  72 & $i=1$; $\forall (\tan \beta , m_0)$;
                                        LSP $\equiv \tilde{\chi}_1^0$;
                                        GUT rel.    
                                      & LEP$_{II}$~\cite{LEPMSSM} \\
                   & $\Rp$-SUSY & 103 & \LLE, \UDD; $\forall$ MSSM
		                      & LEP$_{II}$~\cite{LEPRPV02} \\
                   &            & 100 & \LQD; $\forall$ MSSM 
		                      & LEP$_{II}$~\cite{LEPRPV02} \\
                   &  GMSB      & 150 & LSP/NLSP $\equiv \tilde{G} / \tilde{\chi}_1^0$; 
		                $\beta(\tilde{\chi}_1^0 \rightarrow \gamma \tilde{G}) = 100 \%$
		                      & Tevatron$_{I}$~\cite{D0gx98} \\
                   &            &  95 & LSP/NLSP $\equiv \tilde{G} / \tilde{\tau}_1$;
		                        $\forall M_{\tilde{G}}$
		                      & LEP$_{II}$~\cite{DELPHIgs} \\
   \hline
   \multicolumn{5}{||l||}{{\it Gauginos (Strong sector)}}  \\
%                              ------------------------
   \hline
% gluinos:
  $\tilde{g}$      &  mSUGRA    & 190 &  jets + \Emiss$_T$ final states         
                                      &  Tevatron$_I$~\cite{PDG2000} \\
                   &            & 180 &  dilepton final states                  
		                      &  Tevatron$_I$~\cite{PDG2000} \\
   \hline
   \multicolumn{5}{||l||}{{\it Sfermions}}  \\
%                              ----------
   \hline
% sleptons:
 \sel, \smu, \stau &   MSSM     &  92 / 85 / 68 & $\tilde{l}_R$; $\Delta M_{\tilde{l}\tilde{\chi}}>10\GeV$
                                        $\tilde{l} \rightarrow l \tilde{\chi}_1^0$; $\forall$ mixing 
                                      &  LEP$_{II}$~\cite{LEPMSSM}          \\
      \sel \hspace*{0.9cm}    &$\Rp$-SUSY  &  69-96 & \LLE, \LQD, \UDD; $\tilde{l}_R$ pair prod.; 
					 $\forall$ MSSM
                                      &  LEP$_{II}$~\cite{LEPRPV02}          \\
 \hspace*{0.3cm} \smu, \stau  &            &  61-87 & \LLE, \LQD, \UDD; $\tilde{l}_R$ pair prod.; 
					 $\forall$ MSSM
                                      &  LEP$_{II}$~\cite{LEPRPV02}          \\
                   &   GMSB     &  77 & \stau$_1$; $\tilde{G}$ is LSP, \stau$_1$ is NLSP; 
		                        $\forall M_{\tilde{G}}$         
		                      & LEP$_{II}$~\cite{ALEPHgs} \\
      \snu         &  MSSM      &  43 &                                         
                                      & LEP$_I$~\cite{PDG2000}              \\
                   &$\Rp$-SUSY & 84-99 & \LLE, \LQD, \UDD; $\tilde{\nu}_e$ pair prod.; 
					$\forall$ MSSM
                                      &  LEP$_{II}$~\cite{LEPRPV02}          \\
                   &           & 64-83 & \LLE, \LQD, \UDD; $\tilde{\nu}_{\mu,\tau}$ pair prod.; 
					$\forall$ MSSM
                                      &  LEP$_{II}$~\cite{LEPRPV02}          \\
  \hline 
% squarks:
     \sqrk         &            & 260 &  jets + \Emiss$_T$ final states         
                                      &  Tevatron$_I$~\cite{PDG2000} \\ 
                   &            & 230 &  dilepton final states                  
		                      &  Tevatron$_I$~\cite{PDG2000} \\ 
     \sqrkt{1}     &          & 90-91 &  $\tilde{t}_1 \rightarrow c \tilde{\chi}_1^0, b l \tilde{\nu}$;
                                         $\Delta M_{\tilde{t}_1\tilde{\chi}}>7\GeV$; 
					 $\forall \theta_{mix}$                                      
                                      &  LEP$_{II}$~\cite{LEPMSSM,LEPSTOP02} \\    
                   &            &  88 &  $\tilde{t}_1 \rightarrow c \tilde{\chi}_1^0$;
                                         $M_{\tilde{\chi}_1^0} < 1/2 M_{\tilde{t}}$; 
					 $\forall \theta_{mix}$                                              
                                      &  Tevatron$_{II}$~\cite{PDG2000} \\    
                   &            & 138 &  $\tilde{t}_1 \rightarrow b l \tilde{\nu}$;
                                         $M_{\tilde{\nu}} < 1/2 M_{\tilde{t}}$; 
					 $\forall \theta_{mix}$                                      
                                      &  Tevatron$_{II}$~\cite{TEVSTOP02} \\
 
     \hline 
  \hline \hline
  \end{tabular}
  \caption {{\it Lower limits at 95\% CL on sparticle masses established at
                 LEP and Tevatron colliders.}} 
  \label{tab:susycons}
\end{center}
\end{table*}
%
%...................................................................................
On one hand, the absence of deviations from Standard Model expectations 
in various sparticle search channels has been used in a global manner to 
establish constraints in the parameter space of more restricted theories 
discussed above, such as the constrained (i.e. complemented by GUT relations) 
MSSM, mSUGRA, GMSB theories with assumptions on the nature of the LSP and 
next-to-LSP sparticles, $\Rp$ versions of the constrained MSSM assuming a 
single dominant new Yukawa coupling, etc. Coherent and {\it comprehensive}
review articles on the restrictions imposed on specific theories
can be found in Ref.~\cite{PDG2000}.
On the other hand, searches for specific sparticles have been used
to derive ``most conservative'' lower mass limits with the intention of
remaining independent from a specific choice of model parameters
as far as possible. In a way these constitute ``utterly unavoidable 
constraints''. A review of the most general existing constraints 
on sparticle masses is provided in Table~\ref{tab:susycons}.

\begin{figure}[htb]
  \begin{center}                                                                
  \vspace*{-0.2cm}
    
  \hspace*{-0.5cm} \epsfig{file=\master/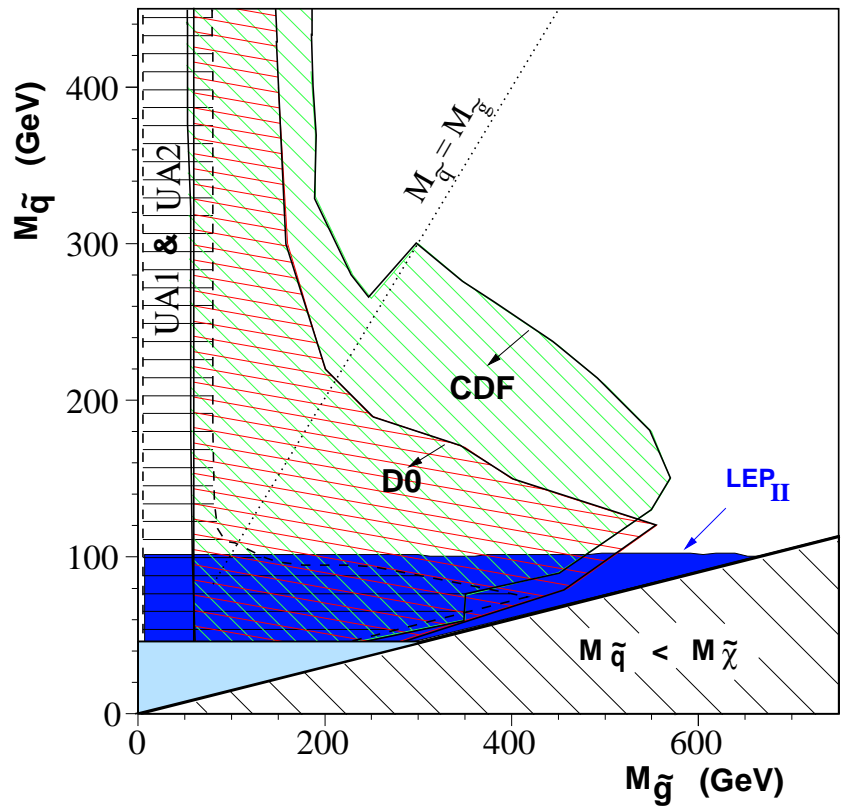,width=0.60\textwidth}
 
  \end{center}

 \caption[]{ \label{fig:sqgl}
            Constraints in the squark mass $M_{\tilde{q}}$ {\it vs.} gluino mass
	    $M_{\tilde{g}}$ plane derived from a (preliminary) 
	    combination~\cite{LEPSTOP02} of ALEPH, DELPHI, L3 and OPAL 
	    results from LEP$_{II}$ collider shown together with results
	    obtained~\cite{TEVQG02} by the D$\emptyset$ and CDF experiments 
	    at Tevatron$_{I}$.} 
\end{figure}

\begin{figure}[htb]
  \begin{center}                                                                
  \begin{tabular}{cc}
    
  \hspace*{-0.2cm} \epsfig{file=\master/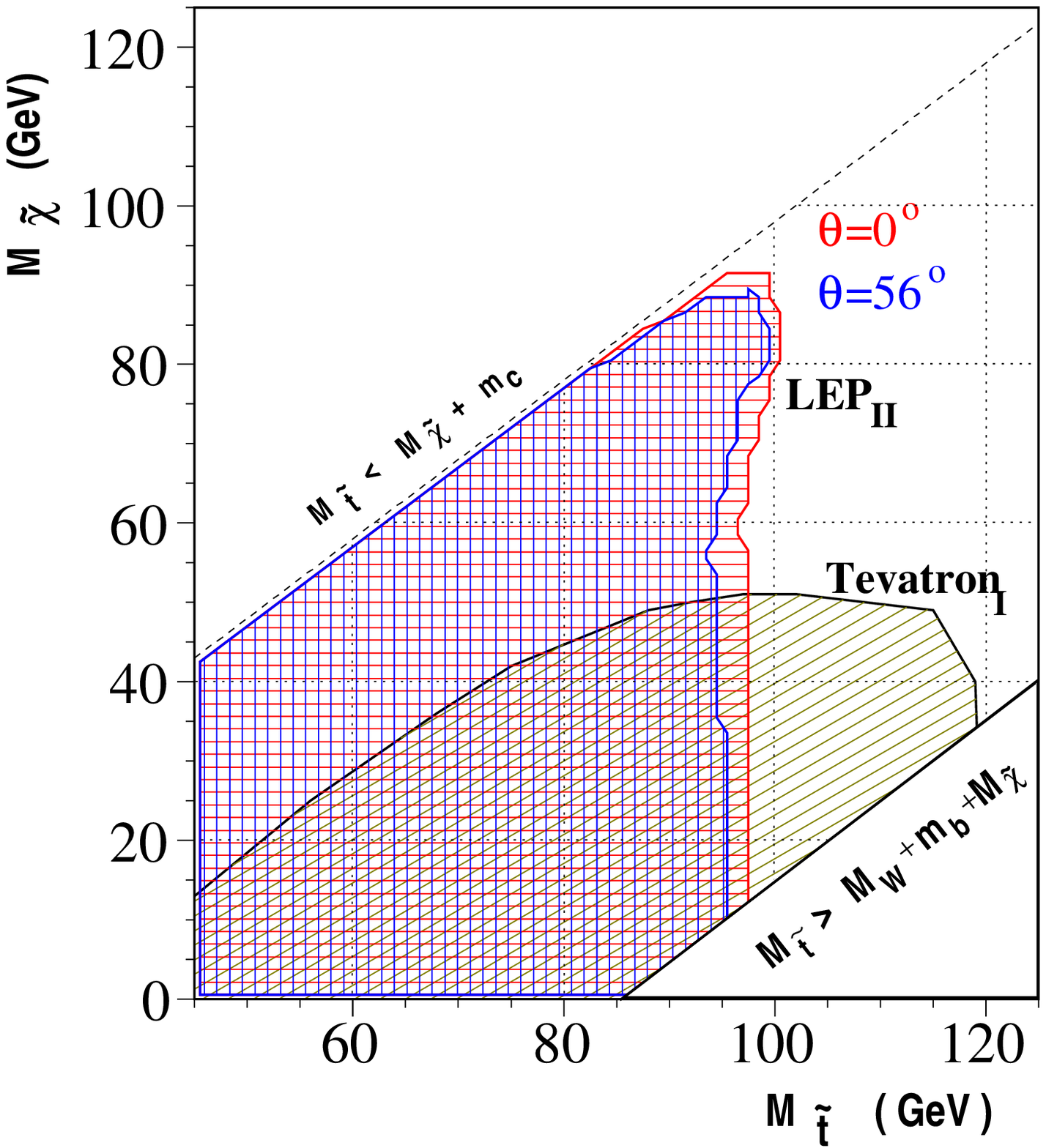,width=0.50\textwidth}
&
  \hspace*{-0.5cm} \epsfig{file=\master/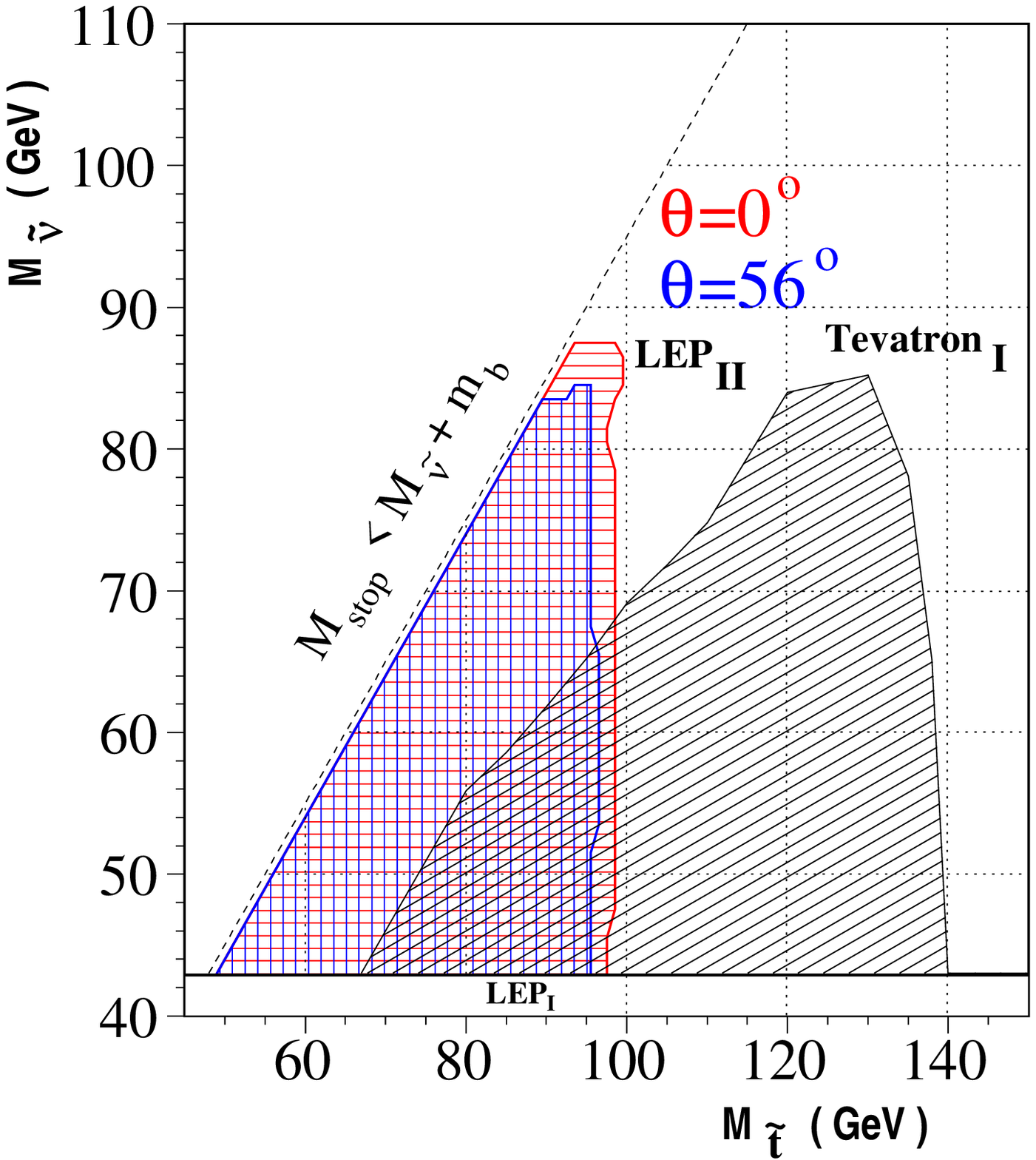,width=0.45\textwidth}

  \end{tabular}
  \end{center}
 \vspace*{-1.0cm}
 
 \hspace*{5.0cm} (a) \hspace*{8.0cm} (b) \\
 
 \vspace*{-0.5cm}

 \caption[]{ \label{fig:stoplim}
            Mass constraints (95\% CL) on the stop $\tilde{t}_1$ eigenstate as a function 
	    of the mass of the lightest supersymmetric particle taken either as
	    a) the neutralino $\tilde{\chi}_1^0$ or b) the sneutrino $\tilde{\nu}_{\tau}$.
	    The excluded domains derived from a (preliminary) combination~\cite{LEPSTOP02} 
	    of ALEPH, DELPHI, L3 and OPAL results from the LEP$_{II}$ collider are shown together
	    with results obtained~\cite{TEVSTOP02} by the D$\emptyset$ and CDF experiments 
	    at Tevatron$_{I}$.} 
\end{figure}

The most general and best constraints are obtained at the LEP$_{II}$ collider,
where remarkably complete analyses of sparticle-pair production have been 
performed including wide parameter scans carried out in the framework of the 
MSSM and mSUGRA models with or without $\Rp$ couplings and for GMSB
models. These are complemented by LEP$_{I}$ results most noticeably 
in the case of GMSB models when the next-to-lightest sparticle (NLSP)
is the \Xo{1}, and also in the case of coloured sparticles with MSSM or
mSUGRA models.
The existing constraints on squark masses $M_{\tilde{q}}$ are summarized in
Fig.~\ref{fig:sqgl} assuming mass degeneracy for five squark flavours. 
The stringent constraints on $M_{\tilde{q}}$ obtained at the Tevatron 
depend on assumptions on the gluino mass $M_{\tilde{g}}$.
%-----------------------------------------------------------------------------------
%------------------------------------------------------------------------------------
Existing constraints on the lightest stop mass eigenstate $\tilde{t}_1$
are summarized in Fig.~\ref{fig:stoplim}.
%-----------------------------------------------------------------------------------
%-----------------------------------------------------------------------------------

For completeness, it should be mentioned that searches for slepton-squark 
pair production through $t$-channel exchange of a \Xo{1}\ have been performed
by the H1 and ZEUS experiments at HERA~\cite{HERAMSSM}. These were sensitive
only up to a sum of masses $M_{\tilde{e}} + M_{\tilde{q}}$ below $150 \GeV$ 
which is now excluded by LEP$_{II}$-alone ``universal'' constraints.
 
% Specific Rp-violating searches:

\begin{figure}[htb]
  \begin{center}                                                                
  \begin{tabular}{cc}
    
  \hspace*{-0.2cm} \epsfig{file=\master/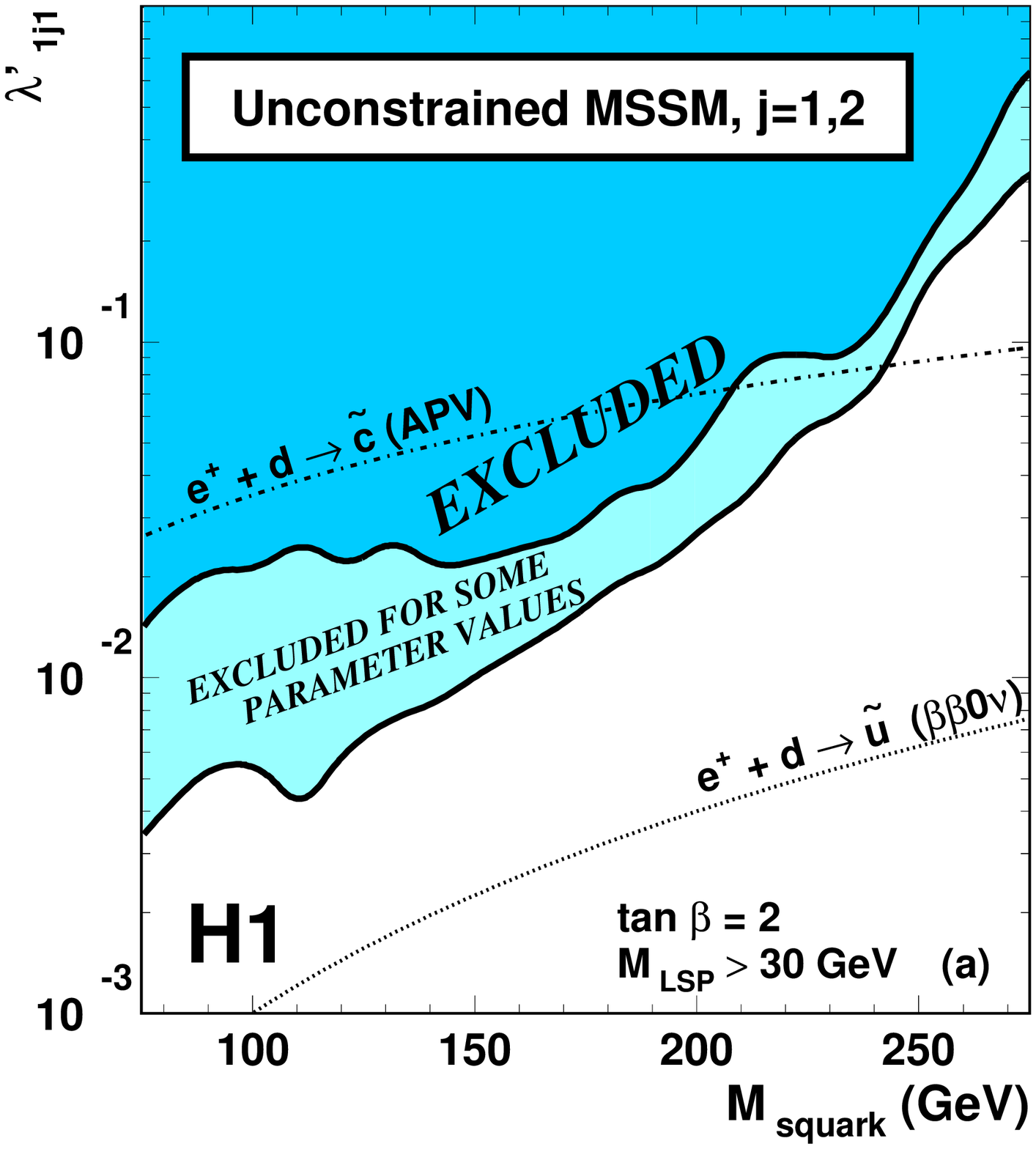,width=0.45\textwidth}
&
  \hspace*{-0.5cm} \epsfig{file=\master/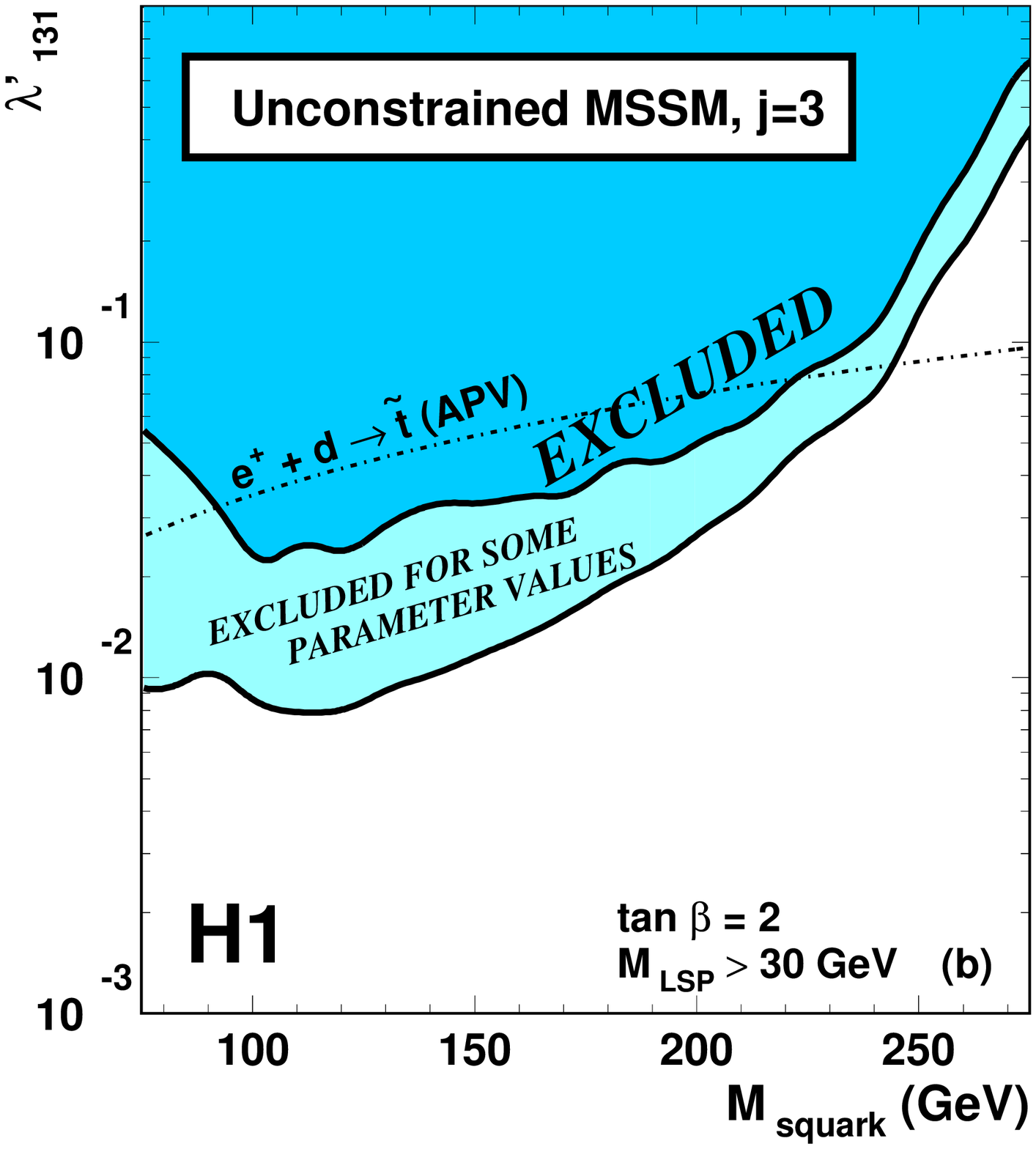,width=0.45\textwidth}

  \end{tabular}
  \end{center}
 \vspace*{-1.0cm}
 
 \hspace*{5.0cm} (a) \hspace*{7.5cm} (b) \\
 
 \vspace*{-0.5cm}

 \caption[]{ \label{fig:lpvsm}
            Upper Limits (95\% CL) on a) the coupling $\lambda'_{1j1}$ with
	    $j=1,2$ and b) $\lambda'_{131}$ as a function of the squark mass
	    for $\tan{\beta} = 2$ in the unconstrained MSSM.
	    The limits are obtained from a scan of the $\mu$ and $M_2$
	    parameters within $-300 < \mu < 300 \GeV$ and 
	    $70 < M_2 < 350 \GeV$ and imposing that the lightest sparticle
	    (LSP) has a mass $M_{LSP}$ above $30 \GeV$.
	    The dark shaded area is excluded for any parameter values.
	    The light shaded area is excluded for some parameters values.
	    The dotted curve is the indirect upper bound~\cite{Dreiner97} on 
	    $\lambda'_{111}$ derived from constraints on neutrinoless 
	    double-beta decays~\cite{Hirsch95,Balysh95}.
	    The dash-dotted curves are the indirect upper 
	    bounds~\cite{Dreiner97} on $\lambda'_{1j1}$ derived from 
	    constraints on atomic-parity violation~\cite{Wood97}.}
\end{figure}

Complementary $\Rp$ SUSY searches have been performed at HERA, LEP and 
Tevatron colliders under the hypothesis of a single dominant 
$\lambda'_{1jk}$ coupling.
The constraints obtained~\cite{H1RPV} by the H1 experiment at HERA from a search
for resonant squark production via $\lambda'_{1jk}$ are shown in
Fig.~\ref{fig:lpvsm}. Similar results were obtained~\cite{ZEUSRPV} 
by the ZEUS experiment.
Also shown in Fig.~\ref{fig:lpvsm} are the best existing indirect 
bounds~\cite{Dreiner97} from low-energy experiments. 
The $\lambda'_{111}$ coupling is seen to be
very severely constrained by the non-observation of neutrinoless double-beta
decay. The most stringent low-energy constraints on $\lambda'_{121}$ and 
$\lambda'_{131}$ come from atomic-parity violation measurements.

\begin{figure}[htb]
  \begin{center}                                                                

  \hspace*{-0.5cm} \epsfig{file=\master/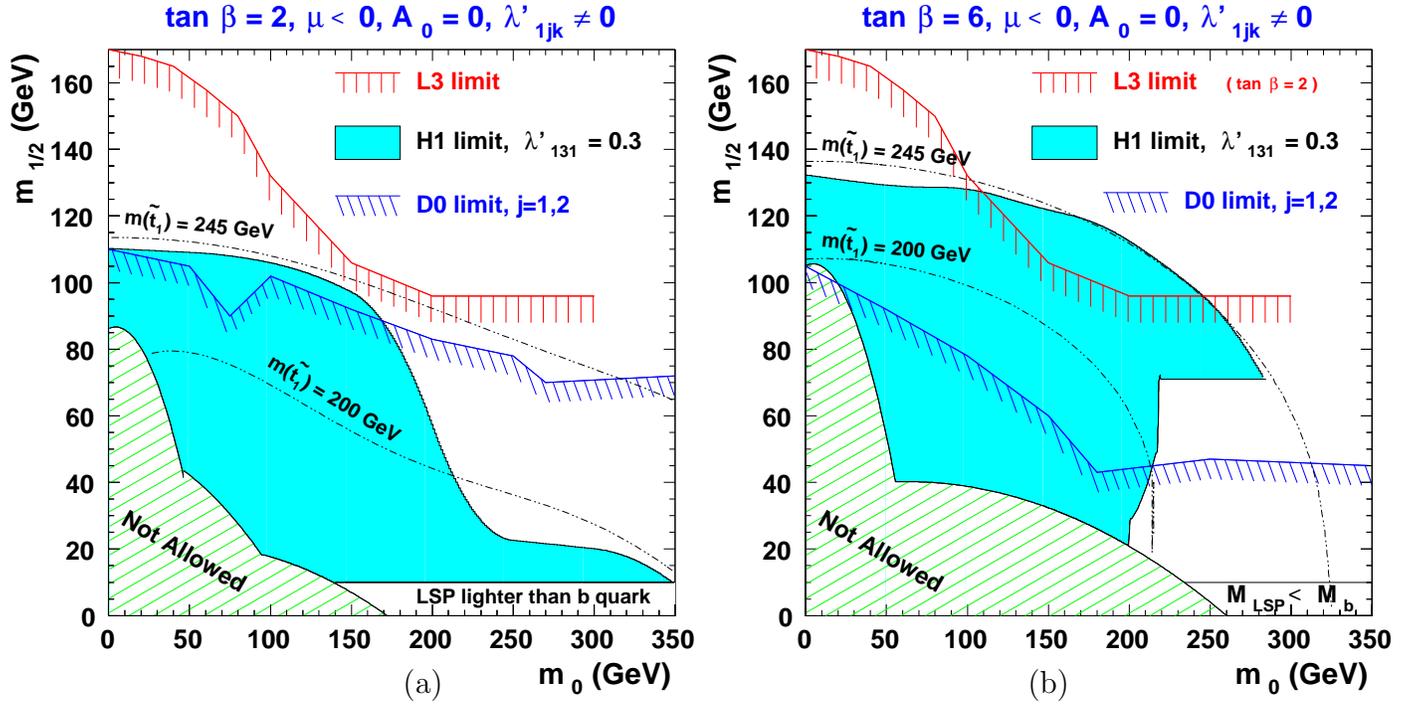,width=1.00\textwidth}

  \end{center}
 \vspace*{-1.0cm}
 
 \hspace*{5.0cm} (a) \hspace*{7.5cm} (b) \\
 
  \vspace*{-0.2cm}
  
  \caption[]{ \label{fig:stophera}
              Constraints on stop production via $\lambda '_{131}$ in
$R$-parity-violating SUSY in the parameter space of Minimal Supergravity.
H1 and D$\emptyset$ limits are shown for a) $\tan{\beta} = 2$ and
b) $\tan{\beta} = 6$.  L3 limit for $\tan{\beta} = 2$ is also shown.} 
\end{figure}

The HERA results analysed in the framework of $\Rp$ mSUGRA are shown in 
Fig.~\ref{fig:stophera}.
%-----------------------------------------------------------------------------------
%-----------------------------------------------------------------------------------
The searches were made under the hypothesis of a single dominant 
$\lambda'_{1jk}$ coupling and the results are presented as excluded domains 
in the parameter space of the model.
 
The constraints from the D$\emptyset$~\cite{D0RPV} experiment at the Tevatron
were obtained from a search for $\tilde{q}$ pair production through gauge 
couplings. 
The analysis profits from an approximate mass degeneracy implicitly 
extended to five $\tilde{q}$ flavours 
($\tilde{d}$,$\tilde{u}$,$\tilde{s}$,$\tilde{c}$,$\tilde{b}$) and both 
(partners) chiralities ($\tilde{q}_L$,$\tilde{q}_R$).
The $\Rp$ couplings are assumed to be significantly smaller than the gauge 
couplings, so that direct $\Rp$ decays are suppressed and each squark rather
decays back into a quark and the LSP through gauge couplings.
The only effect of the $\Rp$ couplings is to make the LSP unstable. The analysis
is further restricted to $\Rp$ coupling values $\gsim 10^{-3}$ to guarantee 
a negligible decay length of the LSP.
In the domains considered, the LSP is almost always the lightest 
neutralino $\Xo{1}$. The $\Xo{1}$ decays via $\lambda'_{1jk}$ into a
first-generation lepton ($e$ or $\nu_e$) and two quarks. The analysis is restricted
to $j=1,2$ and $k=1,2,3$ and, in practice, the D$\emptyset$ selection of 
event candidates requires like-sign di-electrons accompanied by multiple jets.

The constraints from the $L3$~\cite{L3RPV} experiment at LEP were obtained from a search for 
pair production through gauge couplings of neutralinos 
($e^+e^- \rightarrow \tilde{\chi}_m^0 
\tilde{\chi}_n^0$ with $m=1,2$ and $n=1, \ldots ,4$), charginos ($e^+e^- 
\rightarrow \tilde{\chi}_1^+ \tilde{\chi}_1^-$) and scalar leptons ($e^+e^- 
\rightarrow \tilde{l}_R^+ \tilde{l}_R^-$, $\tilde{\nu} \tilde{\nu}$).
The $\Rp$ couplings contribute here again in opening new decay modes for 
the sparticles. A negligible decay length of the sparticles through these
decay modes is ensured by restricting the analysis to coupling values 
$\gsim 10^{-5}$.
All possible event topologies (multijets and lepton and/or missing energy)
resulting from the direct or indirect sparticle decays involving
the $\lambda'_{ijk}$ couplings have been considered in the $L3$ analysis. 
%----------------------------------------------------------------------------
%--------------------------------------------------------------------

The hatched region marked ``not allowed'' in Fig.~\ref{fig:stophera} 
corresponds to points where the radiative electroweak symmetry breaking 
does not occur; or where the LSP is the sneutrino; or that lead to 
tachyonic Higgs or sfermion masses.

For the set of mSUGRA parameters with $\tan \beta = 2$, the Tevatron
experiment excludes squarks with masses $M_{\tilde{q}} < 243 \GeV$ 
(95 \% CL) for any value of $M_{\tilde{g}}$ and a finite value 
($\gsim 10^{-3}$) of $\lambda'_{1jk}$ with $j=1,2$ and $k=1,2,3$. 
The sensitivity decreases for the parameter set with a larger value of 
$\tan \beta$ due in part to a decrease of the photino component of the 
LSP, which implies a decrease of the branching fraction of the LSP into 
electrons, and in part to a softening of the final-state particles for 
lighter charginos and neutralinos.
The best sensitivity at $\tan \beta = 2$ is offered by LEP for any of 
the $\lambda'_{ijk}$ couplings.
HERA offers a best complementary sensitivity to the coupling 
$\lambda'_{131}$ which allows for resonant stop production via 
positron-quark fusion  $e^+ d \rightarrow \tilde{t}_1$.
The HERA constraints (shown here for a coupling
of electromagnetic strength, i.e. $\lambda'_{131} = 0.3$) 
extend beyond LEP and Tevatron constraints towards larger $\tan \beta$.
 % Search for Supersymmetry 

% Forbidden Lepton and Quark Flavour Changing Processes:
\clearpage
%%%%%%%%%%%%%%%%%%%%%%%%%%%%%%%%%%%%%%%%%%%%%%%%%%%%%%%%%%%%%%%%%%%%%%%%%%%%%%
\section{Forbidden Lepton and Quark Flavour-Changing Processes}
\label{sec:flavours}
%%%%%%%%%%%%%%%%%%%%%%%%%%%%%%%%%%%%%%%%%%%%%%%%%%%%%%%%%%%%%%%%%%%%%%%%%%%%%%
\label{sec:FLAVOURS}

%=============================================================================
\subsection{Lepton-Flavour Violation}
\label{sec:LFV}
%=============

In the Standard Model, lepton flavours are separately conserved in every 
reaction. 
However, the model does not provide a fundamental motivation for this
exact additive conservation of electron, muon and tau numbers.
It is regarded in the model as resulting from an accidental symmetry.

The observation of lepton-flavour violation (LFV) could provide essential 
guidance beyond the realm of the Standard Model.
This has motivated extensive LFV searches in the charged-lepton sector
for the last 30 years in, for instance, largely dedicated experiments
using $\mu$ nuclear capture and rare or forbidden decays of $\mu$, 
$\tau$, or $K$, $B$ and $D$ mesons.
In the neutrino sector, the strong implication of neutrino mixing to
explain recent observations~\cite{Fukuda98} suggests that lepton 
flavour is violated there. 
However, a minimal extension of the Standard Model that incorporates 
neutrino masses and mixings predicts a rate of LFV in the charged-lepton 
sector far too small to be detected in current and planned experiments.
This is due to the smallness of the neutrino masses. 
On the other hand, many proposed extensions of the Standard Model,
for example in GUT theories~\cite{GUT-GG, PatiSalam}, 
entail LFV at more fundamental levels and thus predict LFV rates
that could be detected in collider experiments and in low-energy
processes.
%----------------------------------------------------------------------------
\begin{figure}[htb]
  \begin{center}                                                                
  \begin{tabular}{cc}
  \vspace*{-0.2cm}
    
  \hspace*{-1.0cm} \epsfig{file=\master/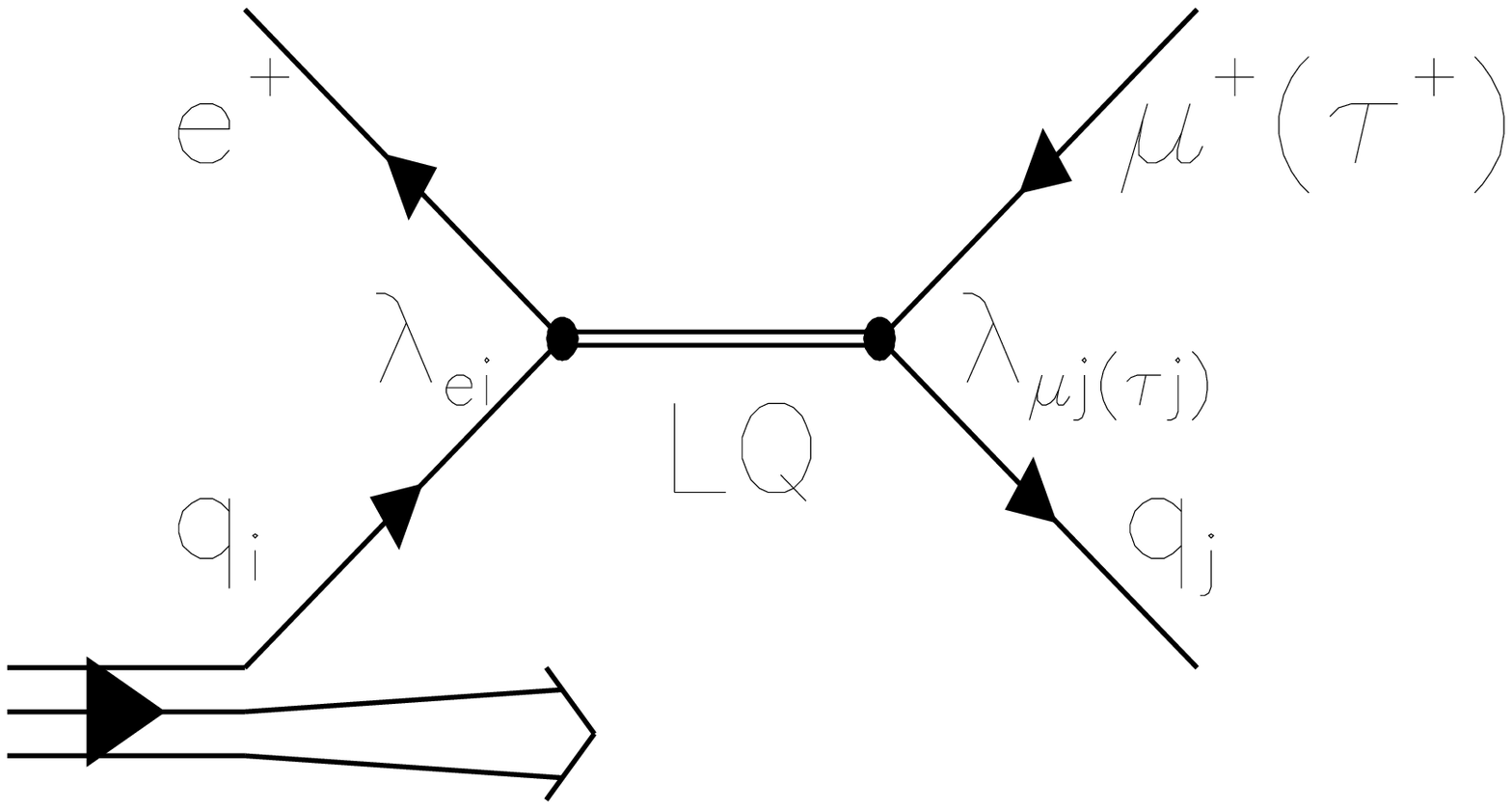,width=0.42\textwidth}
&
  \hspace*{-1.5cm} \epsfig{file=\master/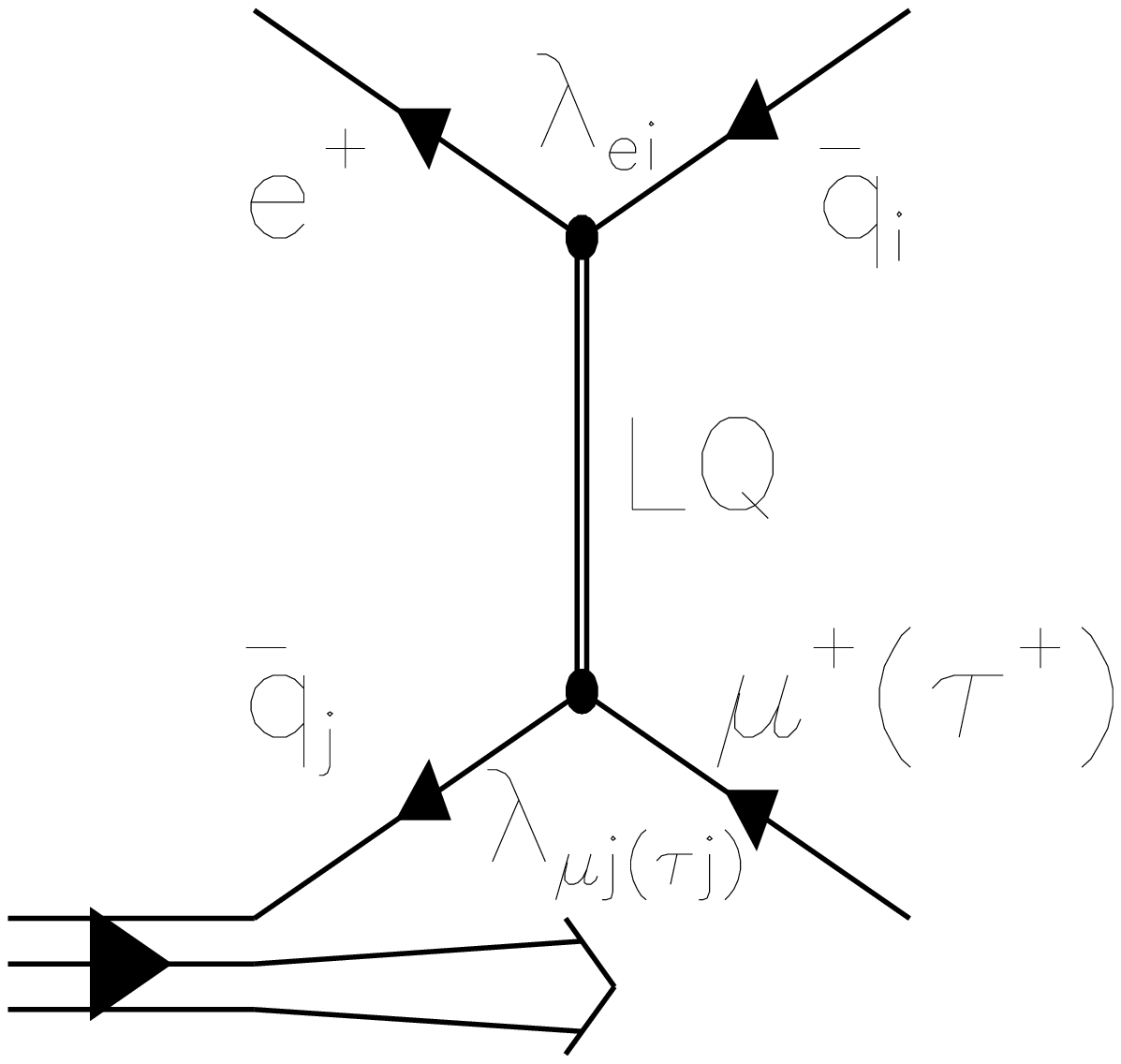,width=0.42\textwidth}

  \end{tabular}
  \end{center}
 \vspace*{-2.5cm}
 
 \hspace*{5.0cm} (a) \hspace*{6.5cm} (b) \\
 
 \vspace*{-0.5cm}
  
 \caption[]{ \label{fig:diaglfv}
            $ep \to \mu X (\tau X)$ process mediated by a LFV LQ.
            a) $s$-channel production;
            b) $u$-channel exchange.}
\end{figure}
%-----------------------------------------------------------------------------

Figure~\ref{fig:diaglfv} shows an example of an LFV process, 
$ep \to \mu(\tau)X$, which could be observed in $ep$ collisions.
The process is mediated by LQ exchange with couplings involving specific 
combinations of lepton and quark generations.
In Fig.~\ref{fig:diaglfv}, the LQ possesses two non-vanishing Yukawa couplings,
a $\lambda_{ei}$ to an electron and a quark of generation $i$, and a 
$\lambda_{\mu j (\tau j)}$ to a muon (tau) and a quark of generation $j$.  

Both the H1~\cite{H19497lq} and ZEUS~\cite{ZEUS9497lfv} experiments have
searched, in $e^+p$ data taken at $\sqrt{s}=300\GeV$, for events with a 
high-transverse-momentum muon or tau balancing a hadronic jet. 
ZEUS also reported preliminary search results~\cite{ZEUS9899lfv} from
$e^\pm p$ data taken at $\sqrt{s}=318\GeV$.
No outstanding LQ candidates were found and the null results were 
interpreted in terms of exclusion limits for two different LQ mass regions
described below.
The exclusion limits at HERA are derived in a framework that differs from 
the minimal BRW model introduced in section~\ref{sec:lqmodels} only by a
relaxing of the diagonality requirement to allow for combinations of 
lepton and quark generations.
In a similar framework, an exhaustive review of the contributions of 
leptoquarks to rare or forbidden lepton and mesons decays and 
to various precision electroweak tests has been performed by Davidson, Bailey 
and Campbell~\cite{Davidson94}. Further discussions concerning contributions
to $\mu \rightarrow e \gamma$, $\mu \rightarrow 3e$ and $\mu - e$ conversion in
nuclei can be found in~\cite{Gabrielli99}. 

\begin{figure}[htb]
  \begin{center}                                                                
  \begin{tabular}{c}
  \vspace*{-0.2cm}
    
  \hspace*{-1.0cm} \epsfig{file=\master/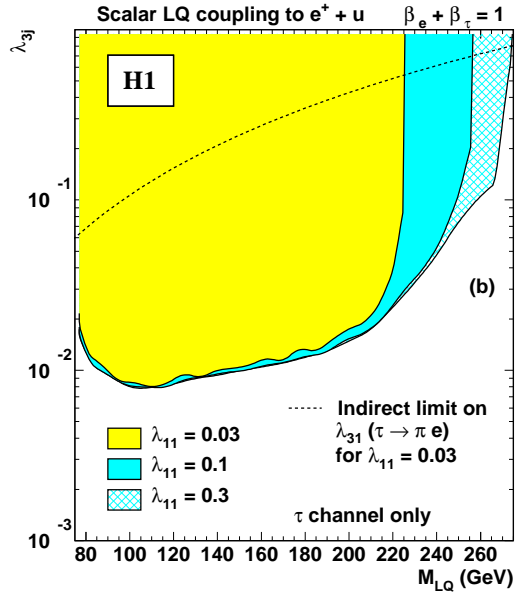,width=0.42\textwidth}

  \end{tabular}
  \end{center}
 \vspace*{-0.5cm}
  
 \caption[]{ \label{fig:h1lfv}
          Limits on the leptoquark coupling $\lambda_{3j}$ (to a $\tau$ and
          a quark of generation $j$),
          for several fixed values of coupling $\lambda_{11}$ (to an electron and a 
          1st-generation quark), as a function of the leptoquark mass.}
\end{figure}

At HERA, the $s$-channel resonant production of leptoquarks dominates
for a low-mass assumption, $M_{LQ} < \sqrt{s}$.
For this case, upper limits on $\lambda_{\tau j}$ coupling ($j=1,2$)
are shown in Fig.~\ref{fig:h1lfv} as a function of a scalar LQ mass 
for several fixed values of $\lambda_{e 1}$.
The limits cover masses up to $270 \GeV$ and explore a mass-coupling
range beyond indirect constraints from rare $\tau$ decays. 
The best indirect constraint for $\lambda_{31}$ comes from the upper 
limit on the branching ratio $\beta_{\tau \rightarrow \pi^0 e}$ which 
could be affected through  $\tau \rightarrow d + LQ^*; LQ^* \rightarrow e + d$.
No low-energy process constrains the coupling $\lambda_{32}$. 
More stringent indirect constraints exist for leptoquarks coupling 
to $e^+ d$ pairs (such as the $\tilde{S}_{1/2,L}$ in BRW model) for 
which the couplings $\lambda_{31}$, $\lambda_{32}$ or $\lambda_{33}$ are 
constrained respectively by $\tau \rightarrow \pi^0 e$, 
$\tau \rightarrow K^0 e$ and $B \rightarrow \tau e X$.
%----------------------------------------------------------------------------
%-----------------------------------------------------------------------------

For a high-mass assumption, $M_{LQ} \gg \sqrt{s}$, the contributions
to $ep \to \mu(\tau)X$ from virtual LQ exchanges in the $s$- and 
$u$-channels have a cross-section proportional to 
($\lambda_{ei}\lambda_{\mu j(\tau j)}/M_{LQ}^2)^2$. 
Note that an initial-state quark that couples to the electron participates
in the $s$-channel for $F=0$ leptoquarks in $e^+p$ collision, while an
initial-state antiquark that couples to the muon or tau must be involved
in a $u$-channel process, as shown in Fig.~\ref{fig:diaglfv}.
For F=2, the quark and antiquark exchange their roles.
Limits derived for high LQ masses~\cite{ZEUS9497lfv} are summarized in 
Table 9 for an $e\rightarrow\tau$ transition involving any quark-generation  
combinations and for all $F=0$ scalar and vector LQ species.
Limits were also obtained~\cite{H19497lq, ZEUS9497lfv} at HERA for 
the $e \rightarrow \mu$ transition.
As can be seen in Table 9, the low-energy limits are quite stringent if
first-generation quarks only are involved.
However, HERA offers a higher sensitivity for many coupling products
involving heavy quarks. 
For these cases, improved sensitivities from rare $B$- and $\tau$-decays 
are expected from $B$-factories in the coming years.

A search for LFV processes in $B^0$-decays has been performed by the
CDF experiment~\cite{CDFBtomue} at the Tevatron. 
The $B^0_d \rightarrow \mu^{\pm} e^{\mp}$ and $B^0_s \rightarrow \mu^{\pm} e^{\mp}$
decays each probe the existence of two types of leptoquarks.
For instance, a leptoquark contributing to $B^0_s \rightarrow \mu^{\pm} e^{\mp}$ 
{\it either} couples to $e-s$ and $\mu-b$ pairs {\it or} couples to  $e-b$ and 
$\mu-s$ pairs.
CDF first establishes the best upper limits on the branching ratios for the
decays $B^0_d \rightarrow \mu^{\pm} e^{\mp}$ 
($\beta < 4.5 \times 10^{-6}$ at 95\% CL)
and $B^0_s \rightarrow \mu^{\pm} e^{\mp}$ 
($\beta < 8.2 \times 10^{-6}$ at 95\% CL).
These results are then interpreted as a lower limit on the mass $M_{V}^{PS}$ 
of Pati-Salam~\cite{PatiSalam} bosons which are vector LFV leptoquarks 
with non-chiral couplings to quarks and leptons. 
In such a theory for lepton-quark unification, the leptoquarks couple
to fermion pairs with a strong coupling $\alpha_s(M_{LQ})$.
The CDF limits on the $B^0$ decays correspond to a lower limit~\cite{CDFBtomue} 
of $M_{V}^{PS} > 20 \TeV$. 
A review of the contributions of Pati-Salam leptoquarks to rare or forbidden 
$K$, $\pi$ and $B$ decays can be found in~\cite{Valencia94}. 
%-----------------------------------------------------------------------------
\begin{figure}[tb]
      Table 9: {\it Limits (95\% CL upper limit)
      on $\lambda_{eq_\alpha}\lambda_{\tau q_\beta}/M_{LQ}^2$
      $(TeV^{-2})$ for $F$=0 LFV leptoquarks mediating the 
      $eq_\alpha \leftrightarrow \tau q_\beta$ transition (bold numbers 
      in the bottom of each  cell).  Each row corresponds to a 
      $(q_\alpha, q_\beta)$ generation combination and each column 
      corresponds to a leptoquark species.
      The numbers in the middle of each cell are the best limit from 
      low-energy experiments.  The cases where the ZEUS limit is
      more stringent are enclosed in a box. The * shows the cases where 
      only the top quark can participate.
      Similar tables exist for $F$=2 LQs, and for the $e$-$\mu$
      transition. In addition, H1 has similar results.}
\begin{center}
\vskip 0.1 in
\epsfig{file=\master/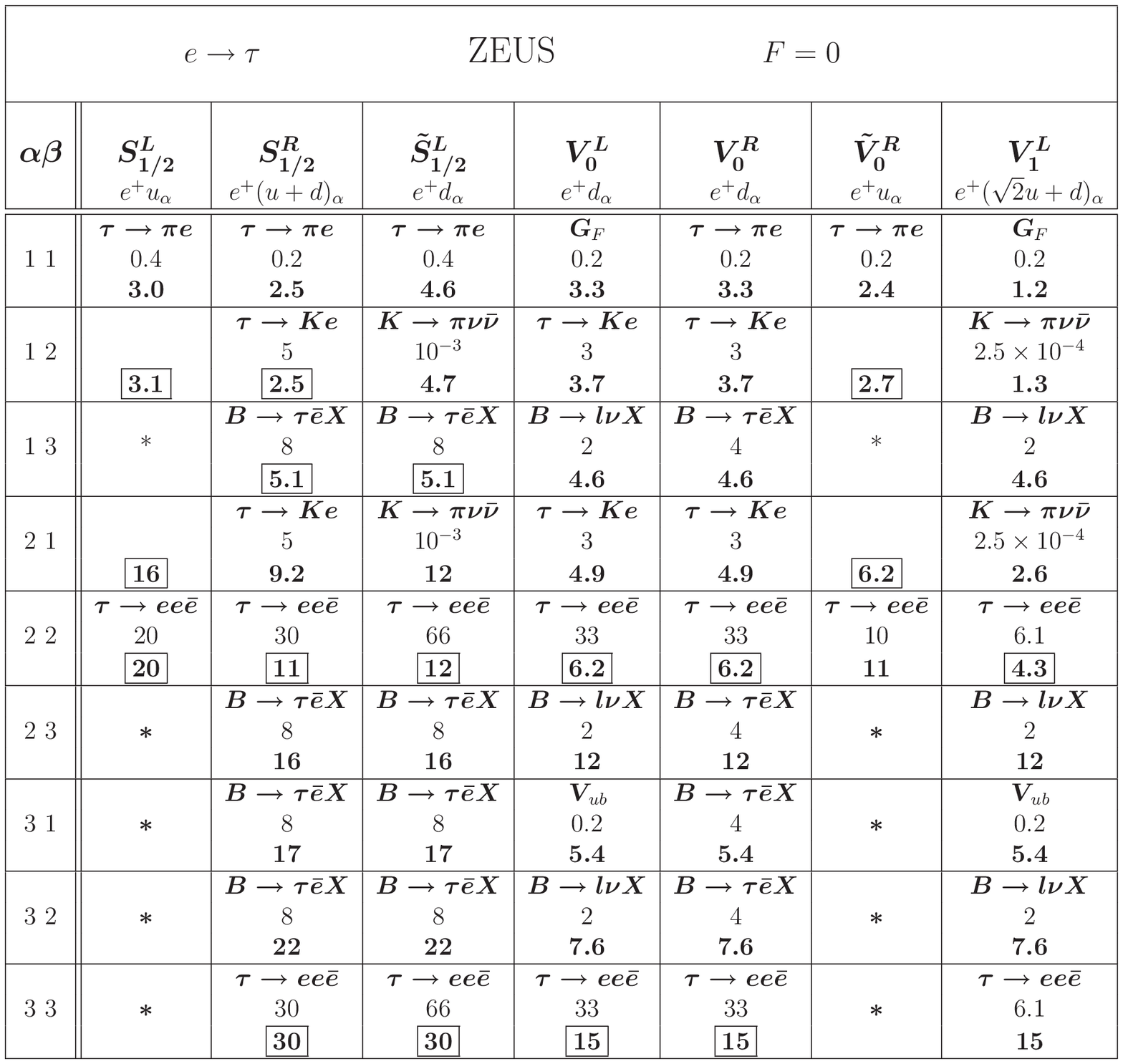,width=14cm}
\end{center}
\end{figure}
\setcounter{table}{9}
%-----------------------------------------------------------------------------

%=============================================================================
\subsection{Flavour-Changing Neutral Currents}
\label{sec:FCNC}
%%=============

In the Standard Model, inter-generation transitions between quarks
can happen only via charged currents, i.e. with the $W$ boson and
off-diagonal elements of the Cabibbo-Kobayashi-Maskawa matrix~\cite{CKM}.
In contrast, neutral currents are flavour diagional.
Flavour-changing neutral currents (FCNC), i.e. the transition between 
quarks of the same charge but of different generations, are not contained 
at tree level and can happen only from higher-order loop contributions.  
These contributions vanish in the limit of degenerate quark masses 
(GIM suppression~\cite{Glashow70}), and therefore, a sizeable (but still 
very small) rate can arise only when the top quark appears in the loop.  
This is the case for instance in the FCNC process $b \rightarrow s \gamma$
which was first observed by the CLEO experiment~\cite{CLEO93} and has been
used to set stringent constraints on physics beyond the 
Standard Model~\cite{AfterLEPb,bsgSUSY,bsgXTRA,bsgTECHNI,bsgHIGGS}.
In constrast, the GIM suppression is very strong for FCNC decays connecting
charge $+2/3$ quarks because of the relative smallness of the mass of the
charge $-1/3$ quarks involved in the higher-order loops. 
Therefore, no detectable rate is predicted in the Standard Model for 
FCNC processes between the top and charm or up quarks; for example, 
the decay branching ratios $\beta( t \rightarrow c \gamma, c Z^0 )$ are 
predicted to be $\sim 10^{-13} \,\, - \,\,  10^{-12}$.

However, considerable enhancements are expected for FCNC processes in the
top sector~\cite{FCNCBSM, Obraz1998, HAN1999} in various new models such as 
models with two or more Higgs doublets, supersymmetric models with or 
without $R$-parity conservation, or models with a composite top quark.
Thus, the top-quark phenomenology could be sensitive to physics beyond the 
Standard Model leading to FCNC processes. Such processes are less tightly 
constrained in the top sector compared to the lighter quarks and this
sector can be tested at current energy-frontier colliders.

\begin{figure}[htb]
  \begin{center}                                                                
  \begin{tabular}{ccc}
  \vspace*{-0.2cm}
    
  \hspace*{-1.0cm} \epsfig{file=\master/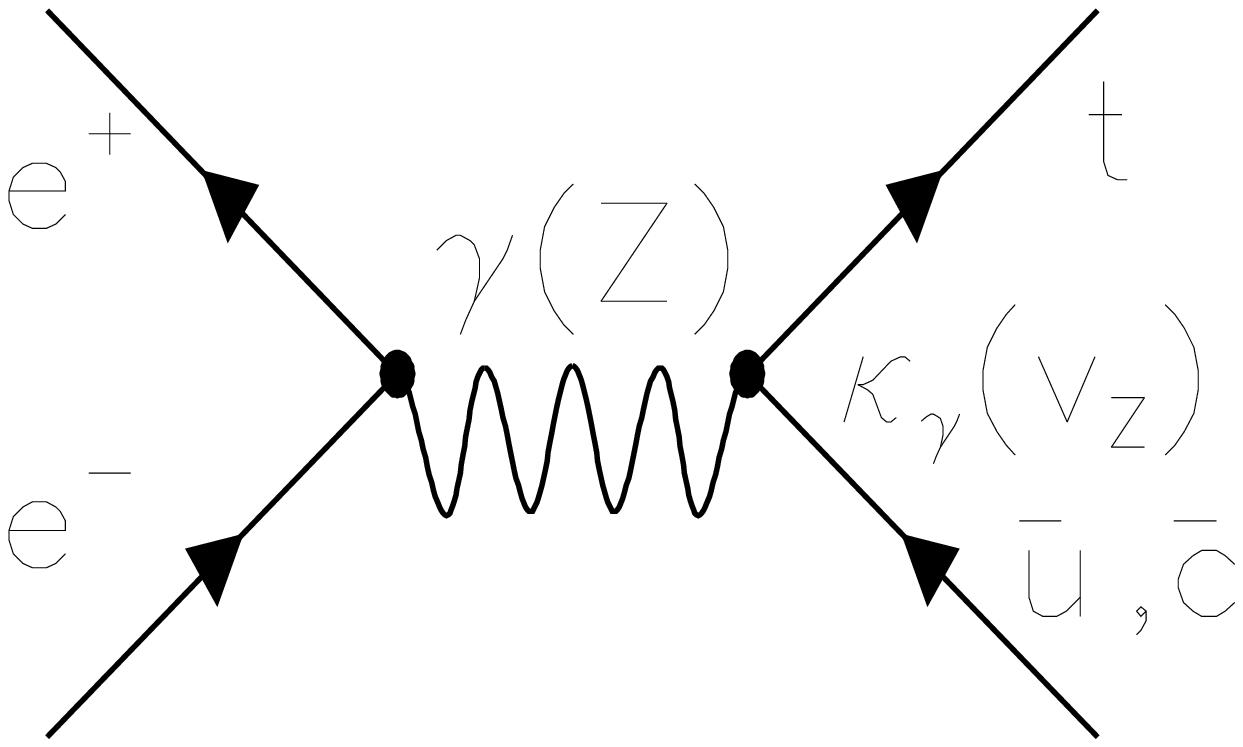,width=0.42\textwidth}
&
  \hspace*{-1.5cm} \epsfig{file=\master/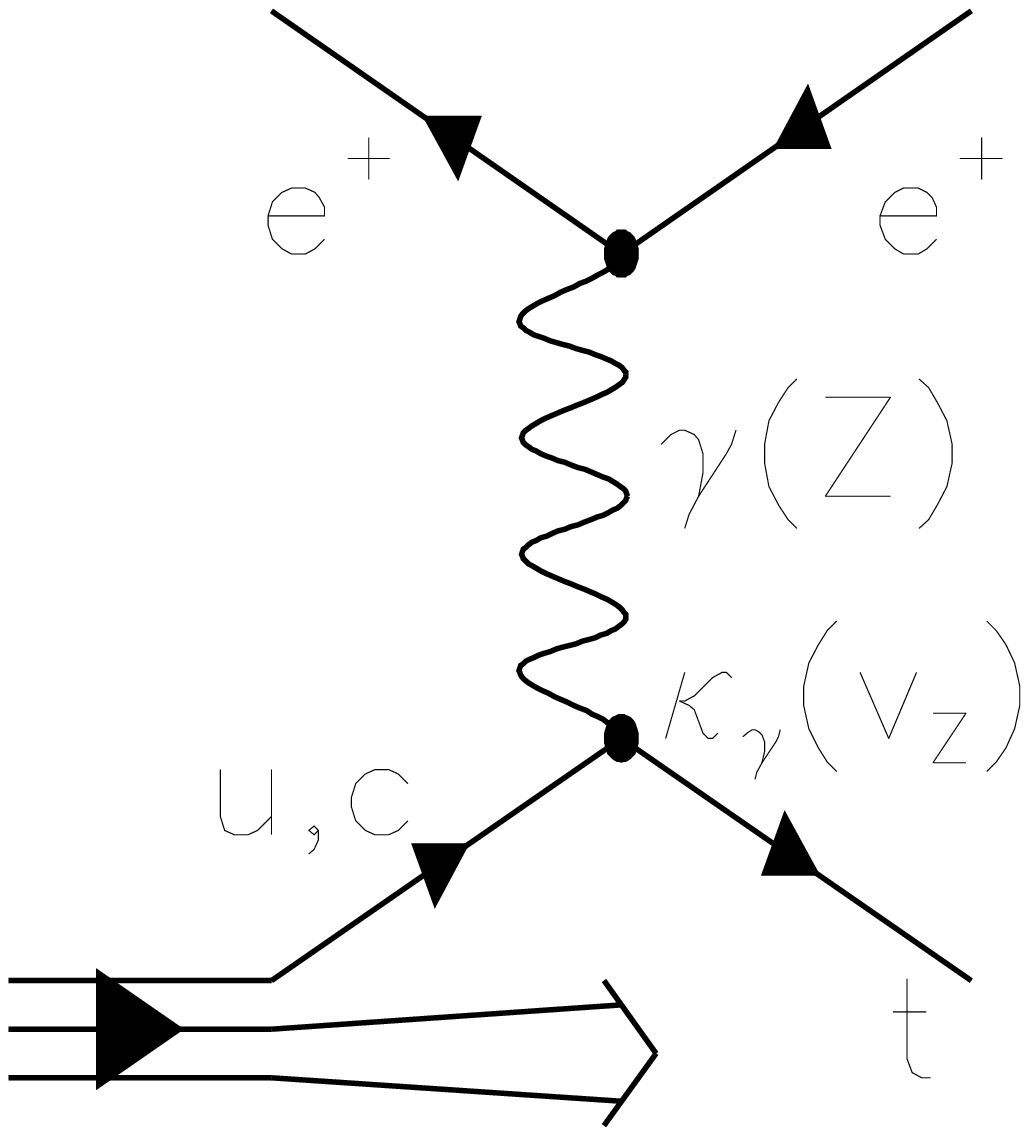,width=0.42\textwidth}
&
  \hspace*{-2.5cm} \epsfig{file=\master/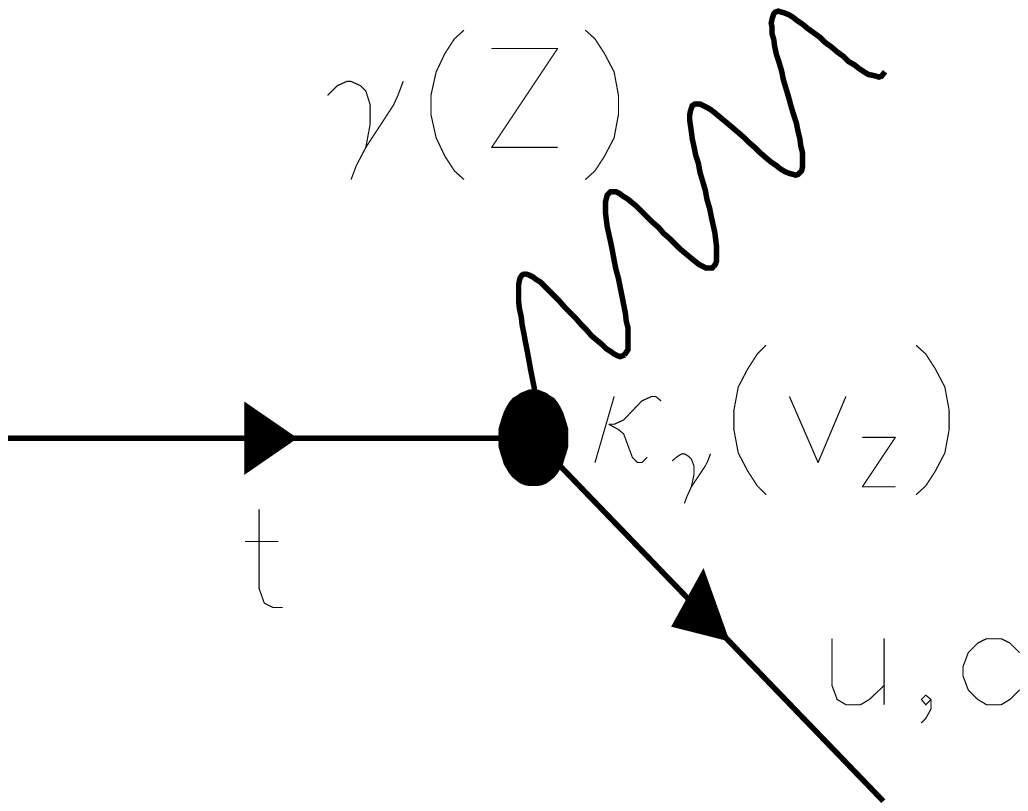,width=0.42\textwidth}

  \end{tabular}
  \end{center}
 \vspace*{-2.5cm}
 
 \hspace*{3.0cm} (a) \hspace*{6.5cm} (b) \hspace*{4.0cm} (c)\\
 
 \vspace*{-0.5cm}
  
 \caption[]{ \label{fig:diagfcnc}
            Top FCNC processes at high-energy colliders.
            a) single-top production in $e^+e^- \to tq$;
            b) single-top production in $ep \to etX$;
            c) FCNC top decay $t \to \gamma (Z) q$.}
\end{figure}

In $e^+e^-$ and $ep$ collisions, single-top production can be
searched for, and in $pp$ collisions, rare decays of produced
top quarks, $t \to \gamma q$ and $t \to Zq$, can be used to
explore the anomalous top FCNC couplings (Fig.~\ref{fig:diagfcnc}).
%----------------------------------------------------------------------------
%-----------------------------------------------------------------------------

In the absence of a specific predictive theory, the
most general effective Lagrangian was proposed in Ref.~\cite{HAN1999} to 
describe FCNC top interactions involving electroweak bosons:
$$  \sum_{U = u,c}
 \left[ 
 \frac{e e_U}{\Lambda} \bar{t} \sigma_{\mu \nu} q^{\nu} 
 (\kappa_{\gamma,U} - i\tilde\kappa_{\gamma,U}\gamma_5) U A^{\mu}
  + \frac{g}{2 \cos \theta_W} \bar{t}
 \{ \gamma_{\mu} (v_{Z,U} - a_{Z,U} \gamma_5) +
 i \frac{1}{\Lambda} \sigma_{\mu \nu} q^{\nu}
 (\kappa_{Z,U} - i\tilde\kappa_{Z,U}\gamma_5) \} 
 U Z^{\mu}\right] $$
where $\sigma_{\mu \nu} = (i/2) \left[ \gamma^{\mu}, \gamma^{\nu} \right]$,
$\theta_W$ is the Weinberg angle, $q$ the four-momentum of
the exchanged boson, 
$e$ and $g$ denote the gauge couplings relative to $U(1)$ and
$SU(2)$ symmetries, respectively, 
$e_U$ denotes the electric charge of up-type quarks, 
$A^{\mu}$ and $Z^{\mu}$ the fields of the photon and $Z$ boson,
and $\Lambda$ denotes the characteristic mass scale of the new interaction.
By convention, $\Lambda$ is set to $m_t$.
Only magnetic operators allow FCNC $t q \gamma$ couplings denoted by
$\kappa_{\gamma, q}$, while $q-t$ transitions
involving the $Z$ boson may also occur via vector (or axial-vector)
interactions with $v_{Z,q} (a_{Z,q})$ coupling
due to the non-vanishing $Z$ mass. 
The collider results so far have been expressed based on the
simplified Lagrangian in Ref.~\cite{Obraz1998}, which
dealt with only $\kappa_{\gamma,q}$ and $v_{Z,q}$ and derived limits\footnote{These two
couplings are often denoted as $\kappa_\gamma$ and $\kappa_Z$ in LEP papers.}
$\kappa_{\gamma,q} < 0.42$ and $v_{Z,q} < 0.73$ from
the CDF experimental results~\cite{CDFFCNC} on radiative top decays:
$BR ( t \rightarrow q \gamma) < 3.2 \%$ 
and $BR ( t \rightarrow q Z) < 33 \%$ at $95 \%$ CL.

Electron-proton collisions at HERA are most sensitive to the $\kappa_{\gamma,u}$
coupling, leading to a $u$-quark in the proton changing to a top
quark with a $t$-channel photon exchange with the electron
(see Fig.~\ref{fig:diagfcnc}b).
The process involving the $Z$-boson is much suppressed due to the
large mass in the $t$-channel propagator.  The anomalous coupling to the
$c$-quark is also suppressed by the small charm density in the proton.

\begin{figure}[htb]
 \vspace*{-1.0cm}
  \begin{center}                                                                
  \epsfig{file=\master/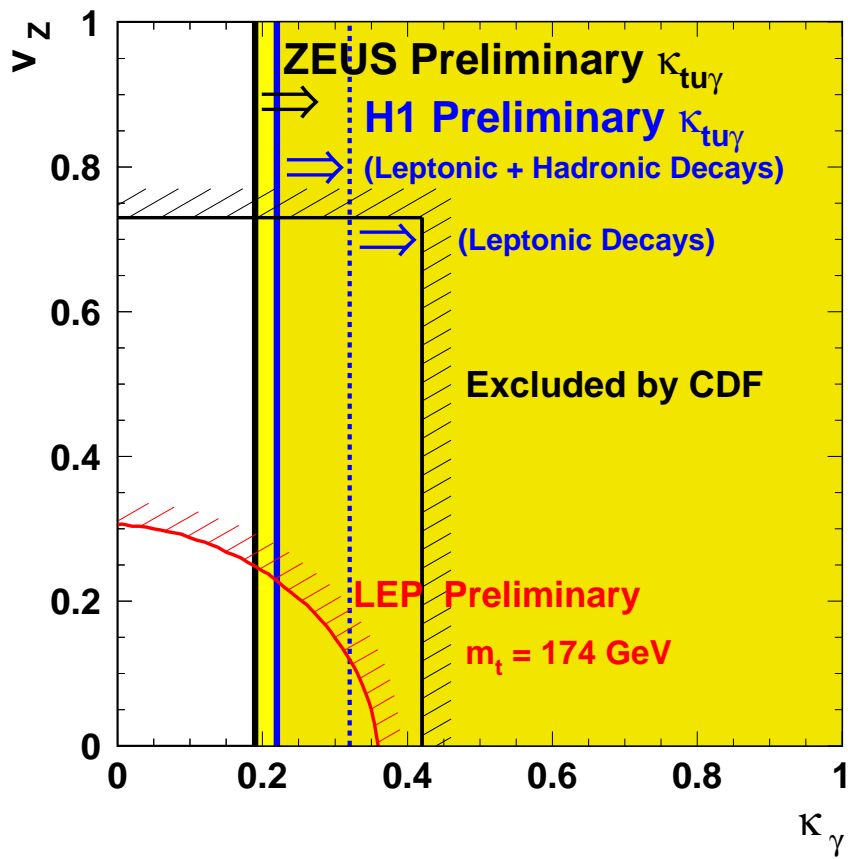,width=0.60\textwidth}

  \end{center}

 \vspace*{-0.5cm}
  
 \caption[]{ \label{fig:fcncresult}
            Top FCNC searches at high-energy colliders.
For H1 and ZEUS limits, the shaded regions to the right of the vertical
lines are excluded.  For CDF (LEP), the hatched regions outside
the curve (rectangle) are excluded, respectively.}
\end{figure}

The single-top production at HERA will yield a high-transverse-momentum
$W$ boson accompanied by an energetic hadron jet coming from the other
decay product, the $b$-quark.  When the $W$ decays leptonically, the
event topology will contain an energetic isolated lepton and large
missing transverse momentum, as well as large hadronic transverse momentum.
For the hadronic decays of $W$, the topology will be a three-jet event with
a resonant structure in dijet and three-jet invariant masses.
Both the H1~\cite{H1FCNCa,H1FCNCb} and ZEUS~\cite{ZEUSFCNC} collaborations 
derived limits on $\kappa_{\gamma,u}$ as shown in Fig.~\ref{fig:fcncresult}.
The H1 limits based on leptonic decay channels only~\cite{H1FCNCa} are 
less stringent due to a slight excess of isolated-lepton
events observed in the data~\cite{H1lepton}.

The figure also compares limits from LEP~\cite{LEPFCNC} and the
Tevatron~\cite{CDFFCNC, Obraz1998}.  They are sensitive to both
$\gamma$ and $Z$ couplings, and both to $u$- and $c$-quark couplings,
since they appear in the final state.  Since the LEP centre-of-mass
energies are close to the threshold, the dependence on the top-mass
uncertainty ($\delta m \simeq \pm 5\rm GeV$) is sizeable, up to 25\% in the
coupling limit, while the corresponding HERA uncertainty is about
10\%~\cite{BelyaevFCNC}.

%----------------------------------------------------------------------------
%-----------------------------------------------------------------------------

% Extended Gauge Groups and Extra Generations
\clearpage
%%%%%%%%%%%%%%%%%%%%%%%%%%%%%%%%%%%%%%%%%%%%%%%%%%%%%%%%%%%%%%%%%%%%%%%%%%%%%%
\section{Extra Bosons and Fermions in Extended Electroweak Models}
\label{sec:extendedSM}
%%%%%%%%%%%%%%%%%%%%%%%%%%%%%%%%%%%%%%%%%%%%%%%%%%%%%%%%%%%%%%%%%%%%%%%%%%%%%%

The symmetries of the Standard Model could be merely the low-energy remains
of a more fundamental theory at high energy scales where the strong and 
electroweak forces would be described by a single Grand Unified (GUT) gauge 
group. But extensions of the standard electroweak gauge symmetries might
also very well be required already at intermediate energy scales, far
below GUT unification scales.
The new phenomenology expected at colliders for models incorporating 
such an extension of the electroweak sector is discussed in this 
section. 

Searches and direct constraints on new particles established at colliders 
as well as the comparison to indirect constraints from low-energy experiments
are often discussed in the context of general effective models. An example
of such an approach was provided by the Buchm\"uller-R\"uckl-Wyler
classification and effective Lagrangian of section~\ref{sec:leptoq} to
study leptoquarks couplings independently of a fundamental theory, 
i.e. implicitly assuming no other new couplings or new particles 
beyond the Standard Model. Similar effective approaches have been developed
for the search for extra gauge bosons, bileptons or doubly charged particles 
discussed below. In the absence of an obvious candidate for the true 
fundamental theory, this seems a pragmatic approach. 
Now the searches are ultimately motivated by specific (true) fundamental 
theories, the consequences of which should be ideally discussed as a whole.
Among the theories that predict the existence of extra gauge bosons 
and exotic Higgs scalars and possibly exotic quarks or leptons,
are the left-right symmetric models and Standard Model extensions containing
triplets of lepton or quark fields. Such models will serve to motivate 
and guide the searches discussed below, where emphasis is put on extra $W'$ 
gauge bosons (subsection~\ref{sec:extraZW}) which come together with
heavy right-handed neutrinos in some models, on exotic 
doubly charged Higgs scalars (subsection~\ref{sec:doublyQ})
and on vector gauge bileptons (subsection~\ref{sec:bileptons}) which
are associated with exotic quarks in some models.

%===========================================================================
\subsection{New Weak Gauge Bosons and Heavy Neutrinos}
\label{sec:extraZW}
%=================

Extra gauge bosons, often generically denoted as $Z'$ and $W'$,
are predicted in left-right symmetric (LRS) 
models~\cite{Pati74,mohapatra86,martin92,aulakh98} 
and in other extensions of the Standard Model, such as
the $3-3-1$ model~\cite{Frampton92,Montero02}. 
They also appear in theories where a strongly interacting sector is 
responsible for a dynamical electroweak symmetry breaking, as discussed 
in the effective Lagrangian approach of the BESS model in 
section~\ref{sec:bess}.
They are required in the Un-unified Standard 
Model~\cite{Georgi89} where quarks and leptons are classified in two 
distinct $SU(2)$ gauge groups and, in general, in models with
separate $SU(2)$ gauge factors for each generation~\cite{Li1981}.
Extra $Z'$ bosons are futhermore motivated by superstring-inspired models based 
on the $E_6$ gauge group which contains $U(1)$ factors beyond the 
Standard Model. 

In the following we concentrate mainly on $W'$ and related searches. 
The existence of new $W'$ bosons is strongly motivated in particular
in LRS models, and will be used here for the discussion.
LRS models based on the gauge group $SU(2)_L \times SU(2)_R \times U(1)_{B-L}$ 
provide a simple extension beyond $SU(2)_L \times U(1)_Y$. 
Such models are themselves motivated for example by $SO(10)$ 
Grand Unified Theories.

In LRS models, the left-handed fermions transform as doublets under 
$SU(2)_L$ and are invariant under $SU(2)_R$ and {\it vice versa} for
right-handed fermions. The models elegantly restore the symmetry for quarks 
and leptons to weak interactions. They furthermore provide a natural
framework to discuss the origin of parity violation and to understand the 
smallness of the neutrino mass (via a seesaw mechanism).
Supersymmetric LRS models based on the extended electroweak gauge group 
$SU(2)_L \times SU(2)_R \times U(1)_{B-L}$ have attracted further 
attention as they offer the possibility to avoid unwanted $R$-parity-violating
interactions by gauge 
symmetries~\cite{mohapatra86,martin92,aulakh98}.
Such interactions are arguably one of the most problematic features
of supersymmetric theories relying on the Standard Model electroweak 
group and in which trilinear $\Rp$ interactions are present unless 
the {\it ad hoc} assumption of an additional discrete symmetry 
is invoked.
In supersymmetric LRS models, there are no such $L$- or $B$-violating
trilinear interactions admissible~\cite{mohapatra86} in the starting 
theory~\footnote{Since there are no massless gauge boson in Nature
                 that couples to $B-L$, the $U(1)_{B-L}$ gauge 
		 symmetry must be broken. In general, $L$-violating 
		 (hence $\Rp$) interactions will be induced in the
		 low-energy effective theory through the spontaneous 
		 or dynamical breaking of $U(1)_{B-L}$ by the 
		 vacuum~\cite{martin96,Chacko98}.}
and, for instance, the proton becomes automatically protected 
from fast decays.

As a result of the additional $SU(2)_R$ symmetry, LRS models predict
the existence of three additional gauge-boson fields coupling to 
right-handed fermions, two charged $W^{\pm}_R$ and a neutral $Z^{0}_R$. 
These appear along with a massive right-handed neutrino $N^i_R$ for
each generation $i$. 
After spontaneous symmetry breaking, the bosons coupling to left- and 
right-handed fermions mix to form physical mass eigenstates.
For the charged bosons $W^-_L \, - \, W^-_R$ (and $W^+_L \, - \, W^+_R$)
one has:
$$ W_1 =  \cos \xi W_L + \sin \xi W_R \,\,\, , \,\,\,
   W_2 = -\sin \xi W_L + \cos \xi W_R \,\,\, ;$$
where $W_1$ is identified as the known $W$ boson and 
$W_2 \equiv W'$ is a new particle.
The expression for the most general Lagrangian that describes the 
interaction of such bosons with Standard Model fermions can be found 
in~\cite{PDG2001}. The Lagrangian contains a phase $\omega$ reflecting
a possible complex mixing parameter in the $W_L - W_R$ mass matrix,
and the $SU(2)_{L,R}$ gauge couplings $g_{L,R}$ with $g_{L} = g_{R}$
if parity invariance is imposed.

Reviews of indirect constraints as a function of the free parameters
$\xi$, $g_{R}/g_{L}$ and $M_{W_2}$ can be found 
in~\cite{Langacker89,Cvetic,PDG2001}. 
Taking into account experimental results on $B-\bar{B}$ mixing,
$b$ decays, neutrinoless double-$\beta$ decays, the $K_L-K_S$ mass
difference, muon decays, etc., a most conservative limit of 
$M_{W_R} ( g_L / g_R ) > 300 \GeV$ was obtained 
in Ref.~\cite{Langacker89} in a model allowing for a non-diagonal
mixing matrix $V^R$ for right-handed quarks and some fine tuning.

\begin{figure}[htb]
  \begin{center}                                                                
  \vspace*{-1.2cm}
  \begin{tabular}{ccc}
  \vspace*{-0.2cm}
    
\hspace*{-0.6cm}  \epsfig{file=\master/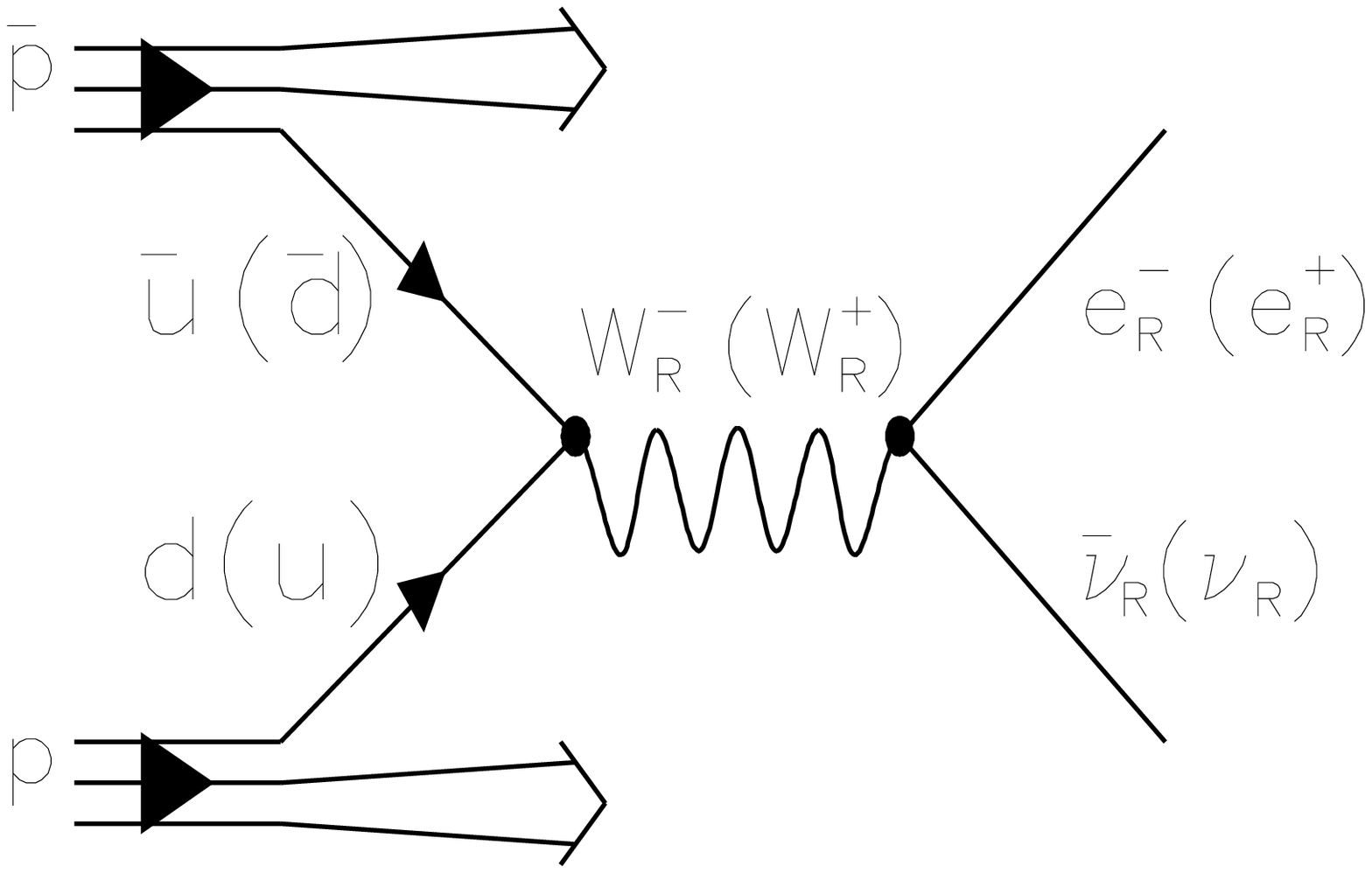,width=0.40\textwidth}
&
\hspace*{-1.4cm} \epsfig{file=\master/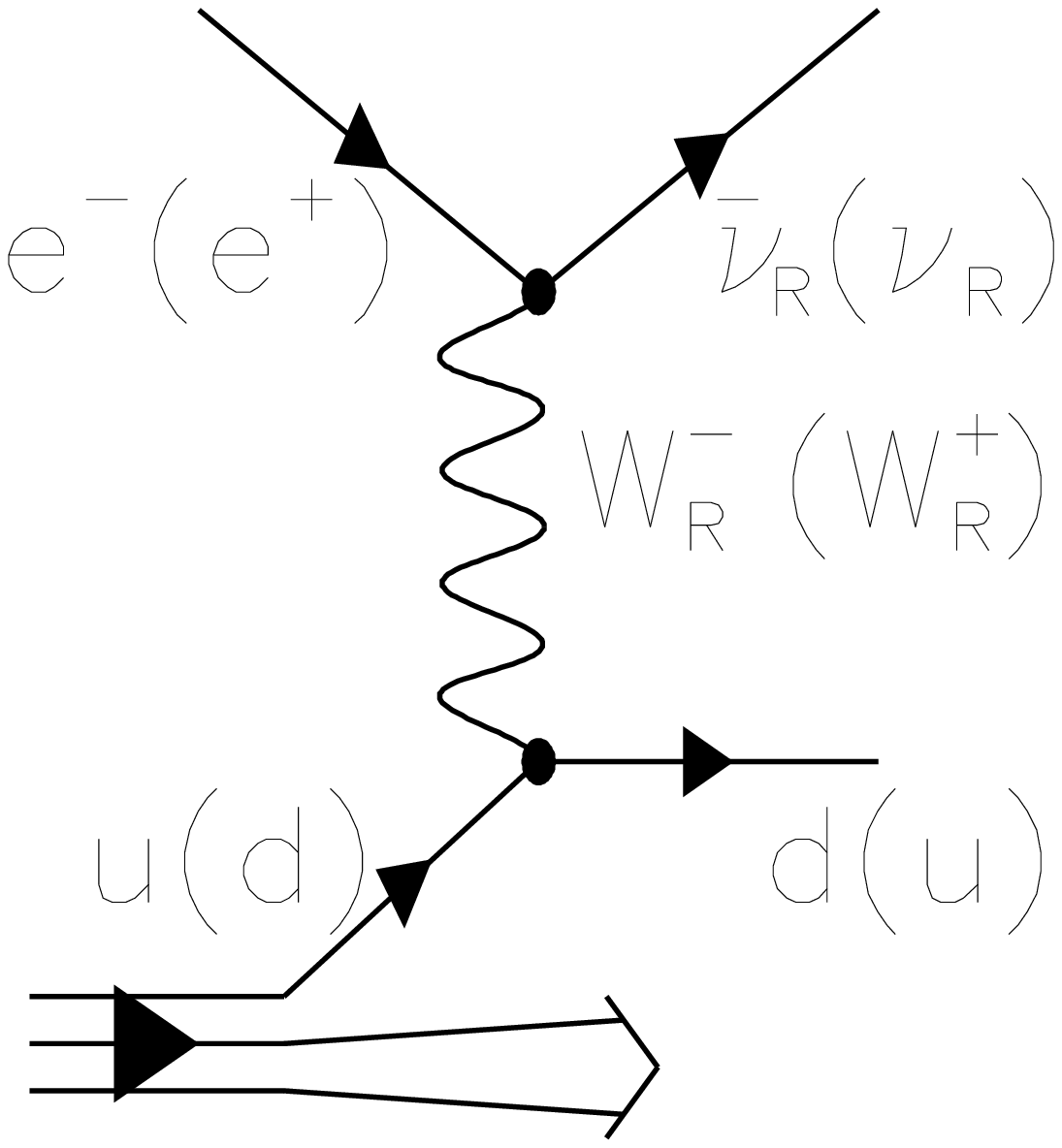,width=0.40\textwidth}
&
\hspace*{-2.2cm} \epsfig{file=\master/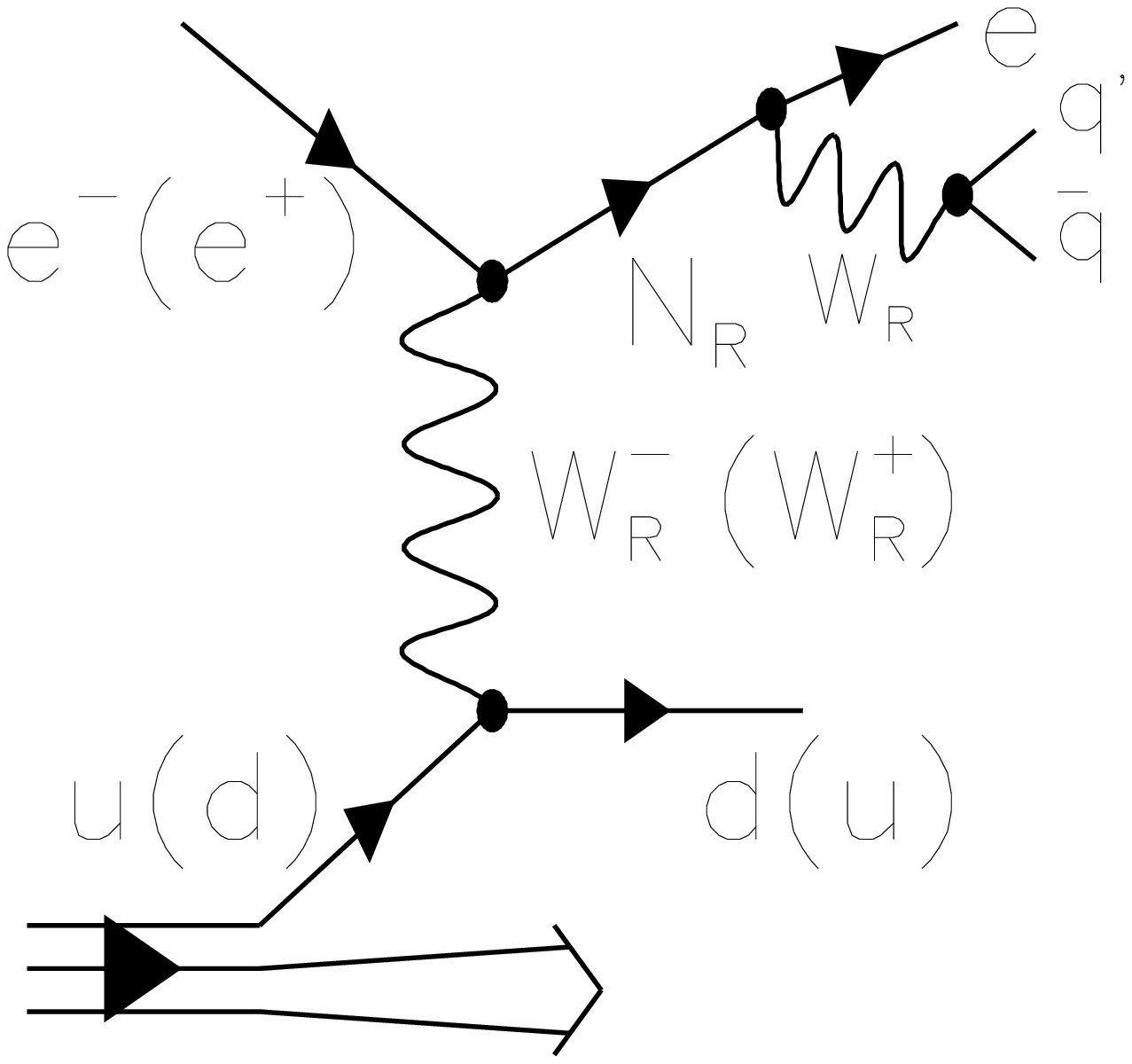,width=0.40\textwidth}

  \end{tabular}
  \end{center}
 \vspace*{-1.0cm}
 
 \hspace*{3.0cm} (a) \hspace*{5.5cm} (b) \hspace*{5.0cm} (c)\\
 
 \vspace*{-0.5cm}
  
 \caption[]{ \label{fig:wrdiags}
            Typical diagrams for $W'$ production at a) $p\bar{p}$ and b,c) $ep$
	    colliders.
            a) Single production via $q' \bar{q}$ fusion; 
            b) $t$-channel virtual exchange in charged current 
	              processes with $M_{\nu_R} \ll M_{W'}$;    
	    c) $t$-channel virtual exchange with subsequent decay
	              of a heavy Majorana neutral lepton $N_R$.} 
\end{figure}

Diagrams for the production or exchange of a heavy $W'$ at colliders
are shown in Fig.~\ref{fig:wrdiags}. At $p\bar{p}$ colliders, the dominant
$W'$ production proceeds via $\bar{u}d$ (${W'}^{-}$) or $u\bar{d}$ (${W'}^{+}$)
fusion as shown in Fig.~\ref{fig:wrdiags}a. At an $ep$ collider, the exchange
of a right-handed $W'$ might interfere with standard charged current 
processes  (Fig.~\ref{fig:wrdiags}b for $M_{\nu_R} \ll M_{W'}$) 
or allow for the production of a heavy and unstable right-handed neutral
lepton $N_R$ (Fig.~\ref{fig:wrdiags}c). 
%----------------------------------------------------------------------------
%----------------------------------------------------------------------------- 

The best direct (but model dependent) $W'$ constraints have been obtained at 
hadronic colliders, first by the UA2 experiment~\cite{UA2WR} at CERN and 
then by the CDF~\cite{CDFWRLNU,CDFWRQQ} and
D$\emptyset$~\cite{D0WRLNU,D0WRQQ} experiments at the Tevatron.
The searches rely on single $W'$ production 
(Fig.~\ref{fig:wrdiags}a) through $\bar{q} q'$ fusion processes, 
e.g. $\bar{u} d \rightarrow {W'}^{-}$ or $\bar{d} u \rightarrow {W'}^{+}$.

Various possible decay modes of the $W'$, motivated for instance by LRS 
models, have been considered separately in the analyses.
This includes SM-like leptonic decays $W' \rightarrow l_R \nu_R$ where the 
right-handed neutrino is assumed to be light ($M_{\nu_R} \ll M_{W'}$) and escapes
detection~\cite{CDFWRLNU,D0WRLNU}.
Alternatively, $W'$ is assumed to initiate a decay chain 
$W' \rightarrow l_R N_R \,;\, N_R \rightarrow l \bar{q} q'$
involving a heavy and unstable right-handed neutrino $N_R$~\cite{D0WRLNU}.
For a very heavy $N_R$ with $M_{N_R} \gg M_{W'}$, the analyses rely
on hadronic decays $W' \rightarrow \bar{q} q'$~\cite{CDFWRQQ,D0WRQQ}.

In LRS models, and in general in extended gauge models where the new 
bosons belong to gauge groups different from those of standard bosons, 
non-vanishing $W' W Z^0$ couplings only occur through mixing after 
symmetry breaking. The mixing angles are expected to be small,
typically of ${\cal{O}}(M_W/M_{W'})^2$, such that the branching 
ratio for $W' \rightarrow W Z^0$ is small~\cite{Altarelli89}.
However, it has been further argued in Ref.~\cite{Altarelli89} that large 
branching ratios in that mode are possible in models with 
a strongly interacting scalar sector or in models with non-linear 
realization of the electroweak interactions in the limit where
the Higgs mass becomes infinite.
A search for singly produced $W'$ decaying into $W Z^0$ has been
recently performed by CDF~\cite{CDFWRWZ}.

Model-dependent assumptions have to be made to translate the 
experimental observations into $W'$ constraints.
These concern mainly: {\it i)} the value of the
the coupling constant $g_R$ of the $W'$ to right-handed fermions,
which enters in the production cross-section; {\it ii)} the $L-R$ 
mixing angle $\xi$, which will determine for instance the relative 
contribution of $W' \rightarrow W Z^0$ decays; {\it iii)} the mass(es) 
and nature (Dirac or Majorana) of the right-handed neutrinos; 
{\it iv)} the values of the elements of the ``CKM'' mass mixing matrix 
for right-handed quarks, which also affect the production cross-section.

\begin{figure}[htb]
  \begin{center}

  \epsfig{file=\master/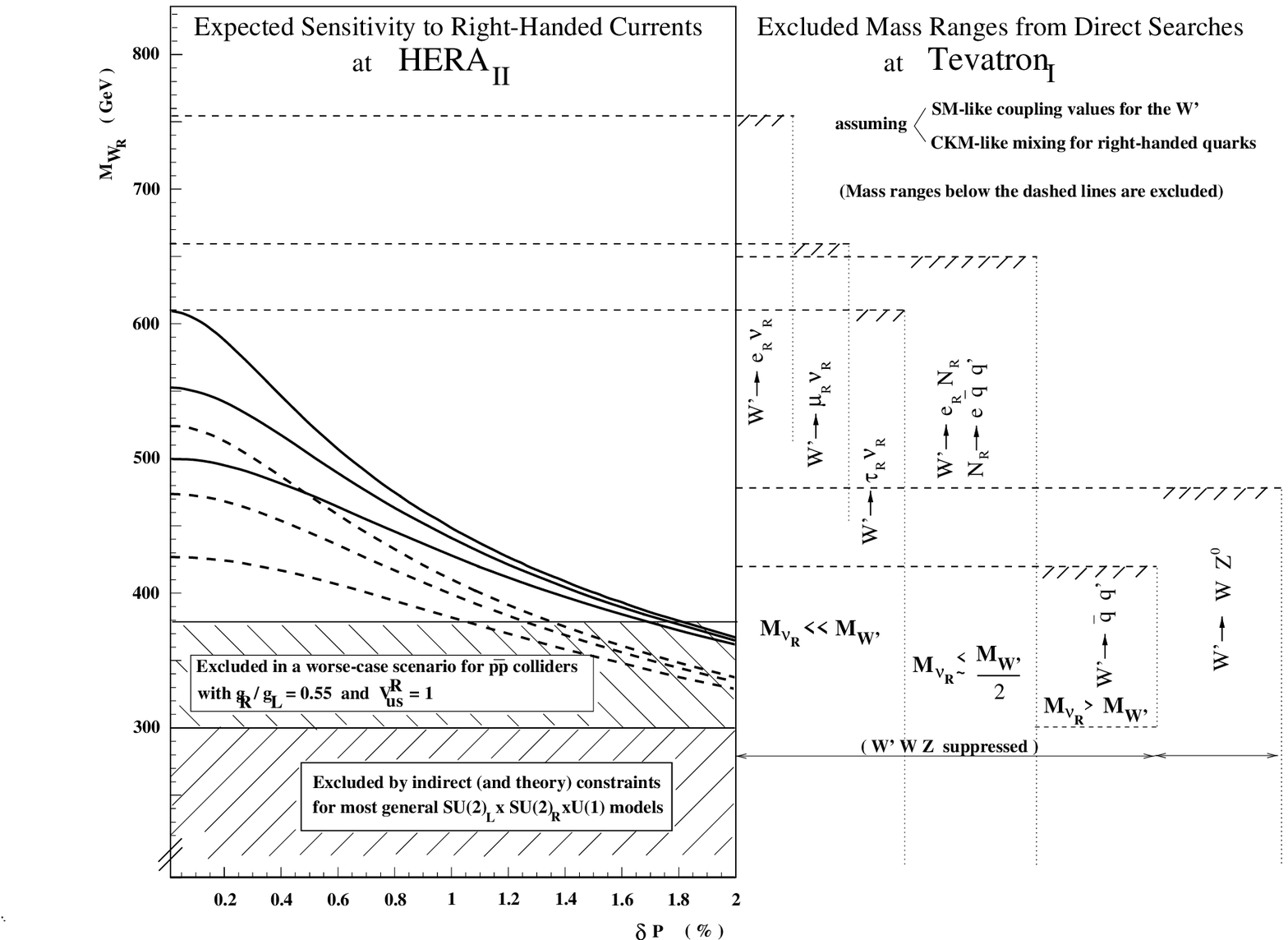,width=1.00\textwidth}

  \end{center}

 \vspace*{-0.5cm}
  
 \caption[]{ \label{fig:wright}
             Direct and indirect constraints on a new heavy right-handed
	     boson $W_R$.
	     The hatched domain at $M_{W_R} ( g_L / g_R ) < 300 \GeV$ 
	     is excluded by a most conservative model taking into account
             indirect measurements~\cite{Langacker89}.	     
	     The curves represent the expected exclusion lower limits 
	     (95 \% CL) obtainable at HERA$_{II}$ by total cross-section 
	     measurements as a function of the precision $\delta P$ on 
	     the lepton-beam polarisation for a polarisation level of
	     $P = 70 \%$ (solid) and $P = 50 \%$ (dashed), and for 
	     integrated luminosities of $0.5 \fb^{-1}$ (lower), 
	     $1 \fb^{-1}$ (middle) and $2 \fb^{-1}$ (upper).  
	     Model-dependant lower limits from direct searches at
	     hadronic colliders are also shown.} 
\end{figure}

The $W'$ mass constraints derived from the Tevatron searches in
various (possibly mutually exclusive) decay modes are displayed
in Fig.~\ref{fig:wright}. The constraints are given at 95\% CL and
assuming SM-like coupling values to ordinary fermions and CKM-like values
%----------------------------------------------------------------------------
%-----------------------------------------------------------------------
for the mass mixing matrix of right-handed quarks.

The $W'$ mass is seen to be severely restricted for leptonic decays 
$W' \rightarrow l_i \nu'_i$ involving any of the three
lepton generations $i$ and assuming $M_{\nu'_i} \ll M_{W'}$.
In the limit where $M_{\nu'_i} \sim 0$, masses in the range
$[ 100 < M_{W'} < 754 \GeV ]$ are excluded for $W' \rightarrow e \nu$
decays, $[ 100 < M_{W'} < 660 \GeV ]$ for $W' \rightarrow \mu \nu$,
and $[ 100 < M_{W'} < 610 \GeV ]$ for $W' \rightarrow \tau \nu$.
For $M_{\nu'_i} \lsim M_{W'}/2$, masses in the range 
$[ 200 < M_{W'} < 650 \GeV ]$ are excluded.

Coupled with the most general indirect constraints for lowish $M_{W'}$ 
discussed above, these direct collider constraints seem at first glance
to completely exclude a possible signal at the HERA$_{II}$ $ep$ collider. 

The sensitivity to new right-handed charged currents at HERA$_{II}$ with 
polarized lepton beams is shown in Fig.~\ref{fig:wright} assuming parity 
invariance and for various technical assumptions on the level of 
lepton-beam polarisation and integrated luminosities. For a precision on 
the polarisation measurement which should routinely reach $1.0\%$ and 
(ambitiously) is aiming for $0.5\%$, and for an integrated luminosity 
of $1 \fb^{-1}$ shared between opposite helicities, $W_R$ masses reaching
$400$ to $500 \GeV$ (depending on the polarisation level) will be
probed in charged current processes with $M_{\nu_R} \ll M_{W'}$.

In the case of a heavy right-handed neutrino $N_R$ decaying rapidly
inside the detector, a drastic reduction of the Standard Model 
background (hence an improved discovery reach) becomes possible 
at HERA by considering the decay $N_R \rightarrow e \bar{q} q'$. 
For $M_{\nu_R} < M_{W'}$, this decay involves either a virtual 
$W_R$ or, through mixing, a real ordinary $W$.
Production and decay of Dirac neutrinos will lead only to 
$\Delta L = 0$ processes where the charged lepton in the final state
carries the same charge as the incident lepton beam.
In contrast, both positively and negatively charged leptons would
occur with equal rates in the decay of Majorana neutrinos.

The production and decay of Majorana neutrinos at HERA via $W_L$
has been studied in Ref.~\cite{Buchmuller91}. The process involves
the boson mixing parameter $\xi$ and a leptonic mass-mixing
matrix $V_L$. The total cross-section varies with the square 
of the strength $(V \xi)_{eN}$ between the incident $e_R$ and 
the produced heavy neutrino $N_R$. 
For $(V \xi)_{eN} = 0.1$, the existence of Majorana neutrinos
with masses up to $M_N \simeq 200 \GeV$ could be probed at 
HERA$_{II}$ for an integrated luminosity of $1 \fb^{-1}$. 
Irrespective of $V_L$ and $\xi$, Majorana neutrinos could also be 
produced at HERA via $W_R$ exchange as in the diagram of 
Fig.~\ref{fig:wrdiags}c. The sensitivity expected in this channel 
was studied in the $M_{W_R}$ {\it vs.} $M_{N_R}$ plane in
Ref.~\cite{Buchmuller92}. Masses up to $M_{W_R} = 700 \GeV$
could be probed for heavy neutrinos with masses up to 
$M_{N_R} \simeq 120 \GeV$. The mass reach for $M_{W_R}$ decreases
with increasing $M_{N_R}$ and masses up to only $M_{W_R} = 450 \GeV$ 
could be probed for $M_{N_R} \simeq 180 \GeV$.
Thus, the mass reach at HERA$_{II}$ for right-handed currents accompanied
by the production of a heavy Majorana neutrino decaying via
$N_R \rightarrow e \bar{q} q'$ appears to be already well covered 
by Tevatron$_I$ searches.

The stringent collider constraints on $M_{W'}$ from leptonic decays
can be partly evaded if $M_{N_R} \gg M_{W'}$ but then masses in the 
range $[ 300 < M_{W'} < 420 \GeV ]$ are nevertheless excluded via 
the $W' \rightarrow \bar{q} q'$ channel. If this in turn is turned off
because $W' \rightarrow W Z^0$ decays dominate, then masses in the 
range $[ 200 < M_{W'} < 480 \GeV ]$ remain excluded.

A way to partly evade $p \bar{p}$ collider constraints while possibly
preserving the sensitivity of charged current processes at an $ep$ 
collider consists of allowing for a non-standard quark mass-mixing matrix 
in the right-handed sector~\cite{Rizzo94}. As an illustration of an 
extreme case, one can consider for example a mixing matrix with 
$V^R_{us} = 1$ (thus with $V^R_{ud} = 0$ for a unitary $V^R$). 
This leads to a suppression of the production in $p \bar{p}$ collisions,
which depends dominantly on a product of the $u$ and $d$ valence-quark
densities in the proton. In contrast, the ${W'}^{+}$ exchanged with a 
valence quark in an $e^+ p$ interaction or the ${W'}^{-}$ exchanged in an
$e^- p$ interaction can only couple respectively to the $d$ or 
$u$ quark. The ${W'}^-$ contribution to charged currents in $e^- p$
collisions would thus remain largely unaffected in this extreme 
case. Such an extreme case was considered in a D$\emptyset$ 
analysis~\cite{D0WRLNU} which excludes masses in the range
$[ 200 < M_{W'} < 380 \GeV ]$ for $V^R_{us} = 1$ and 
$g_R / g_L = 0.55$. This is labelled as a ``worse-case scenario''
for $p \bar{p}$ colliders in Fig.~\ref{fig:wright}. It helps
in emphasizing the necessity for extremely good precision 
$\delta P$ on the lepton-beam polarisation at the $ep$ collider.

For completeness, and although it appears to be very difficult
if not impossible to avoid the $M_{W_R} > 300 \GeV$ constraints 
from precision measurements and rare or forbidden 
decays~\cite{Langacker89}, it should be remarked that none of
the existing direct searches carried out at CERN or Tevatron 
covered the range $M_{W_R} \lsim 100 \GeV$. 
This is even more so in the case where $M_{N_R} > M_{W_R}$
if the $W_R$ decays only via $W_R \rightarrow \bar{q} q'$.
In that case a ``hole'' has been left between the di-jet coverage 
of the UA2 experiment, which reaches $M_{W_R} \simeq 250 \GeV$,
and the range covered by the CDF analysis which starts at 
$M_{W_R} \simeq 300 \GeV$.
The sensitivity at HERA to heavy Majorana neutrinos in the
case $M_{W_R} \lsim 100 \GeV$ and $M_{N_R} > M_{W_R}$ was
discussed in Ref.~\cite{Sciulli96} and a preliminary analysis made
by ZEUS showed~\cite{ZEUSWR96} for an integrated luminosity
of about $10 \pb^{-1}$ that only $M_{N_R}$ values below the
the top quark mass could be probed at HERA for $g_R = g_L$ 
and $W_R$ masses up to the common $W$ mass.

\subsection{Doubly Charged Higgs Scalars}
\label{sec:doublyQ}
%=================

%
% Introduction:
%

We have discussed in subsection~\ref{sec:extraZW} the case 
of the additional gauge bosons predicted by LRS models. Here we would 
like to focus on a possible consequence of the extension of the Higgs 
sector required by such models, namely the prediction of doubly charged
Higgs physical states~\cite{Pati74,Mohapatra80,Mohapatra81}. 
Doubly charged Higgs bosons are also present in other 
scenarios~\cite{Gelmini81,Georgi85,Gunion90,Godbole95,Gunion95} containing  
triplet Higgs fields but not necessarily incorporating
left-right symmetry.

In most LRS models, new additional triplets of Higgs scalar bosons are 
introduced which act solely in the leptonic 
sector~\cite{Rizzo82,Gunion96,Huitu99}.
Two different Higgs multiplets are needed to preserve the left-right 
symmetry. They connect to either left- or right-handed lepton chiral 
states. The so-called ``right-handed'' Higgs field is responsible for 
$SU(2)_R$ symmetry breaking and gives the heavy mass to the right-handed
Majorana neutrinos needed in the seesaw mechanism.
As a general feature of LRS models~\cite{Pati74,Mohapatra80,Mohapatra81}, 
the Higgs multiplets contain doubly charged elements, leading to the existence 
of two physical doubly charged Higgs particles, labelled 
$\Delta^{--}_L$ and $\Delta^{--}_R$. 
Actually, it has been shown on very general grounds that large classes 
of supersymmetric LRS models do necessarily~\cite{Chacko98} 
contain such doubly charged Higgs fields and that the physical
states tend to be very light~\cite{Huitu94,Chacko98}.
This remains true whether or not the $SU(2)_R$ weak scale is in the 
superheavy range and holds even when $R$-parity gets spontaneously 
broken~\cite{Chacko98}.
The doubly charged Higgs bosons are members of the so-called ``left-'' and 
``right-handed'' triplets $(\Delta^0,\Delta^{-},\Delta^{--})_{L,R}$
and carry the quantum number $\mid B-L \mid = 2$.

%
% General aspects on couplings (production and decays)
%

The doubly charged Higgs boson could in principle couple to ordinary EW 
bosons and/or to other Higgs bosons but, in a likely scenario, the 
couplings to lepton pairs could determine the relevant 
phenomenology. Indeed, trilinear $\Delta L = 2$ couplings of the type 
$W W \Delta^{--}$ which could allow for single production via 
$t$-channel $W W$ fusion are not necessarily present in the 
theory~\cite{Gunion95,Gunion96}.
For the $W_L W_L \Delta^{--}_L$, the coupling must be in any case
vanishingly small given the constraints set by the electroweak 
$\rho$ parameter which involves the mass ratio of ordinary weak 
bosons~\cite{Maalampi02}.
For the $W_R W_R \Delta^{--}_R$, the coupling strength depends
on the scale of the left-right symmetry breaking and will
be suppressed for a very heavy $W_R$~\cite{Huitu97}. 
Moreover, in particular in the framework of supersymmetric LRS 
models, the decay $\Delta^{--}_R \rightarrow W_R^- W_R^-$ of a
real doubly charged Higgs boson might very well be closed because of
a heavy or superheavy $W_R$.
In addition, bosonic decays of the type 
$\Delta^{--} \rightarrow \Delta^{-} W_R^-$ or 
$\Delta^{--} \rightarrow \Delta^{-} \Delta^{-}$ 
are possible but also likely to be disallowed.
Finally, given that the doubly charged Higgs boson cannot couple to quark 
pairs because of charge conservation, a real doubly charged Higgs boson may 
very well be left with only like-sign lepton-pair decays.
Of course, couplings to the $\gamma$ and $Z^0$ are always present.

Since the $\Delta_L$ and $\Delta_R$ triplets are not involved in the 
mass-generation mechanism for the ordinary charged leptons, the 
doubly charged Higgs boson couples to ordinary charged leptons 
independently of their mass ! 
Hence, there are no mass-suppression effects for light leptons and 
the decay branching ratio into like-sign charged leptons of each
of the three generations could be in principle similar.  
In detail the couplings $k^i_L$ of the $\Delta^{--}_L$ are arbitrary
and can be treated as free parameters. The couplings $k^i_R$ of the 
$\Delta^{--}_R$ are proportional to the mass $M(N^i)$ of the new heavy 
right-handed Majorana neutrinos which are introduced for each 
generation $i$. These could be almost degenerate but not 
necessarily so. While heavy Majorana neutrinos could be beyond the
reach of colliders, the $\Delta^{--}_R$ could still retain a large 
coupling to like-sign lepton pairs.
The relative $k^i$ values will fix the (unknown) relative branching ratio 
into like-sign lepton pairs of a given generation.

%
% Essential Aspects on Indirect Constraints :
%

A very important property of the doubly charged Higgs boson when considering
indirect constraints~\cite{Swartz89} is that they naturally avoid some 
of the most sensitive tests of lepton-flavour conservation. The reason 
is that such tests often involve initial-state hadrons where they
can only contribute as higher-order corrections. 
The indirect constraints can be parametrized in terms of the mass
$M_{\Delta^{--}}$ of the scalar and a coupling constant $h_{ij}$ where
$i,j = e, \mu, \tau$. They have been discussed 
in~Refs.~\cite{Swartz89,Gunion89,Lusignoli89,Barenboim97}.

The off-diagonal products $h_{ij} h_{i'j'}$ with either $i \neq j$ or
$i' \neq j'$ suffer from stringent constraints for the first- and 
second-generation charged leptons from forbidden 
$\mu \rightarrow e^+ e^- e^-$ and $\mu \rightarrow e \gamma$ 
decays~\cite{Barenboim97}. 
Assuming that only purely diagonal couplings are non-vanishing,
the existing constraints are remarkably mild. Constraints involving
$ h_{ee} $ come from a possible virtual $\Delta^{--}$ exchange contribution 
to Bhabha scattering in $e^+e^-$ collisions (see below)
which yields~\cite{Swartz89,Barenboim97} 
$$ h_{ee}^2 \lsim 9.7 \times 10^{-6} \GeV^{-2} M_{\Delta^{--}}^2 $$
and from the search for muonium ($\mu^+ e^-$) to anti-muonium
($\mu^- e^+$) conversion which yields
$$ h_{ee} h_{\mu\mu} \lsim 5.8 \times 10^{-5} \GeV^{-2} 
                                M_{\Delta^{--}}^2 \, . \,$$
For the coupling $h_{\mu\mu}$ alone, avoiding possible extra 
contribution to $(g-2)_{\mu}$ yields
$$ h_{\mu\mu}^2 \lsim 2.5 \times 10^{-5} \GeV^{-2} M_{\Delta^{--}}^2\, . \,$$				
No stringent constraints involving the $\tau$ lepton have been 
established.
 
%
% Production and phenomenology at Colliders:
%

\begin{figure}[htb]
  \begin{center}                                                                
  \begin{tabular}{cccc}
  \vspace*{-0.2cm}
    
\hspace*{-1.4cm} \epsfig{file=\master/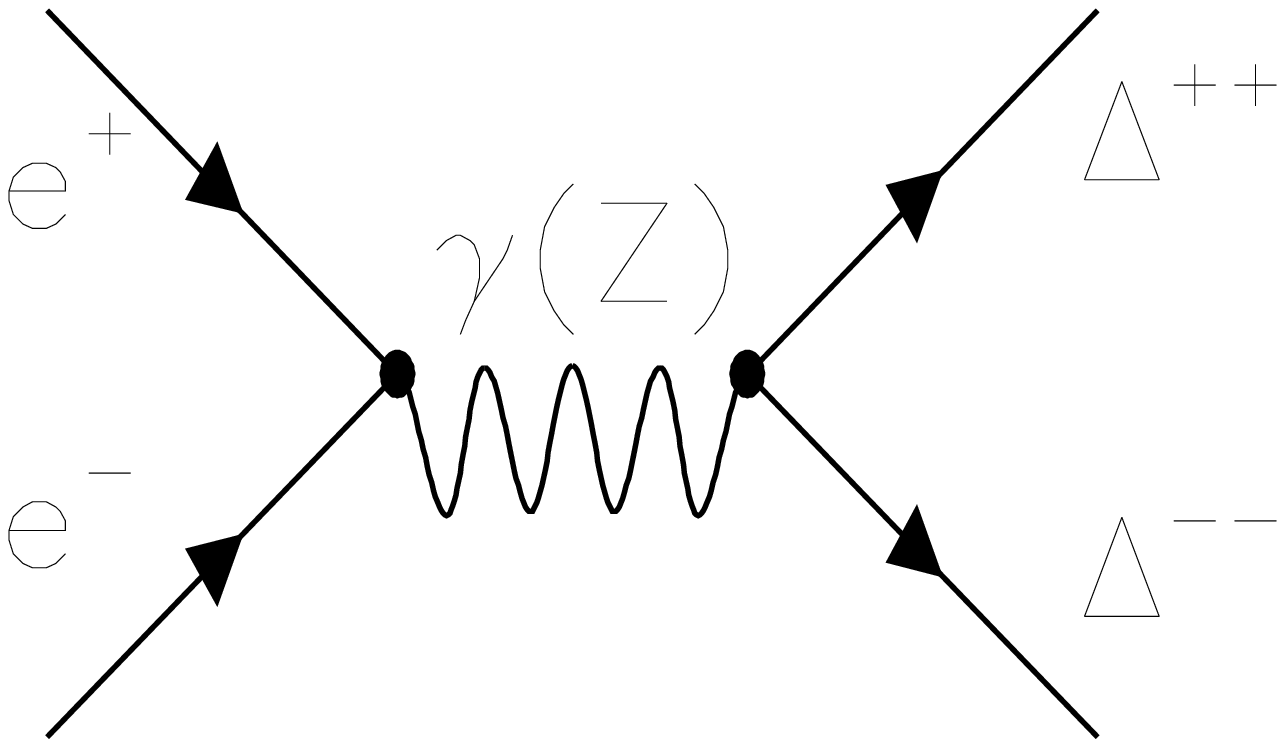,width=0.38\textwidth}
&
\hspace*{-2.7cm} \epsfig{file=\master/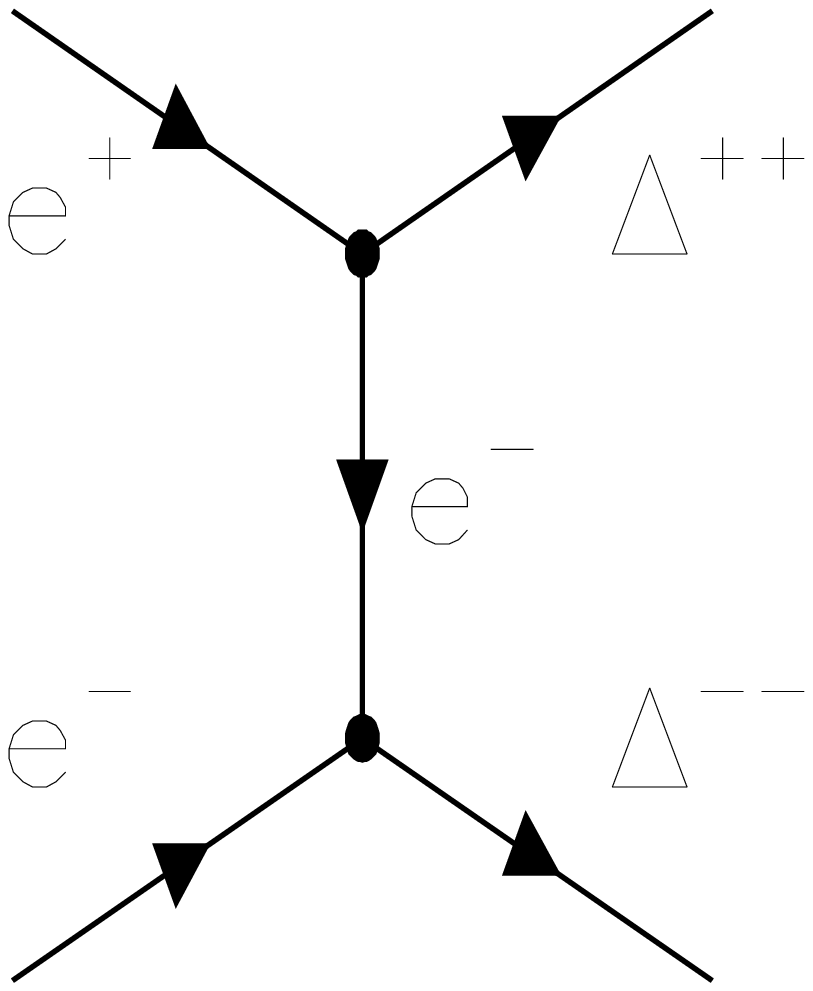,width=0.38\textwidth}
&
\hspace*{-3.1cm} \epsfig{file=\master/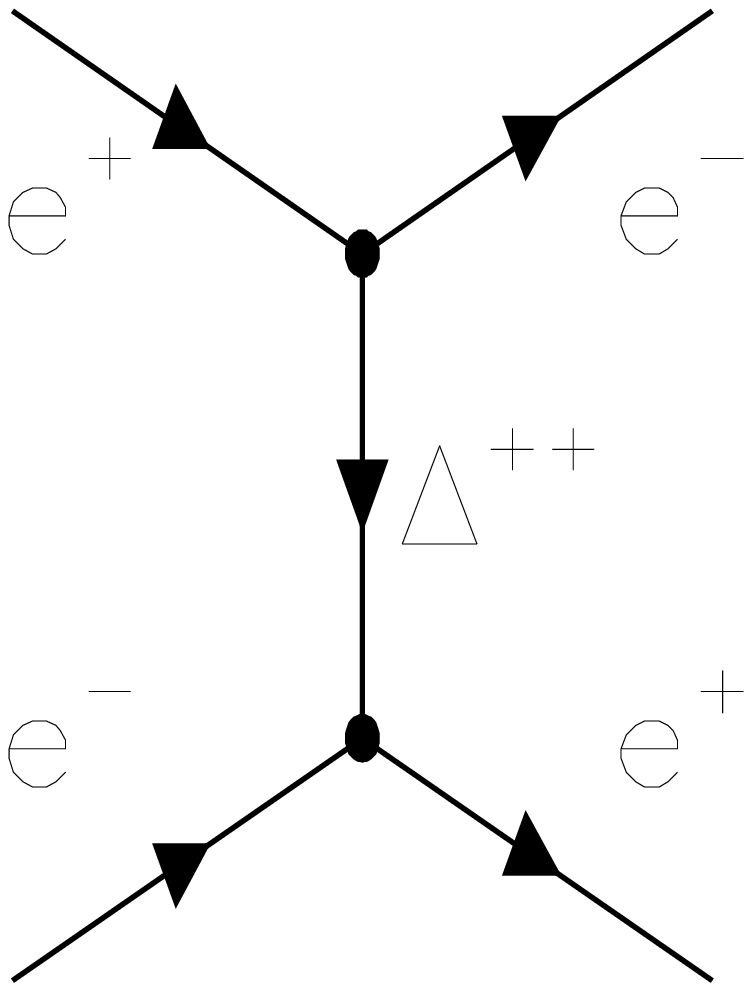,width=0.38\textwidth}
&
\hspace*{-2.9cm} \epsfig{file=\master/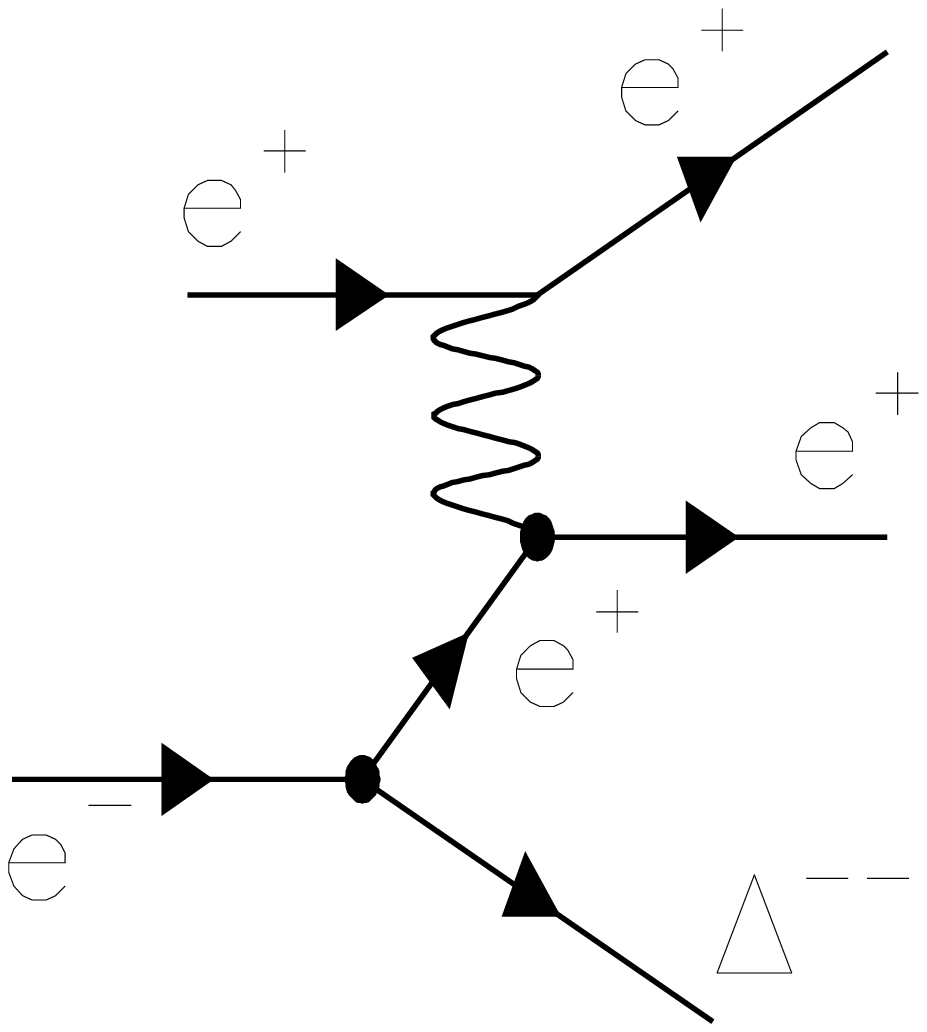,width=0.38\textwidth}

  \end{tabular}
  \end{center}
 \vspace*{-2.3cm}
 \hspace*{2.5cm} (a) \hspace*{4.0cm} (b) \hspace*{3.5cm} (c) \hspace*{3.5cm} (d) \\  
 \vspace*{-0.5cm}
 \caption[]{ \label{fig:diagdelta}
            Typical diagrams for doubly charged Higgs boson production at $e^+e^-$
	    colliders.
            a) $s$-channel pair production; 
            b) $u$-channel pair production;    
	    c) $t$-channel virtual exchange;
	    d) single production.} 
\end{figure}

At $e^+e^-$ colliders the $\Delta^{--}$ scalars can be pair 
produced through their $Z \Delta^{++} \Delta^{--}$ or 
$\gamma \Delta^{++} \Delta^{--}$ couplings
(see Fig.~\ref{fig:diagdelta}a).
Pair production can also proceed through $u$-channel 
exchange (Fig.~\ref{fig:diagdelta}b) involving only couplings
to electron-positron pairs.
For pair production, the kinematic reach is of course 
restricted to $M_{\Delta^{\pm\pm}} < \sqrt{s_{ee}}/2$. 
The $\Delta^{\pm\pm}$ can be exchanged in the $t$-channel 
(Fig.~\ref{fig:diagdelta}c), thus providing an anomalous 
contribution to ``Bhabha'' scattering.
Finally, single production is possible via diagrams involving a 
$\gamma^* e \rightarrow \Delta^{\pm\pm} e$ sub-process as seen
for example in Fig.~\ref{fig:diagdelta}d.
%----------------------------------------------------------------------------
%----------------------------------------------------------------------------- 
The relevant phenomenology has been discussed in 
Refs.~\cite{Rizzo82,Lusignoli89,Grifols89,Dutta98}.
Single production of doubly charged Higgs bosons in $\gamma e$ collisions 
at linear colliders has been discussed in 
Refs.~\cite{Rizzo83,Barenboim97,Godfrey02}.
Resonant production in $e^-e^-$ collisions has been discussed in
Ref.~\cite{Gunion95}. The case for future $\gamma \gamma$ colliders
has been discussed in Ref.~\cite{Chakra98}.

\begin{figure}[htb]
  \begin{center}                                                                
  \begin{tabular}{ccc}
  \vspace*{-0.5cm}
    
 \hspace*{-1.0cm} \epsfig{file=\master/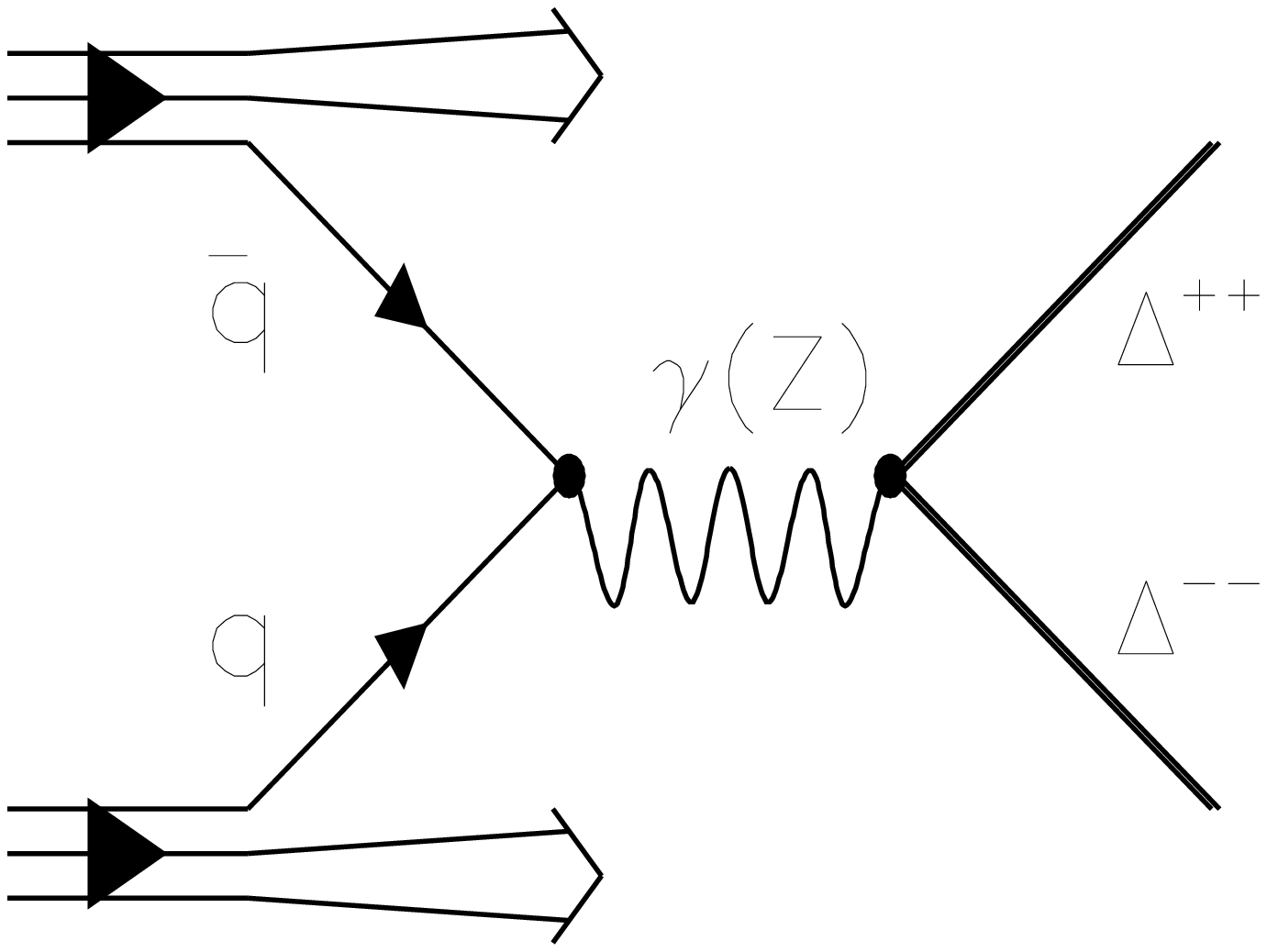,width=0.40\textwidth}
&
 \hspace*{-1.6cm} \epsfig{file=\master/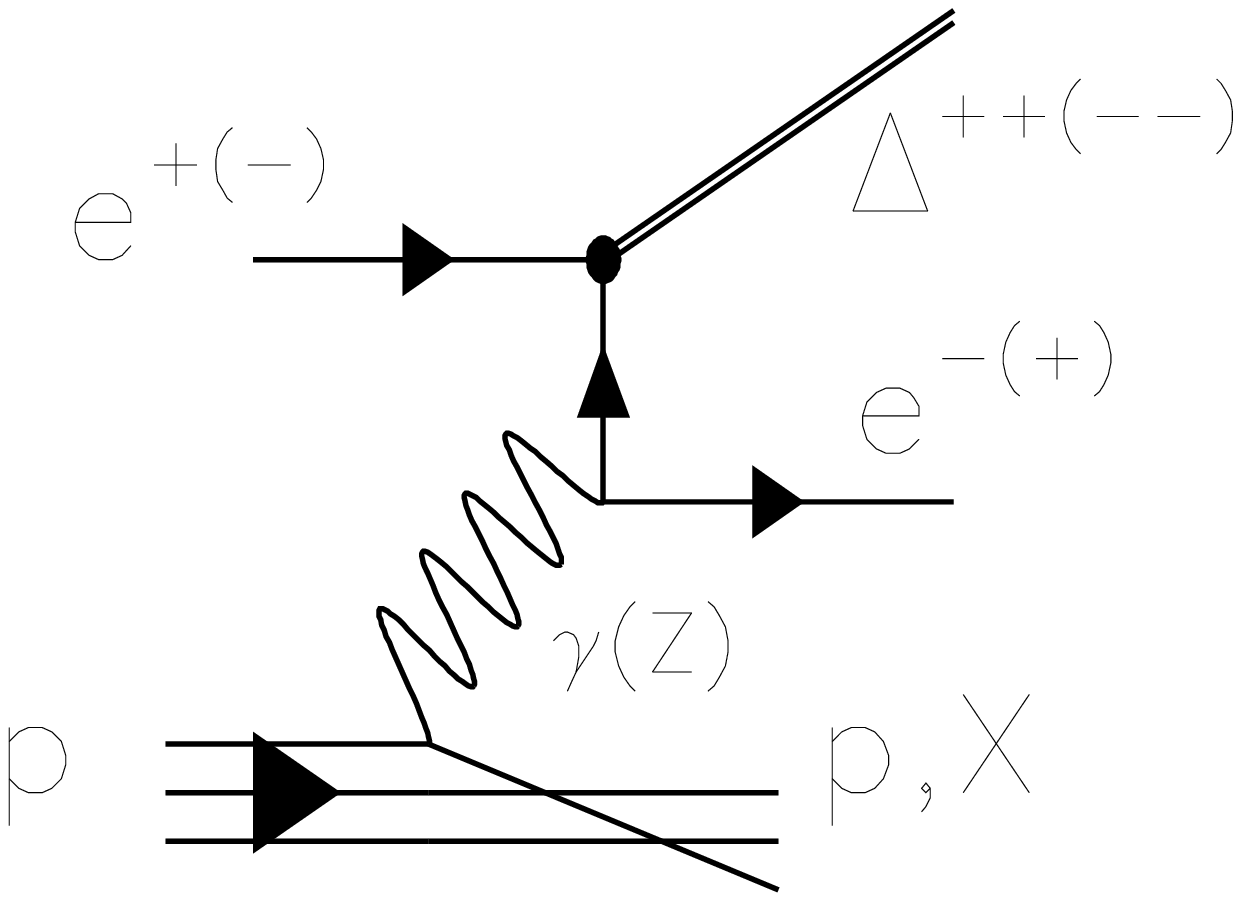,width=0.44\textwidth}
&
 \hspace*{-2.6cm} \epsfig{file=\master/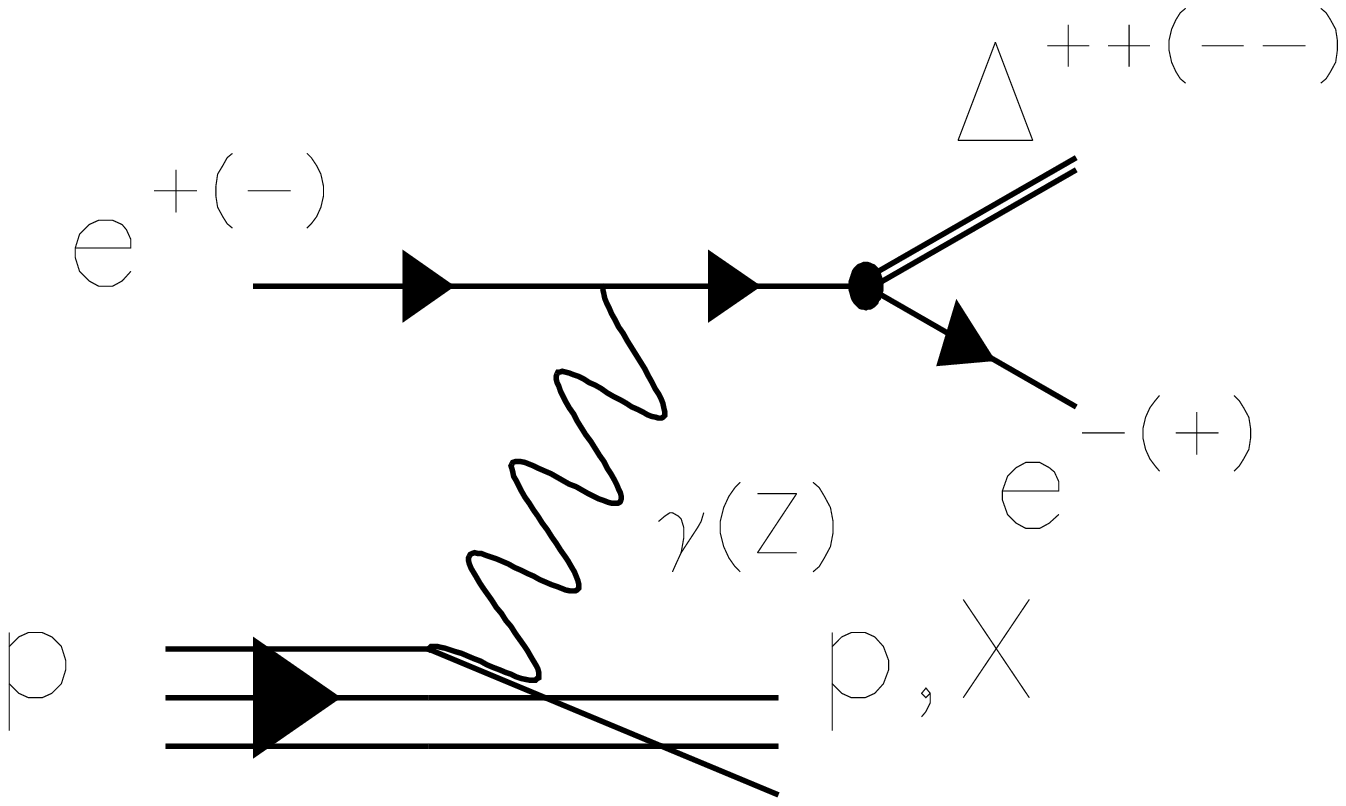,width=0.44\textwidth}

  \end{tabular}
  \end{center}
 \vspace*{-2.3cm}
 \hspace*{3.0cm} (a) \hspace*{5.0cm} (b) \hspace*{5.0cm} (c) \\  
 \vspace*{-0.5cm}
  
 \caption[]{ \label{fig:diagppdta}
            Typical diagrams for doubly charged Higgs boson production at hadronic
	    and lepton-hadron colliders.
            a) pair production in $p \bar{p}$ collisions; 
            b,c) single production in $e p$ collisions.} 
\end{figure}

At the Tevatron $p \bar{p}$ collider, a doubly charged Higgs boson could be pair 
produced via its coupling to $\gamma/Z^0$ electroweak bosons in 
the reactions
$p \bar{p} \rightarrow \gamma/Z^0 \, X
           \rightarrow \Delta^{--} \Delta^{++} \, X$
(see Fig.~\ref{fig:diagppdta}a).
Such production reaching larger $\Delta^{\pm\pm}$ masses will be
possible at a $p p$ collider like the LHC, where it requires an 
anti-quark carrying a large momentum fraction of the proton.
Single production via $WW$ fusion at hadronic colliders could very well 
be suppressed by vanishingly small trilinear couplings.
The relevant phenomenology has been discussed 
in~Refs.~\cite{Gunion89,Grifols89,Gunion96,Huitu97,Dutta98,Datta00,Maalampi02}.
%----------------------------------------------------------------------------
%----------------------------------------------------------------------------- 

With a production cross-section estimated~\cite{Grifols89} to be below 
${\cal{O}}(10^{-2}) \pb$ for $\Delta^{\pm\pm}$ masses above
$45 \GeV$, the mass reach for pair production via photon-photon fusion 
at the HERA $e p$ collider is completely covered by LEP.
On the other hand, HERA allows for single production of doubly charged 
scalars through the $h_{ee}$ coupling for example by the fusion of 
the incoming electron with an electron provided by a photon radiated 
from the proton (see Fig.~\ref{fig:diagppdta}b), or in ``leptonic 
radiative return'' processes (Fig.~\ref{fig:diagppdta}c). 
The phenomenology for HERA has been discussed 
in Ref.~\cite{Accomando94} but considering only elastic production, 
and thus neglecting possibly equally important 
contributions from quasi-elastic and inelastic processes.
HERA is found to offer an almost background-free search environment
in the reactions 
$e^- p \rightarrow  e^+ p \Delta^{--}$ followed by the decays
$ \Delta^{--} \rightarrow e^- e^- \,;\, (\mu^-  \mu^-, \tau^- \tau^-)$
or for non-diagonal couplings      
$e^- p \rightarrow \mu^+ p \Delta^{--}$ followed by the decays
$ \Delta^{--} \rightarrow e^- \mu^- \,;\, (e^- \tau^-, \mu^- \tau^-)$.
A most promising signal at high masses would be three leptons with 
two of them of the same sign at large invariant mass values.

%
% Existing Constraints :
%

\begin{figure}[h]
  \begin{center}                                                                
  \vspace*{-1.0cm}
    
  \epsfig{file=\master/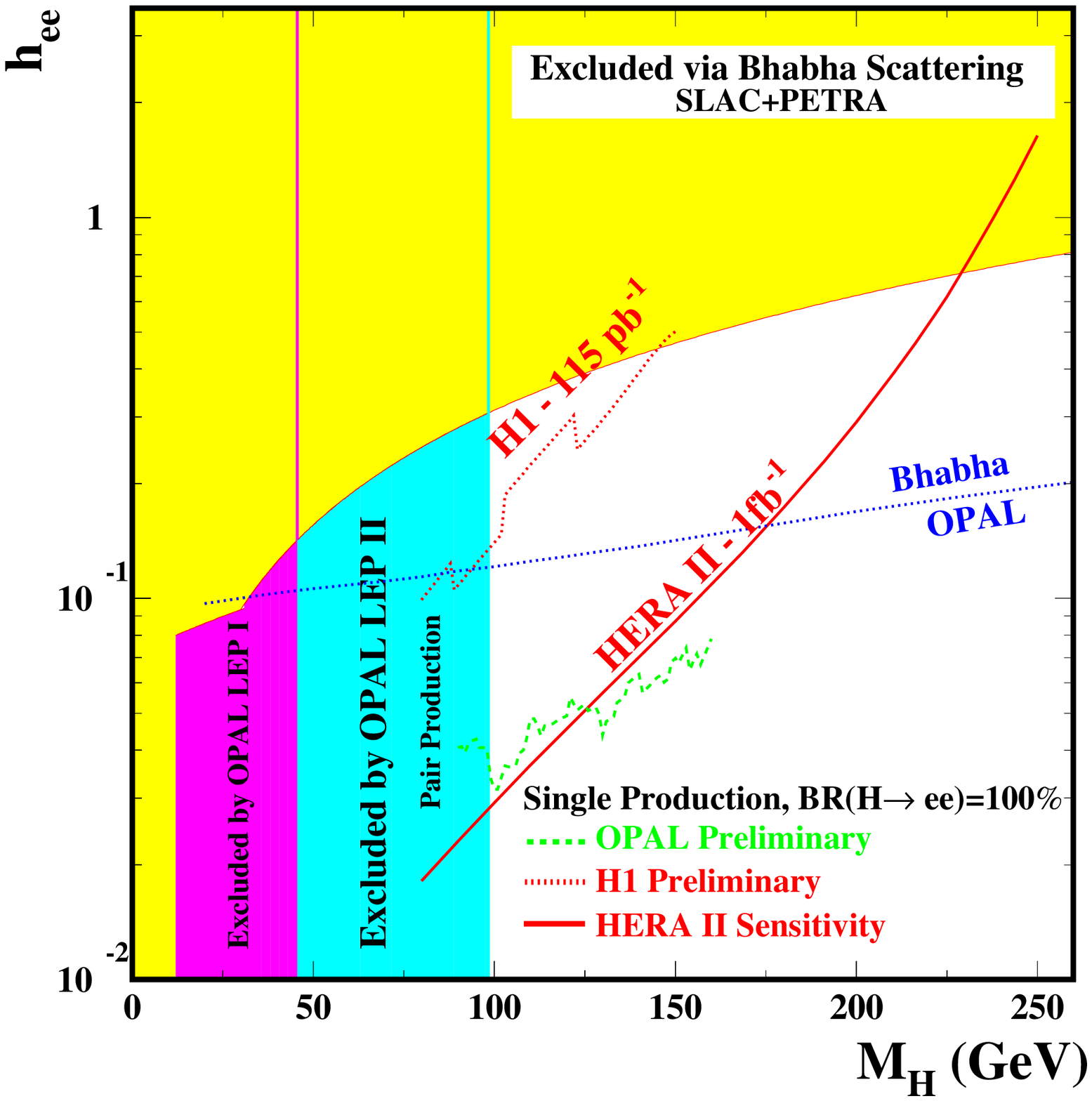,width=0.70\textwidth}

  \end{center}
 \vspace*{-0.5cm}
  
 \caption[]{ \label{fig:dqconstraints}
            Existing direct and indirect constraints on doubly charged Higgs
	    bosons in the coupling $h_{ee}$ {\it vs.} mass $M_{\Delta}$ plane.
	    Recent collider results from OPAL searches~\cite{OPAL02DQ,OPALNEW02DQ} at 
	    LEP$_{II}$ and from a direct search by H1~\cite{H1DQ2002} at
	    HERA$_{I}$ are shown together with estimated prospects for 
	    HERA$_{II}$. } 
\end{figure}

A summary of existing direct and indirect constraints on the
doubly charged Higgs boson is shown in~Fig.~\ref{fig:dqconstraints}
in the case of the $h_{ee}$ coupling as a function of
its mass.
%----------------------------------------------------------------------------
%--------------------------------------------------------------------------- 
Masses $M_{\Delta^{\pm\pm}} \lsim 45.6 \GeV$ have been excluded by the
OPAL experiment analyzing $Z^0$ decays at LEP$_I$~\cite{OPAL92DQ}.
As seen on~Fig.~\ref{fig:dqconstraints}, this was extended by OPAL to 
$M_{\Delta^{\pm\pm}} \lsim 98.5 \GeV$ (95\% CL)
in a search~\cite{OPAL02DQ} for pair production in the $s$-channel 
using LEP$_{II}$ data at centre-of-mass energies between $189 \GeV$ 
and $209 \GeV$. Similar results were derived by OPAL for any relative
values of the $h_{ee}$, $h_{\mu\mu}$ and $h_{\tau\tau}$ couplings, assuming a
$100 \%$ decay branching fraction into charged-lepton pairs.
A similar lower limit on $M_{\Delta^{\pm\pm}}$ has been obtained 
recently by the DELPHI experiment at LEP$_{II}$~\cite{DELPHIDQ} in 
the $\Delta^{\pm\pm} \rightarrow \tau^\pm \tau^\pm$ channel.
A search for single production of doubly charged Higgs bosons~\cite{OPALNEW02DQ} has 
been performed very recently by OPAL which also considered indirect effects
on measurements of Bhabha scattering at LEP$_{II}$. The resulting constraints
are shown in~Fig.~\ref{fig:dqconstraints}.

The constraints obtained at HERA$_I$ by the H1 experiment~\cite{H1DQ2002}
are also shown on~Fig.~\ref{fig:dqconstraints} together with an estimation of
the prospects for HERA$_{II}$.
For a coupling of electromagnetic strength, $h_{ee} = e$ with
$e \equiv \sqrt{4 \pi \alpha}$, doubly charged Higgs bosons which would
decay with $100\%$ branching into like-sign and like-flavour charged
leptons have been probed at HERA$_I$ for masses $M_{\Delta} \gsim 130 \GeV$.
The sensitivity at HERA$_{II}$ will be competitive with the one
of LEP$_{II}$, extending the mass reach to $M_{\Delta} \gsim 200 \GeV$.

%==========================================================================
\subsection{Bileptons}
\label{sec:bileptons}
%=================

The doubly charged Higgs scalars discussed above can also be seen as 
a special kind of particle (the scalars) among those coupling to
like-sign lepton pairs and generically called ``bileptons''.
Another kind of bileptons that have received considerable attention 
in the literature are vector 
bileptons~\footnote{There is currently no agreed convention in the
literature for the usage of the name ``bilepton''. In a recent general
classification~\cite{Davidson98}, the name has been used to designate 
any bosons coupling to pairs of leptons including e.g. electroweak 
bosons carrying lepton number $L = 0$.
Here we rather reserve the name to scalar or vector bosons carrying 
$\mid L \mid = 2$. Some authors have used the nomenclature
``bilepton'' or ``dilepton'' to designate more specifically vector bosons
carrying $\mid L \mid = 2$.}.

Vector bileptons originally appeared in extensions of the Standard Model 
such as the ``lepton-triplet theories''~\cite{Wilczek77} and in GUT 
theories such as $SU(15)$~\cite{Frampton90a}.
But most of the recent discussions have focused on a chiral theory
where the three fermion generations and family structure of the Standard 
Model is embedded in a $SU(3)_C \times SU(3)_L \times U(1)_Y$ gauge 
group~\cite{Frampton92,Montero02}. 
The symmetries of this $3-3-1$ group are assumed to break down to those 
of the Standard Model at low energies.
This spontaneous breaking requires the existence of four new massive 
gauge vector bosons carrying lepton number $L = \pm 2$
and an additional neutral gauge boson $Z'$~\cite{Frampton92}.
The bileptons appear in doublets formed of doubly charged and 
singly charged members ($X^{\pm\pm},X^{\pm}$).

At high energies, the particle content in the leptonic sector is exactly 
the same as in the Standard Model but there is a symmetry among the 
$l^-_i$, $l^+_i$, and $\nu_i$ of each generation $i$, and these come in 
triplets of $SU(3)_L$. The quark sector is more complex and the
quark content must be enlarged beyond ordinary quarks.
In particular, at least one of the quark generation must be treated 
differently from the other two. 
As a generic feature, models containing doubly charged bileptons
require the existence of one exotic quark of electric charge $Q_{em}= +5/3$ 
in either one of the generations and two $Q_{em}= -4/3$ 
quarks~\cite{Wilczek77,Frampton92,Kim81,Pisano92,Montero02} in the other two
generations.
This special treatment of the quarkonic sector is required to ensure anomaly 
cancellation for exactly three fermion generations. 
Turning the argument around, it has been seen as a virtue of such models 
(in contrast to the Standard Model)
that the number of fermion generations, which is related to the number 
of quark colours through the requirement of anomaly cancellation,
is thus ``predicted'' to be exactly three. 

A comprehensive review of existing indirect constraints on bileptons
has been recently performed~\cite{Davidson98} in a general approach
considering the Lagrangian for all possible bilepton-lepton-lepton
couplings consistent with electroweak symmetries. The allowed bilepton
states in such an approach are listed in Table~\ref{tab:cdbil}.
% ------------------ TABLE : L=2 Bileptons  -------------------------
\begin{table*}[htb]
  \renewcommand{\doublerulesep}{0.4pt}
  \renewcommand{\arraystretch}{1.2}
 \vspace{-0.1cm}

\begin{center}
  \begin{tabular}{||c|c|c|c|c|c||}
    \hline \hline
     \multicolumn{6}{||c||}{{\large $|\rm L|$=2 \,\, Bileptons }} \\ \hline
     Type  & Spin & $T_3$  &$Q_{em}$& Coupling to & siblings \\
     \hline \hline
  $L_{1}$  &  0   &   0    &    1   & $l_L \nu_L$ &                  \\ \hline
  $\tilde{L}_{1}$  
           &  0   &   0    &    2   & $l_R l_R$   &                  \\ \hline
  $L_{2}^{\mu}$  
           &  1   & $-1/2$ &    1   & $e_R \nu_L$ &  
	                                ($X^-$,$X^{--}$) {\it doublet in} \\
           &      & $+1/2$ &    2   & $e_R e_L$   & 3-3-1 {\it models} \\ \hline
  $L_{3}$  
           &  0   & $-1$ &    0   & $\nu_L \nu_L$ & 
{\it ``Left-handed''} ($\Delta^0$,$\Delta^-$,$\Delta^{--}$) {\it Higgs triplet} \\
           &      &   0  &    1   & $e_L \nu_L$   &  
{\it in Left-Right symmetric models}                           \\
           &      & $+1$ &    2   & $e_L e_L$     &                  \\ \hline
    \hline
  \end{tabular}
  \caption {\small \label{tab:cdbil}
               Bileptons with leptonic number $|L|=2$ in the 
	       Cuypers-Davidson effective model~\cite{Davidson98}
	       from a most general lepton-number conserving and
	       renormalisable Lagrangian consistent with 
	       $SU(2)_L \times U(1)_Y$ symmetries.
	       The bileptons are grouped in isospin families.
	       The allowed states can be distinguished by their
	       spin, the third component $T_3$ of their weak isospin, 
	       and their electric charge $Q_{em}$.
	       The bilepton-lepton-lepton couplings are left as free
	       parameters in the model. The coupling matrix for 
	       the $L_{1}$ isosinglet is by construction 
	       antisymmetric in flavour space while those of the
	       $\tilde{L}_{1}$ isosinglet and $L_{3}$ 
	       isotriplet are symmetric. The isodoublet $L_{2}^{\mu}$
	       is left with an arbitrary $3 \times 3$ coupling matrix.}

\end{center}
\end{table*}
% ------------------------------------------------------------------------

\begin{figure}[htb]
  \begin{center}
\vspace*{-2.0cm}
  \begin{tabular}{ccc}
  \vspace*{-0.5cm}
    
 \hspace*{-1.0cm} \epsfig{file=\master/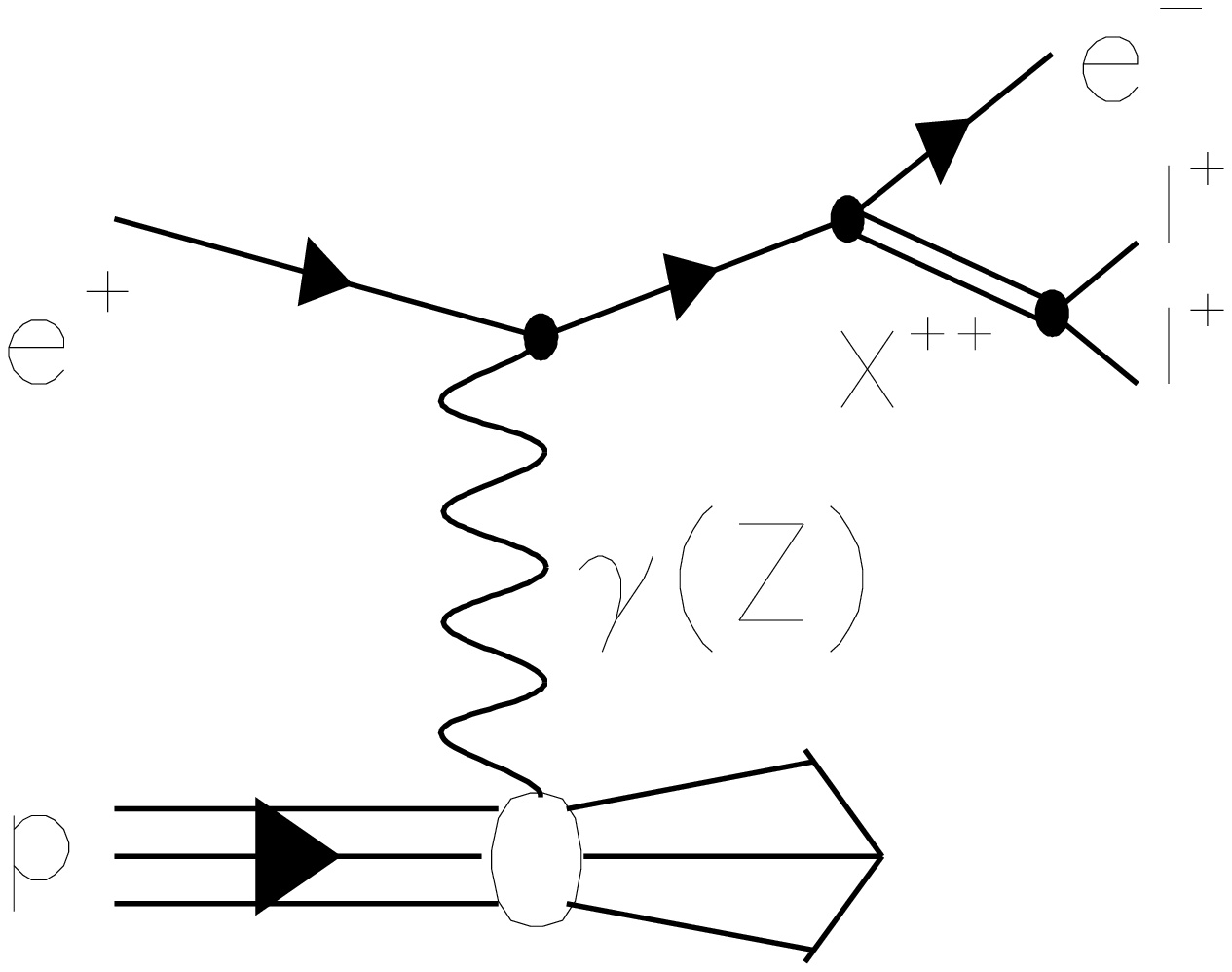,width=0.40\textwidth}
&
 \hspace*{-1.6cm} \epsfig{file=\master/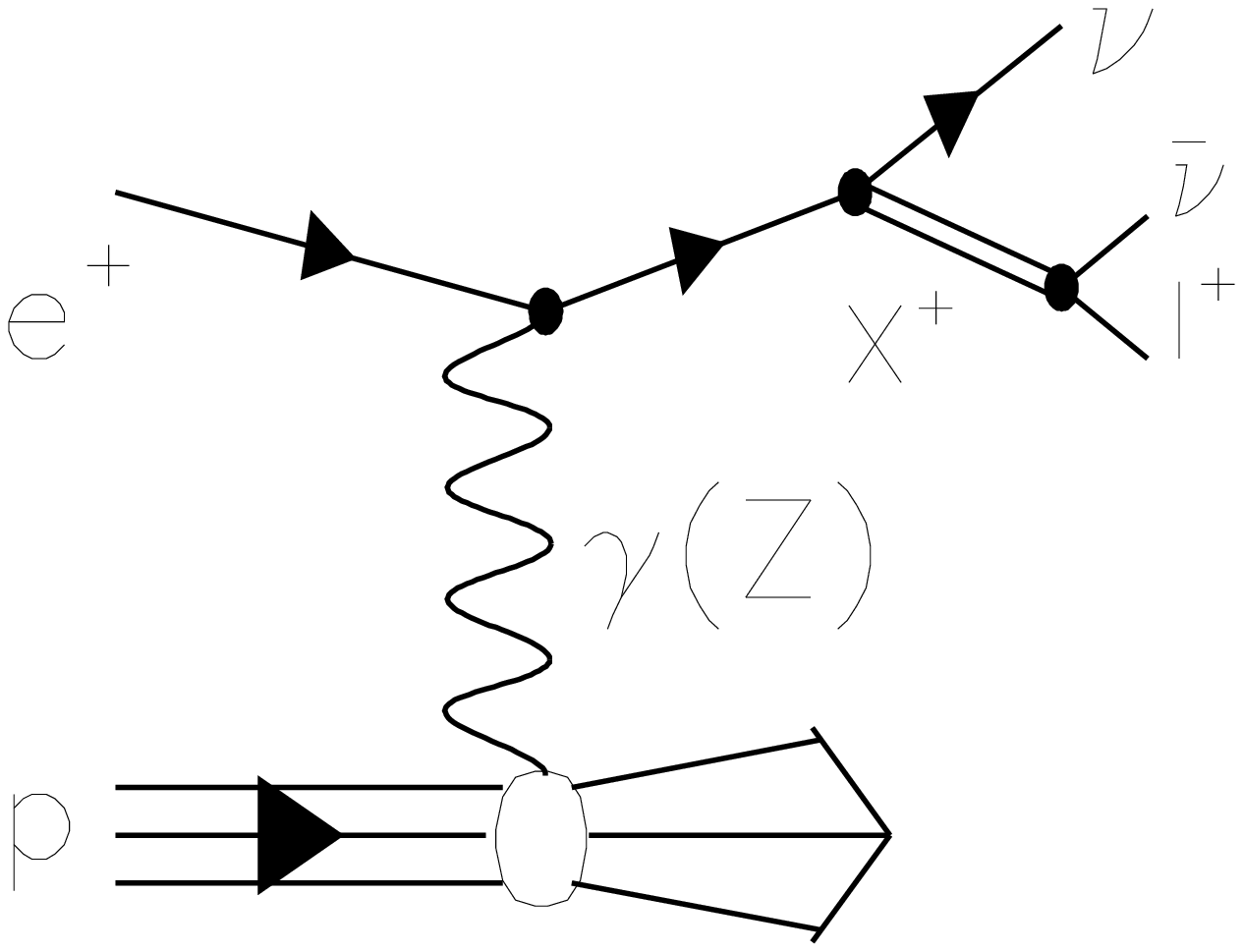,width=0.44\textwidth}
&
 \hspace*{-2.6cm} \epsfig{file=\master/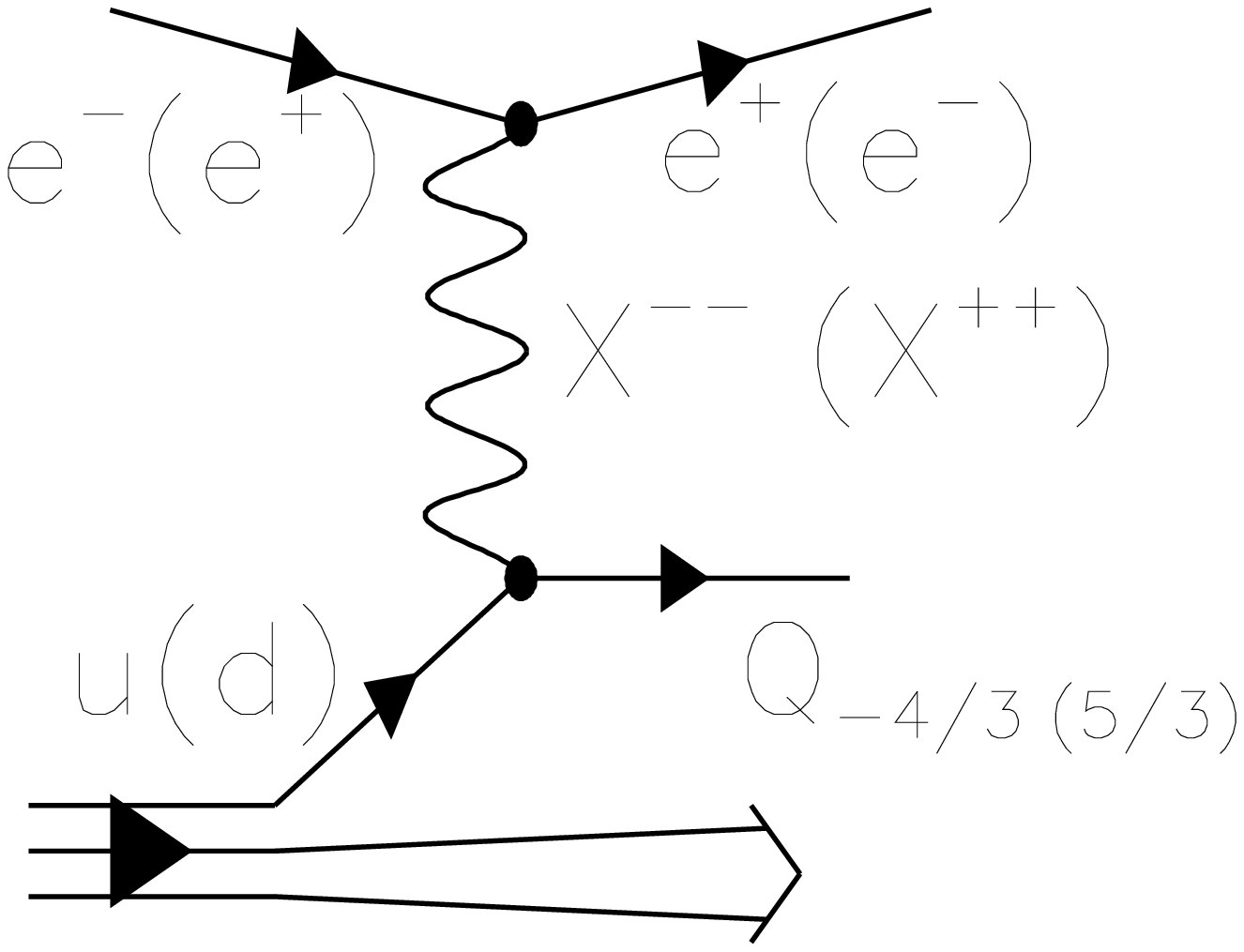,width=0.44\textwidth}

  \end{tabular}
  \end{center}
 \vspace*{-1.3cm}
 \hspace*{3.0cm} (a) \hspace*{5.0cm} (b) \hspace*{5.0cm} (c) \\  
 \vspace*{-0.5cm}
  
 \caption[]{ \label{fig:diagepdlp}
            Typical diagrams for bilepton production at lepton-hadron 
	    colliders.
            a) single production of a doubly charged $X^{++}$;
            b) single production of singly charged $X^{+}$;	    
            c) $t$-channel exchange of a $X^{--}$ ($X^{++}$) bilepton 
	              in $e^-p$ ($e^+p$) leading to the creation of an exotic 
		      quark of electric charge $-4/3$ ($+5/3$).} 
\end{figure}

The phenomenology aspects of vector-bilepton production has been studied
in the literature for $e^+e^-$~\cite{Rizzo92,Lepore94},
$p \bar{p}$~\cite{Dion99} and $ep$~\cite{Agrawal92,Coutinho99} colliders.
As in the case of doubly charged Higgs scalar bosons, the doubly charged
vector bileptons $X^{\pm\pm}$ could be a source of spectacular multi-lepton 
events. 
A diagram for single production of a doubly charged $X^{--}$ at an
$ep$ collider is shown in Fig.~\ref{fig:diagepdlp}a.
%----------------------------------------------------------------------------
%-----------------------------------------------------------------------------
The production of a singly charged bilepton at the $ep$ collider could also
lead in principle to striking event topologies as seen in 
Fig.~\ref{fig:diagepdlp}b. In case real production of a dilepton would
turn out to be inaccessible at the HERA$_{II}$ collider, striking event
topologies could still be expected from a $t$-channel exchange of a
doubly charged bilepton leading to the creation of an exotic quark as shown
in Fig.~\ref{fig:diagepdlp}c.
In 3-3-1 models containing an exotic quark with $Q_{em}= -4/3$ in the first
generation~\cite{Frampton92}, the dominating process would be 
$e^- p \rightarrow e^+ X$ via the exchange of a $X^{--}$.
In contrast, 3-3-1 models containing the exotic quark with $Q_{em}= 5/3$ in
the first generation~\cite{Pisano92} prefer $e^+ p \rightarrow e^- X$ via the 
exchange of a $X^{++}$.
Such processes have been studied in Refs.~\cite{Sasaki95,Coutinho99}. 
Sizeable cross-sections are expected at HERA$_{II}$
for the exchange of doubly charged bileptons with masses from
$\simeq 200$ up to $1000 \GeV$ associated with exotic quarks 
with masses below $\simeq 200$ down to $100 \GeV$, respectively.
 
Having dared to contemplate the possibility of creating exotic quarks
at $ep$ colliders in processes involving the virtual exchange of 
bileptons, it is only fair to mention that the possible existence
of exotic quarks and leptons is a subject by itself which has been
thoroughly reviewed recently in a very general context~\cite{Frampton00}.
Direct searches for a fourth-generation quark of charge $Q_{em} = -1/3$ 
have been performed by the D$\emptyset$ and CDF experiments
at the Tevatron~\cite{D0CDFQQ} with a sensitivity to masses reaching
$\sim 200 \GeV$.

% Excited states of fermions, compositeness
\clearpage
%%%%%%%%%%%%%%%%%%%%%%%%%%%%%%%%%%%%%%%%%%%%%%%%%%%%%%%%%%%%%%%%%%%%%%%%%%%%%%
\section{Excited States of Fermions}
\label{sec:FSTAR}
%%%%%%%%%%%%%%%%%%%%%%%%%%%%%%%%%%%%%%%%%%%%%%%%%%%%%%%%%%%%%%%%%%%%%%%%%%%%%%

\begin{figure}[htb]
 \vspace*{-1.0cm}
  \begin{center}                                                                
  \begin{tabular}{ccc}
  \vspace*{-0.2cm}
    
  \hspace*{-1.0cm} \epsfig{file=\master/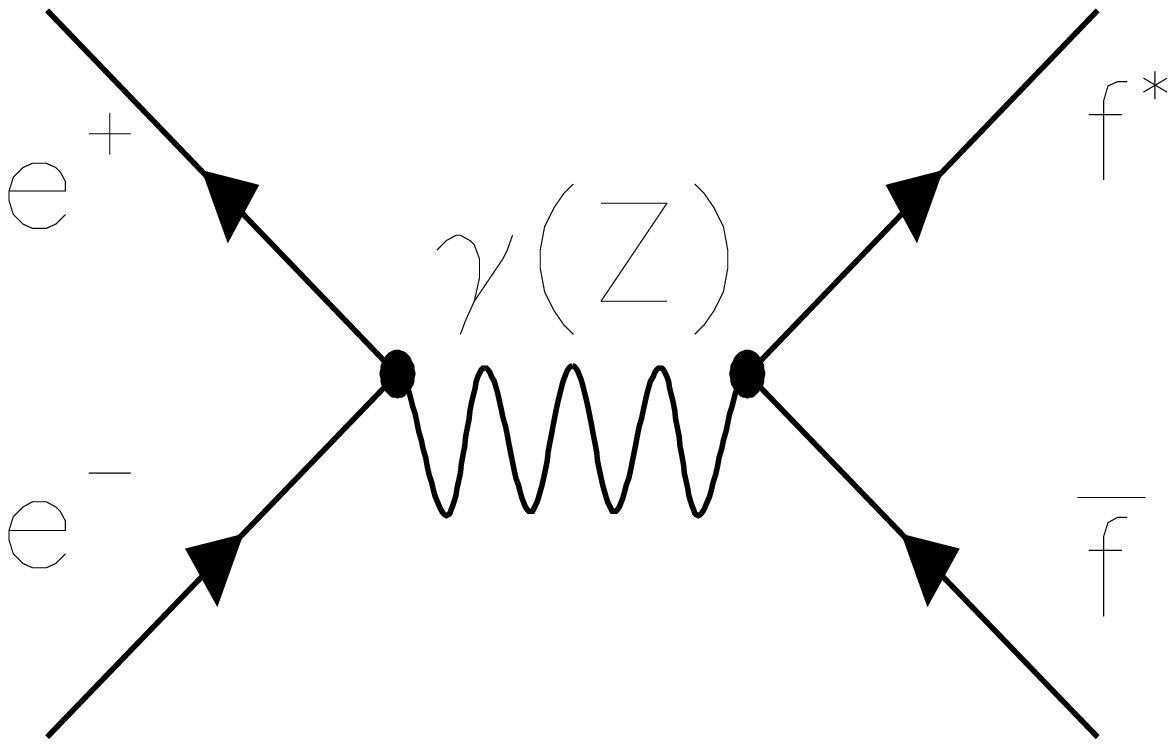,width=0.42\textwidth}
&
  \hspace*{-2.0cm} \epsfig{file=\master/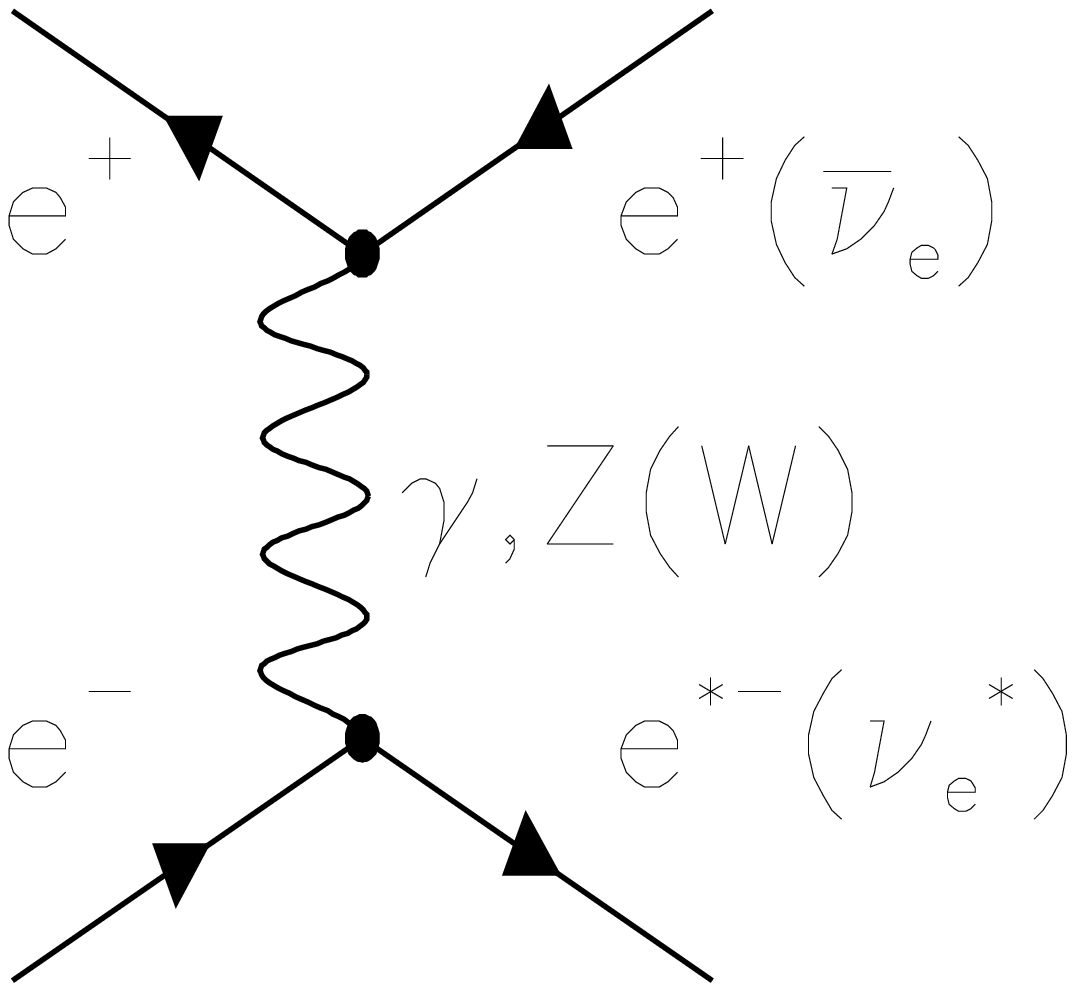,width=0.42\textwidth}
&
  \hspace*{-2.0cm} \epsfig{file=\master/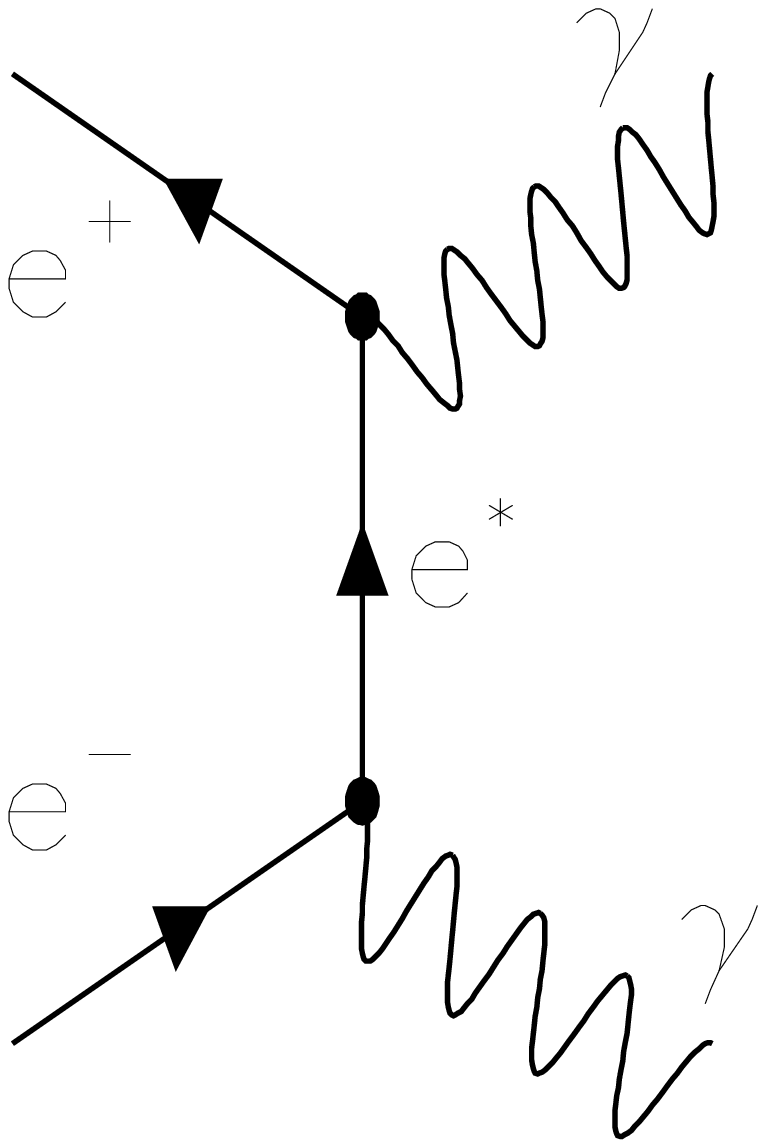,width=0.42\textwidth}

  \end{tabular}
  \end{center}
 \vspace*{-2.8cm}
 \hspace*{3.0cm} (a) \hspace*{5.5cm} (b) \hspace*{5.5cm} (c) \\  
 \vspace*{-0.5cm}
  
 \caption[]{ \label{fig:fstar3}
            a) single $f^*$ production from $s$-channel
            $\gamma,Z$ at LEP;
            b) single $e^*(\nu^*)$ production from $t$-channel
            process at LEP;
            c) virtual $e^*$ exchange in $e^+e^- \to \gamma\gamma$.}
\end{figure}

\begin{figure}[htb]
  \begin{center}                                                                
  \begin{tabular}{cc}
  \vspace*{-0.2cm}
    
  \hspace*{-1.0cm} \epsfig{file=\master/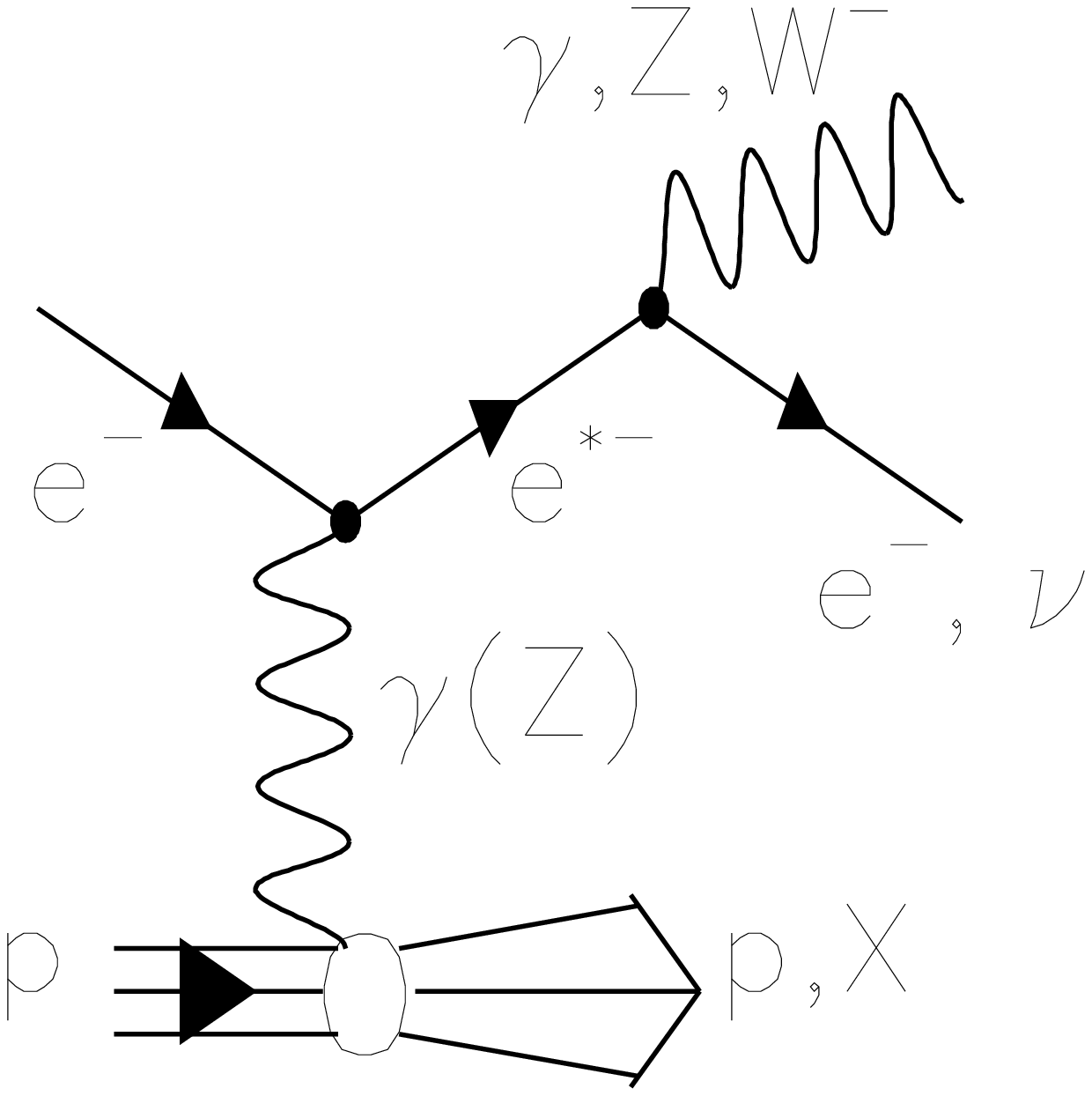,width=0.42\textwidth}
&
  \hspace*{-1.0cm} \epsfig{file=\master/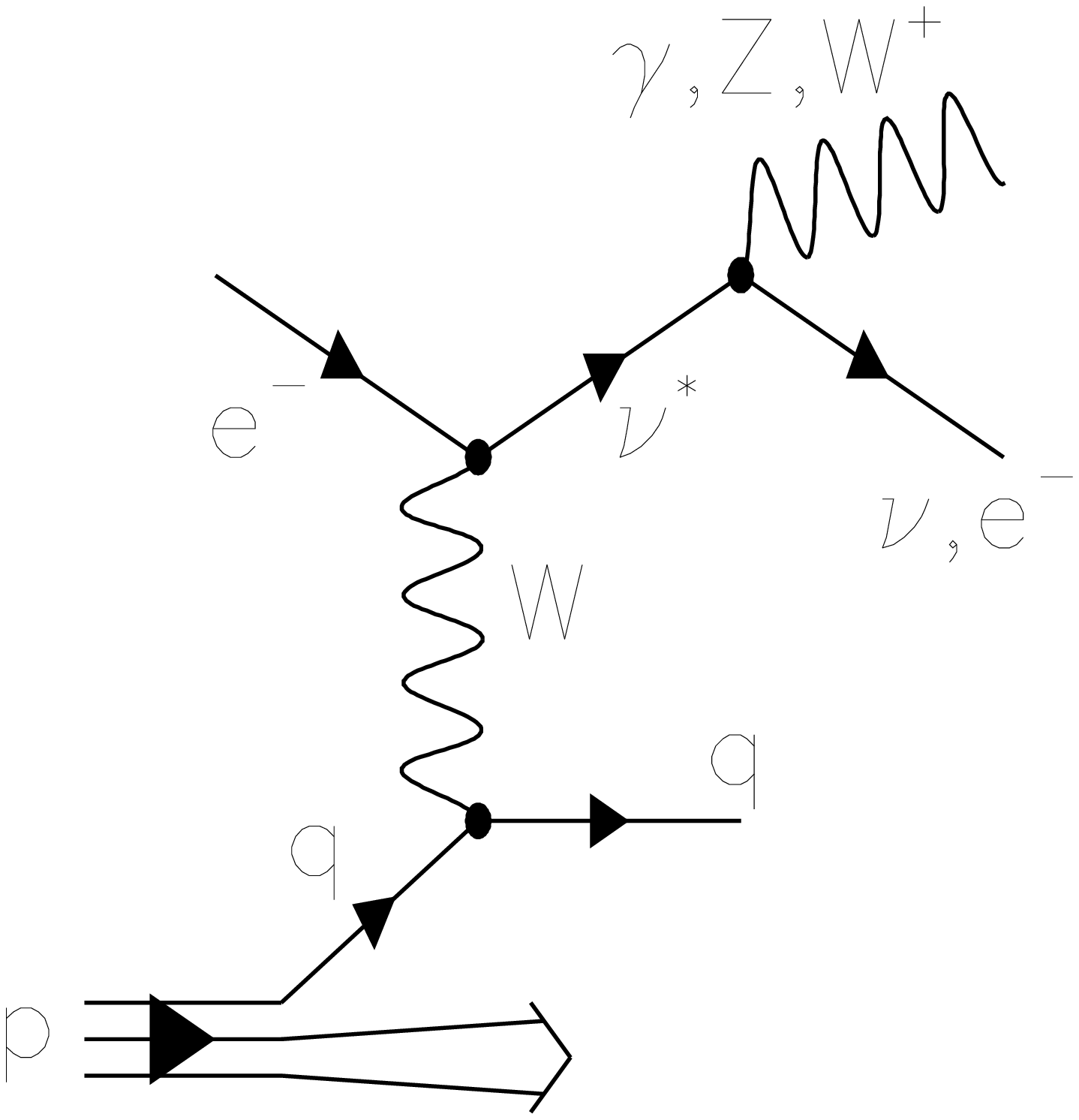,width=0.42\textwidth}

  \end{tabular}
  \end{center}
 \vspace*{-1.3cm}
 \hspace*{3.5cm} (a) \hspace*{7.0cm} (b) \\  
 \vspace*{-0.5cm}
  
 \caption[]{ \label{fig:fstar1}
            a) $e^*$ production and decay at HERA;
            b) $\nu^*$ production and decay at HERA.}
\end{figure}

\begin{figure}[htb]
  \begin{center}                                                                
  \begin{tabular}{cc}
    
  \hspace*{-1.0cm} \epsfig{file=\master/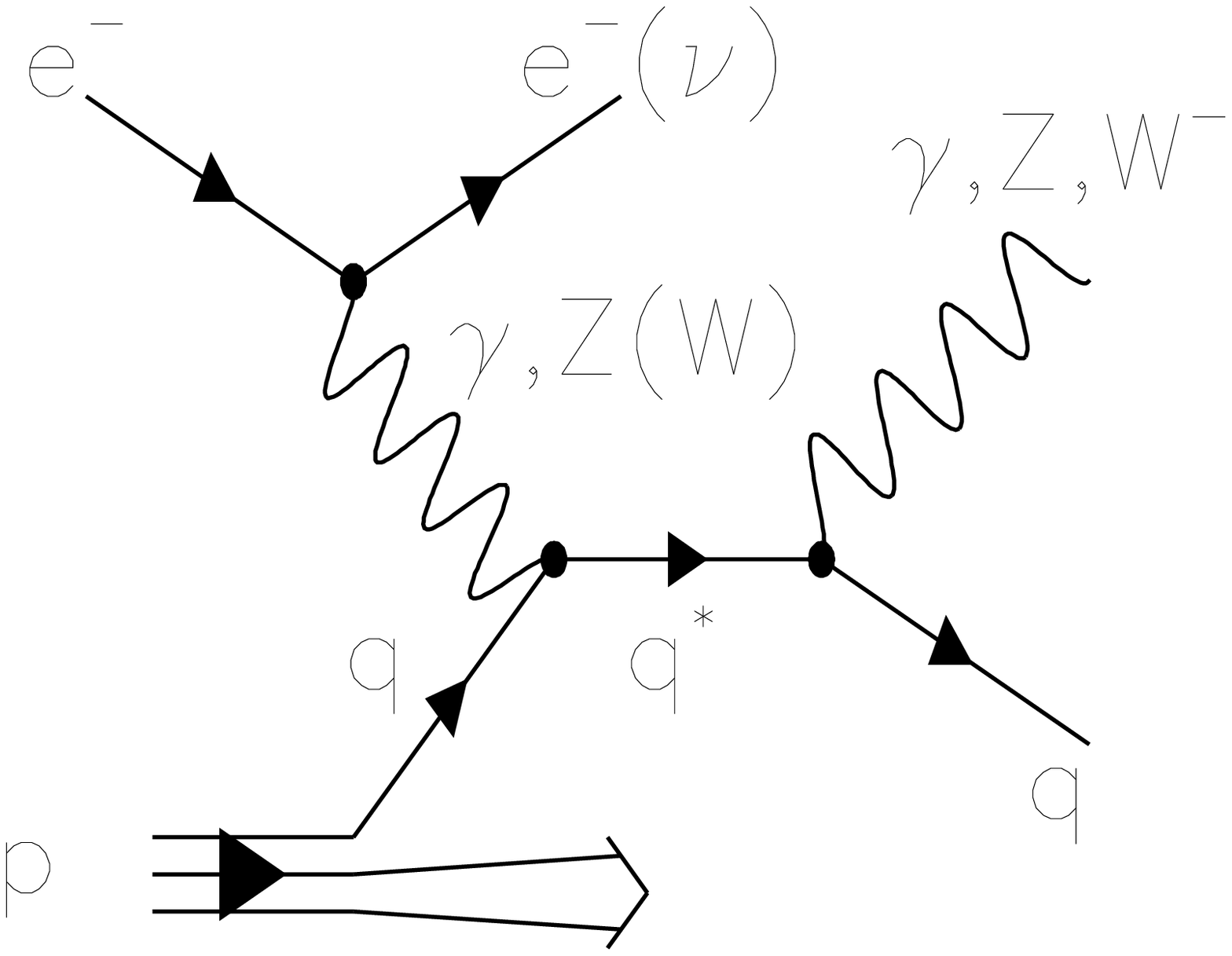,width=0.42\textwidth}
&
  \vspace*{-0.2cm}
  \hspace*{-0.0cm} \epsfig{file=\master/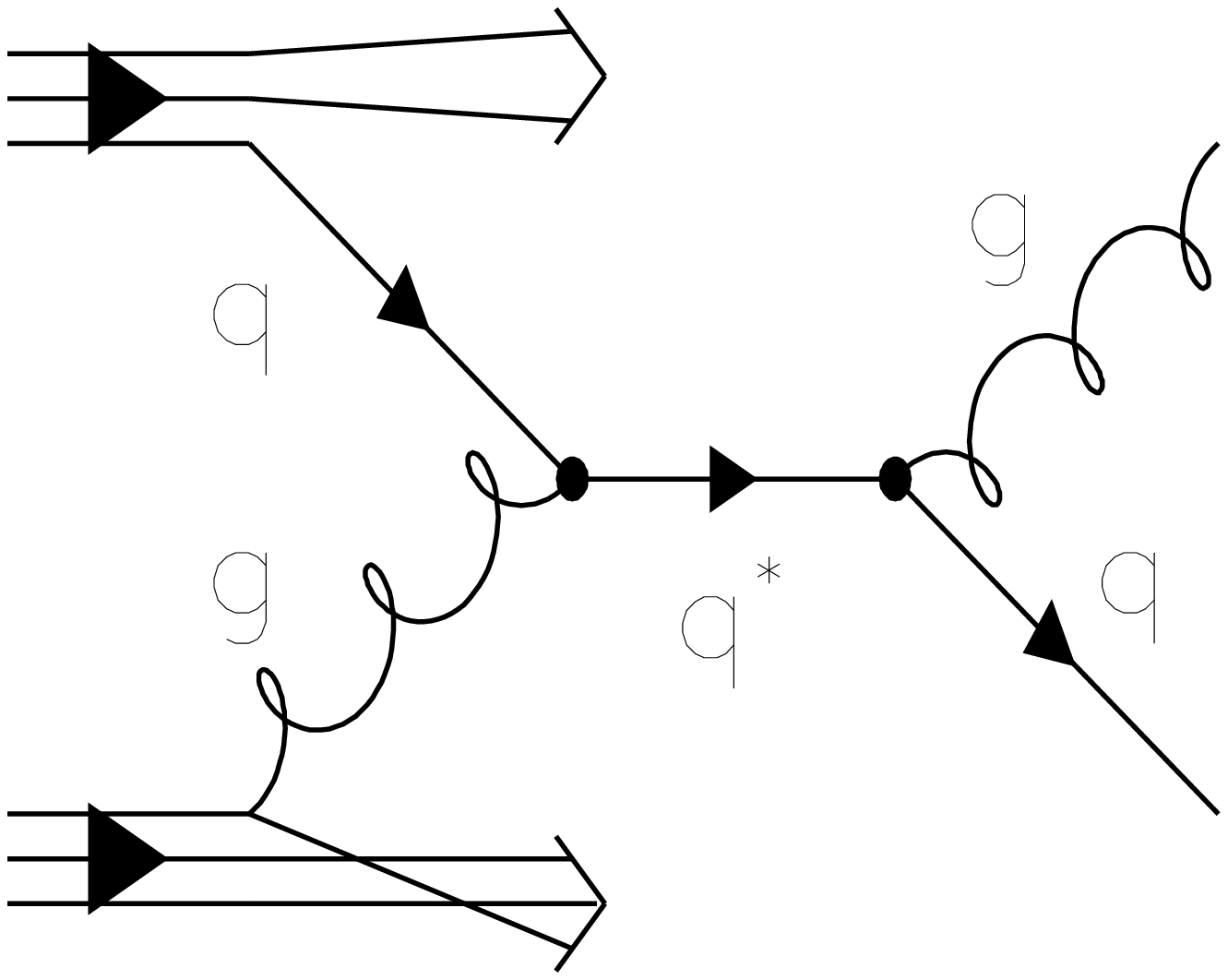,width=0.42\textwidth}

  \end{tabular}
  \end{center}
 \vspace*{-2.3cm}
 \hspace*{3.5cm} (a) \hspace*{7.0cm} (b) \\  
 \vspace*{-0.5cm}
  
 \caption[]{ \label{fig:fstar2}
            a) $q^*$ production and decay at HERA;
            b) $q^*$ production at Tevatron.}
\end{figure}

The wide spectrum of ``elementary constituents of matter'', i.e.
the repetition of lepton and quark multiplets over three generations,
leads one to speculate that they may not be the ultimate elementary
particles but rather composite objects consisting of more fundamental
entities.  In this hypothesis, it is possible that excited states of
fermions exist, at a mass scale comparable to the dynamics of the new
``binding force''.  They may be produced at energy-frontier colliders
and would decay back ``radiatively'' into an ordinary fermion and a
gauge boson (photon, W, Z or gluon).
(Figs. \ref{fig:fstar3}, \ref{fig:fstar1}, \ref{fig:fstar2})

The magnetic transition between the ordinary and excited fermions
was formulated in the literature~\cite{Hagiwara85, Baur90, Boudjema93}
in the following Lagrangian:

\begin{equation*}
\label{equ:exflagr} 
{\mathcal L}_{f^*f} = \frac{1}{2\Lambda}\bar{f^*_R}\sigma^{\mu\nu}
\left[g  f  \frac{\vec\tau}{2}\vec W_{\mu\nu} + 
      g' f' \frac{Y}{2}            B_{\mu\nu} +
      g_sf_s\frac{\lambda^a}{2}    G_{\mu\nu}^a \right]f_L + h.c.,
\end{equation*}

where $\vec W_{\mu\nu},B_{\mu\nu}$ and $G_{\mu\nu}^a$ are the field-strength
tensors of the SU(2), U(1) and SU(3) gauge fields, $\vec\tau, Y$ and
$\lambda^a$ are the corresponding gauge-group generators, and $g, g'$ and
$g_s$ are the gauge coupling constants, respectively.
$\Lambda$ is the compositeness scale and $f, f'$ and $f_s$ are weight
parameters associated with the three gauge groups that are determined by
the unknown composite dynamics.

Once produced, excited fermions $f^*$ can decay back to the ground 
state by radiating a boson. For colour-neutral $f^*$,
in the limiting case $f = -f'$ ($f = f'$), the coupling  $\gamma e e^*$
($\gamma \nu \nu^*$) vanishes and the decay must involve a $Z$ or $W$
boson. For excited quarks, the decay $q \rightarrow q g$ will generally 
dominate if $ \mid f_s \mid \simeq \mid f \mid \simeq \mid f' \mid$.

In $e^+e^-$ collisions, the dominant contribution to the pair production 
of charged excited fermions is $s$-channel 
$\gamma$ and $Z$ exchange in reactions 
$e^+ e^- \rightarrow l^* \bar{l^*} \,;\, \nu^* \bar{\nu^*}$.
%-----------------------------------------------------------------------------
%-----------------------------------------------------------------------------
In the case of excited neutrinos, only the $Z$ exchange contributes.
Single production, described by the above Lagrangian, also proceeds
through $s$-channel $\gamma$ and $Z$ exchange in reactions 
$e^+ e^- \rightarrow l l^* \,;\, \nu \nu^*$ (see
Fig.~\ref{fig:fstar3}a). 
For single production of $e^*$ ($\nu^*$), important additional contributions 
come from $t$-channel $\gamma$ and $Z$ ($W$) exchange (Fig.~\ref{fig:fstar3}b). 
Furthermore, $t$-channel exchange of a virtual $e^*$ will give additional
contribution to $e^+e^- \to \gamma\gamma$ events, allowing LEP2 data to
constrain the $e^*$ mass domain above its centre-of-mass energy (Fig.~\ref{fig:fstar3}c).
%----------------------------------------------------------------------------
%-----------------------------------------------------------------------------

At HERA, excited fermions could be produced via $t$-channel exchange of
gauge bosons as shown in Figs.~\ref{fig:fstar1},~\ref{fig:fstar2}.
The $e^*$ production has a significant contribution from (quasi-)elastic
production, $ep \to e^*p (e^*N)$.  The $\nu^*$ production is a charged
current reaction, resulting in much larger production cross-section
for $e^-p$ collisions compared to $e^+p$.
%-----------------------------------------------------------------------------
%-----------------------------------------------------------------------------

The Tevatron has a large discovery potential
for excited quarks, provided they have considerable SU(3) coupling
strength ($f_s$), such that production via quark-gluon fusion (see
Fig.~\ref{fig:fstar2}) becomes significant.  The signal would be an
enhancement in the dijet invariant-mass distribution.

Usually experimental constraints are derived by assuming certain
relations between $f, f'$ and $f_s$, by which the decay branching
ratios of the excited fermions are fixed and limits are set on
the single quantity $f/\Lambda$.
%----------------------------------------------------------------------------
\begin{figure}[htb]
  \begin{center}                                                                
  \begin{tabular}{cc}

  \vspace*{-0.2cm}

  \hspace*{-0.5cm} \epsfig{file=\master/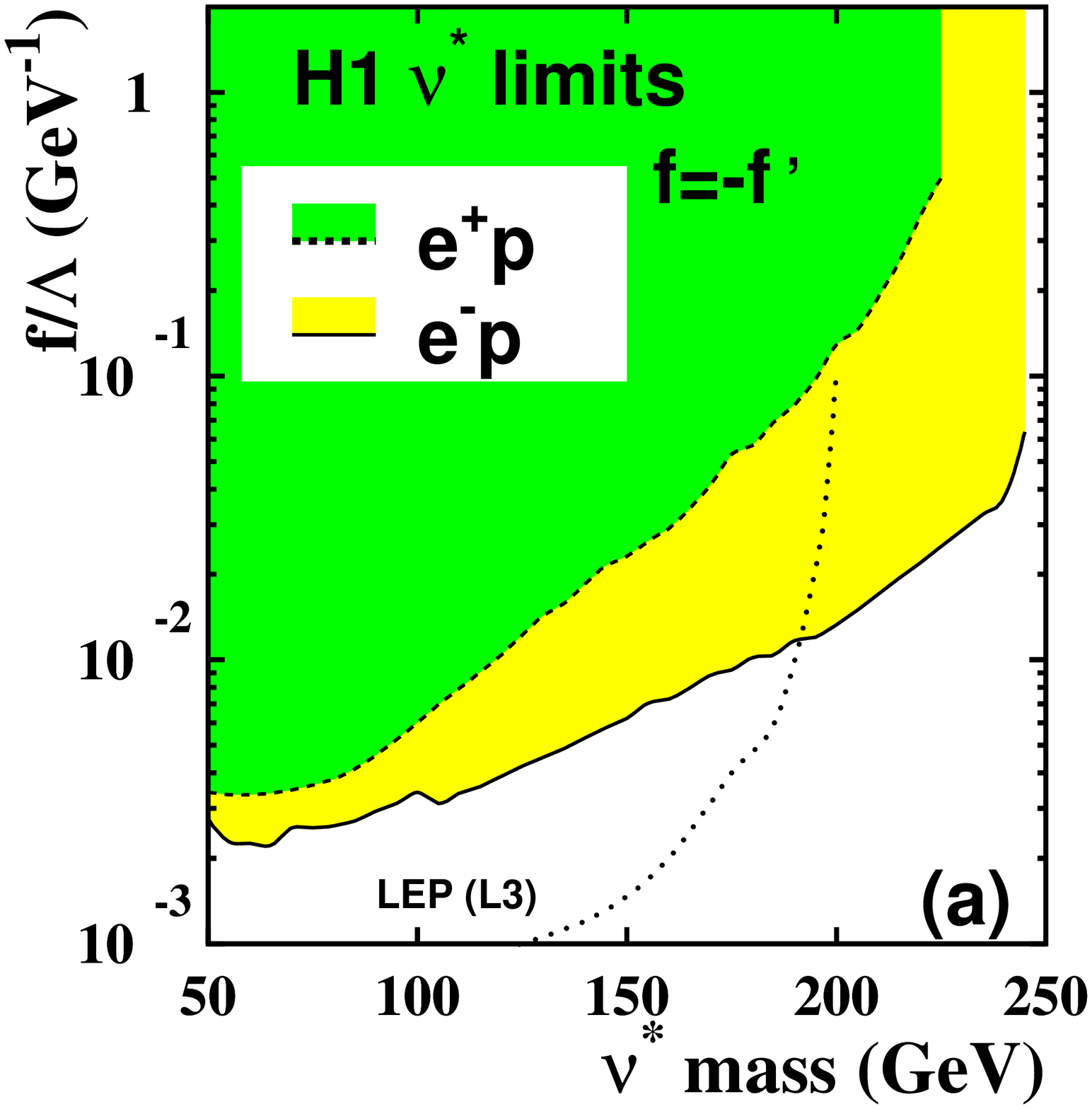,width=0.50\textwidth}
 &
  \hspace*{-0.5cm} \epsfig{file=\master/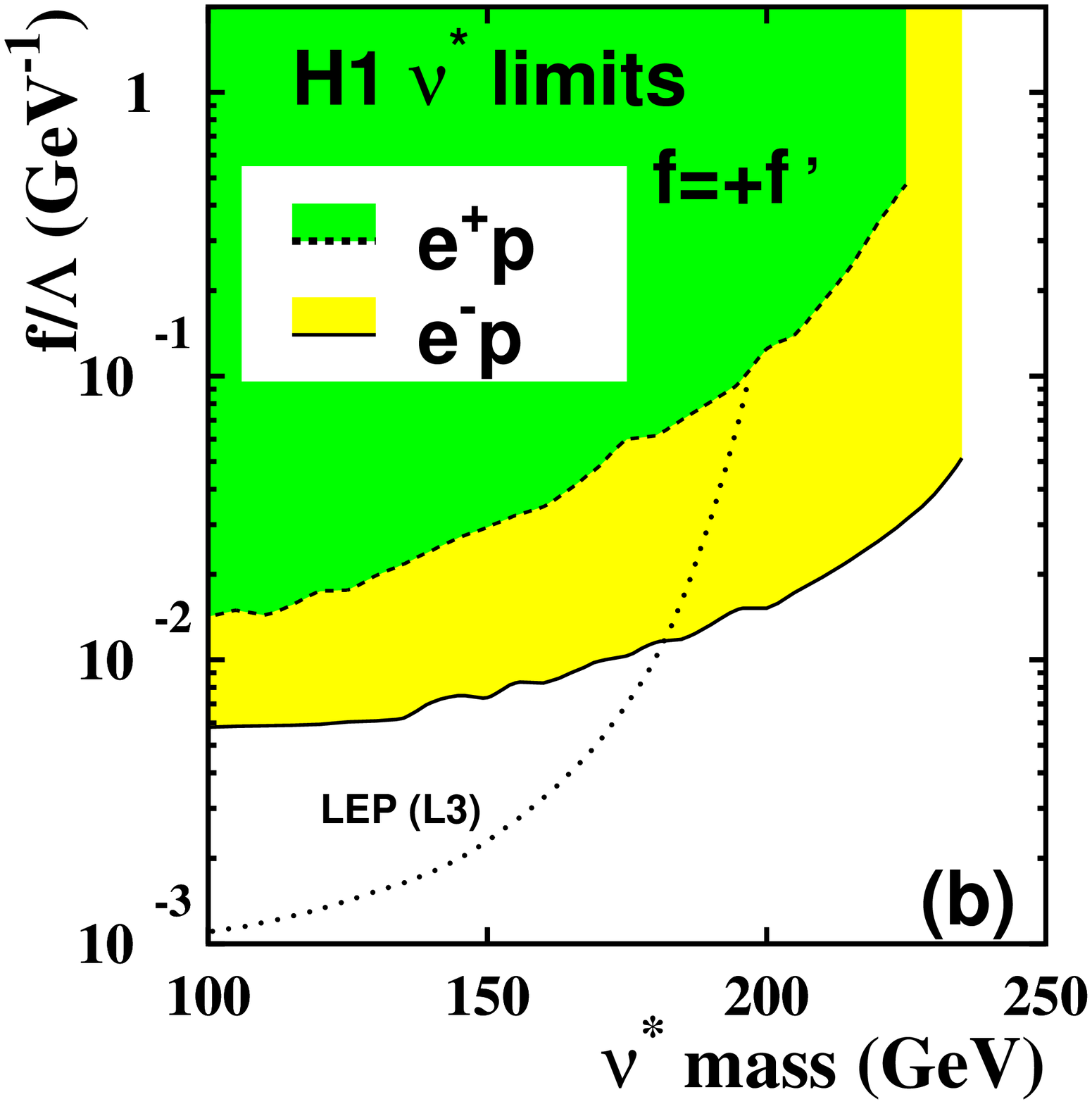,width=0.50\textwidth}

  \end{tabular}
  \end{center}
 
  \caption[]{ \label{fig:nustarlim}
            a)  Constraints on excited neutrinos for $f = -f'$.
	    b) Constraints on excited neutrinos for $f = +f'$.} 
\end{figure}
%--------------------------------------------------------------------

The limits on $\nu^*$ from H1~\cite{H19899nustar} and L3~\cite{L3nustar}
are shown in Fig.~\ref{fig:nustarlim}.
Corresponding results from ZEUS can be found in Ref.~\cite{ZEUSfstar}.
The $f=+f'$ case has a vanishing branching ratio for the experimentally 
clean decay mode $\nu^*\to\nu\gamma$, thus giving worse limits.
The different sensitivities between $e^-p$ and $e^+p$ data are evident.
It can be seen that HERA offers a higher sensitivity than LEP
for $\nu^*$ masses above $\simeq 200 \GeV$.
%----------------------------------------------------------------------------
\begin{figure}[htb]
  \begin{center}                                                                
\begin{tabular}{cc}

  \hspace*{-0.5cm} \epsfig{file=\master/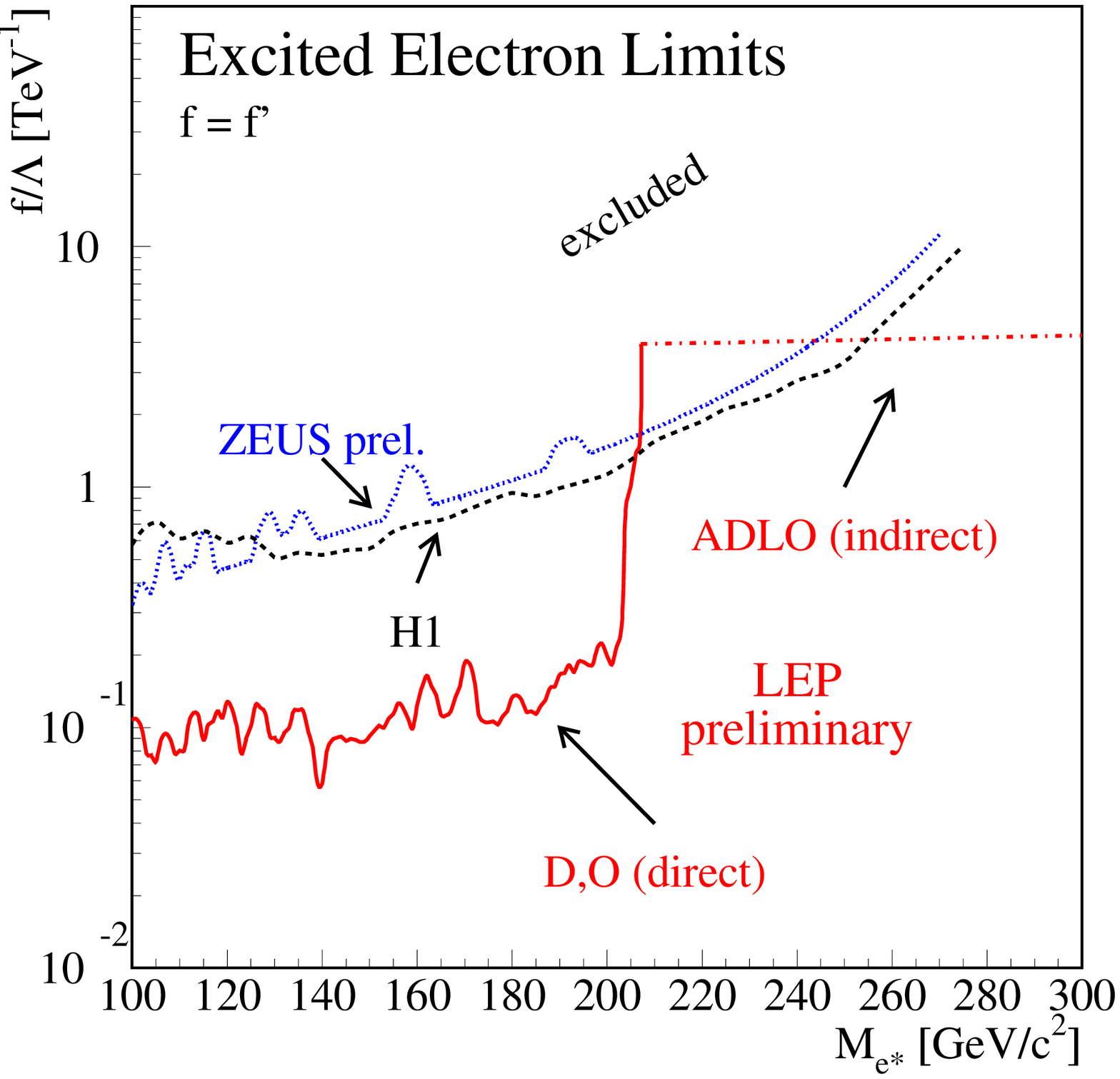,width=0.55\textwidth}
&
  \hspace*{-1.3cm} \epsfig{file=\master/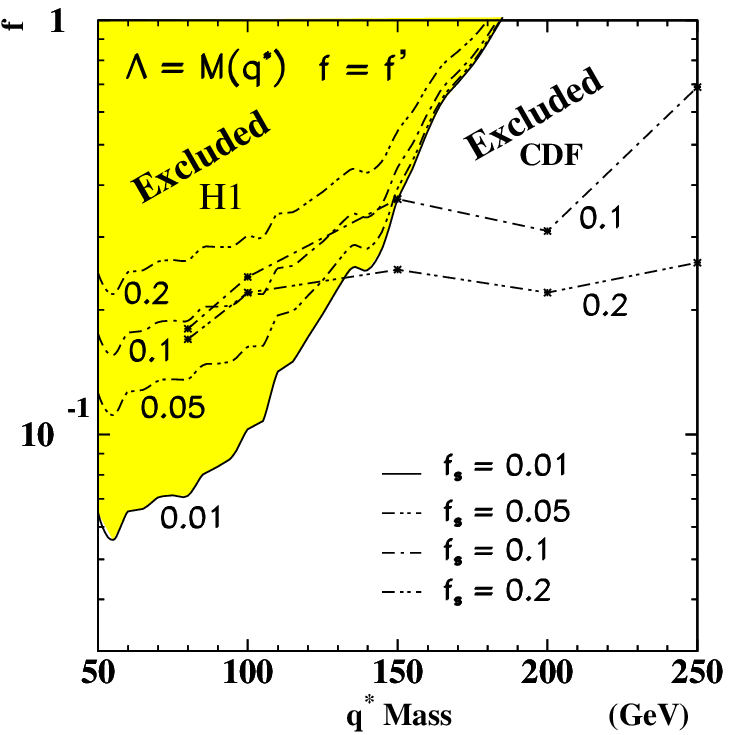,width=0.50\textwidth}

\end{tabular}
  \end{center}
  \vspace*{-0.2cm}
   
 \vspace*{-0.3cm}
 \hspace*{5.0cm} (a) \hspace*{8.0cm} (b) \\  
 \vspace*{-0.5cm}
  \caption[]{ \label{fig:eqstarlim}
           a)  Constraints on excited electrons from HERA and LEP.
 	   b)  Constraints on excited quarks from H1 and CDF.} 
\end{figure}
%----------------------------------------------------------------------------

The latest limits on $e^*$ from HERA~\cite{HERA2002estar} and
LEP~\cite{LEPFSTAR} are compared in Fig.~\ref{fig:eqstarlim}a.
Due to the indirect sensitivity of LEP to $e^*$ in the $\gamma\gamma$ final 
state, majority of the region excluded by HERA at high mass is also excluded 
by LEP. 
Nevertheless, the expected ten-fold luminosity increase at HERA$_{II}$ 
will allow to cover an unexplored domain for $e^*$ masses 
up to about $270 \GeV$.
In Fig.~\ref{fig:eqstarlim}b, limits on $f$ from
H1~\cite{H19497fstar} and CDF~\cite{CDFqstar} are shown for $q^*$ under the 
assumption $f=+f'$ and $\Lambda=M(q^*)$, and for different values of $f_s$.
It can be seen that HERA and the Tevatron have complementary sensitivities;
as long as $f_s$ is not very small, Tevatron sensitivity reaches to
very high $q^*$ masses (up to $760 \GeV$ for $f=f'=f_s=1$), while HERA has
a better sensitivity when $f_s$ is small or vanishing, i.e. excited quarks
are produced and decay predominantly with electroweak couplings.
        % Search for excited states of fermions

% Contact Interactions (compositeness, leptoquarks, ...) 
\clearpage
%%%%%%%%%%%%%%%%%%%%%%%%%%%%%%%%%%%%%%%%%%%%%%%%%%%%%%%%%%%%%%%%%%%%%%%%%%%%%%
\section{Contact Interactions}
\label{sec:contact}
%%%%%%%%%%%%%%%%%%%%%%%%%%%%%%%%%%%%%%%%%%%%%%%%%%%%%%%%%%%%%%%%%%%%%%%%%%%%%%

% Introduction/general CI concept:

\begin{figure}[htb]
  \vspace*{-0.1cm}

  \begin{center}
  \begin{tabular}{ccc}
     \mbox{\epsfxsize=0.25\textwidth 
           \epsffile{\master/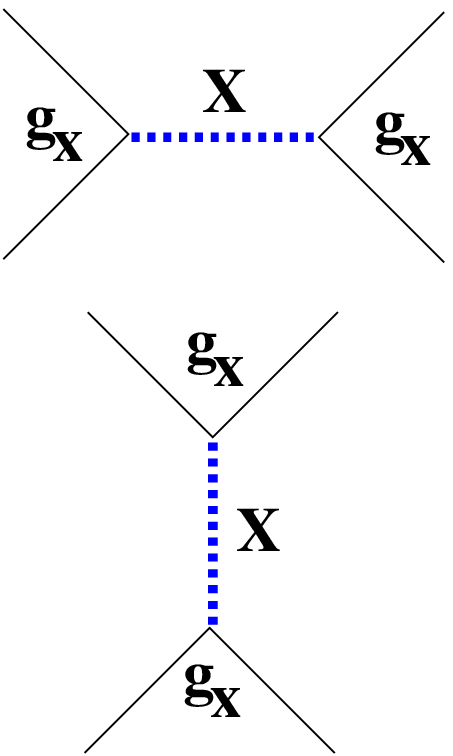}}
   &
     \raisebox{40pt}{
     \mbox{\epsfxsize=0.20\textwidth 
           \epsffile{\master/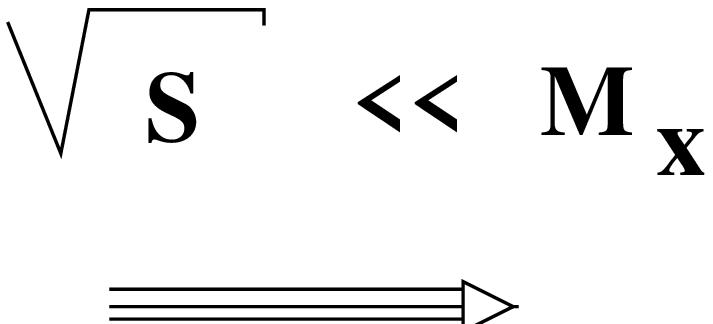}}}
   &
     \raisebox{10pt}{
     \mbox{\epsfxsize=0.40\textwidth 
           \epsffile{\master/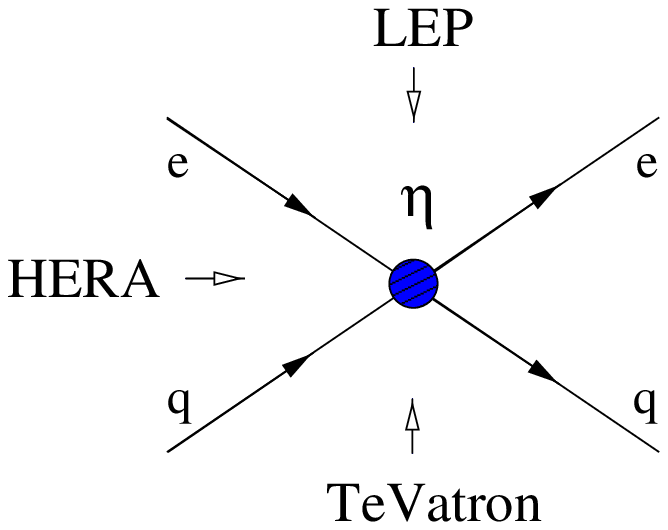}}} \\  
  \end{tabular}
  \end{center}
     \vspace*{-0.3cm}

 \caption[]{ \label{fig:treecidiag}
 {\small Schematic representation of the transition from a virtual
         exchange to a contact interaction. The tree-level exchanges
  	 through couplings $g_X$ of a heavy $X$ particle with mass 
	 $M_X$ from some underlying (renormalizable) theory ``contract'' 
	 to an point-like effective four-fermion (non-renormalizable)
	 contact interaction with couplings $\eta \propto g_X^2 / M_X^2$.
	 The complementarity through crossed diagrams for probing similar
	 contact $eeqq$ interactions at $ep$ (HERA), $e^+e^-$ (LEP) 
	 and $p\bar{p}$ (Tevatron) colliders is manifest.}}
\end{figure}

Effects from very heavy particles $X$, with masses $M_X$ much larger than 
the centre-of-mass energies $\sqrt s$ available at a given collider, 
could still be detected in experiments through virtual exchange. 
For sufficiently heavy $X$ particles, the propagators in the $s$-, $t$- or 
$u$-channel exchange diagram ``contract'' to an effective point-like
four-fermion contact interaction (CI), analogous to Fermi's proposal for a 
four-fermion interaction to explain $\beta$-decay, which subsequently
was understood to be mediated 
by the heavy $W$ particle in the true underlying theory. 
The transition from a tree-level exchange to a contact interaction is
illustrated in Fig.~\ref{fig:treecidiag}.
% ----------------- FIGURE: Contact Interaction diagrams  ------------------
%---------------------------------------------------------------------------
The concept of contact interaction finds an obvious application in the
search for a compositeness of ordinary leptons and quarks but is in essence
much more general and is applicable to inclusive searches for various kinds
of new virtual phenomena. 
Yet, the searches at $ep$ and $p\bar{p}$ colliders have been essentially 
restricted to neutral current (NC)-like processes in which the new particle field associated 
to large mass scales interferes with the ordinary $\gamma$ and $Z^0$ fields
of the Standard Model.

% CI Lagrangian and assumptions:

The effect in NC processes of new physics at scales $M_X \gg \sqrt s$ 
can be described~\cite{Eichten83,Ruckl83,Haberl91} by adding a 
four-fermion term ${\cal{L}}_{CI}^{NC}$ to the Standard Model 
interaction ${\cal{L}}_{SM}^{NC}$. The ${\cal{L}}_{CI}^{NC}$ term
itself can in general be decomposed into three different products of
fermion bilinears containing scalar-scalar, vector-vector and 
tensor-tensor Lorentz-invariant structures. General expressions
for these can be found in Ref.~\cite{Haberl91} where it is otherwise
argued on a model-dependent basis that scalar-scalar and tensor-tensor
terms are more severely constrained. Pragmatically, experimental analyses 
at colliders have considered the effect of vector-vector terms.

The effective Lagrangian for $eeqq$ vector contact interactions can be
written as~\cite{Haberl91}:
\begin{equation*}
\label{equ:cilagr} 
\begin{array}{rr}
\mathcal L = \mathcal L_{SM} +
\displaystyle \sum_{q}\{\eta^q_{LL}(\bar e_L \gamma_\mu e_L)(\bar q_L \gamma^\mu q_L)
+ \eta^q_{LR}(\bar e_L \gamma_\mu e_L)(\bar q_R \gamma^\mu q_R)\\
+ \eta^q_{RL}(\bar e_R \gamma_\mu e_R)(\bar q_L \gamma^\mu q_L)
+ \eta^q_{RR}(\bar e_R \gamma_\mu e_R)(\bar q_R \gamma^\mu q_R)\}
\end{array}
\end{equation*}
where the subscripts $L$ and $R$ denote the left- and right-handed
helicity projections of the fermion fields.
The values of each $\eta_{ij}$ parameter depend on the
chiral structure and dynamics of the new interaction.
To quantify the experimental sensitivity, it is conventionally
assumed that each $\eta$ takes the values $\epsilon g^2/\Lambda^2$,
where $\Lambda$ is the mass scale of the new interaction.
It is moreover conventional to assume a `strong' value of 
$g^2 = 4 \pi$, as was originally motivated in compositeness models.
The constant $\epsilon$ is either +1, -1 or 0 for each chirality 
combination, which defines the ``model'' of the new 
interaction. An $\epsilon = +1$ or $\epsilon = -1$ correspond
to different signs of interference with respect to the Standard Model.
The chirality structure of the CI model can be adjusted to avoid the 
severe constraints~\cite{Langacker91,Barger98} coming in particular from 
atomic-parity violation (APV)~\cite{Wood97,Bennett99,Deandrea97}.
These are cancelled in particular if 
$\eta^q_{LL} + \eta^q_{LR} - \eta^q_{RL}- \eta^q_{RR} = 0$
is satisfied for the quarks $q$, as 
realized for instance in the so-called $VV$, $AA$ and $VA$ compositeness
models with the mixtures
$VV = LL + LR + RL + RR$, $VA = LL - LR + RL - RR$, and
$AA = LL - LR - RL + RR$.
Isospin invariance is generally assumed in the analysis, which imposes
$\eta^u_{RL} = \eta^d_{RL}$ for all $u$-type and $d$-type quarks.
The SU(2)-conserving CI scenarios with $\eta^u_{LL} \neq \eta^d_{LL}$
would also induce $e\nu qq'$ CI signals.

\begin{figure}[htb]
  \begin{center}                                                                
  \begin{tabular}{ccc}
  \vspace*{-0.2cm}
    
  \hspace*{-1.0cm} \epsfig{file=\master/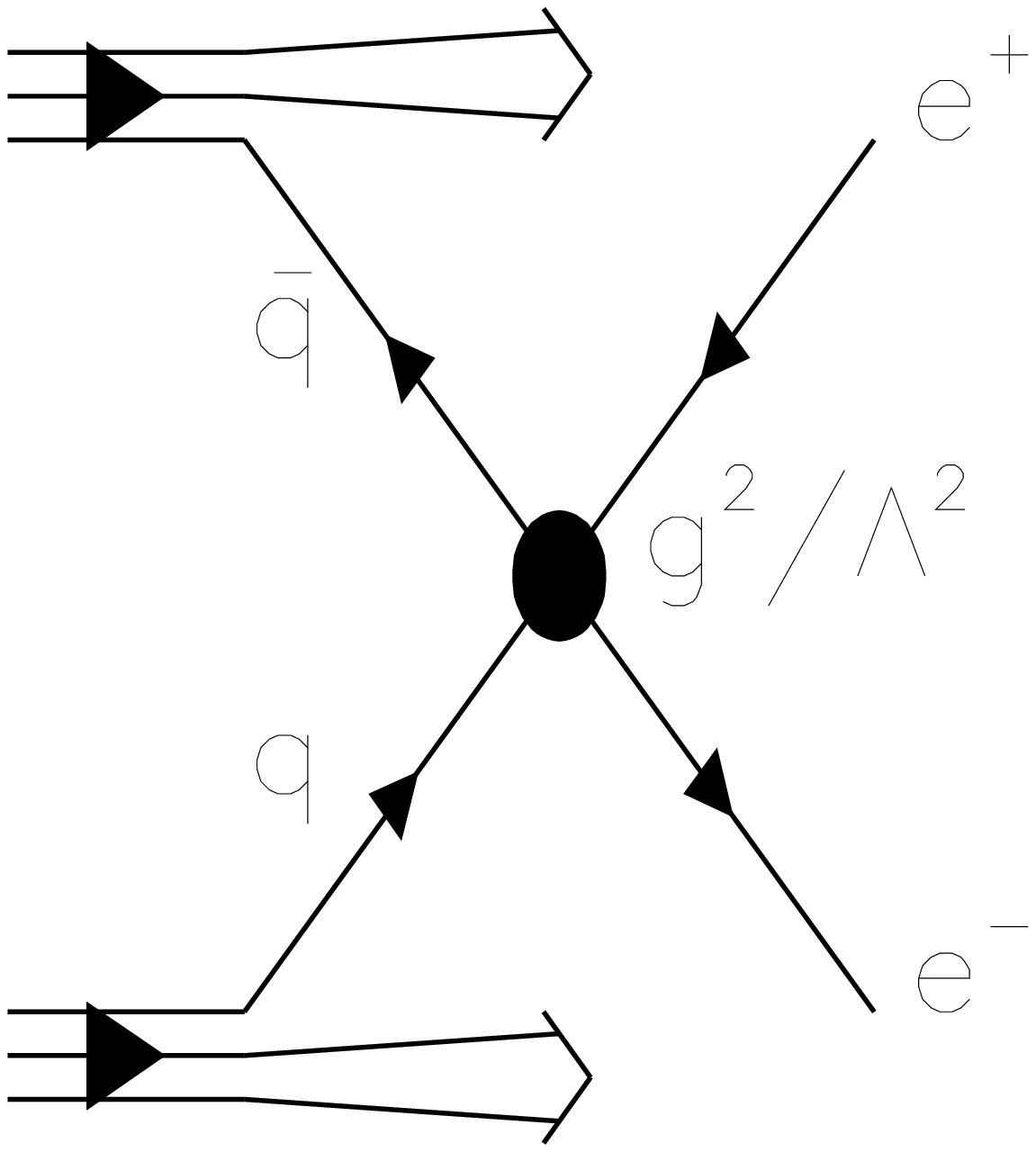,width=0.42\textwidth}
&
  \hspace*{-2.0cm} \epsfig{file=\master/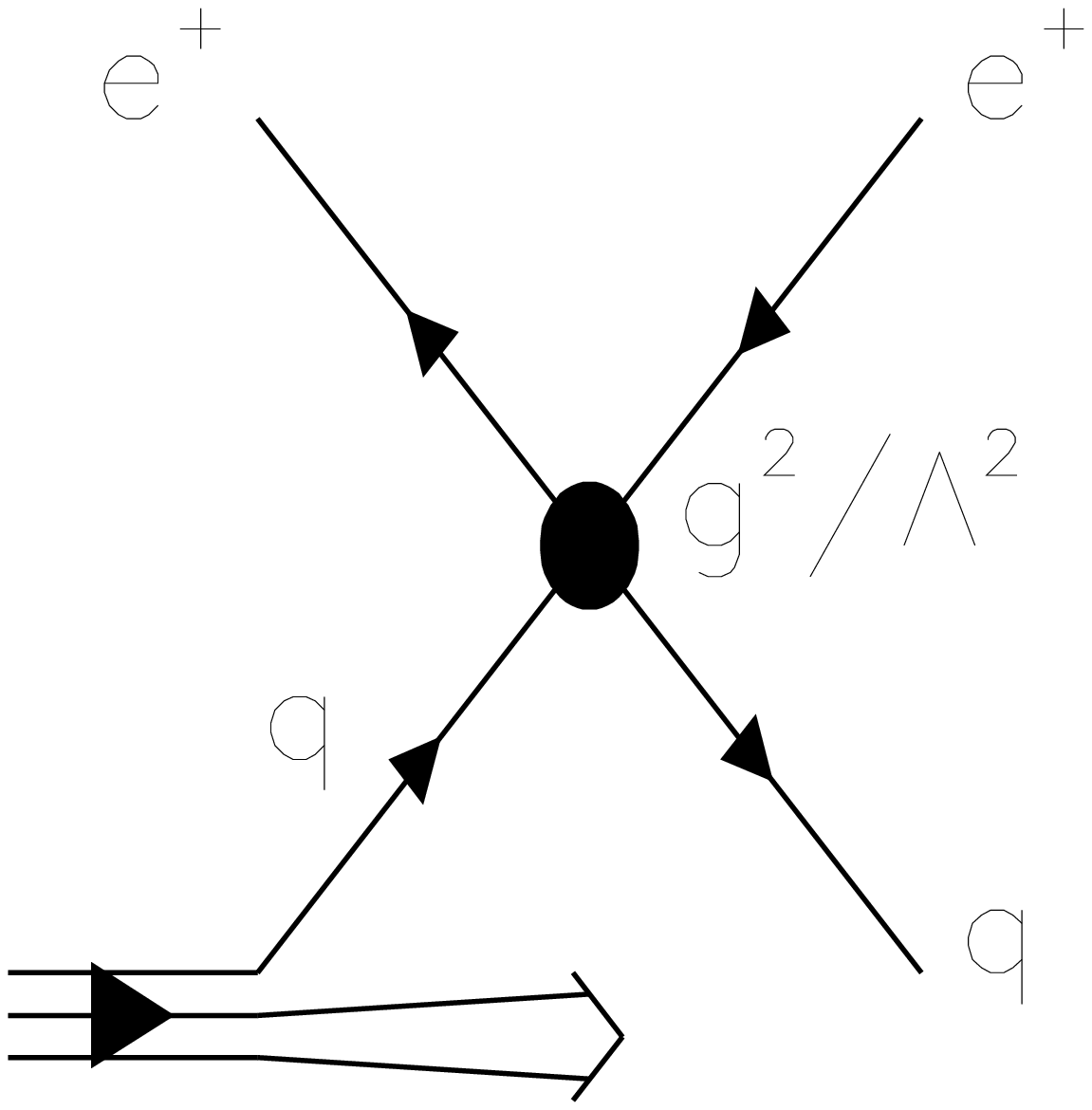,width=0.42\textwidth}
&
  \hspace*{-2.0cm} \epsfig{file=\master/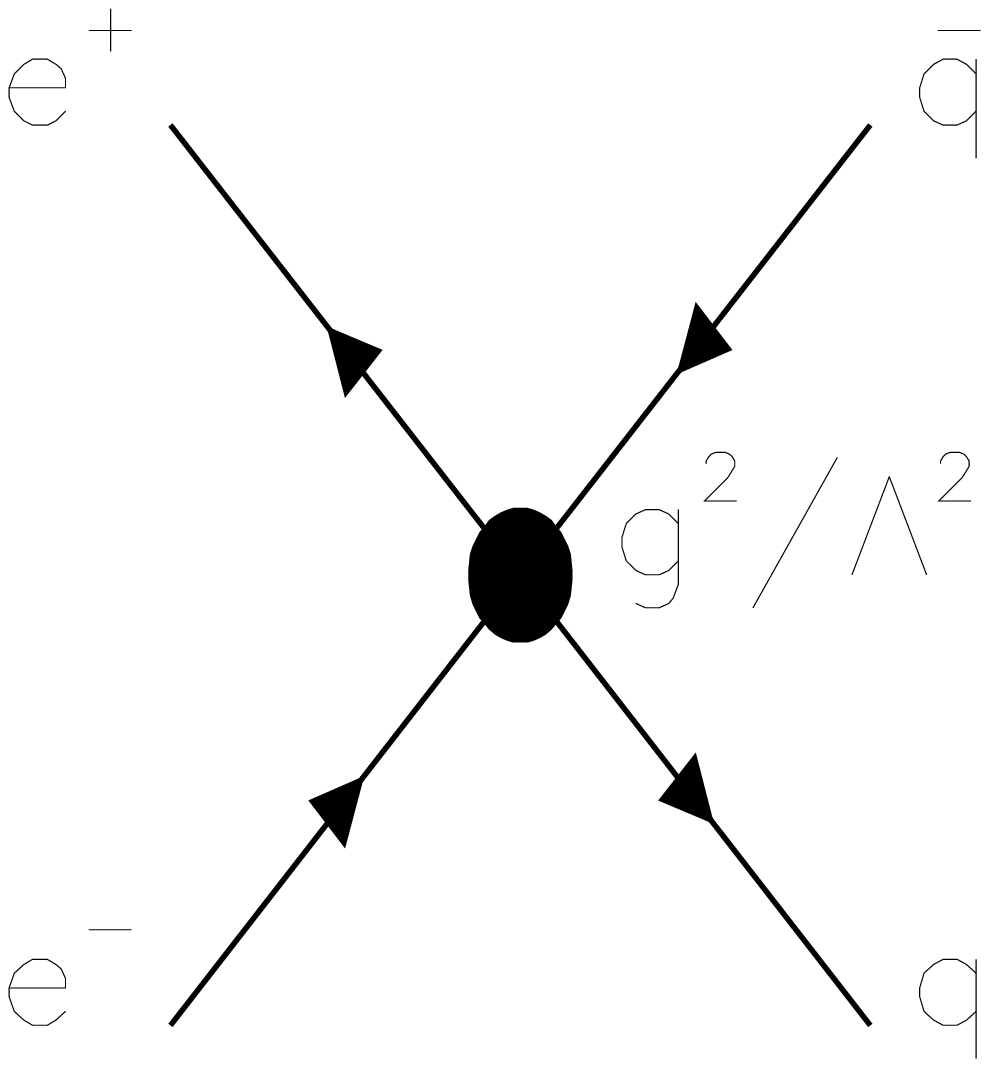,width=0.42\textwidth}

  \end{tabular}
  \end{center}
 \vspace*{-1.8cm}
 \hspace*{3.0cm} (a) \hspace*{5.5cm} (b) \hspace*{5.5cm} (c) \\  
 \vspace*{-0.5cm}
  
 \caption[]{ \label{fig:cidiag}
            Probing $eeqq$ contact interactions in a) $p\bar p$,
            b) $ep$ and c) $e^+e^-$ colliders.}
\end{figure}

\begin{figure}[htb]
  \begin{center}                                                                
  \begin{tabular}{cc}
  \vspace*{-0.2cm}
    
  \hspace*{-1.0cm} \epsfig{file=\master/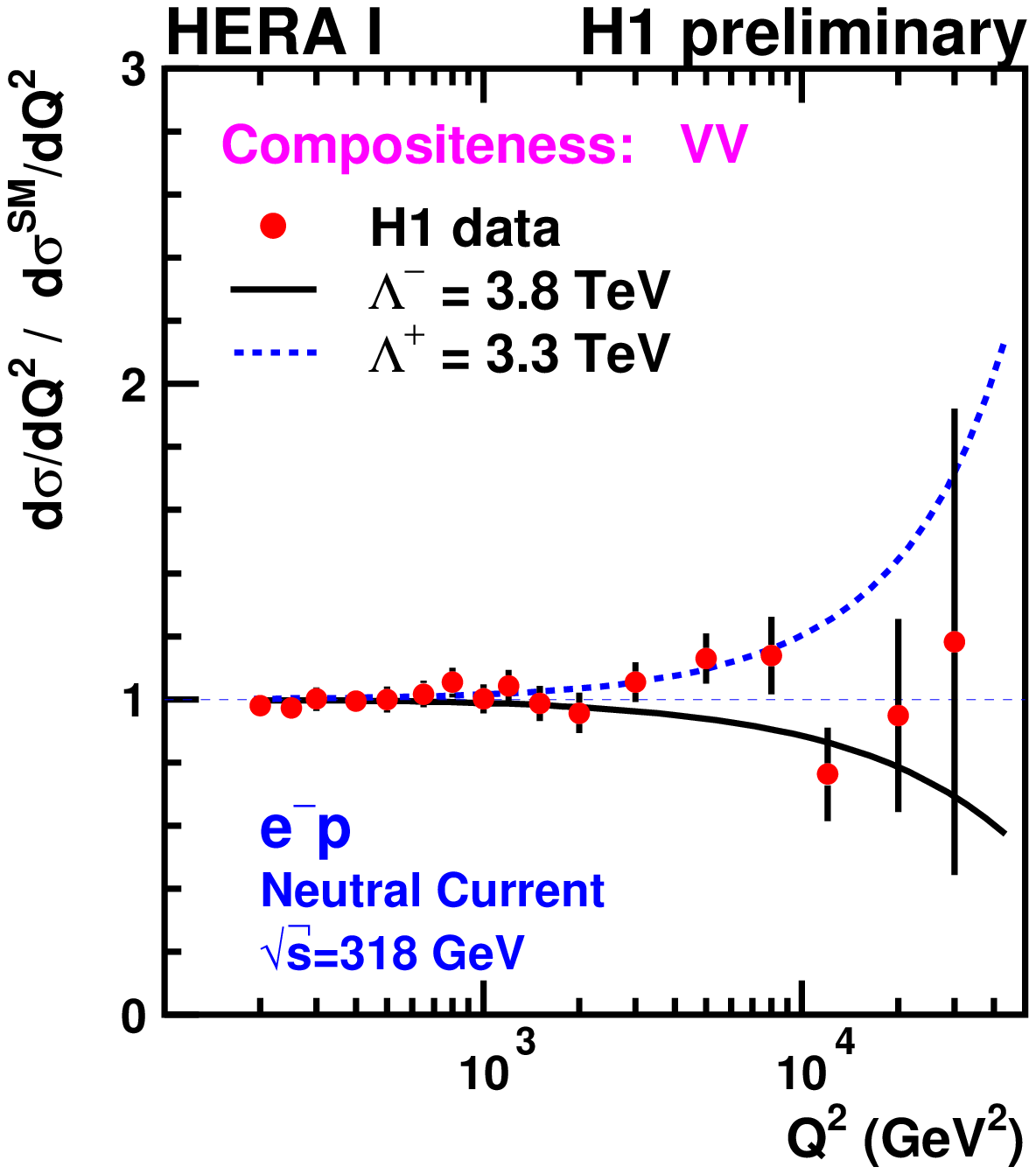,width=0.42\textwidth}
 &
  \hspace*{-1.0cm} \epsfig{file=\master/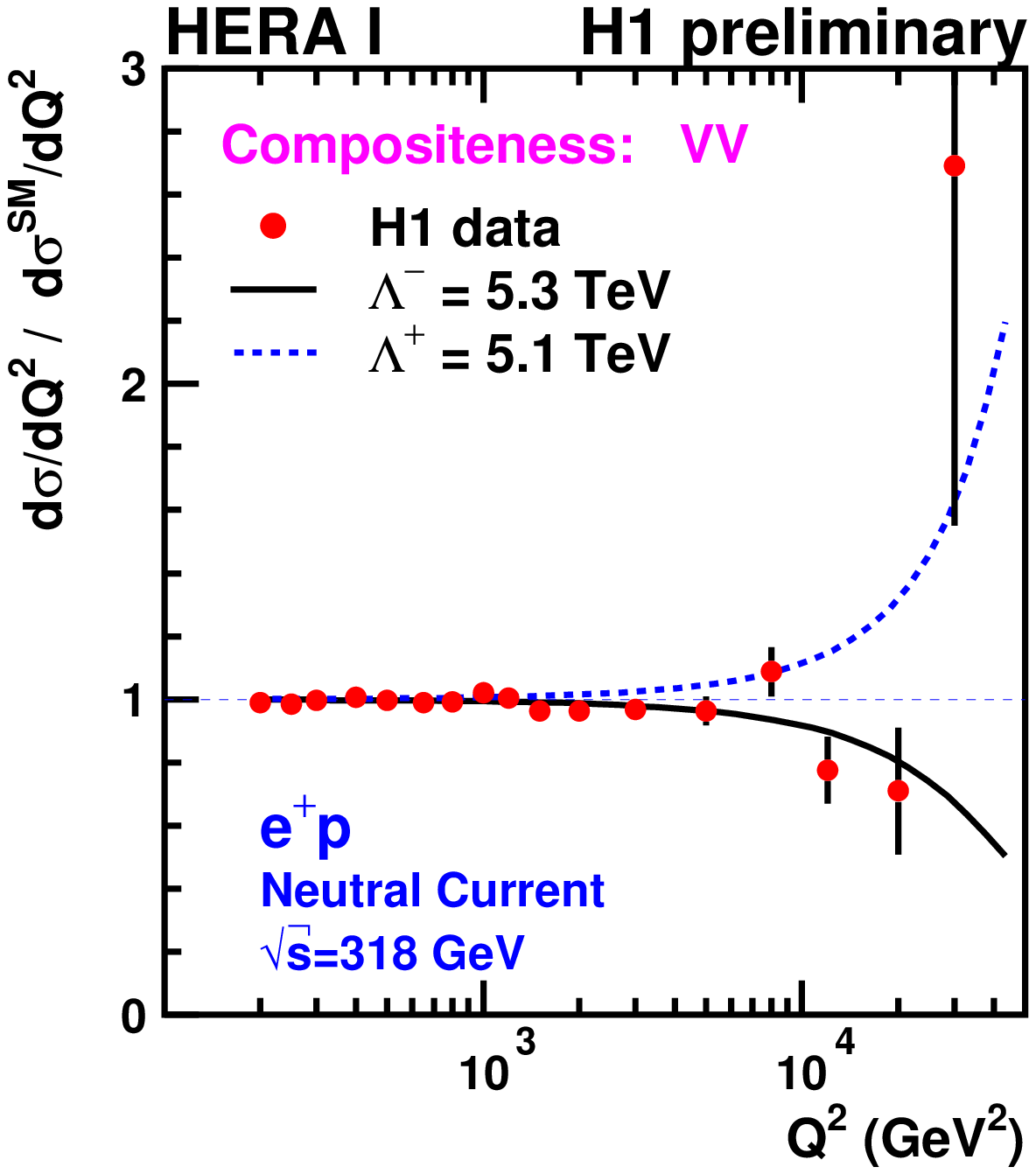,width=0.42\textwidth}

  \end{tabular}
  \end{center}
 \vspace*{-0.3cm}
 \hspace*{5.0cm} (a) \hspace*{6.5cm} (b)\\  
 \vspace*{-0.5cm}
  
 \caption[]{ \label{fig:h1vv}
             Ratio of a) $e^-p$ and b) $e^+p$ NC DIS cross-sections,
             measured by H1, divided by the SM prediction.
             The curves are VV CI models corresponding to 95\% CL exclusion
             obtained from each data set.}
\end{figure}

\begin{figure}[htb]
  \begin{center}                                                                
  \begin{tabular}{c}
  \vspace*{-0.2cm}
    
  \hspace*{-1.0cm} \epsfig{file=\master/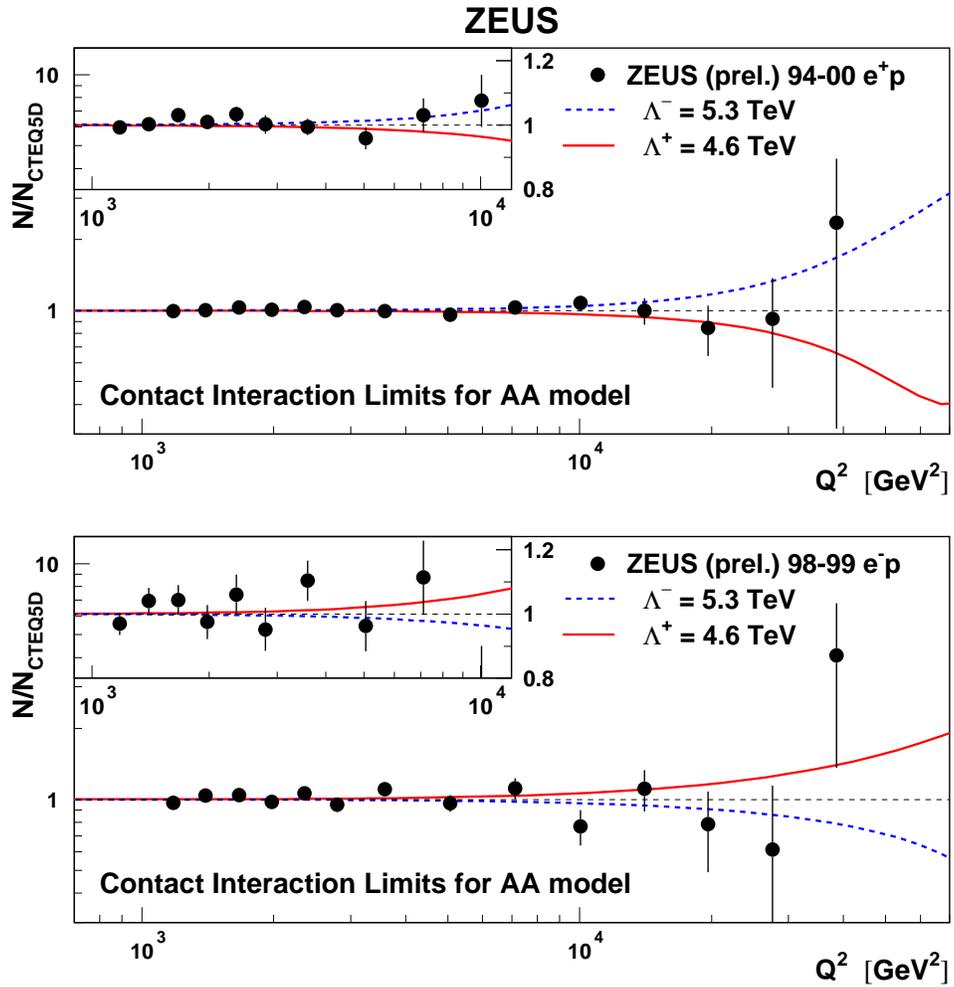,width=0.74\textwidth}

  \end{tabular}
  \end{center}
 \vspace*{-0.5cm}
  
 \caption[]{ \label{fig:zeusaa}
             Ratio of $e^+p$ ({\it top}) and $e^-p$ ({\it bottom}) NC DIS events
             observed by ZEUS, divided by the SM prediction.
             The curves are AA CI models corresponding to 95\% CL exclusion
             obtained from the combined data set.}
\end{figure}

Examples of diagrams for $eeqq$ contact interactions at each of the three
existing types of high-energy colliders are shown in Fig.~\ref{fig:cidiag} 
with emphasis on the different initial- and final-state particles. 
In addition, four-quark (four-lepton) interactions can be probed at Tevatron 
(LEP).
At HERA, the $eeqq$ contact interaction (CI) would modify the 
NC DIS cross-sections at high $Q^2$.
The pure-CI term would increase the cross-section at the highest $Q^2$,
while the SM-CI interference could act either constructively or
destructively in the intermediate $Q^2$ region.
Figure \ref{fig:h1vv} shows preliminary $e^-p$ and $e^+p$ cross-sections
measured by H1 as a ratio to the SM prediction~\cite{H1CI}. 
Since no significant
deviation is found, fits to CI models were made to obtain 95\% CL
exclusion limits on $\Lambda$ for both $\epsilon=1$ ($\Lambda^+$) and 
$\epsilon=-1$ ($\Lambda^-$) cases.
Similarly Fig.~\ref{fig:zeusaa} shows the $Q^2$ distribution of NC DIS events
from ZEUS, again as a ratio to SM prediction~\cite{ZEUSCI}.
At the Tevatron, searches for $llqq$ CI are made in Drell-Yan dilepton
production.  The presence of CI would alter the cross-sections at large
masses.  Both CDF~\cite{CDFDY} and D$\emptyset$~\cite{D0DY} have obtained limits
for $eeqq$ terms from di-electron data.  CDF~\cite{CDFDY} also looked at
di-muon data which constrained $\mu\mu qq$ terms.
At LEP2~\cite{ALEPHCI, DELPHICI, L3CI, OPALCI}, 
measurements of hadronic cross-sections constrainted
$eeqq$ CI terms.  

Table~\ref{tab:tab-ci} summarizes the limits from three
colliders for various $eeqq$ CI models with different chiral structures.
Except for the purely chiral models (LL, LR, RL and RR), all models
in the table respect the above mentioned condition imposed by
APV experiments.
Each row of the table corresponds to two models, depending on
the overall interference sign with respect to the SM.
Note that the limits from LEP were obtained under the assumption that all
quark flavours participate in the CI reaction, while the HERA and Tevatron
results are sensitive to first-generation quarks, which dominate the high-$x$
region of the proton structure function.

The Tevatron $\bar p p$ experiments are also sensitive to
$qqqq$ CI models by comparing hadronic-jet production with
the QCD predictions.  Limits on $\Lambda$ as high as 2.4~TeV have
been obtained from observables such as dijet mass or angular
distributions~\cite{TeVqq}.
Recently CDF has also looked for $qq'e\nu$-type CI effects in the
high-mass $e\nu$ final state and derived a lower limit of
2.81~TeV on $\Lambda$~\cite{CDFqqen}.

\begin{figure}[htb]
  \begin{center}                                                                
  \begin{tabular}{c}
  \vspace*{-0.2cm}
    
  \hspace*{-1.0cm}
  \epsfig{file=\master/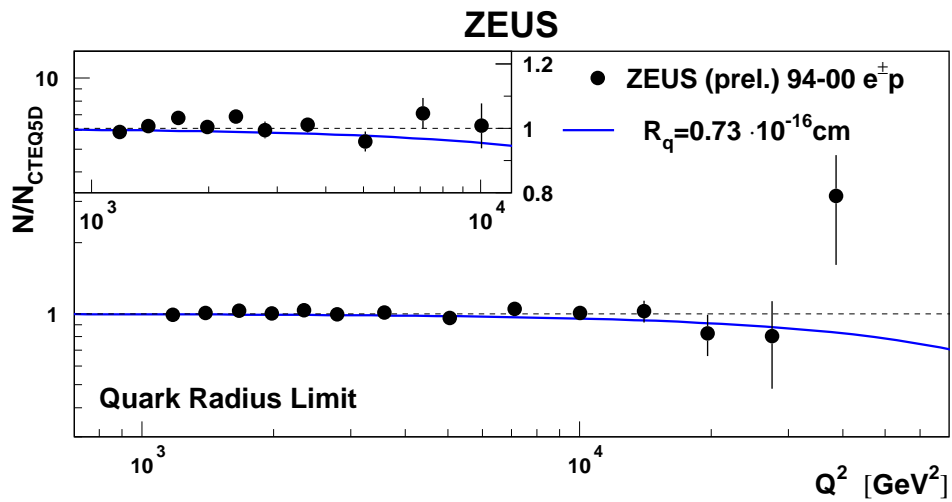,width=0.74\textwidth}

  \end{tabular}
  \end{center}
 \vspace*{-0.5cm}
  
 \caption[]{ \label{fig:zeusrq}
             Ratio of $e^\pm p$ NC DIS events
             observed by ZEUS, divided by the SM prediction.
             The curve corresponds to the quark form factor at which
             the limit is set.}
\end{figure}

The results above could also be interpreted in terms of ``radius'' of
the quark, with the classical form-factor approximation.
For example, the high-$Q^2$ NC DIS cross-section will decrease by a factor
$(1-R_e^2/6Q^2)^2(1-R_q^2/6Q^2)^2$ under the assumption of non-zero root-mean-square
electroweak radii of electron and quark, respectively.
Assuming a point-like electron ($R_e$=0), limits on $R_q$ of
$0.82\cdot10^{-16}$cm and $0.73\cdot10^{-16}$cm have been obtained
from H1~\cite{H1CI} and ZEUS~\cite{ZEUSCI} data, respectively.
Figure~\ref{fig:zeusrq} shows an example from ZEUS.
CDF~\cite{CDFDY} gives a limit of $0.79\cdot 10^{-16}$cm from the Drell-Yan
results\footnote{derived from the quoted limit of $0.56\cdot10^{-16}$cm
which assumes $R_e$=$R_q$} and L3~\cite{L3CI} gives
$0.42\cdot10^{-16}$cm, but the latter assumes that all produced flavours
are composite.
%----------------------------------------------------------------------------
%----------------------------------------------------------------------------
%
%----------------------------------------------------------------------------
%----------------------------------------------------------------------------
%----------------------------------------------------------------------------
%------------------------------------------------------------------------------
%       general CI
%------------------------------------------------------------------------------

%\begin{sidewaystable}[tbp]
\begin{table}[tbp]
  \begin{center}
   \begin{tabular}{|c@{~[}r@{,}r@{,}r@{,}r|cc|cc|cc|cc|cc|cc|cc|}
\hline
  \multicolumn{5}{|c|}{{95\% CL limit [TeV]}} & 
  \multicolumn{2}{c|}{ZEUS} & 
  \multicolumn{2}{c|}{H1} & \multicolumn{2}{c|}{D$\emptyset$} & 
  \multicolumn{2}{c|}{CDF} & \multicolumn{2}{c|}{ALEPH} & 
  \multicolumn{2}{c|}{L3} & \multicolumn{2}{c|}{OPAL} \\
\hline
 \multicolumn{5}{|r|}{Coupling structure}
 & &  & &  & & & & & & & & & & \\
  Model & 
 $\epsilon_{_{LL}}$ & $\epsilon_{_{LR}}$  &
                 $\epsilon_{_{RL}}$ &  $\epsilon_{_{RR}}$] &
   $\Lambda^{-}$  &  $\Lambda^{+}$ & $\Lambda^{-}$  &  $\Lambda^{+}$ & 
   $\Lambda^{-}$  &  $\Lambda^{+}$ & $\Lambda^{-}$  &  $\Lambda^{+}$ & 
   $\Lambda^{-}$  &  $\Lambda^{+}$ & $\Lambda^{-}$  &  $\Lambda^{+}$ & 
   $\Lambda^{-}$  &  $\Lambda^{+}$ \\
\hline                   %   ZEUS    %    H1     %   D$\emptyset$      %    CDF    %   ALEPH   %    L3     %   OPAL    
LL &+1 & 0 & 0 & 0]     &     &     & 2.3 & 2.8 & 4.2 & 3.3 & 3.7 & 2.5 & 6.2 & 5.4 & 2.8 & 4.2 & 3.1 & 5.5 \\
LR & 0 &+1 & 0 & 0]     &     &     & 1.8 & 3.2 & 3.6 & 3.4 & 3.3 & 2.8 & 3.3 & 3.0 & 3.5 & 3.3 & 4.4 & 3.8 \\
RL & 0 & 0 &+1 & 0]     &     &     & 1.9 & 3.2 & 3.7 & 3.3 & 3.2 & 2.9 & 4.0 & 2.4 & 4.6 & 2.5 & 6.4 & 2.7 \\
RR & 0 & 0 & 0 &+1]     &     &     & 2.3 & 2.8 & 4.0 & 3.3 & 3.6 & 2.6 & 4.4 & 3.9 & 3.8 & 3.1 & 4.9 & 3.5 \\
\hline
VV &+1 & +1 & +1 & +1]  & 7.0 & 6.5 & 5.4 & 5.1 & 6.1 & 4.9 & 5.2 & 3.5 & 7.1 & 6.4 & 5.5 & 4.2 & 7.2 & 4.7 \\
AA &+1 &$-1$&$-1$& +1]  & 5.3 & 4.6 & 3.9 & 2.5 & 5.5 & 4.7 & 4.8 & 3.8 & 7.9 & 7.2 & 3.8 & 6.1 & 4.2 & 8.1 \\
VA &+1 &$-1$&+1  &$-1$] & 3.4 & 3.3 & 2.9 & 2.9 &     &     &     &     &     &     &     &     &     &     \\ 
LL$-$LR&+1 &$-1$& 0& 0] & 4.0 & 2.7 &     &     & 4.5 & 3.9 &     &     &     &     &     &     &     &     \\
LL+RL  & +1 & 0 &+1& 0] & 4.7 & 4.7 &     &     &     &     &     &     &     &     &     &     &     &     \\
LL+RR  & +1 & 0 & 0&+1] & 4.3 & 4.2 & 3.7 & 3.8 & 5.1 & 4.2 &     &     & 7.4 & 6.7 & 3.7 & 4.4 & 4.4 & 5.4 \\
LR+RL  &  0 &+1 &+1& 0] & 5.6 & 5.6 & 4.1 & 4.3 & 4.4 & 3.9 &     &     & 4.5 & 2.9 & 5.2 & 3.1 & 7.1 & 3.4 \\
LR+RR  &  0 &+1 & 0&+1] & 4.8 & 4.8 &     &     &     &     &     &     &     &     &     &     &     &     \\
RL$-$RR&  0 &0 &+1&$-1$]& 2.6 & 3.9 &     &     & 4.3 & 4.0 &     &     &     &     &     &     &     &     \\ 
\hline
    \end{tabular}
  \end{center}
  \caption{Relations between couplings 
  $[\epsilon_{LL},\epsilon_{LR}, \epsilon_{RL}, \epsilon_{RR}]$ for the 
         compositeness models and the {95\%} CL lower limits on
         the compositeness scale, $\Lambda$, resulting from 
         HERA, Tevatron and LEP2 experiments.
         Each row of the table represents
         two $eeqq$ CI scenarios corresponding to
%         $\eta<0$ ($\Lambda^{-}$) and $\eta>0$ ($\Lambda^{+}$).
         the coupling structure defined in the leftmost column ($\Lambda^{+}$)
         and another with all $\epsilon$'s negated ($\Lambda^{-}$).
        The same coupling structure applies
        to $d$ and $u$ quarks.}
  \label{tab:tab-ci}
%\end{sidewaystable}
\end{table}

%------------------------------------------------------------------------------
%       Lepto quarks
%------------------------------------------------------------------------------
The leptoquarks described in section~\ref{sec:leptoq}, when
much heavier than the centre-of-mass energy of the collider,
can influence the SM processes via virtual effects with
$s$-, $u$- (for HERA) or $t$-channel exchange (at LEP and Tevatron).
Their effect at the low-energy limit can be expressed as a CI model
in which $g/\Lambda$ is replaced with $\lambda/M$, the ratio
between the Yukawa coupling and the leptoquark mass, and the
coefficients $\epsilon^q_{ij}$ take particular constant values
depending on the leptoquark species, as in Table~\ref{tab:tab-lq}.
The table summarizes the limits from HERA~\cite{H1CI, ZEUSCI} and
LEP2~\cite{ALEPHCI, L3CI, OPALLQ}.
%-------------------------------------------------------------------------------
%       Lepto quarks
%-------------------------------------------------------------------------------
\begin{table}[btp]
\begin{center}
\begin{tabular}{|cl|cc|ccc|c|c|}
\hline
\multicolumn{9}{|c|}{ 95\% CL limit  $M_{LQ}/\lambda_{LQ}$ [TeV] } \\
\hline
\multicolumn{2}{|c|}{CI Model}  & \multicolumn{5}{|c|}{Collider Experiments} & Low Energy & Global Fits \\
\hline
LQ   & Coupling Structure & ZEUS & H1 & ALEPH & L3 & OPAL & (APV and                 & \\
type &                    &      &    &       &    &      & $\pi \rightarrow e \nu$) & \\
\hline
%                                      % ZEUS   %  H1    % ALEPH  %  L3    % OPAL   & APV / pienu & 
$S_{\circ}^L$ & 
$\epsilon^{u}_{_{LL}}=+\frac{1}{2}$    & {0.75} & {0.72} & {0.64} & {1.24} & 0.64   & 3.5 & 3.7 \\
$S_{\circ}^R$ & 
$\epsilon^{u}_{_{RR}}=+\frac{1}{2}$    & {0.69} & {0.67} &        & {0.96} &        & 2.8 & 3.9 \\
$\tilde{S}_{\circ}^{R}$ & 
$\epsilon^{d}_{_{LL}}=+\frac{1}{2}$    & {0.31} & {0.33} & {0.22} & {0.26} & {0.59} & 3.0 & 3.6 \\
$S_{1/2}^L$ & 
$\epsilon^{u}_{_{LR}}=-\frac{1}{2}$    & {0.91} & {0.87} & {0.06} & {0.18} & {0.46} & 2.8 & 3.5 \\
$S_{1/2}^R$ & 
$\epsilon^{d}_{_{RL}}=\epsilon^{u}_{_{RL}}=-\frac{1}{2}$
                                       & {0.69} & {0.37} &        & {0.35} & {0.63} & 2.1 & 2.1 \\
$\tilde{S}_{1/2}^{L}$ & 
$\epsilon^{d}_{_{LR}}=-\frac{1}{2}$    & {0.50} & {0.43} &        &        & {0.37} & 3.0 & 3.8 \\
$S_{1}^{L}$ & 
$\epsilon^{d}_{_{LL}}=+1, \; \epsilon^{u}_{_{LL}}=+\frac{1}{2}$  
                                       & {0.55} & {0.48} & {0.77} & {0.64} & {0.93} & 2.5 & 2.4 \\
\hline
$V_{\circ}^L$ & 
$\epsilon^{d}_{_{LL}}=-1$              & {0.69} & {0.77} & {1.09} & {1.79} &        & 4.3 & 8.1 \\
$V_{\circ}^R$ & 
$\epsilon^{d}_{_{RR}}=-1$              & {0.58} & {0.64} & {0.38} & {0.41} & {0.45} & 2.2 & 2.3 \\
$\tilde{V}_{\circ}^{R}$ & 
$\epsilon^{u}_{_{RR}}=-1$              & {1.03} & {1.00} & {0.89} & {0.89} & {1.10} & 2.0 & 1.9 \\
$V_{1/2}^L$ & 
$\epsilon^{d}_{_{LR}}=+1$              & {0.49} & {0.42} & {0.41} & {0.61} & {0.66} & 5.4 & 2.1 \\
$V_{1/2}^R$ & 
$\epsilon^{d}_{_{RL}}=\epsilon^{u}_{_{RL}}=+1$         
                                       & {1.15} & {0.94} & {0.48} & {0.54} & {0.60} & 2.2 & 7.5 \\
$\tilde{V}_{1/2}^{L}$ & 
$\epsilon^{u}_{_{LR}}=+1$              & {1.26} & {1.02} & {0.29} & {0.45} & {0.55} & 2.0 & 2.1 \\
$V_{1}^{L}$ & 
$\epsilon^{d}_{_{LL}}=-1, \; \epsilon^{u}_{_{LL}}=-2 $  
                                       & {1.42} & {1.38} & {1.50} & {1.21} & {1.53} & 6.6 & 7.3 \\
\hline
\end{tabular}
\end{center}
  \caption{Lower limits (95\% CL) on the ratio $M_{LQ}/\lambda_{LQ}$ of the 
           leptoquark mass $M_{LQ}$ to the Yukawa coupling $\lambda_{LQ}$ for various 
	   contact-interaction models. The models are defined by the coefficients 
	   $\epsilon^{q}_{ij}$ and correspond to the interactions of scalar leptoquarks 
	   $S$ (upper table part) and vector leptoquarks $V$ (lower table part)
	   of different types in the limit $M_{LQ} \gg \sqrt{s}$. 	   
           Results are summarized for the CI analysis of HERA~\cite{H1CI, ZEUSCI} and
	   LEP2~\cite{ALEPHCI, L3CI, OPALLQ} data.	   
           Low energy constraints~\cite{Hewett97b} derived from 
	   Atomic Parity Violation (APV) measurements and precision tests of 
	   lepton universality in $\pi \rightarrow l \nu_l$ decays 
	   are also summarized. 
	   Results from ``global fits''~\cite{Zarnecki2000,Barger2000} (see text) 
	   are shown for comparison.}
  \label{tab:tab-lq}
\end{table}
%-----------------------------------------------------------------------------
Also listed in Table~\ref{tab:tab-lq} are low energy constraints derived in 
Ref.~\cite{Hewett97b} from precision measurements of APV 
and of lepton universality in $\pi \rightarrow l \nu_l$ decays.  
The constraints from low energy measurements alone are seen to exclude a domain 
already beyond the reach of actual HERA data for leptoquarks in the CI limit 
($M_{LQ} \gg \sqrt{s_{ep}}$). As was shown in Ref.~\cite{Zarnecki2002}, this 
will remain so at  HERA$_{II}$ even when considering integrated luminosities ${\cal{L}}$
approaching $1 \fb^{-1}$ in a single experiment, given that the mass reach only
improves in powers of ${\cal{L}}^{1/4}$.
Combining~\cite{Zarnecki2000,Barger2000} all existing data from colliders 
and low energy experiments leads to so-called ``global fit'' constraints shown
for comparison in Table~\ref{tab:tab-lq} from the analysis of Ref.~\cite{Zarnecki2000}.
As was shown in Ref.~\cite{Zarnecki2002}, only the full integrated luminosity expected
in the lifetime of Tevatron$_{II}$ will allow a single collider experiment to start
to compete in sensitivity with these existing bounds.

      % Search for contact interactions       

% Extra-dimensions and other exoticas
\clearpage
%%%%%%%%%%%%%%%%%%%%%%%%%%%%%%%%%%%%%%%%%%%%%%%%%%%%%%%%%%%%%%%%%%%%%%%%%%%%%%
\section{Large Compactified Extra Dimensions}
\label{sec:xtra}
%%%%%%%%%%%%%%%%%%%%%%%%%%%%%%%%%%%%%%%%%%%%%%%%%%%%%%%%%%%%%%%%%%%%%%%%%%%%%%

It has been realized recently that the problem of the hierarchy between
the electroweak scale and the Planck scale, two seemingly fundamental 
scales in Nature, could be solved in theories with extra dimensions.
Viable scenarios have been constructed in ($4+n$)-dimensional string-inspired 
theories such as the Arkani-Hamed, Dimopoulos and Dvali (ADD) 
model~\cite{Arkani98} with $n \ge 2$ ``large'' compactified extra 
dimensions, or the Randall--Sundrum model~\cite{Randall99} with 
$n =1$ ``small'' and (so-called) ``warped'' extra spatial dimension. 
In such quantum-gravity models, the gravitational force is expected to 
become comparable to the gauge forces close to the weak scale, eventually 
leading to (model dependent) effects in the $\TeV$ range observable at high 
energy 
colliders~\cite{Hewett99,Giudice99,Allanach00,Besancon01,Hewett02,Giudice02}. 
The phenomenology and results discussed below are based on the ADD scenario.

In the ADD scenario, a gravitational ``string'' scale, $M_s$, in ($4+n$)
dimensions is introduced close to the weak scale. It is related to the
usual Planck scale, $M_p \sim 10^{19} \GeV$ (which is no longer fundamental
but now rather merely the scale of effective four-dimensional gravity),
via a relation $M_p^2 = R^n M_s^{2+n}$, where $R$ is a characteristic
(large) size of the $n$ compactified extra dimensions.
The graviton is allowed to propagate in these extra dimensions. 
Their finite size $R$ implies that the graviton will appear in our 
familiar 4-dimensional universe as a ``tower'' of massive Kaluza-Klein
excitation states. 
The effects from virtual graviton exchange are expected to depend only
weakly on the number of extra dimensions, while in constrast direct graviton
emission is expected to be suppressed by a factor 
$(M_s)^{n-2}$~\cite{Hewett99,Giudice99}.
The virtual exchange of Kaluza-Klein towers between Standard Model 
particles leads to an effective contact interaction with a coupling 
coefficient $\eta_G = \lambda/M^4_S$~\cite{Giudice99}.
Of the spin-$0$, $1$ and $2$ states of the Kaluza-Klein towers, only the 
spin-2 gravitons interact in the ADD scenario with the Standard Model 
fields of our familiar universe~\cite{Hewett99,Hewett02}.

The contributions of virtual graviton exchange to deep inelastic scattering
in $ep$ collisions have been derived in Ref.~\cite{H1xtra} by applying crossing 
relations to the cross-sections given in Ref.~\cite{Giudice99} for $e^+e^-$ 
collisions. At the parton level, the differential cross-sections for
the basic processes of elastic $e^+q \to e^+q$ and $e^+g \to e^+g$ scattering
can be written as 
\begin{eqnarray*}
  \frac{\mathrm d \sigma (e^+q \rightarrow e^+q)}{\mathrm d t} & = &
  \frac{\mathrm d \sigma^{SM}}{\mathrm d t} +
  \frac{\mathrm d \sigma^{G}} {\mathrm d t} +
  \frac{\mathrm d \sigma^{\gamma G}}{\mathrm d t} +
  \frac{\mathrm d \sigma^{Z G}}{\mathrm d t} \ , \\[.4em]
  \frac{\mathrm d \sigma^{G}}{\mathrm d t} & = &
  \frac {\pi\, \lambda^2}{32\, M_S^8} \, \frac{1}{s^2} %\,
  \left\{32\,u^4 + 64\,u^3 t + 42\,u^2 t^2 + 10\,u\,t^3 + t^4\right\} \ , \\[.4em]
  \frac{\mathrm d \sigma^{\gamma G}}{\mathrm d t} & = &
  -\frac{\pi\, \lambda}{2\, M_S^4} \,
   \frac{\alpha\, e_q}{s^2} \, \frac{(2\, u + t)^3}{t} \ , \\[.4em]
  \frac{\mathrm d \sigma^{Z G}}{\mathrm d t} & = &
  \frac{\pi\, \lambda}{2\, M_S^4} \,\frac{\alpha}{s^2\sin^2 2\,\theta_W} 
  \left \{ v_e v_q\, \frac{(2\,u + t)^3}{t-m_Z^2} 
         - a_e a_q\, \frac{t\,(6\,u^2 + 6\,u\,t + t^2)}{t-m_Z^2} \right \} 
      \, , \ \ \ \\[.4em]
  \frac{\mathrm d \sigma(e^+g \rightarrow e^+ g)}{\mathrm d t} & = &
  \frac{\pi\, \lambda^2}{2\, M_S^8} \, \frac{u}{s^2} %\,
  \left\{2\, u^3 + 4\, u^2 t + 3\, u\, t^2 + t^3 \right\},
\end{eqnarray*}
where the contributions of the Standard Model (SM), of the pure graviton (G)
exchange and of $\gamma G$ and $Z G$ interference have been distinguished.
Here $s$, $t = - Q^2$ and $u$ are the Mandelstam variables,
$e_q$ is the quark charge and $v_f$ and $a_f$ are the vector and axial-vector 
couplings of the fermions to the $Z$.
The corresponding cross-sections for $e^+ \bar{q}$ scattering are obtained
by replacing $v_q \to -v_q$ and $e_q \to -e_q$ in the expressions above.
For $e^-q$ scattering, the interference of the graviton exchange with
$\gamma$ and $Z$ exchange behaves oppositely to that in $e^+q$ scattering.
Integral expressions for the inclusive $e^+p$ cross-section obtained by
integrating over parton distributions in the proton are given in 
Ref.~\cite{H1xtra}. The gravitational effects arising from the 
gluon contribution are expected to be small, ${\cal{O}}(1\%)$, compared
to those coming from quarks and antiquarks.

\begin{figure}[htb]

  \begin{center}                                                                
  \begin{tabular}{cc}
  \vspace*{-0.2cm}

  \hspace*{-1.0cm} \epsfig{file=\master/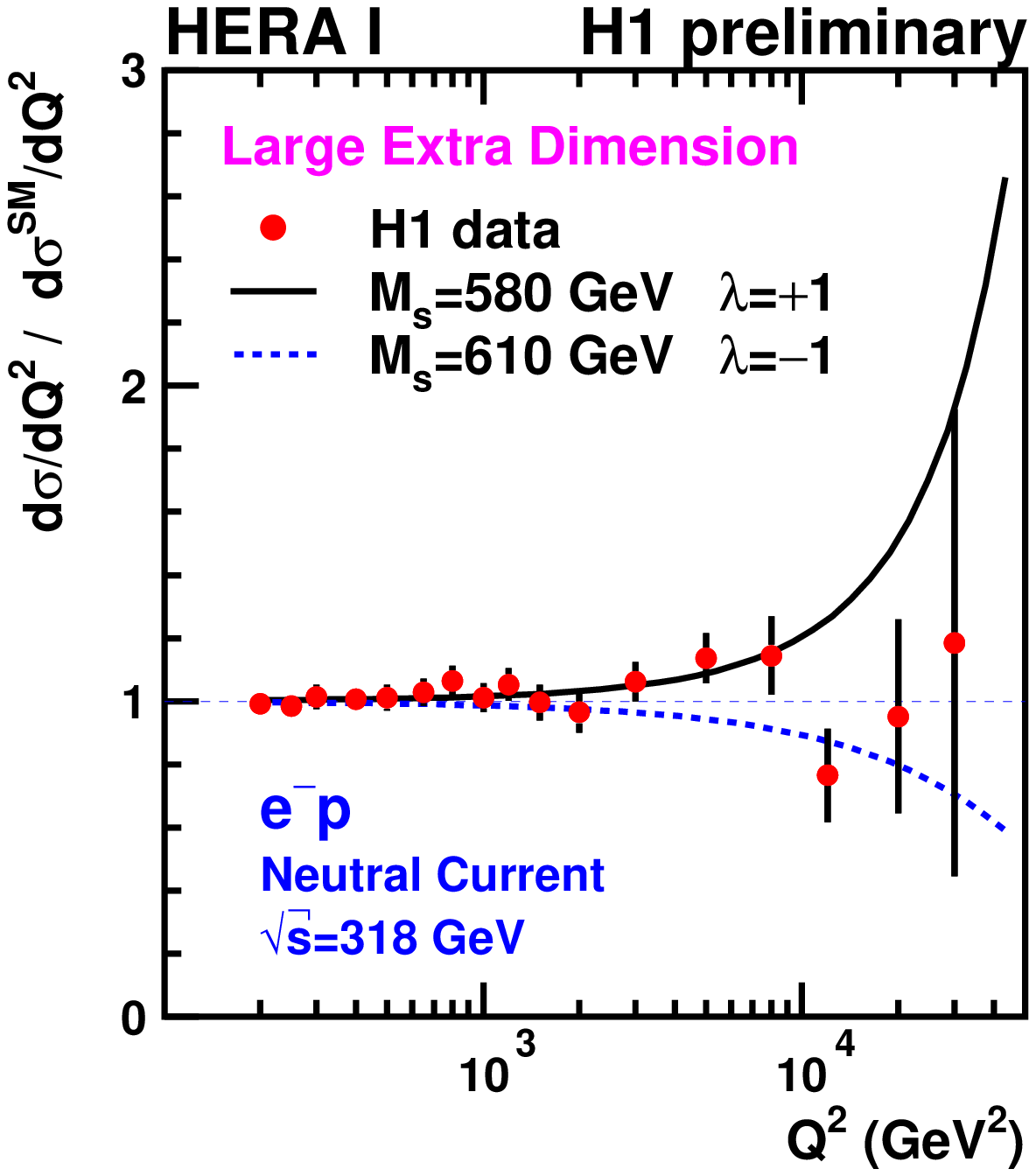,width=0.42\textwidth}
 &
  \hspace*{-1.0cm} \epsfig{file=\master/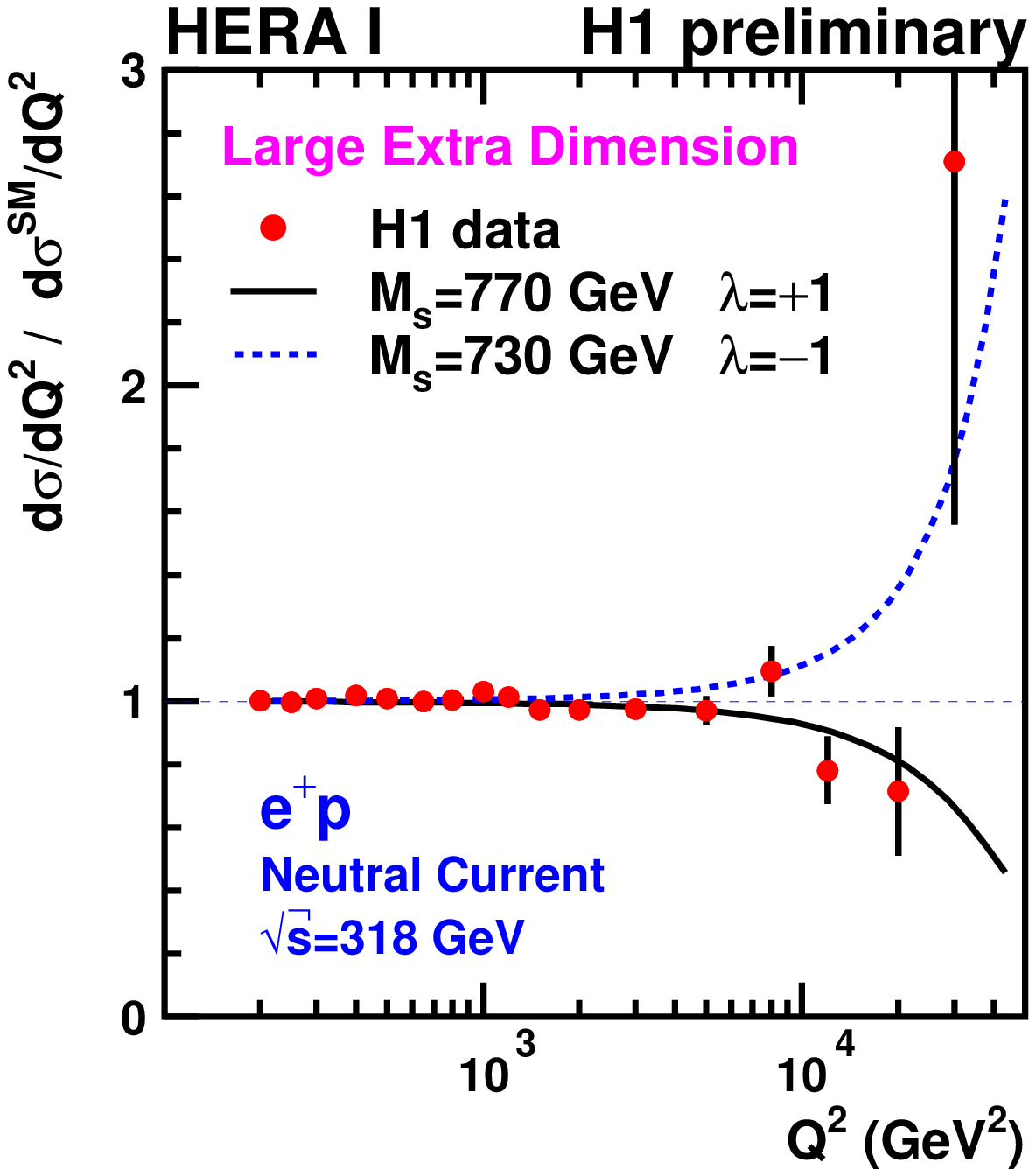,width=0.42\textwidth}
    
  \end{tabular}
  \end{center}
 \vspace*{-0.5cm}
 \hspace*{5.0cm} (a) \hspace*{6.5cm} (b)\\  
 \vspace*{-0.5cm}
  
 \caption[]{ \label{fig:h1kaluza}
             Comparison of the deep inelastic neutral current differential 
	     a) $e^-p$ and b) $e^+p$ cross-section measured by H1 with 
	     expectations from the Standard Model; expected effects from the 
	     exchange of Kaluza-Klein towers of gravitons for values of the 
	     string scales derived as the 95\% CL lower limit are also shown.
	     A combined analysis of the $e^-p$ and $e^+p$ data yields a
	     a lower limit on $M_S$ of $0.83 \TeV$ for $\lambda = +1$ and
	     $0.79 \TeV$ for $\lambda = -1$.} 
\end{figure}

\begin{figure}[htb]
  \begin{center}                                                                
 \vspace*{-1.0cm}

  \epsfig{file=\master/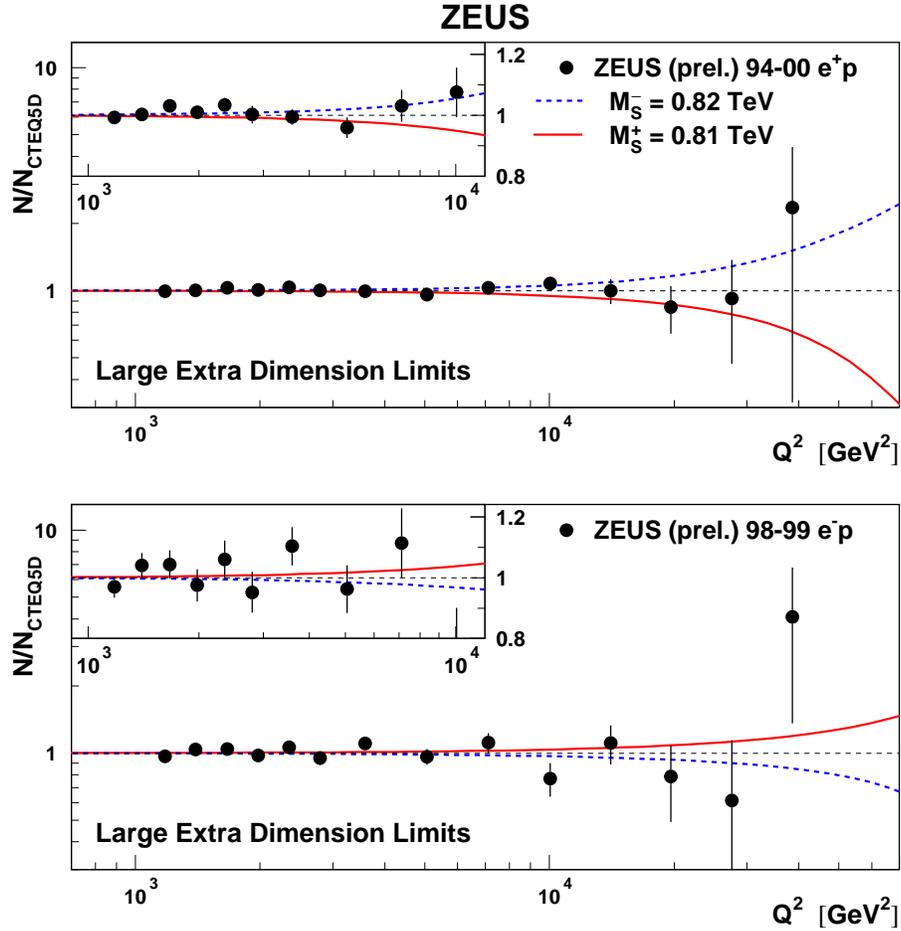,width=0.70\textwidth}

  \end{center}

 \vspace*{-0.5cm}
  
 \caption[]{ \label{fig:zkaluza}
             Comparison of high-$Q^2$ deep inelastic neutral current events
	     measured by ZEUS with expectations from the Standard Model; 
	     curves show the expected effects from the exchange of Kaluza-Klein
	     towers of gravitons for values of the string scales derived
	     as the 95\% CL lower limit from combined $e^+p$ and $e^-p$ data.} 
\end{figure}

Results from HERA experiments on the search for virtual graviton exchange
in theories with large extra dimensions are given in 
Figs.~\ref{fig:h1kaluza}~\cite{H1CI} and~\ref{fig:zkaluza}~\cite{ZEUSCI}.
The high-$Q^2$ NC DIS events are presented as a ratio to the SM prediction, 
together with the effect of Kaluza-Klein graviton exchange at a mass scale 
excluded at 95\% CL.
Here the coupling $\lambda$, which depends on the full theory and is 
expected to be of order unity, has been fixed by convention to 
$\lambda = \pm 1$. A combined analysis of the $e^-p$ and $e^+p$
data yields very simliar lower limits on $M_S$ for both $\lambda = +1$
and $\lambda = -1$ of about $0.8 \TeV$.
%----------------------------------------------------------------------------
%-----------------------------------------------------------------------------
%----------------------------------------------------------------------------
%-----------------------------------------------------------------------------

Searches for virtual effects from theories with large extra dimensions have 
also been reported from Tevatron~\cite{xtraTEV} and 
LEP experiments~\cite{xtraLEP}.
The $D\emptyset$ experiment at the Tevatron considered di-electron and 
di-photon invariant mass and angular distributions and compared these to
Standard Model expectation. The analysis made use of the entire sample of 
data collected by $D\emptyset$ at Tevatron$_{I}$, which corresponds 
to an integrated luminosity of $\approx 130 \pb^{-1}$.
No deviation from expectation was observed, and a lower limit on the mass 
scale $M_S$ of $1.2 \TeV$ (95\% CL) was 
obtained~\footnote{The results given here are expressed in the formalism 
of Giudice, Rattazzi and Wells~\cite{Giudice99}. Collider results are also
often discussed in the formalisms of Hewett~\cite{Hewett99} or that of
Han, Lykken and Zhang~\cite{Han99} where a different definition of the
coupling coefficient $\eta_G$ is used.}.
The L3 and OPAL experiments at LEP have searched for effects from virtual 
exchange of gravitons in fermion-pair production 
$e^+ e^- \rightarrow f \bar{f}$ in analyses making use of an integrated 
luminosity of $\approx 180 \pb^{-1}$ collected per experiment 
at $\sqrt{s} = 189 \GeV$. The L3 analysis in addition considers
boson-pair production $e^+ e^- \rightarrow \gamma\gamma, WW, ZZ$. 
No deviation from Standard Model expectation was observed and the
LEP$_{II}$ searches yielded lower limits of about $1 \TeV$. 

     % Search for extra-dimensions 

% Conclusions (prospects)
\clearpage
%%%%%%%%%%%%%%%%%%%%%%%%%%%%%%%%%%%%%%%%%%%%%%%%%%%%%%%%%%%%%%%%%%%%%%%%%%%%%%
\section{Summary}
\label{sec:conclusion}
%%%%%%%%%%%%%%%%%%%%%%%%%%%%%%%%%%%%%%%%%%%%%%%%%%%%%%%%%%%%%%%%%%%%%%%%%%%%%%

In this paper we have reviewed the status of searches for physics beyond 
the Standard Model of electroweak and strong interactions at LEP$_{I}$,
LEP$_{II}$, Tevatron$_{I}$ and HERA$_{I}$ colliders. We also have presented
new avenues for discoveries at the (upgraded) Tevatron$_{II}$ and HERA$_{II}$ 
colliders.

Leptoquark colour-triplet bosons are seen to be ideally suited to searches 
at $ep$ and $\bar{p}p$ colliders. Such scalar or vector bosons are predicted
in various unification theories, as a consequence of the symmetry between the
leptonic and quarkonic sectors.
The CDF and D$\emptyset$ experiments at the Tevatron collider offer the best
discovery mass reach for leptoquark bosons of all three generations 
if they appear in the context of minimal models with interactions 
invariant under  $SU(3) \times SU(2) \times U(1)$ gauge groups.
In a more general context where leptoquarks could have a low decay branching
fraction into final states containing first- or second-generation charged
leptons, a complementary sensitivity is offered by the H1 and ZEUS 
experiments at the HERA collider. At HERA$_{II}$, leptoquarks with masses 
approaching $300 \GeV$ could be discovered for Yukawa coupling values 
corresponding to an interaction of electromagnetic strength.
We have seen that HERA also offers a sensitivity beyond existing low-energy 
indirect constraints if leptoquarks are allowed to mediate lepton-flavour 
violating transitions.

Leptoquark-like composite objects appear as bound states of fundamental
``preons'' in some compositeness theories. They also appear as ``technipions''
$\pi_{LQ}$ in some specific Technicolour theories which address the problem 
of electroweak symmetry breaking via a dynamical mechanism. We have discussed
stringent constraints established at the Tevatron on the technihadrons 
of Technicolour theories. We mentioned the possible interest of $t$-channel 
$\pi_{LQ}$ production for HERA$_{II}$.

The relatively low energy scale at which electroweak symmetry breaking occurs
when compared to characteristic Grand Unification energy scales is accommodated 
naturally in supersymmetric (SUSY) theories. The search for the SUSY partners
of ordinary particles has constituted a major theme in high energy physics
over the past decades. We have discussed how the collider phenomenology 
depends on assumptions made for the parameters of specific SUSY models 
and on the chosen (a priori unknown) mechanism responsible for the breaking 
of the supersymmetry. A review of the best (and least model-dependent) existing
lower limits on sparticle masses was presented. The most stringent constraints
on gaugino-higgsinos and sleptons have been established at the LEP collider
for a very wide range of parameters of either the Minimal SUSY Standard Model, 
Minimal Supergravity models or Gauge Mediated SUSY Breaking models.
If the $R$-parity quantum number which distinguishes ordinary particles
($R_p = +1$) from supersymmetric particles ($R_p = -1$) is exactly
conserved in Nature, the best opportunity for a discovery will be provided
by squark and gluino searches at the Tevatron$_{II}$ collider.
The discovery mass reach for sfermions at colliders can be considerably enlarged
by single sparticle real production or virtual exchange involving $\Rp$ Yukawa 
couplings. We have discussed with some emphasis the case of the
lepton-number violating couplings $\lambda'$ which could allow for resonant
squark production at HERA through lepton-quark fusion or resonant slepton
production at the Tevatron through quark-antiquark fusion.

$R$-parity violation is a possible source of flavour-changing neutral currents
(FCNC) which are also predicted for instance in various models incorporating
an extended Higgs sector. We have seen that HERA and Tevatron experiments 
have access, with large integrated luminosities, to possible FCNC processes 
beyond the reach of LEP$_{II}$ via anomalous magnetic couplings of the top 
to lighter up or charm quarks.

We have discussed prominent models that rely on an extension of the
standard electroweak gauge symmetries with emphasis on left-right symmetric 
models and on models containing triplets of lepton or quark fields.
These models predict the existence of extra gauge bosons such as an extra 
$W'$ which couples to right-handed quarks and of exotic Higgs particles such 
as a doubly charged scalar coupling to lepton pairs.
We have seen that very stringent constraints are established at the Tevatron
on the $W'$ mass for a wide range of values of the model parameters which 
include the coupling constant $g_R$ of the $W'$ to right-handed fermions, the 
mixing angle $\xi$ between the $W_L$ and $W_R$ states, and the mass(es) 
and nature (Dirac or Majorana) of some new right-handed neutrinos.
A complementary sensitivity could be offered by the HERA$_{II}$ collider
only for very drastic choices of the quark mass-mixing matrix in the 
right-handed sector and provided that very high precision can
be obtained for the lepton-beam polarisation.
We have argued that doubly charged Higgs bosons $\Delta^{\pm\pm}$
could remain accessibly light even in left-right symmetric models with 
Majorana neutrinos and $W'$ bosons beyond the reach of colliders.
We have shown that the $\Delta^{\pm\pm}$ would lead to striking event
topologies in particular at the HERA collider.

We discussed the possibility of creating excited states of fermions
via magnetic transitions from the ground state of leptons or quarks 
in compositeness models.
Excited electrons and neutrinos could be discovered at the
HERA$_{II}$ collider while excited quarks could be discovered 
at Tevatron$_{II}$ collider. The HERA$_{II}$ and Tevatron$_{II}$
colliders were shown to offer a comparable sensitivity via four-fermion
processes to compositeness (or in general virtual exchange of very heavy 
particles) with characteristic mass scales in the TeV range.
We have furthermore shown that inclusive measurements could be used to
set stringent constraints on models with extra compactified dimensions.

  % Conclusions 

% Acknowledgements
% \clearpage
%%%%%%%%%%%%%%%%%%%%%%%%%%%%%%%%%%%%%%%%%%%%%%%%%%%%%%%%%%%%%%%%%%%%%%%%%%%%%%
\section*{Acknowledgements}

We wish to thank L.~Bellagamba, B.~Foster, P.~Schleper, A.~Sch\"oning 
and A.F.~Zarnecki for useful discussions and suggestions and for reading 
sections of this manuscript. 
We want to extend our gratefulness to other members of the 
H1 and ZEUS Collaborations of the HERA collider in DESY 
for their support and encouragements.
We are especially grateful
to D.~Acosta for discussions concerning leptoquarks and
experimental results from collider experiments at the Tevatron,
to S.~Davidson for discussions concerning bileptons, 
to B.~Olivier for his help in revisiting the phenomenology of doubly 
charged Higgs bosons, and to E.~Perez for discussions concerning supersymmetry 
and theories with extra dimensions.
The participation of M.~Kuze in the ZEUS Collaboration was supported by
the Japanese Ministry of Education, Culture, Sports,
Science and Technology (MEXT) and its grants for Scientific Research.
The participation of Y.~Sirois in the H1 Collaboration was supported
by the French Institut National de Physique Nucl\'eaire
et de Physique des Particules (IN2P3) of the Centre National pour la Recherche
Scientifique (CNRS).
  % Acknowledge

%================Bibliography ========================

\end{document}